\documentclass[12pt,a4paper,fleqn,openright,twoside]{book} 
\usepackage{graphicx}
\usepackage{amssymb}
\usepackage{amsmath}
\usepackage[T1]{fontenc}
\usepackage[cp 1250]{inputenc}
\usepackage{textcomp}
\usepackage{amsfonts}
\usepackage{epsfig}
\usepackage{cite}
\usepackage{esint}
\usepackage{fancyhdr}
\usepackage{color}
\usepackage{longtable}
\usepackage{mathrsfs}

\graphicspath{}

\title{{\bf Hyperaccreting Neutron-Star Disks, Magnetized Disks and Gamma-Ray Bursts
}}
\author{Dong Zhang}
\date{May, 2009}
\begin{document}
\thispagestyle{empty}
\newpage
\thispagestyle{empty}
\begin{center}
{\large Astronomy \& Astrophysics  }\\
{\large Department of Astronomy  }\\
{\large Nanjing University}\\
\end{center}
\begin{center}\end{center}
\begin{center}\end{center}

\begin{center}
{\Large{\bf Hyperaccreting Neutron-Star Disks}}
\end{center}
\begin{center}
{\Large{\bf Magnetized Disks}}
\end{center}
\begin{center}
{\Large{\bf and Gamma-Ray Bursts}}
\end{center}

\vspace{0.2cm}
\begin{center}
{by}
\end{center}

\vspace{0.2cm}
\begin{center}\end{center}
\begin{center}
{\large \textbf{Dong Zhang}}
\end{center}
\begin{center}\end{center}
\begin{center}\end{center}
\begin{center}
{\normalsize \textbf{Thesis for Master of Science} }\\
{\normalsize supervised by }
\end{center}
\begin{center}
{\large Professor Zi-Gao Dai}
\end{center}
\begin{center}
{\normalsize Nanjing, China\\}
\end{center}
\begin{center}\end{center}

\begin{center}
{\vspace{0.5cm}\normalsize May, 2009}
\end{center}
\begin{center}
\begin{figure}[h]
\hspace{6.7cm} \vspace{-1.5cm}
\parbox{0.10\textwidth}{\centerline{\epsfig{file=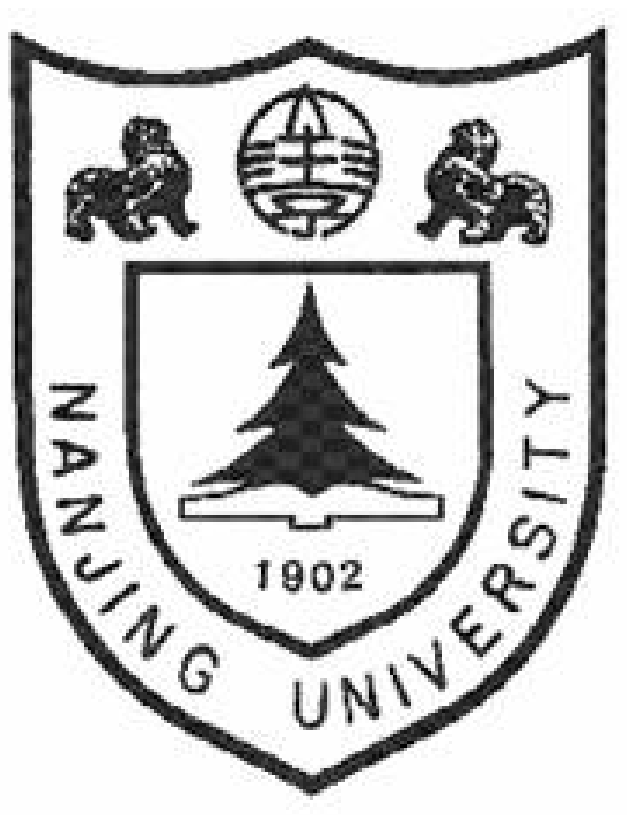,width=0.12\textwidth}}}
\end{figure}
\end{center}

\newpage
\pagestyle{plain}
\thispagestyle{empty}
\null\vfil \textit{ {\large
\begin{center}
For the memory of my Grandfather
\end{center}}
}
\vfil\null
\clearpage

\newpage
~ \\
\thispagestyle{empty}
\begin{center}
\LARGE{
\textbf{Abstract}
}
\end{center}

\hspace{\parindent}
~\\
~\\
~\\
~\\
\indent Gamma-ray bursts (GRBs) have been an enigma since their
discoveries forty years ago. What are the nature of progenitors and
the processes leading to formation of the central engine capable of
producing these huge explosions? My thesis focus on the
hyperaccreting neutron-star disks cooled via neutrino emissions as
the potential central engine of GRBs. I also discuss the effects of
large-scale magnetic fields on advection and convection-dominated
accretion flows using a self-similar treatment. This thesis is
organized as follows:

In Chapter 1, we first present a brief review on the theoretical
models of GRB for forty years. Although GRBs were first considered
from supernovae at cosmological distances, Galactic models become
more popular in 1980s, particularly after the detections of
cyclotron absorption and emission lines by the Konus and Ginga
detectors, which showed that neutron stars with magnetic fields
$\sim 10^{12}$ G are the most possible sources of GRBs. Galactic
GRBs were purposed to be produced from stellar flares or giant
stellar flares from main sequence or compact stars, from starquakes
of neutron stars, from accretion processes and thermonuclear burning
onto neutron stars, from collapses of white dwarfs and so on. It was
not until the mid-1990s that the observation results by BATSE
strongly suggested that GRBs originate at cosmological distance. The
main Cosmological GRB models include the mergers of neutron star
binaries and black hole-neutron star binaries, the ``collapsar"
model (i.e., the collapses of massive star), the magnetar model, and
the quark stars and strange stars model. We distinguish GRB
progenitors and its direct central engines. Furthermore, we discuss
two GRB progenitor scenarios in detail: the collapsar scenario for
long-duration bursts, and compact start merger scenario for
short-hard bursts.Observation evidence supports the hypothesis that
long-duration GRBs are supernova-like phenomena occurring in star
formation region related to the death of massive stars. Their energy
is concentrated in highly relativistic jet. We investigate various
collapsar formation scenarios, the requirements of collapsar angular
momentum and metallicity for generating GRBs, and the relativistic
formation, propagation and breakout in collapsars. Mergers of
compact objects due to gravitational radiation as the sources of
short-hard GRBs have been widely studied for years. We present the
neutron star-neutron star, black hole-neutron star and white
dwarf-white dwarf binaries as the most possible progenitors of short
bursts. We also introduce the numerical simulations results of
merges of the three types of binaries. In addition to the above
progenitor scenarios, we also mention the ``supranova" and magnetar
scenario.

Most GRB progenitor scenarios lead to a formation of a stellar-mass
black hole and a hyperaccretion disk around it. This accreting black
hole system is commonly considered as the direct central engine of
GRBs. In Chapter 2 we discuss the physical properties of accretion
flows around black holes. Accretion flows may be identified in three
cases: cooling-dominated flows, advection-dominated or
convection-dominated flows, and advection-cooling-balance accretion
flows. On the other hand, accretion flows can be classified by their
cooling mechanisms: no radiation, by photon radiation, or by
neutrino emission. We first investigate the structure of the normal
accretion disks, which could be cooled by photon radiation
effectively, and then we discuss the similar structure of advection
and convection-dominated flows. Next we describe the main
thermodynamical and microphysical processes in neutrino-cooled
disks, and show the properties of such neutrino-cooled flows around
black holes. Moreover, we discuss the two mechanisms for producing
relativistic jets from the accreting black holes: neutrino
annihilation and MHD processes, especially the Blandford-Znajek
process.

Chapter 3-5 are my works on hyperaccretion neutron star-disks and
magnetized accretion flows. As mentioned in Chapter 2, it is usually
proposed that hyperaccretion disks surrounding stellar-mass black
holes, with an accretion rate of a fraction of 1 $M_{\odot}$
s$^{-1}$ are central engines of GRBs. In some models, however,
newborn compact objects are introduced as neutron stars rather than
black holes. Thus, hyperaccretion disks around neutron stars may
exist in some GRBs. Such disks may also occur in Type II supernovae.
In Chapter 3 we study the structure of a hyperaccretion disk around
a neutron star. Because of the effect of a stellar surface, the disk
around a neutron star must be different from that of a black hole.
Clearly, far from the neutron star, the disk may have a flow similar
to the black hole disk, if their accretion rate and central object
mass are the same. Near the compact object, the heat energy in the
black-hole disk may be advected inward to the event horizon, but the
heat energy in the neutron star disk must be eventually released via
neutrino emission because the stellar surface prevents any heat
energy from being advected inward. Accordingly, an energy balance
between heating and cooling would be built in an inner region of the
neutron star disk, which could lead to a self-similar structure of
this region. We therefore consider a steady-state hyperaccretion
disk around a neutron star, and as a reasonable approximation,
divide the disk into two regions, which are called inner and outer
disks. The outer disk is similar to that of a black hole and the
inner disk has a self-similar structure. In order to study physical
properties of the entire disk clearly, we first adopt a simple
model, in which some microphysical processes in the disk are
simplified, following Popham et al. and Narayan et al. Based on
these simplifications, we analytically and numerically investigate
the size of the inner disk, the efficiency of neutrino cooling, and
the radial distributions of the disk density, temperature and
pressure. We see that, compared with the black-hole disk, the
neutron star disk can cool more efficiently and produce a much
higher neutrino luminosity. Finally, we consider an elaborate model
with more physical considerations about the thermodynamics and
microphysics in the neutron star disk (as recently developed in
studying the neutrino-cooled disk of a black hole), and compare this
elaborate model with our simple model. We find that most of the
results from these two models are basically consistent with each
other.

In Chapter 4 we further study the structure of such a hyperaccretion
neutron-star disk based on the two-region (i.e., inner \& outer)
disk scenario following Chapter 3, and calculate the neutrino
annihilation luminosity from the disk in various cases. We
investigate the effects of the viscosity parameter $\alpha$, energy
parameter $\varepsilon$ (measuring the neutrino cooling efficiency
of the inner disk) and outflow strength on the structure of the
entire disk as well as the effect of emission from the neutron star
surface boundary emission on the total neutrino annihilation rate.
The inner disk satisfies the entropy-conservation self-similar
structure for the viscosity parameter $\varepsilon\simeq 1$ and the
advection-dominated structure for $\varepsilon<1$. An outflow from
the disk decreases the density and pressure but increases the
thickness of the disk. Moreover, compared with the black-hole disk,
the neutrino annihilation luminosity above the neutron-star disk is
higher, and the neutrino emission from the boundary layer could
increase the neutrino annihilation luminosity by about one order of
magnitude higher than the disk without boundary emission. The
neutron-star disk with the advection-dominated inner disk could
produce the highest neutrino luminosity while the disk with an
outflow has the lowest. As a result, the neutrino annihilation above
the neutron-star disk may provide sufficient energy to drive GRBs
and thus observations on GRB-SN connection could constrain the
models between hyperaccreting disks around black holes and neutron
stars with outflows.

In Chapter 5, we study the effects of a global magnetic field on
viscously-rotating and vertically-integrated accretion disks around
compact objects using a self-similar treatment. We extend Akizuki \&
Fukue's work (2006) by discussing a general magnetic field with
three components ($r, \varphi, z$) in advection-dominated accretion
flows (ADAFs). We also investigate the effects of a global magnetic
field on flows with convection. For these purposes, we first adopt a
simple form of the kinematic viscosity $\nu=\alpha
c_{s}^{2}/\Omega_{K}$ to study magnetized ADAFs: a vertical and
strong magnetic field, for instance, not only prevents the disk from
being accreted but also decreases the isothermal sound speed. Then
we consider a more realistic model of the kinematic viscosity
$\nu=\alpha c_{s}H$, which makes the infall velocity increase but
the sound speed and toroidal velocity decrease. We next use two
methods to study magnetized flows with convection, i.e., we take the
convective coefficient $\alpha_{c}$ as a free parameter to discuss
the effects of convection for simplicity. We establish the
$\alpha_{c}-\alpha$ relation for magnetized flows using the
mixing-length theory and compare this relation with the
non-magnetized case. If $\alpha_{c}$ is set as a free parameter,
then $|v_{r}|$ and $c_{s}$ increase for a large toroidal magnetic
field, while $|v_{r}|$ decreases but $|v_{\varphi}|$ increases (or
decreases) for a strong and dominated radial (or vertical) magnetic
field with increasing $\alpha_{c}$. In addition, the magnetic field
makes the $\alpha_{c}-\alpha$ relation be distinct from that of
non-magnetized flows, and allows the $\rho\propto r^{-1}$ or
$\rho\propto r^{-2}$ structure for magnetized non-accreting
convection-dominated accretion flows with $\alpha+g\alpha_{c}< 0$
(where $g$ is the parameter to determine the condition of convective
angular momentum transport).

Finally, we give an outlook in Chapter 6.

\newpage
\chapter*{Acknowledgements}
\markboth{Acknowledgements}{Acknowledgements}
\addcontentsline{toc}{chapter}{Acknowledgements}

First and foremost, I would like to record my gratitude to my
advisor Professor Zi-Gao Dai for his supervision for my three-year
graduate study at Nanjing University. His broad knowledge in high
energy astrophysics especially in the areas of Gamma-ray bursts and
compact objects inspired me to explore the nature of Gamma-ray
Bursts. Professor Dai always gave me lots of suggestions during my
research, and discussed with me on many topics, from Gamma-ray
bursts central engines to theoretical physics. Without his
invaluable encouragement and patient guidance, I could never
accomplish my research in the topic of hyperaccreting neutron-star
disks.

Moreover, I should express my earnest thanks to Professor Daming
Wei, Yong-Feng Hang , Xiang-Yu Wang and Tan Lu in Nanjing GRB Group.
Their valuable ideas, suggestions and comments in the group
discussion seminar each week are extremely helpful for me to improve
my work and learn new knowledge in theoretical astrophysics. I also
acknowledge my current and former colleagues who have contributed to
our group, especially to Yun-Wei Yu, Fa-Yin Wang, Xue-Feng Wu,
Xue-Wen Liu, Lang Shao, Tong Liu, Yuan Li, Si-Yi Feng, Rongrong Xue,
Zhi-Ping Jin, Ming Xu, Hao-Ning He, Cong-Xin Qiu, Wei Deng, Tingting
Gao, Lei Xu, Ting Yan, Haitao Ma and Lijun Gou. I was and still am
fortunate to communicate with them and learn lots of interesting
topics from them.

It is a pleasure of me to pay tribute to the faulty in both the
Departments of Astronomy and Physics at Nanjing University,
especially to the professors who have taught me various graduate
courses and discussed problems with me: Zi-Gao Dai, Qiu-He Peng,
Peng-Fei Chen, Xiang-Dong Li, Xin-Lian Luo, Yu-Hua Tang, Yang Chen,
Jun Li, Zhong-Zhou Ren, Ren-Kuan Yuan, Ji-Lin Zhou, Tian-Yi Huang
and Xiang-Yu Wang. Special thanks also to several professors outside
Nanjing University who gave many useful suggestions on my work:
Professor Kwong-Sang Cheng at the University of Hong Kong, Professor
Feng Yuan at the Shanghai Astronomical Observatory and Professor
Ye-Fei Yuan at the USTC.

Many thanks to my friends and classmates: Meng Jin,  Cun Xia, Yan
Wang, Bo Ma, Chao Li, Yunrui Zhou, Jun Pan, Wenming Wang, Xiaojie
Xu, Bing Jiang, Jiangtao Li, Changsheng Shi, Fanghao Hu, Yang Guo,
Yuan Wang, Tao Wang, Zhenggao Xiong, Qing-Min Zhang, Fangting Yuan,
Xian Shi, Xin Zhou, Shuinan Zhang, Peng Jia and Xiang Li. It is my
honor to share friendship and passion with all of you.

Last but most importantly, I should express my appreciation to my
mother and father. You are always the strongest support behind me.

I dedicate this thesis to my grandfather, I will always remember
you.

     \normalsize
     \def\contentsname{\Large \mbox{} \hspace{6cm} Contents}
     \tableofcontents
     \pagestyle{myheadings}
     \markboth{Contents}{Contents}
     \newpage
     \clearpage
     \normalsize
\pagestyle{fancy}
\fancyhead{}
\fancyfoot{}
\renewcommand{\headrulewidth}{0.9pt}
\fancyhead[RO]{\textbf{\sffamily{{{\thepage}}~}}}
\fancyhead[LO]{\bf\footnotesize{{\nouppercase{\rightmark}}}}
\fancyhead[RE]{\textbf{\sffamily{{{\thepage}}~}}}
\fancyhead[LE]{\bf\footnotesize{{\nouppercase{\rightmark}}}}
\advance\headheight by 5.3mm
\advance\headsep by -3mm
%
%

\chapter{GRB Progenitors}
\label{chap1} \setlength{\parindent}{2em}

\section{Detectors and Observations}

Theoretical astrophysics is always based on the observation events
and data. In \S 1.1 we list the detectors of Gamma-ray bursts (GRBs)
and their main contributions. The theories of GRBs are based on
these observations.

\textbf{$\blacksquare$Vela Satellites}

The total number of Vela Satellites is 12, six of the Vela Hotel
design, and six of the Advanced Vela design, launched from October
1963 to April 1970. These military satellites were equipped with
X-ray detectors, $\gamma$-ray detectors and neutron detectors, and
the advanced ones also with the silicon photodiode sensors. The
first flash of gamma radiation sinal was detected by Vela 3 and Vela
4 on July 2, 1967. Further investigations were carried out by the
Los Alamos Scientific Laboratory, lead by Ray Klebesadel. They
traced sixteen gamma-ray bursts between 1969 July and 1972 July,
using satellites Vela 5 and Vela 6, and reported their work in 1973
(Klebesadel et al. 1973).

\textbf{$\blacksquare$ First IPN (Inter-Planetary Network)}

The investigation of the new phenomenon gamma-ray bursts become a
fast growing research area after 1973. Beginning in the mid
seventies, second generation gamma-ray sensors started to operate.
By the end of 1978, the first Inter-Planetary Network had been
completed. In addition to the Vela Satellites, gamma-ray burst
observations were conducted from Russian Prognoz 6,7, the German
Helio-2, NASA's Pioneer Venus Orbiters, Venera 11 and Venera 12
spacecrafts. A group from Leningard using the KONUS experiment
aboard Venera 11 and Venera 12 did significant work for gamma-ray
burst survey. The important results of the first IPN projects
include:

(1) First survey of GRB angular and intensity distribution by KONUS
experiments (Matzet et al. 1981).

(2) Discovery of first SGR (SGR 0526-66, Mazet et al. 1979, Cline et
al. 1980).

(3) Cyclotron and annihilation lines observations for many bursts
(Mazets et al. 1981). The absorption and emission lines observation
raised lots of debates until the discovery of cyclotron absorption
lines by Ginga in 1988.

\textbf{$\blacksquare$ Ginga}

Ginga was an X-ray astronomy satellite launched on February 1987.
Cyclotron features in the spectra of three GRBs (Murakami et al.
1988, Fenimore et al. 1988, Yoshida et al. 1991) were reported and
interpreted as photon scattering process near the neutron star
surface with strong magnetic fields $\sim 1.7\times10^{12}$ G.
Neutron-Star model become popular in those years. Moreover, Ginga
data also provided the early evidence of the existence of X-ray
Flashes (XRFs, Strohmayer et al. 1998), which were discussed in
detailed in the BeppoSAX era.

\textbf{$\blacksquare$ Compton Gamma-Ray Observatory (CGRO)}

The Compton Gamma Ray Observatory (CGRO) is part of NASA's Great
Observatories\footnote{The others are the Hubble Space Telescope,
the Chandra X-ray Observatory, and the Spitzer Space Telescope.}.
CGRO carried a complement of four instruments which covered
observational band from 20keV to 30GeV, and the Burst and Transient
Source Experiment (BATSE) is one of them. BATSE become the most
ambitious experiment to study GRBS before BeppoSAX. It had detected
a total of 2704 GRBs from April 1991 to June 2000, which provided a
large sample for GRBs statistical work. Several significant results
concerning the characteristics of GRBs were made based on the
crucial data from BATSE. The cosmological GRB orgin was first
established, although the debate between galactic and cosmological
origin still continued until BeppoSAX. Moreover, the bimodality of
GRB durations was established by analyzing the distribution of
$T_{90}$ for hundreds of BATSE-observed-GRBs.

(1) The several BATSE catalogs had confirmed the apparent isotropy
of the GRB spatial distribution (Meegan et al. 1992). Then the
cosmological origin of GRBs began to be accepted by most
astronomers. The energy of GRBs is about $10^{52}$ergs.

(2) BATSE data showed that GRBs were separated into two classes:
short duration bursts ($T_{90}<$2s) with predominantly hard spectrum
and long duration bursts ($T_{90>}$2s) with softer spectrum
(Kouveliotou et al. 1993). Next two different classes of progenitors
were suggested for this distinction: mergers of neutron star
binaries (or neutron star and black hole systems) for short bursts,
and collapse of massive stars for long bursts. however, some others
believe there are three kinds of GRBs also based on the BASTE data,
which should reflect three different types of progenitors (Mukherjee
et al. 1998).

\textbf{$\blacksquare$ BeppoSAX}

BeppoSAX was an Italian-Dutch satellite for X-ray and gamma-ray
astronomy. It was launched in 1996 and ended its life in 2003. This
instrument had led to the discoveries of many new features of
gamma-ray bursts and greatly accelerated people's understanding
about GRBs:

(1) The discovery of the X-ray afterglow of GRBs, which opened the
afterglow era with multi-wavelength observations and confirmed the
cosmological origin of GRBs. BeppoSAX first detected GRB 970228 with
its X-ray afterglow (Costa et al. 1997), and later optical afterglow
was also detected by the ground-based telescopes. Next, Keck II
detected the spectrum of optical afterglow of GRB 970508, registered
by BeppoSAX, and determined a redshift of $z=0.835$, which confirmed
a cosmological distance (Metzger et al. 1997).

(2) The connection between GRBs and Type Ib/c supernove (SNe Ib/c).
In 1998 BeppoSAX with the ground-based optical telescopes provided
the first clue of a possible connection between GRB 980425 and a new
supernova SN 1998 bw (Galama et al. 1998).

(3) The confirmation of a new sub-class of GRBs--X Ray Flashes
(XRFs), which emit the bulk of their energy around a few Kev, at
energies significantly lower than the normal GRBs (Heise et al.
2001). Furthermore, BeppoSAX first reported the detection of the
X-ray-rich GRBs (Frontera et al. 2000).

(4) The discovery of GRBs with bright X-ray afterglows but no
detectable optical afterglows, i.e. the 'dark bursts'. The next
instrument HETE-2 made a more detailed detection on 'dark bursts'.

(5) The discovery of the possible correlation between the isotropic
equivalent energy radiated by GRBs and the peak energy, such as
Amati relation (Amati et al. 2002) and Ghirlanda relation (Ghirlanda
et al. 2004), provided the new possible standard candles to measure
the Universe with parameters.

\textbf{$\blacksquare$ HETE-2}

The first High Energy Transient Explorer (HETE-2) failed during the
launch in 1996 and lost the opportunity to explore the significant
characteristics of GRBs before BeppoSAX. However, the second HETE,
HETE-2, with multiwavelength instruments, had got many achievements.
It first confirmed a event of GRB-SN connection (GRB 030329 \& SN
2003dh, Stanek et al. 2003). It first discovered a short-hard GRB
with an optical counterpart (GRB 050709) and studied the afterglow
properties of short bursts together with Swift. Moreover, it studied
the phenomena XRFs more detailed than BeppoSAX.

\textbf{$\blacksquare$ Swift}

Swift is a multi-wavelength space-based observatory dedicated to the
study of GRBs. It contains three instruments work together to
observe GRBs and their afterglows in the gamma-ray, X-ray,
ultraviolet and optical wavebands: the BAT (Burst Alert Telescope)
detects GRBs events in the energy range between 15 keV to 150 keV,
and computes its coordinates in the sky; the XRT (X-ray telescope)
takes imagines and performs spectral analysis of the GRB X-ray
afterglows in the energy range from 0.2 to 10 keV; and the UVOT
(Ultriviolert/Optical Telescope) is used to detect an optical
afterglow after Swift has slewed toward a GRB. This satellite was
launched in November 2004 and is still in operation. The major
breakthroughs from Swifts up to this date include:

(1) established a completely new view of the early X-ray afterglow
with several well-defined phases: after the prompt emission there is
a rapid decline phase, which is followed by a shallow decay phase
combined with X-ray flares, and then is the normal decay phase.
Moreover, Swift has detected the existence of the achromatic breaks,
causing some issues of the jet break interpretation of the break
phase of X-ray afterglows.

(2) detected GRBs with high-redshift, e.g., GRB 050904 with $z$=6.29
(Price et al. 2006), GRB 080913 with $z$=6.7 (Schady et al. 2008).
These detections provide a unique way of probing the Universe when
the first stars were formed. Combined with the observation by Hubble
and Spitzer, Swift

(3) first observed the afterglows of short hard GRBs (GRB 050509b,
Hjorth et al. 2005).

(4) the discovery of GRB 060614 called for a rethink of the GRB
classification. GRB 060614 is a burst with $T_{90}=104$s other
properties like short bursts. The classification of GRBs should
based on their afterglows, host galaxies, not only their durations
of prompt emission (e.g., Zhang 2006).

\begin{figure}
\centering\resizebox{1.0\textwidth}{!} {\includegraphics{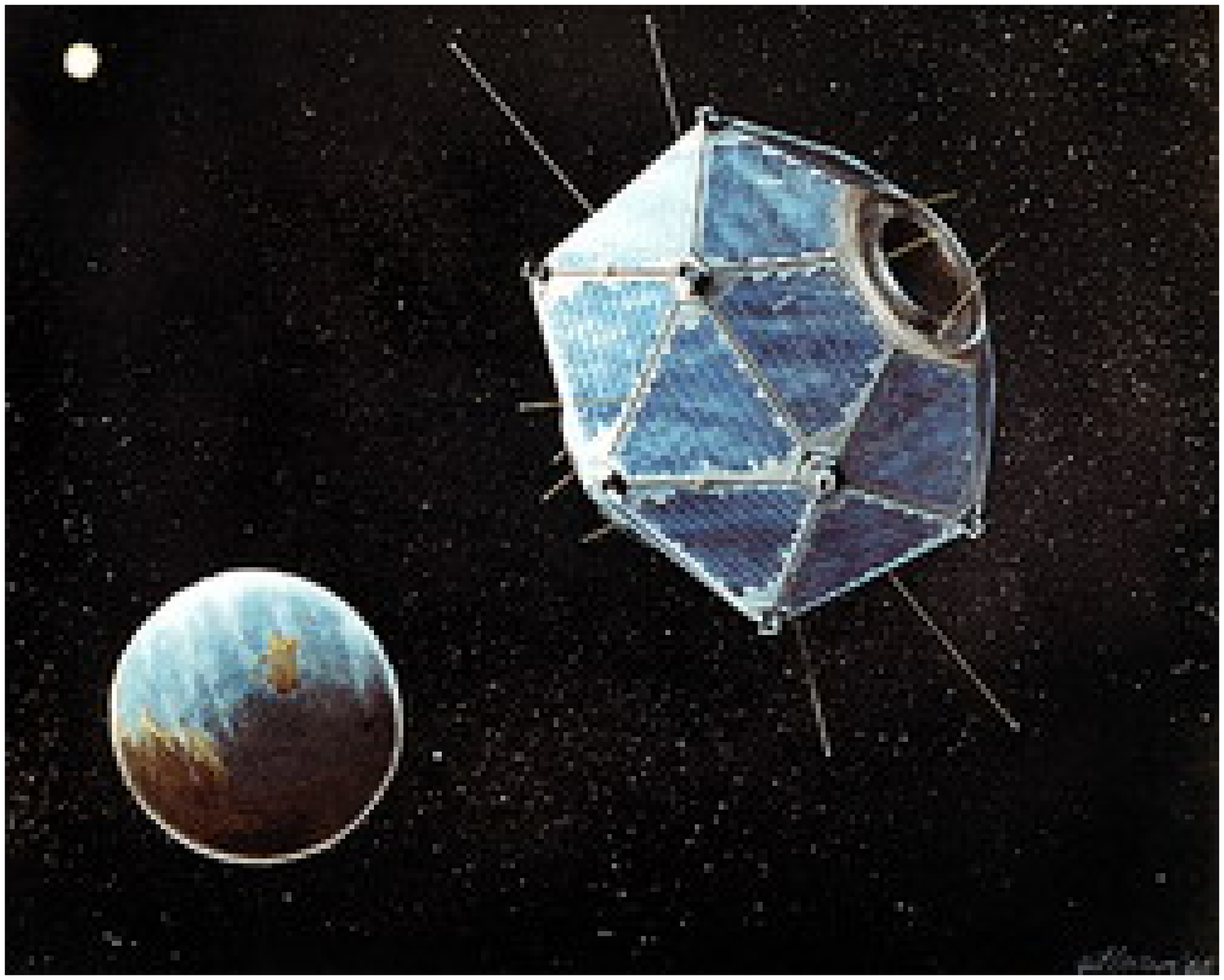}
\includegraphics{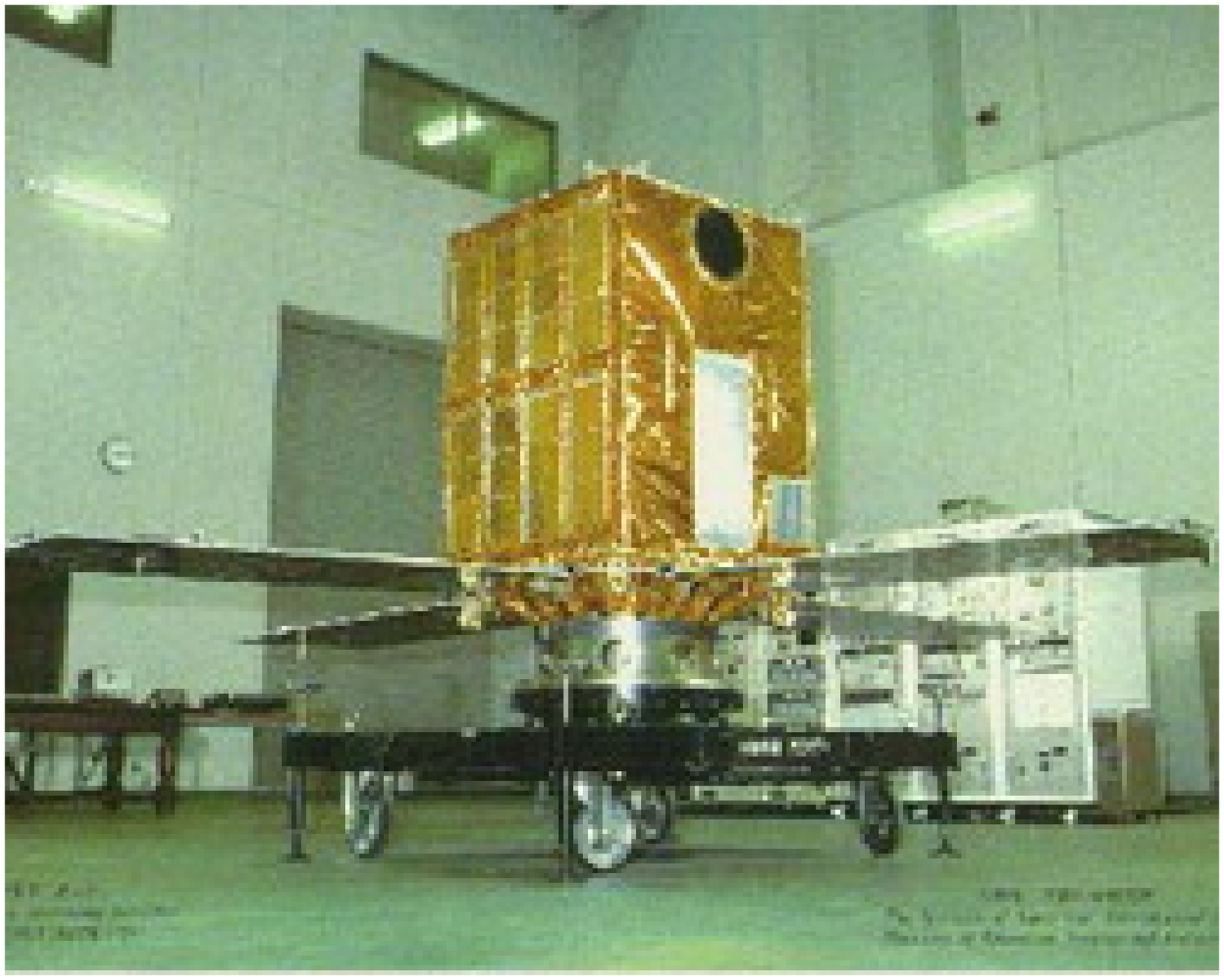}\includegraphics{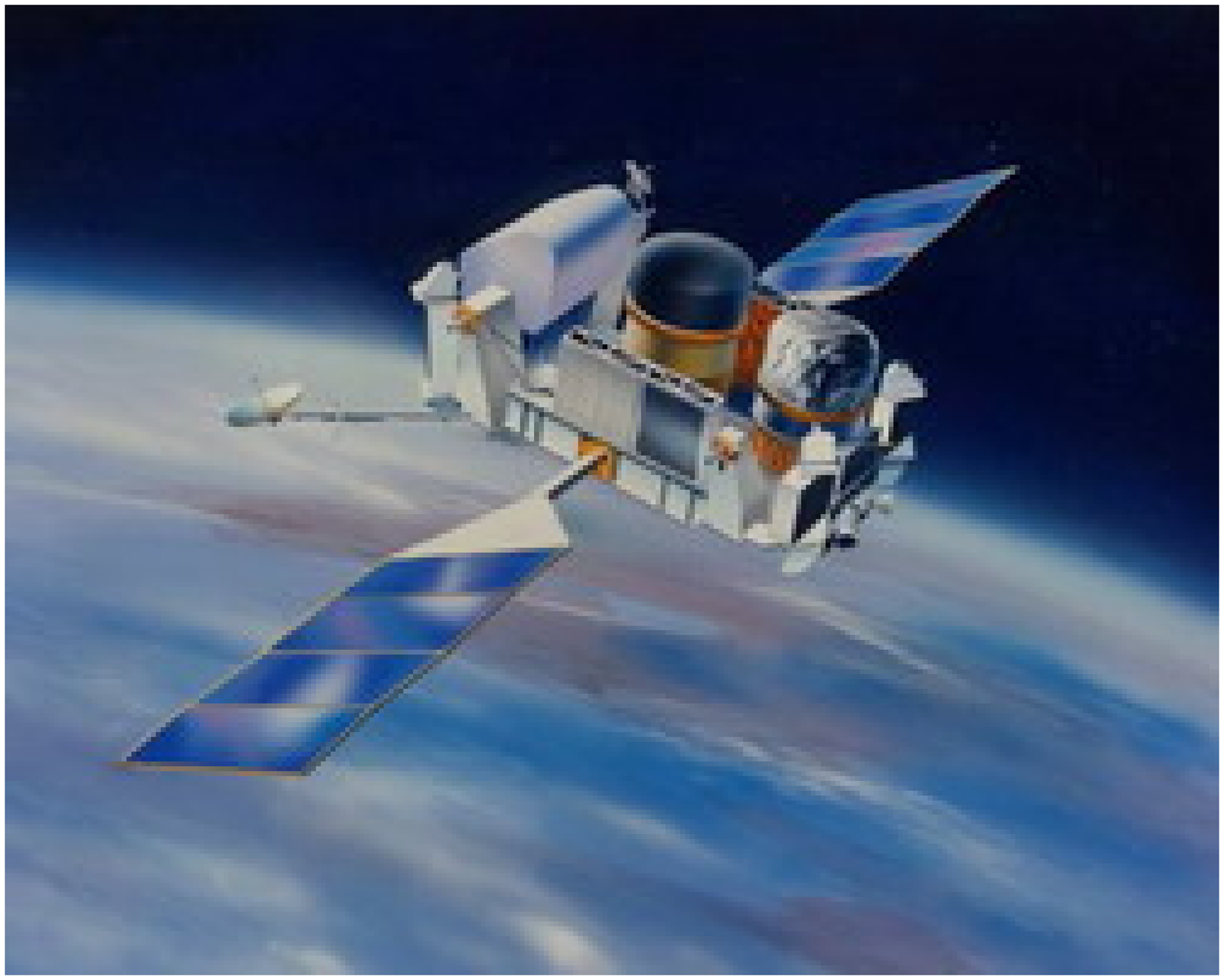}}\\
\centering\resizebox{1.0\textwidth}{!} {\includegraphics{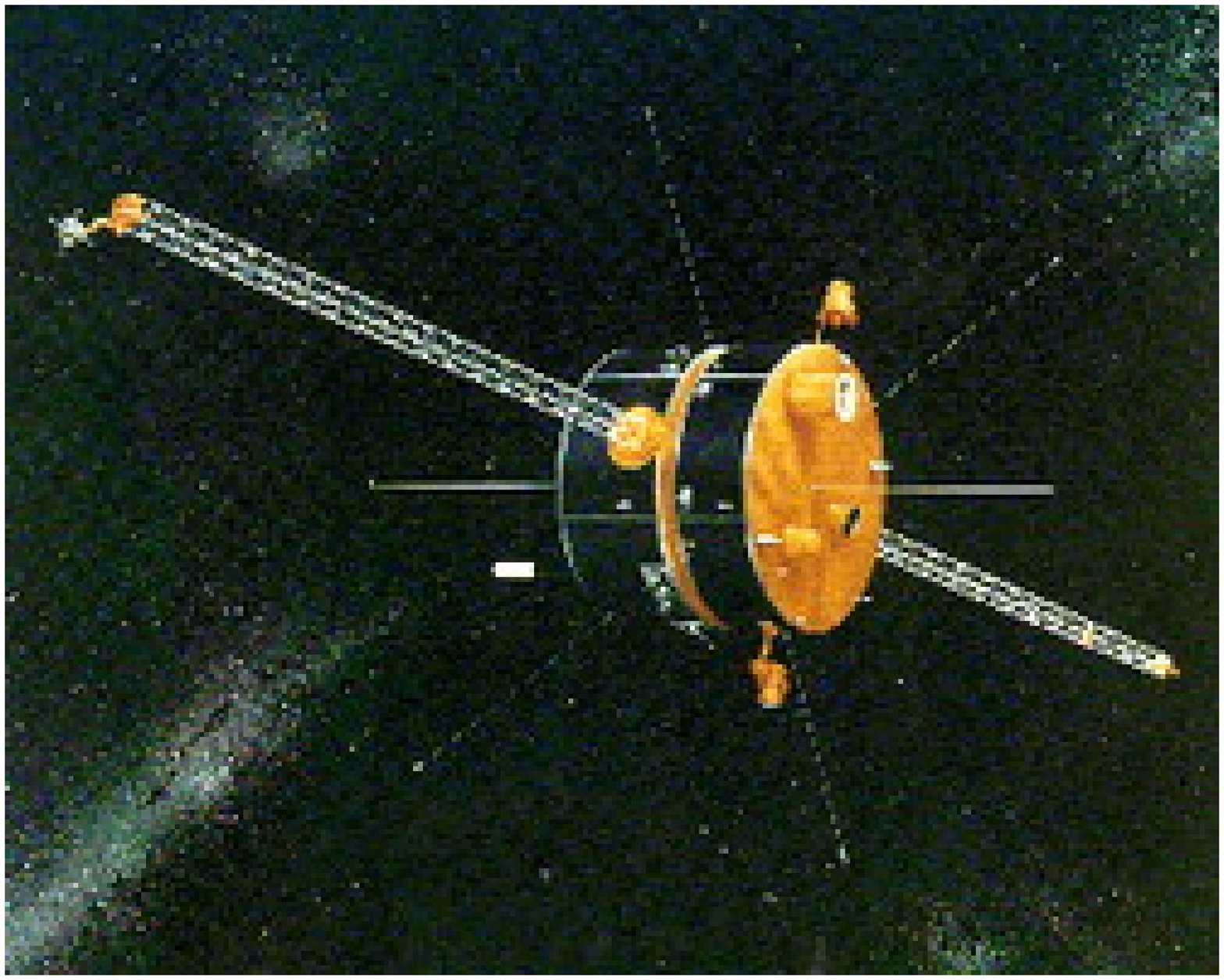}
\includegraphics{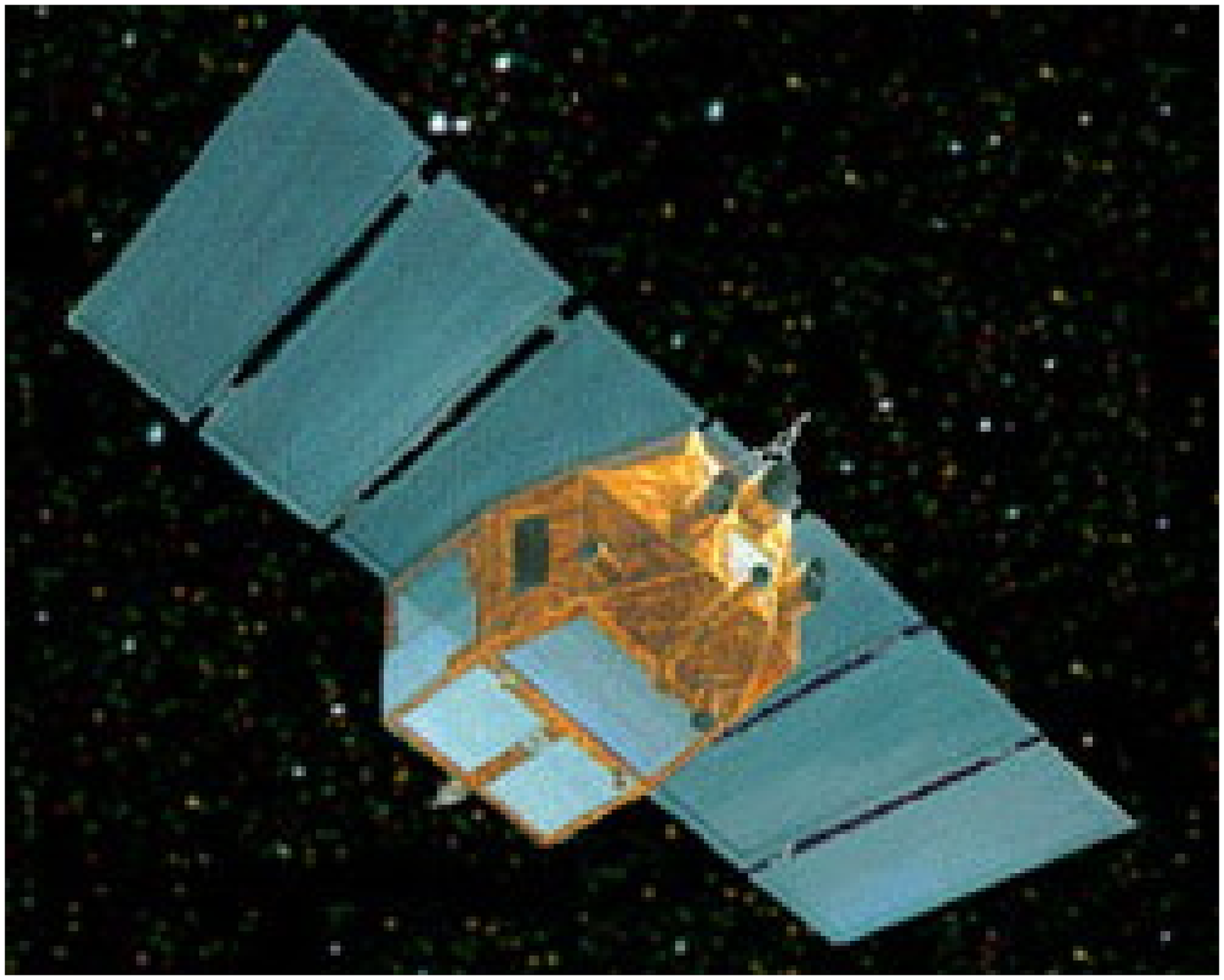}\includegraphics{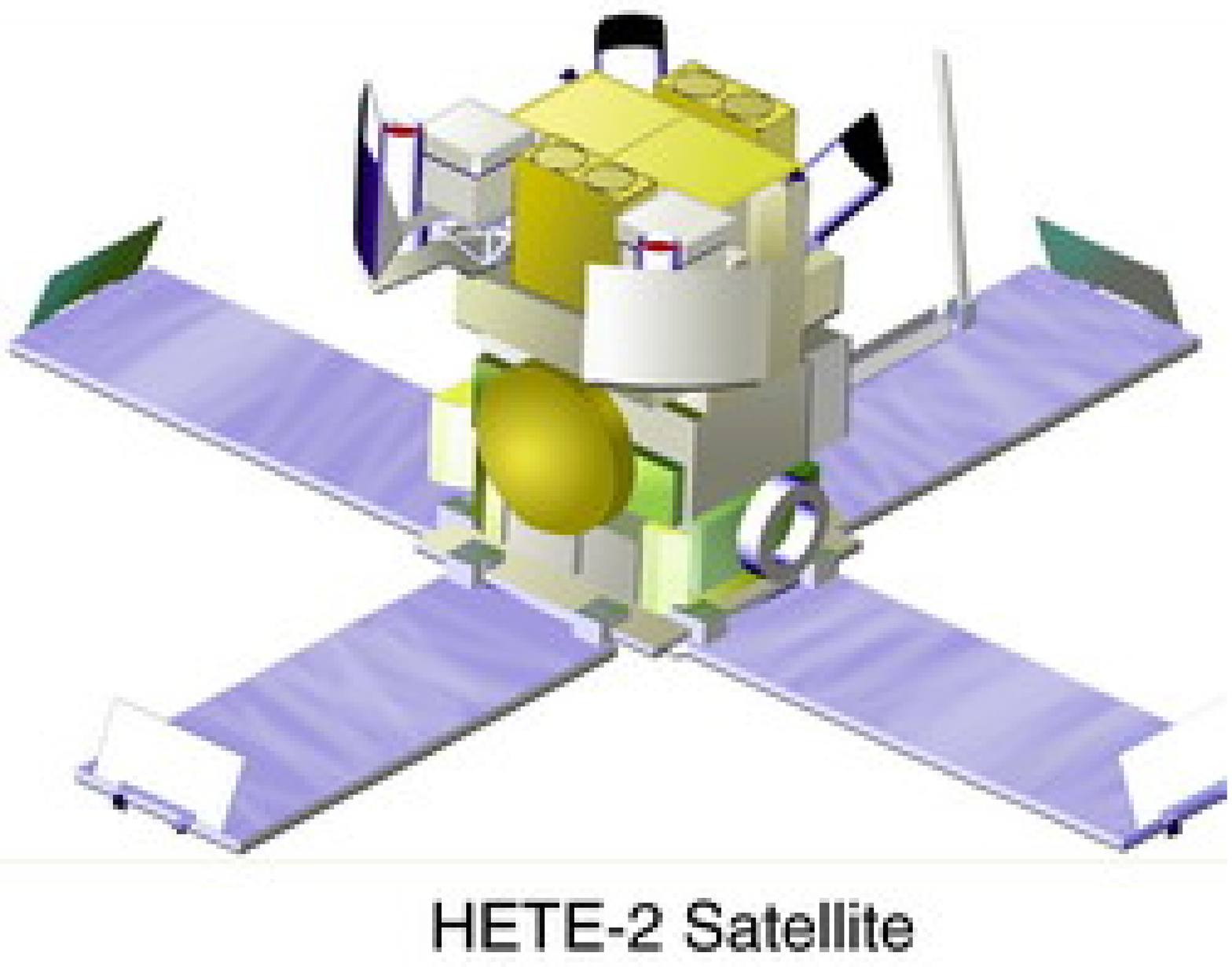}}\\
\centering\resizebox{1.0\textwidth}{!} {\includegraphics{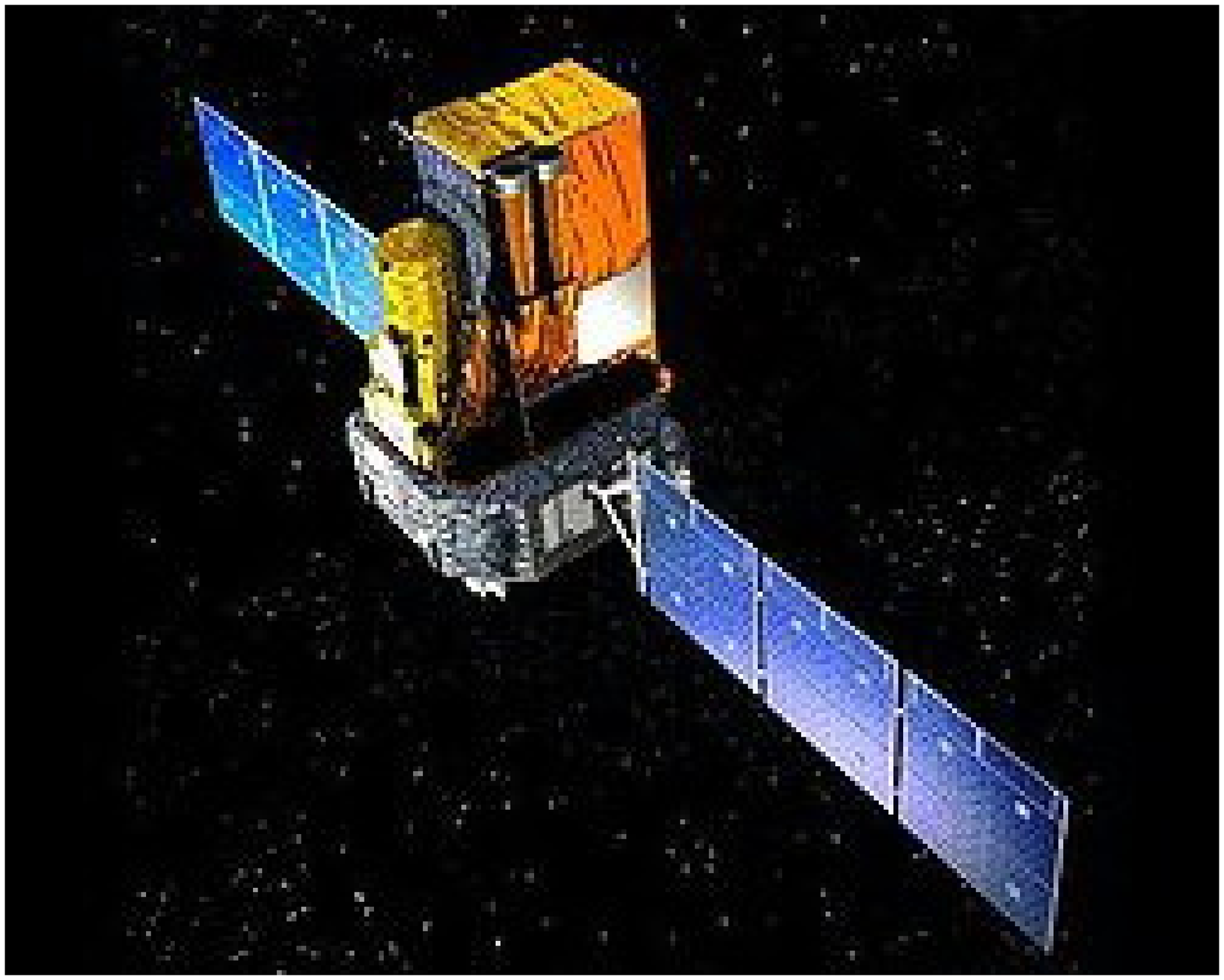}
\includegraphics{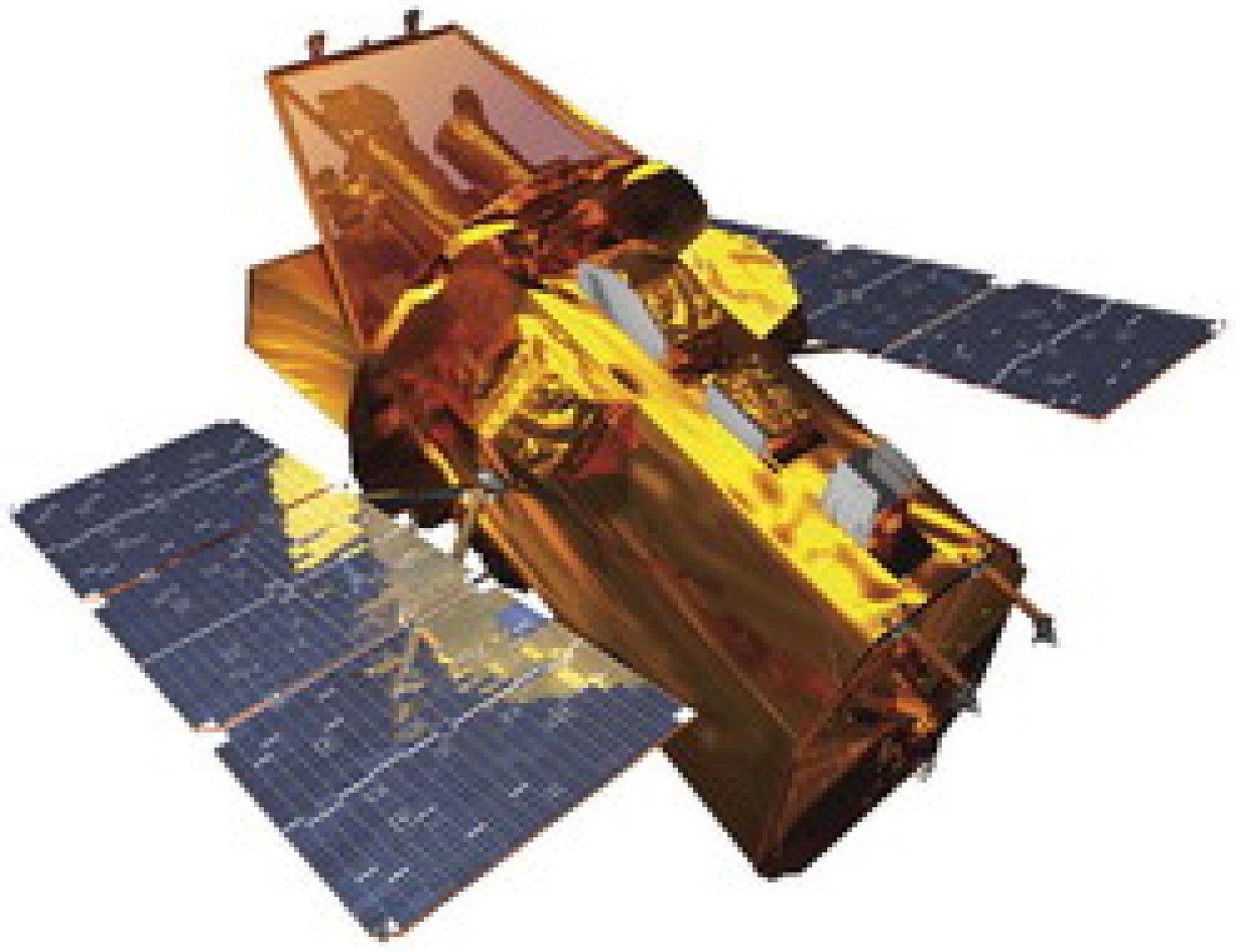}\includegraphics{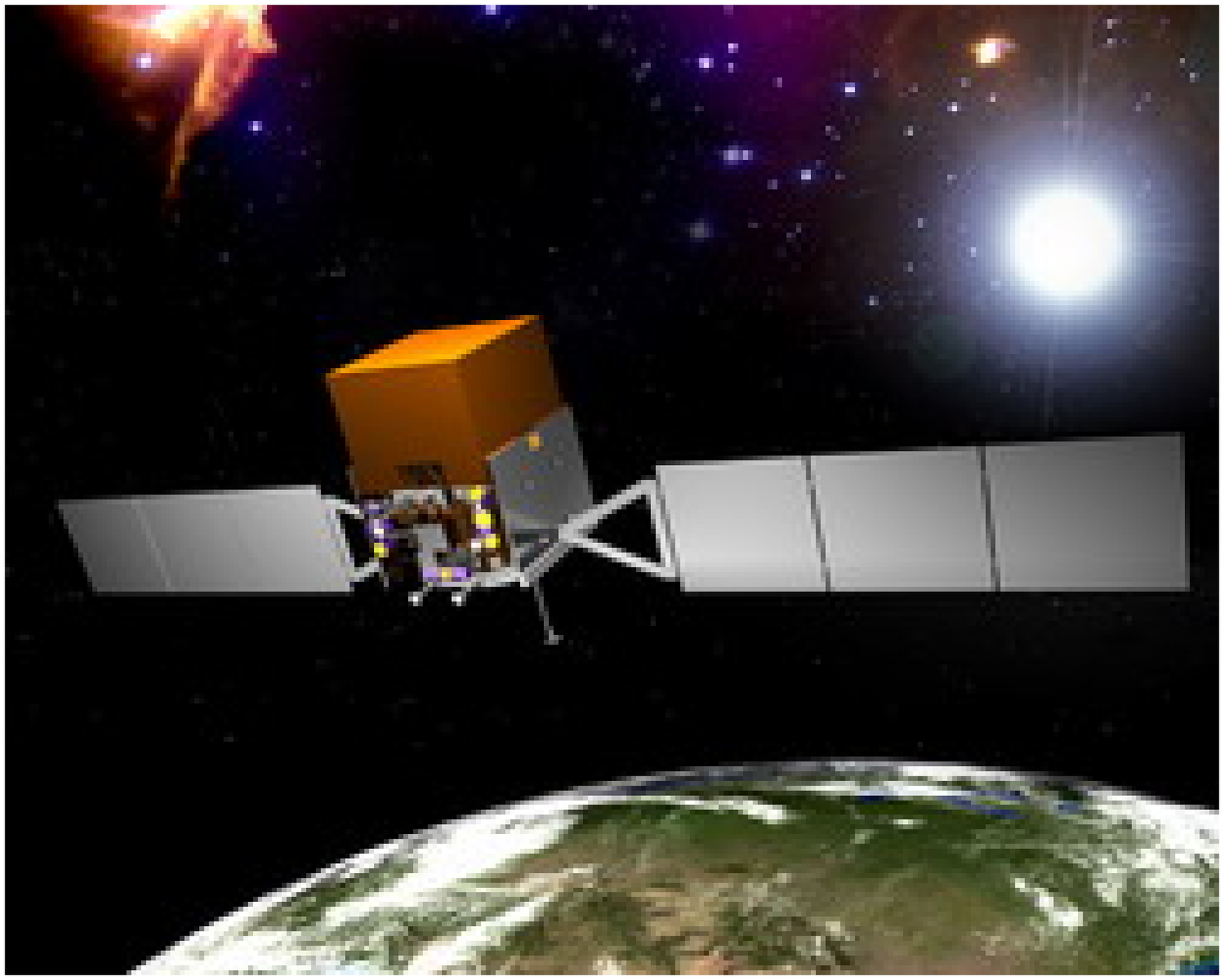}}
\caption{Some former and current GRB detectors. First row (left to
right): Vela 5B (1969-1979), Ginga (1987-1991), CGRO/BATSE
(1991-2000); second row: Konus-Wind (1994- ), BeppoSAX (1996-2003),
HETE-2 (2000- ); third row: INTEGRAL (2002- ), Swift (2004- ), Fermi
(2008- ).}\label{fig111}
\end{figure}

\textbf{$\blacksquare$ Current Missions}

In addition to the Swift satellite, there are also many present
missions to study GRBs. The present long-baseline GRB IPN, i.e., the
third IPN, includes Ulysses in 1992, Konus-wind in 1994, Rossi XTE
in 1995, HETE-2 in 2000, Mars Odgssey in 2001, RHESSI in 2002,
INTEGRAL in 2002, Swift in 2004, MESSENGER in 2004, Suzaku in 2005,
AGILE in 2007 and Fermi in 2008. On the other hand, BATSE-CGRO
(1991-2000), SROSS-C2 (1994-2001?), SZ-2 (2001), NAER (1996-2001),
BeppoSAX (1996-2003) and SZ-2 (2001) have ceased operations today.
Here I list some significant missions in this subsection.

\begin{itemize}
\item Konus-Wind

The joint Russian-American Konus-Wind experiment is presently
carried out onboard the NASA's Wind spacecraft, which was launched
in November 1994 and mainly to study the 'solar wind'. The Konus
experiment on Wind satellite, provides omnidirectional and
continuous coverage of the entire sky in the hard X- and gamma-ray
domain. Konus also provides the only full-time near-Earth vertex in
the present IPN project. Konus-Wind has detected thousands of GRBs
since 1994 until now.

\item INTEGRAL

The European Space Agency's International Gamma-Ray Astrophysics
Laboratory (INTEGRAL), launched in 2002, is a successor to CGRO. It
can similarly determine a coarse position by comparing gamma counts
from one side to another. It also possesses a gamma-ray telescope
with an ability to determine positions to under a degree. The
science achievements of INTEGRAL has covered many areas in gamma-ray
astronomy, the observations of GRBs made by INTEGRAL over the last
years includes the discovery of the nearby low energy GRB 031203
(Gotz et al. 2003).

\item RHESSI

RHESSI was launched in 2002 to perform solar studies. However, its
gamma instrument could detect bright gamma sources from other
regions of the sky, and produce coarse positions through
differential detectors. For instance, RHESSI observed 58 bursts in
2007 and 34 in the first 9 months in 2008.

\item AGILE

This X-ray and Gamma ray astronomical satellite, launched in April
2007 and equipped with scientific instruments capable of imaging
distant celestial objects in the X-ray and Gamma ray regions of the
electromagnetic spectrum, is adding data to the third IPN.

\item Fermi

The Fermi Gamma-ray Space Telescope is a space observatory being
used to perform gamma-ray astronomy observations from low Earth
orbit. Its main instrument is the Large Area Telescope (LAT), with
which astronomers mostly intend to perform an all-sky survey
studying of gamma radiation high-energy sources and dark matter.
Another instrument, the Gamma-ray Burst Monitor (GBM), is being used
to study GRBs.
\end{itemize}

\section{A Review of GRB Models}

Gamma-ray bursts have been an enigma since their discoveries forty
years ago. What are the nature of progenitors and the processes
leading to formation of the central engine capable of producing
these huge explosion? The first GRB model was published (Colgate
1968) even before the discovery of these burst events by Vela in
1969. This model suggested that transient prompt gamma-rays and
X-rays would be emitted as Doppler-shifted Planck radiation from the
relativistically expanding outer layers of supernovae. Klebesadel et
al. (1973) discussed the possibility that GRBs are associated with
some particular supernovae which are not bright in the optical band
(Thorne 1968) in the first observation paper of GRBs. However, it
was not until 25 years later when BeppoSAX could detect the
connection between GRB 980425 and SN 1998bw. Hawking (1974) showed
that black holes would create and emit particles like photons and
evaporate themselves. In this point of view, gamma-ray bursts could
provide the direct evidence of the existence of black holes.
However, today we know this physical opinion could not explain most
GRBs, after we accumulated sufficient observation evidence of
bursts.

Although GRBs were first considered at cosmological distances from
supernovae, Galactic models became more popular at later time,
particularly after the detections of cyclotron absorption and
emission lines by Konus and Ginga (Mazets et al. 1981; Murakami et
al. 1988). More than one hundred GRB Galactic models had been
published before 1990s. The early observations (1st IPN, Ginga,
etc.) led to considering neutron stars with strong magnetic fields
as ideal candidates for GRBs sources. Moreover, the neutron star
models were also reinforced by the discoveries of SGR.

However, on the other hand, Paczy\'{n}ski and a few other
astronomers pointed out again the idea that gamma-ray bursts could
be at cosmological distances like quasars, with redshifts around 1
to 2 (Paczy\'{n}ski 1986; Goodman 1986). The energy needed was
comparable to the energy typically released by a supernovae. Their
cosmological model was based on the evidences of GRBs isotropy
spatial distribution and deviation of the intensity distribution
from the -3/2 law, both of which were still controversial in 1980s,
but crucially confirmed by BATSE-CGRO in 1990s. The observation
results by BATSE in mid-1990s strongly suggested that the bursts
originate at cosmological distance. Thanks to BeppoSAX and the
discovery of afterglows with redshifts, the cosmological nature of
GRBs has been well established since 1997.

BeppoSAX with the multi-wavelength observations on the afterglows
raised researchers interests in explaining the lightcurve and
spectrum using various radiation processes during and after bursts;
and the models of GRB progenitors and central engines became less
than before. Currently, the most widely-accepted model for the
origin of most observed long GRBs is called the ``collapsar" model
(Woosley 1993), in which the core of an extremely massive,
low-metallicity, rapidly-rotating star collapses into a compact
object. On the other hand, short hard GRBs are commonly believed to
be caused by star-neutron star (NS-NS) or neutron star-black hole
(NS-BH) mergers.

In this section, I give a brief review of the GRB models for forty
years since 1969. I still first introduce the Galactic models of
GRBs, which have been demonstrated as wrong for years. However, we
should keep in mind once there was a period of time most researchers
agreed the idea that GRBs came from the Milky Way, and we need to
know the reason why they hold on this idea. Their idea was based on
the observation results in 1970s and 1980s. Astrophysics theoretical
models are always based on the observations, although usually
observations could be incomplete or even have mistakes. Also, the
physical processes and radiation mechanisms pointed out in these
former models might explain other phenomena even if they cannot
explain GRBs today.

Table \ref{tab111} lists part of the history models of GRBs since
Colagate (1968). The models before 1993 (especially Galactic Models
before the BATSE era) has been listed in Nemiroff (1993). We mainly
focus on the model which discuss the origin of GRB energy and the
objects to produce GRBs, not the energy propagation ways and
radiation mechanisms in GRB production process.

\begin{center}
{\footnotesize\begin{longtable}{lllll} \hline \hline
No. & Author & Year & Place & Description \\
\hline
1. & Colgate & 1968 & COS & SN shocks stellar surface in distant galaxy \\
2. & Colgate & 1974 & COS & Type II SN shock brem, IC scat at stellar surface \\
3. & Stecker et al. & 1973 & DISK & Stellar superflare from nearby star \\
4. & Stecker et al. & 1973 & DISK & Superflare from nearby WD \\
5. & Harwit et al. & 1973 & DISK & Relic comet perturbed to collide with old galactic NS \\
6. & Lamb et al. & 1973 & DISK & Accretion onto compact objects from flare in companion \\
7. & Zwicky & 1974 & HALO & NS chunk contained by external pressure \\
8. & Grindlay et al. & 1974 & SOL & Iron dust grain up-scatters solar radiation \\
9. & Brecher et al. & 1974 & DISK & Directed stellar flares on nearby stars \\
10. & Schlovskii & 1974 & DISK & Comet from system's cloud strikes WD/NS \\
11. & Bisnovatyi- et al. & 1975 & COS & Absorption of neutrino emission from SN \\
12. & Bisnovatyi- et al. & 1975 & COS & Thermal emission when small star heated by SN shock \\
13. & Bisnovatyi- et al. & 1975 & COS & Ejected matter from NS explodes \\
14. & Pacini et al. & 1974 & DISK & NS crustal starquake glitch \\
15. & Narlikar et al. & 1974 & COS & White hole emits spectrum that soften with time \\
16. & Tsygan & 1975 & HALO & NS corequake excites vibrations, changing E,B-fields \\
17. & Chanmugam & 1974 & DISK & Convetion inside WD produces flares \\
18. & Prilutski et al. & 1975 & COS & Collaspe of supermassive body in AGN \\
19. & Narlikar et al. & 1974 & COS & WH excites syn emission and IC scat \\
20. & Piran & 1975 & DISK & IC scat deep in ergosphere of Kerr BH \\
21. & Fabian & 1976 & DISK & NS crustquake shocks NS surface \\
22. & Chanmugam & 1976 & DISK & flares from magnetized WD via MHD instability \\
23. & Mullan & 1976 & DISK & Thermal radiation from flares near magnetized WD \\
24. & Woosley et al.& 1976 & DISK & Carbon detonation from accreted matter onto NS \\
25. & Lamb et al. & 1977 & DISK & Mag grating of accret disk around NS causes sudden accretion \\
26. & Piran et al. & 1977 & DISK & Instability in accretion onto Kerr BH \\
27. & Dasgupta & 1979 & SOL & Charged intergal rel dust grain enters sol sys. \\
28. & Tsygan & 1980 & DISK & WD/NS surface nuclear burst causes chromospheric flares  \\
29. & Ramaty et al. & 1981 & DISK & NS vibration heat atm to pair produce, annihilate, syn cool \\
30. & Newman et al. & 1980 & DISK & Asteroid from interstellar medium hits NS \\
31. & Ramaty et al. & 1980 & HALO & NS core quake caused by phase transition, vibrations \\
32. & Howard et al. & 1981 & DISK & Asteroid hits NS, B-field confines mass, creates high temp \\
33. & Mitrofanov et al. & 1981 & DISK & Helium flash cooled by MHD waves in NS outer layers \\
34. & Colgate et al. & 1981 & DISK & Asteroid hits NS, tidally disrupts, heated, expelled along B lines \\
35. & van Buren & 1981 & DISK & Asteroid enters NS B-field, dragged to surface collision \\
36. & Kuznetsov & 1982 & SOL & Magnetic reconnection at heliopause \\
37. & Katz & 1982 & DISK & NS flares from pair plasma confined in NS magnetosphere \\
38. & Woosley et al. & 1982 & DISK & Magnetic reconnection after NS surface He flash \\
39. & Fryxell et al. & 1982 & DISK & He fusion runaway on NS B-pole helium lake \\
40. & Hameury et al. & 1982 & DISK & $e$-capture triggers H flash triggers He flash on NS surface\\
41. & Mitrofanov et al. & 1982 & DISK & B induced cyclo res in rad absorp giving rel e-s, IC scat \\
42. & Fenimore et al. & 1982 & DISK & BB X-ray IC scat by hotter overlying plasma \\
43. & Lipunov et al. & 1982 & DISK & ISM matter accum at NS magnetipause then suddenly accrets \\
44. & Baan & 1982 & HALO & Nonexplosive collapse of WD ito rotating, cooling NS \\
45. & Ventura et al. & 1983 & DISK & NS accretion from low mass binary companion \\
46. & Bisnovatyi- et al. & 1983 & DISK & Neutron rich elements to NS surface with quake, undergo fission \\
47. & Bisnovatyi- et al. & 1984 & DISK & Thermonuclear explosion beneath NS surface \\
48. & Ellision et al. & 1983 & HALO & NS corequak \& uneven heating yield SGR pulsations \\
49. & Hameury et al. & 1983 & DISK & B-field contains matter on NS cap allowing fusion \\
50. & Bonazzola et al. & 1984 & DISK & NS surface nuc explosion causes small scale B reconnection \\
51. & Michel & 1985 & DISK & Remnant disk ionization instability causes sudden accretion \\
52. & Liang & 1984 & DISK & Resonant EM absorp during magnetic flares gives hot syn e-s \\
53. & Liang et al. & 1984 & DISK & NS magnetic fields get twisted, recombine, create flares \\
54. & Mitrofanov & 1984 & DISK & NS magnetosphere excited by starquake \\
55. & Epstein & 1985 & DISK & Accretion instability between NS and disk \\
56. & Schlovskii et al. & 1985 & HALO & Old NS in Galactic halo undergoes starquake \\
57. & Tsygan & 1984 & DISK & Weak B-field NS spherically accretes, Comptonizes X-rays \\
58. & Usov & 1984 & DISK & NS flares result of magnetic convective-oscillation instability \\
59. & Hameury et al. & 1985 & DISK & High Landau e-s beamed along B lines in cold atm of NS \\
60. & Rappaport et al. & 1985 & DISK & NS \& low mass stellar companion gives GRB + optical flash \\
61. & Tremaine et al. & 1986 & DISK & NS tides disrupt comet, debris hits NS next pass \\
62. & Muslimov et al. & 1986 & HALO & Radially oscillating NS \\
64. & Sturrock & 1986 & DISK & Flare in the magnetosphere of NS accelerates e-s along B-field \\
65. & \textbf{Paczy\'{n}ski} & \textbf{1986} & \textbf{COS} & \textbf{Cosmo GRBs: rel $e^{-}e^{+}$ opt thk plasma outflow indicated} \\
66. & Bisnovatyi- et al. & 1986 & DISK & Chain fission of superheavy nuclei below NS surface during SN \\
67. & Alcock et al. & 1986 & DISK & SN ejects strange mat lump craters rotating SS companion \\
68. & Vahia et al. & 1988 & DISK & Magnetically active stellar system gives stellar flare \\
69. & Babul et al. & 1987 & COS & energy released from cusp of cosmic string \\
70. & Livio et al. & 1987 & DISK & Oort cloud around NS can explain SGRs \\
71. & McBreen et al. & 1988 & COS & G-wave bkgrd makes BL Lac wiggle across galaxy lens caustic \\
72. & Curtis & 1988 & COS & WD collapse, burns to form new class of stable particles \\
73. & Melia & 1988 & DISK & Be/X-ray binary sys evolves to NS accretion GRB with recurrence \\
74. & Ruderman et al. & 1988 & DISK & $e^{-}e^{+}$ cascades by aligned pulsar outer-mag-sphere reignition \\
75. & Paczy\'{n}ski & 1988 & COS & Energy released from cusp of cosmic string (revised) \\
76. & Murikami et al. & 1988 & DISK & NS \& accretion disk reflection explains GRB spectra \\
77. & Melia & 1988 & DISK & Absorption features suggest separate colder region near NS \\
78. & Blaes et al. & 1989 & DISK & NS seismic waves couple to magnetosphere ALfen waves \\
79. & Trofimenko et al. & 1989 & COS & Kerr-Newman white holes \\
80. & Sturrock et al. & 1989 & DISK & NS E-feild accelerates electrons which then pair cascade \\
81. & Fenimire et al. & 1988 & DISK & Narrow absorption features indicate small cold area on NS \\
82. & Rodrigues & 1989 & DISK & Binary member loses part of crust \\
83. & Pineault et al. & 1989 & DISK & Fast NS wanders though Oort clouds, fast WD bursts only optical \\
84. & Melia et al. & 1989 & DISK & Episodic electrostatic accel and Comp scat from rot high-B NS \\
85. & Trofimenko & 1989 & COS & Different types of white, "grey" holes can emit GRBs \\
86. & \textbf{Eichler et al.} & \textbf{1989} & \textbf{COS} & \textbf{NS-NS binary collide, coalesce} \\
87. & Wang et al. & 1989 & DISK & Cyclo res \& Raman scat fits 20, 40 keV dips, magnetized NS \\
88. & Alexander et al. & 1989 & DISK & QED mag resonant opacity in NA atmosphere \\
89. & Melia  & 1990 & DISK & NS magnetospheric plasma oscillations \\
90. & Ho et al. & 1990 & DISK & Beaming of radiation necessary of magnetized NS \\
91. & Mitrofanov et al. & 1990 & DISK & Interstellar comets pass through dead pulsar's magnetosphere \\
92. & Dermer & 1990 & DISK & Compton scat in strong NS B-field \\
93. & Blaes et al. & 1990 & DISK & Old NS accrets from ISM, surface goes nuclear  \\
94. & Paczynski & 1990 & COS & NS-NS collision causes neutrino collisions, drives super-Ed wind \\
95. & Zdziarski et al. & 1991 & COS & Scat of microwave background photons by rel e-s \\
96. & Pineault & 1990 & DISK & Young NS drifts through its own Oort cloud \\
97. & Trofimenko et al. & 1991 & HALO & White hole supernova gave simulatancous burst of GW from 1987A \\
98. & Melia et al. & 1991 & DISK & NS B-field undergoes resistive tearing, accelerates plasma \\
99. & Holcomb et al. & 1991 & DISK & Alfven waves in non-uniform NS atmosphere accelerate particles \\
100. & Haensel et al. & 1991 & COS & Strange stars emit binding energy in grav rad and collide\\
101. & Blaes et al. & 1991 & DISK & Slow interstellar accret onto NS, e- capture starquakes \\
102. & Frank et al. & 1992 & DISK & Low mass X-ray binary evolve into GRB sites \\
103. & Woosley et al. & 1992 & HALO & Accreting WD collapsed to NS \\
104. & Dar et al. & 1992 & COS & WD accrets to form naked NS, GRB, cosmic rays \\
106. & Hanami & 1992 & COS & NS-planet magnetospheric interaction unstable \\
107. & M\'{e}sz\'{a}ros et al. & 1992 & COS & NS-NS collision produces anisotropic fireball\\
108. & Carter & 1992 & COS & Normal stars tidally disrupted by AGN BH \\
109. & Usov & 1992 & COS & WD collapses to form NS, B-field brakes NS rotation instantly \\
110. & Naryan et al. & 1992 & COS & NS/BH-NS merge gives optically thick fireball \\
111. & Bzainerd & 1992 & COS & Syn emission from AGN jets \\
112. & M\'{e}sz\'{a}ros et al. & 1992 & COS & NS/BH-NS have neutrinos collide to gamms in clean fireball \\
113. & Cline et al. & 1992 & DISK & Primordial BHs evaporating could account for short bursts \\
114. & Frank et al. & 1992 & DISK & Low mass X-ray binary evolves into GRB sites \\
115. & Eichler et al. & 1992 & HALO & Hgih vel halo pulsars accrete after being kicked from disk \\
116. & Eichler et al. & 1992 & HALO & WD mergers yield GRBs \\
117. & Blaes et al. & 1992 & GAL & old NS accretes from mol cloud, R-T instab at crust \\
118. & Melia et al. & 1992 & COS & Crustal adjustments by extragal radio pulsars \\
119. & Schramm et al. & 1992 & COS &  Conversion of NS to strange star close to AGN \\
120. & Hojman et al. & 1993 & HALO & NS popul at MW halo bdry expected by hydro density jump \\
121. & Thompson et al. & 1993 & COS & Sudden NS convection with high B drives e-pairs, gammas \\
122. & Smith et al. & 1993 & DISK & e-beams accel by E-field near NS with high B \\
123. & Fatuzzo et al. & 1993 & COS & Alfven waves accel particles which upscat soft photons \\
124. & Bisnovatyi- & 1993 & GAL & Absorption by cloud of heavy elements around NS \\
125. & McBreen et al. & 1993 & COS & Relativistic jets from cocooned AGN \\
126. & \textbf{Woosley} & \textbf{1993} & \textbf{COS} & \textbf{Spinning W-R star collapse, collaspar model} \\
127. & Kundt et al. & 1993 & GAL & Spasmodic NS accretion causes beamed cooling sparks \\
128. & Cheng et al. & 1993 & GAL & NS glitch reignites magnetosphere of dead pulsar \\
129. & Media et al. & 1993 & COS & NS structural readjustments explain both SGRs and GRBs \\
130. & Thompson & 1994 & COS & Compton upscattering by mildly relativistic Alfv\'{e}n turbulence\\
131. & Cheng et al. & 1996 & COS & Conversion of NSs to stranger stars (detailed) \\
132. & Ma et al. & 1996 & COS & Phase transion of NS, super-giant glitch \\
133. & Katz & 1997 & COS & Different rotation pulsars, BH-thick disk model \\
134. & \textbf{Paczy\'{n}ski} & \textbf{1998} & \textbf{COS} & \textbf{Star-Formation, Collaspar \& BZ effect} \\
135. & Kluzniak & 1998 & COS & Differential rotating magnetar, buoyant instability \\
136. & Dai et al. & 1998 & COS & Differential rotating stranger stars \\
137. & Vietri et al. & 1998 & COS & Supranova model \\
138. & Spruit & 1999 & COS & GW and buoyancy instab in magnetized NS in X-ray binaries \\
139. & MacFadyen & 1999 & COS & Collaspar model (detailed simulation)\\
140. & Popham et al. & 1999 & COS & Numerical model of hyperaccreting BHs \\
141. & Ruderman et al. & 2000 & COS & WD collapses to a different rotating NS (advanced) \\
142. & Wheeler et al. & 2000 & COS & UMHDW jet from NS drives shocks and generates GRBs\\
143. & Ouyed et al. & 2001 & COS & Quark-Nova, accretion onto quark stars \\
144. & Fryer et al. & 2001 & COS & Collapsar occurs for massive metal-deficient first generation stars\\
145. & MacFadyen et al. & 2002 & COS & "Type II collaspar", delayed BH forms by fallback of materials\\
146. & Konigl et al. & 2002 & COS & An association of PWBs with GRBs facilitate collimation \\
147. & Thompson et al. & 2004 & COS & Magnetars spin down affected by neutrino-cooled wind \\
148. & MacFadyen & 2005 & COS & BH disk in binary systems to produce X-ray flares \\
149. & Dar & 2005 & COS & Beamed spike of burst flares of SGRs as SHBs \\
150. & Pazynski & 2005 & COS & Quark stars, surface helps to produce ultrarelativistic outflow\\
151. & Metzger et al. & 2007 & COS & Extended emission of short bursts from protomagnetar spin-down \\
152. & Lu et al. & 2008 & COS & Tidal disruption of a star by an IMBH \\
153. & Zhang et al. & 2008 & COS & Hyperaccreting onto neutron stars as the central engine\\
155. & Kumar et al. & 2008 & COS & Fall-back accret of the stellar envelope produce the X-ray light \\
\hline \hline \caption{A list of GRB models from 1968 to the end of
2008, which includes more than 150 models. The models before 1993
has been listed in Nemiroff (1993). Here ``COS" refers to a
cosmological distance, ``DISK" refers to the disk of our Galaxy,
``HALO" refers to the halo of our Galaxy.}\label{tab111}
\end{longtable}}
\end{center}

\subsection{Galactic Models}

In this subsection, I introduce various types of Galactic GRB
models. A shorter brief review can be found in Vedrenne \& Atteia
(2009). Here I try to give a more detailed review.

\textbf{$\blacksquare$ Stellar Flares}

GRBs were first considered as flares from stellar or compact objects
just after their discovery. Stecker \& Frost (1973) suggested the
possibility that GRBs are due to the stellar superflares from nearby
stars or white dwarfs, as they thought supernovae model showed by
Colgate (1968) was not able to explain the observation data of GRBs.
Then Bercher \& Morrison (1974) followed them to consider the
stellar flares providing the energy of bursts with directed beams of
inverse Compton scattering to provide $\gamma$-rays; Vahia et al.
(1988) compared GRBs with solar hard X-ray flares and suggested GRBs
originate from solar like activity. On the other hand, flares on
compact objects were more commonly suggested,as the they could be
much hotter, denser with several orders of magnitude higher energy
than the stellar flares. Some researchers discussed that GRBs are
thermal (Mullan 1976) or synchrotron radiation (Chanmugan 1974,
1976) from flares on magnetic white dwarfs, which are caused by
convective instabilities under certain conditions. Liang (1984),
Liang et al. (1984) and Usov (1984) discussed flares of neutron
stars with magnetic fields from $10^{12}$ to $10^{14}$G. Neutron
star flares could be the results of twist and recombination of
magnetic fields or convective-oscillation instability.

These authors mentioned above all considered flares are caused by
instabilities (e.g. convective instability) inside normal or compact
objects, and the energy of GRBs is provided by the magnetic fields.
Accreting process will probably also cause flares, but we consider
this situation in the accretion models. However, flare model is
difficult to explain the temporal structure of GRBs with variability
on time scales $\sim0.1$ s, unless the energy propagation and
radiation mechanism are well studied.

\textbf{$\blacksquare$ Starquakes}

Pacini \& Ruderman (1974) first proposed that GRBs are emitted due
to the magnetospheric activity related to glitches from a population
of old, slowly spinning neutron stars. The starquake and glitch
model have greatly benefited from observations of energetic glitches
in the Crab and Vela radio pulsars since they were first observed in
1969 (Pines et al. 1974). The glitches and their theoretical models
allowed a better understanding of the coupling between the
superfluid core and the neutron star crust. Tsygan (1975) discussed
that GRBs from neutron star vibrations excited by corequakes. Fabian
et al. (1976) suggested that based on the local hypothesis, up to
$10^{39}$ ergs of elastic energy might be released in a neutron star
crustquake via sound wave transformation into a surface shock caused
by the crustquake. Later Ramaty et al. (1980) proposed a vibrating
neutron star in the Large Magellani Cloud with a energy of
$~10^{38}$ erg is the orgin of GRB 790505, which was claimed to be
observed in a supernova remnant. Ellison \& Kanzanas (1983)
elaborated on this model by examining the overall energetic and
characteristic of the shocks produced by the neutron star corequake.
The temperature in the shock could be sufficient high to produce
$\gamma$-rays. The starquake models were last refined by Blaes et
al. (1989) by considering the effect of neutron star seismic waves
coupling to the magnetospheric Alfv\'{e}n waves with the limit of
magnetic fields strength.

The lack of gamma-ray bursts observation in coincidence with the
glitches in the radio pulsars (e.g., Vela and Crab pulsar) was a
problem of the starquake model. Pacini \& Ruderman (1974) argued
that gamma-ray bursts might only be detected from the nearby
pulsars. Another problem was that the birth rate of galactic neutron
stars is much less than the GRB rate. As a possible explanation, we
need each neutron star to produce at least $10^{3}$ bursts during
its lifetime. However, no repetitions were detected. Blaes et al.
(1989) discussed that the elastic energy stored by the neutron star
crust, which is sufficient for a single burst, should be replenished
to supple many bursts. However, the elaborate mechanism was not
established.

\textbf{$\blacksquare$ Accretion onto Neutron Stars}

Accretion onto neutron stars was commonly considered as the sources
of GRBs before the BATSE era. Various accretion processes were
suggested with different effects: accretion materials, such as
comets, asteroids, planets around the star, could directly impact
onto the star surface; or accretion materials may form a disk around
the compact object, as usually mentioned in a binary system, or the
disrupted comets, asteroids or planets by tidal force. GRBs could
come from the instability-induced accretion from disks, sudden and
single accretion, or thermonuclear outbursts in the surface layer of
neutron stars via accreting nuclear flues. Here I introduce these
various senecios as follows.

Lamb et al. (1973) suggested the binary environment that GRBs may
originate from occurrences of accretion onto compact objects in
binary systems, where accretion matter could be provided by flares
from stellar companions. The function of stellar flares in the
binary scenario of their work is quite different from that in
Stecker \& Frost (1973), who showed the possible that GRBs are from
flares directly. Later Lamb et al. (1977) studied the magnetospheric
instabilities of neutron star in the case of spherically symmetric
accretion, which can produce bursts variability ($\sim 0.1-1$ s) and
durations ($\sim 1-100$ s). Accretion instabilities were also
studied by Michel (1985) and Epstein (1985). On the other hand,
based on deep sky surveys in the X-ray and optical wavelengths,
Ventura et al. (1983) gave the upper limit of X-ray luminosity of
GRBs for Galactic model ($<10^{31}$ ergs s$^{-1}$), and showed that
the companion of a GRB would have to be a low mass object if GRBs
are from binary systems. Following their work, Rappaport \& Joss
(1985) pointed out that optical flashes may be associated with GRBs
and be detected by ground-based telescopes\footnote{However, X-ray
and optical afterglow had not been detected for decades until the
BeppoSAX era. }. The evolution of low mass X-ray binaries was
discussed by Frank et al. (1992), who refined the binaries with
short orbital periods ($\sim 10$hr) as the possible sites of some
GRBs.

In the binary model, the source of accretion matter is from the
stellar companion, and GRBs are usually produced from the accretion
instability. Another accretion model, which was first proposed by
Harwit \& Salpeter (1973), proposed that the accreted material onto
a neutron star comes from a comets, an asteroid a planet or
interstellar matter, and the sudden accretion causes the gamma-ray
transient event. However, the frequency of comet impact on the
compact object directly is much lower than the GRB events (Guseinov
\& Van\'{y}sek 1974). Tidally disrupted comet and ``rain down
falling" process was then studied by Tremaine \& \.{Z}ytkow (1986).
Pineault and Poisson (1989) considered the scenario that a neutron
star wanders though a comet could, which is not the neutron star's
own, but belongs to another unrelated star. They predicted that GRBs
are provided in the star formation rate, where usually have dense
comet clouds; and comet accreting onto a white drawft may cause a
"optical flash". On the other hand, the collision between an
asteroid and a neutron star was also studied as the possible
mechanism to provide bursts (Newman and Cox 1980; Howard et al.
1981; Colgate and Petschek 1981; van Buren 1981). The asteroid could
interact with the strong magnetic field near the neutron star
surface, then to be disrupted, heated and dragged onto the neutron
star; the kinematic energy of asteroid is radiated via Alfv\'{e}n
wave. However, the probability of asteroid or planet collision is
even much lesser than comet collision.

Besides the accretion instability and sudden collision scenario, the
thermonuclear model was studied first by Woosley \& Taam in 1976.
Magnetized neutron stars may accrete hydrogen and helium at
$10^{-10} M_{\odot} yr^{-1}$ from a companion or dense cloud or
nebula, and accumulate these nuclear fuels in the polar caps of
$\sim 10^{10}$ cm$^{2}$. Compression occurs and fusion reactions
make hydrogen bursts through CNO cycle, and a carbon shell is formed
and cooled via neutrino emission. When the density at the base of
the carbon layer reach $\sim 2\times10^{9}$g cm$^{-3}$, this carbon
layer cannot be cooled efficiently, which leads to carbon
detonations within a timescale of $\sim 1$ms. The energy released in
this thermonuclear runaway is then transported up to the atmosphere
via instabilities and convection mechanisms, and be cooled by
emitting neutrinos and gamma-rays in times $<1$s. Mitrofanov \&
Ostryakov (1981) showed that thermonuclear energy could be
transported via MHD waves (e.g., Alfven wave) in strong magnetic
fields. Later Woosley \& Wallace (1982) considered He flash with the
temperature $\sim 10^{8}$ K on the neutron star surface as the
energy of bursts, and $\gamma$-ray emission comes directly from the
magnetically confined photosphere and from relativistic electrons
accelerated by magnetic recombination. He flash process and energy
transportation in the magnetic fields were next discussed by Fryxell
\& Woosley (1982) and Hameury et al. (1982). Moreover, Blaes et al.
(1990) introduced a pycnonuclear model: GRBs are from neutron stars
accreting interstellar gas at a rate $\sim 10^{10}$ g s$^{-1}$. The
slow accretion rate causes hydrogen and helium to burn through
pycnonuclear reactions, and the metastable layers of crust lying
deep below the stellar surface provides the nuclear energy and
finally produces GRBs.

Thermonuclear model had been previously used with success to explain
type-I X-ray bursts or rapid X-ray transients. For these various
phenomena the accretion rate and the magnetic fields strength of the
neutron star play a decisive role. Similarly, the novae attributed
to white dwarfs are also explained by thermonuclear runaways, after
accumulation of a certain amount of matters. Therefore, the
thermonuclear model was used to be considered as the "better"
explanation for GRBs for years. On the other hand, the absence of
optical or X-ray counterparts of the GRBs also presents the serious
difficulty for this model\footnote{Please note, counterparts
mentioned here do not mean afterglows but the GRB-like events
radiate optical or X-ray transients.}.

\textbf{$\blacksquare$ Accretion onto Black Holes}

Although GRBs were usually considered as associating with outburst
activities of the neutron stars in 1970s and 1980s, accreting black
holes were also pointed out as the possible sources of GRBs in 1970s
by Piran and Shaham (1975, 1977). They proposed that $\gamma$-rays
are produced during the occurrence of instabilities in accretion of
matter onto rapidly rotating black holes, when the infalling X-ray
photons, which are produced in the X-ray binary systems, are Compton
scattered by tangentially moving electrons deep in the ergosphere of
the rotating black hole via the Penrose mechanism. This model
requires GRBs to be associated with X-ray binaries, but no
observation evidence shows such association. On the other hand, the
discovery of cyclotron lines by Konus and Ginga greatly suggested
the magnetized neutron stars as the gamma-ray bursters. It was not
until BATSE ear accreting black hole model was reconsidered in the
collaspar scenario, when neutrino annihilation process or
Blandford-Znajek effect instead of Pernose mechanism are more
commonly studied as the mechanisms to provide the energy and of
GRBs.

\textbf{$\blacksquare$ Collapse of White Dwarfs}

Baan (1982), Woosley and Baron (1992) discussed the scenario that
GRBs are from the collapse of white dwarfs. An accreting white
dwarfs may be pushed to the Chandrasekhar mass and explodes to probe
a Type Ia supernova. However, when the core evolves to a high
density $\gtrsim 10^{10}$g cm$^{-3}$ prior to carbon ar neon
ignition, the white dwarf could collapse directly to a neutron star
without nuclear explosions. Baan (1982) suggested that a GRB is due
the the initial rapid cooling of the surface of the newly formed
neutron star. However, in most cases the cooling lightcurve cannot
explain the temporal proprieties of GRB lightcurves. Woosley and
Baron (1992) discussed the process of neutrino emission and
annihilation during the collapse with wind $0.005M_{\odot}$
s$^{-1}$, and showed that accretion-induced collapse of a white
dwarf cannot be accompanied by a GRB at cosmological distance, while
such collapses might be detected with $\gamma$-ray transient
emissions within our Galaxy. Later accretion-induced collapse model
was reconsidered for GRB model in cosmological distance, as
researchers studied another collapse process: neutron star collapse
to a black hole.

\subsection{Cosmological Models}

Cosmological GRB models have been reviewed by many authors (e.g.,
Piran 2005; M\'{e}sz\'{a}ros 2006; Woosley \& Bloom 2006; Nakar
2007). However, we distinguish the concepts of GRB ``progenitors"
and GRB ``central engines". The GRB progenitors evolve to become GRB
central engines in a short period of time, and the central engines
directly provide the energy of GRBs. For example, a collapsar or a
massive collapse star without immediate supernova explosion is a GRB
progenitor in the collapsar scenario, while the direct central
engine is the new formed accreting black hole system. The compact
object binaries and their mergers are the progenitors of
short-bursts, while the direct central engines are also also the
accreting black holes. In some models, on the other hand, the
progenitors are the same as the central engines, such as the old
magnetar and neutron star model. In this section we focus on the GRB
progenitor models, not the central engines. Therefore we do not
mention the accreting black holes\footnote{The mergers between a
neutron star and a black hole is a exception. The black hole/neutron
star binary is a GRB progenitor.}, but discuss the magnetars.

\textbf{$\blacksquare$ AGN Models}

Based on the fireball scenario and internal shock model, the GRB
short time variability time scale is determined by the activity of
the "inner engine", and $\delta t \sim 1$ms discussed a compact
stellar source. However, large scale models such as AGN models had
also been discussed. McBreen \& Metcalfe (1988) suggested that the
lensed GRB sources are probably BL Lac objects. The fluctuations in
the position of a BL Lac object originating in a stochastic
background of gravitational wave intensifies the image of this
source and give rise to GRBs. Later McBreen et al. (1993) suggested
gravitational lensing as the important tests of the theory that the
GRB sources are located at cosmological distance. However, it is
quite doubtful whether the gravitational wave stochastic fluctuation
effect could explain the radio of GRBs occurrences. Carter (1992)
studied process of a normal stellar being tidally disrupted by AGN
center black hole, which would give rise to GRBs; and Brainerd
(1992) considered synchrotron emission from AGN jets as the
radiation mechanism of GRBs. Yes, there are many similar physical
aspects between AGN and GRB models. However, there is no observation
that GRBs are associated with AGNs, and no repeating and transient
GRBs are detected from the AGN, which might be expected to have
repeating bursts in theory.

\textbf{$\blacksquare$ NS-NS and NS-BH Mergers}

Neutron star binary mergers (NS-NS) or neutron star-black hole
mergers (NS-BH) occur inevitably when binary systems spiral into
each other as a result of damping of gravitational radiation. The
Galactic merger rate is of the order of $10^{-6}-10^{-5}$ yr$^{-1}$,
or 200-3000 Gpc$^{-3}$yr$^{-1}$, which is sufficient to explain the
bursts rate. GRB is suggested not produced during the merger process
directly, but by the accretion process in the new-formed accretion
disk-stellar massive black hole system, which is the common outcome
of both NS-NS and NS-BH mergers and the direct engines of GRBs. The
possibility of two neutron star mergers as a GRB energy source was
first mentioned by Paczy\'{n}ski (1986), Goodman (1986) and Goodman
et al. (1987), who discussed $10^{53}$ergs as the binding energy of
a neutron star(NS), and the neutrino-antineutrino annihilation
process may occur in the merging circumstances. Eichler et al.
(1989) discussed the coalescing neutron stars with nucleosynthesis
of the neutron-rich heavy elements and accompanied neutrino bursts
in detail. Paczy\'{n}ski (1990) considered that neutron star mergers
could produce a super-Eddington wind, accelerate it to a
relativistic velocity, and release energy via neutrino annihilation
during a few seconds. Next Paczy\'{n}ski (1991) also discussed the
binary system of a spinning stellar mass black hole and a neutron or
a strange star at cosmological distances as the possible GRB
progenitor. Narayan et al. (1992) suggested magnetic Parker
instability as well as neutrino annihilation as two different
mechanisms to provide the energy for the bursts, which as the
results of NS-NS or NS-BH mergers. On the other hand, the
Blandford-Znajek mechanism Blandford \& Znajek 1977) for extracting
the spin energy of the stellar massive black hole was also mentioned
as the possible MHD process to provide energy of GRBs (Nakamura et
al. 1992, Hartmann \& Woosley 1995).

However, there are several problems should be pointed out for the
NS-NS (NS-BH) model. Some have been explained but others are still
unanswered. First of all, the idea that GRBs energy are produced via
$\nu\bar{\nu}\rightarrow e^{-}e^{+}\rightarrow\gamma\gamma$
mentioned above was used to be criticized as not an efficient source
of GRBs (e.g. Jaroszynski 1996). M\'{e}sz\'{a}ros \& Rees (1992a,
1992b) suggested that $e^{-}e^{+}$ pair fireball must be
anisotropic, because the merger process and tidal heating generate a
radiation-driven wind, and blow off energy matter before the
$e^{-}e^{+}$ pair plasma acquires a high Lorentz factor except along
the binary rotation axis, where the baryon density is low and an
ultrarelativistic pair plasma could escape. The second problem is
about the typical photon signature time scale of the merger or the
accretion process. Narayan et al. (2001) showed that the accreting
neutrino-dominated accretion disk formed after merger is unlikely to
produce long bursts\footnote{However, if the accretion disk system
is formed in the environment of collaspar, the accretion matter is
provided by the stellar envelope, and the accretion time-scale is
determined by the fallback time-scale of the collapsar. Therefore,
the accretion time-scale for collapsar is much longer than the disk
formed via mergers.}. In other words, compact object mergers could
only explain short GRBs. This idea has been commonly accepted today.
However, an additional problem for the short GRBs is the discrepancy
between the inferred lifetime distribution of the merging systems
and the recent observed lifetime distribution of short bursts. The
type of NS-NS systems in our Galaxy has a typical lifetime of $\sim$
100Myr, while the distribution of short bursts is usually at least
several Gyr old (Nakar 2007).

\textbf{$\blacksquare$ Collapsar Model}

The collapsar model is one of the popular model for long GRBs after
the discovery of GRB-supernovae association. Collapsars are rotating
massive stars whose iron core eventually collapses to form a black
hole and a thick disk around that hole. This model was first
considered as the "failed Type Ib supernovae" (Woosley 1993): stars
heavier than $35\sim40M_{\odot}$ are thought to lose their hydrogen
envelope before dying, and the relatively bigger iron cores prevent
the stars from explosion. The unsuccessful outgoing shock after the
iron core first collapse leads the accreting core further collapses
to a black hole, surrounded by a neutrino-cooled dense disk.A pair
fireball is generated by neutrino annihilation and electron-neutrino
scattering, and provides enough energy to a GRB. However, the
discovery of GRB-supernovae association showed the relation between
GRBs and the collapses of massive stars, and required the "failed
supernovae" to become "successful supernovae (hypernovae)". The
energy of supernovae is considered to be provided by the outflows
from the accretion disks. MacFadyen \& Woosley (1999) discussed the
continued evolution of rotating helium star using a 2-D
hydrodynamics simulation. A compact disk could form when the stellar
special angular momentum $j$ satisfied $3\times10^{16}\leq
j\leq2\times10^{17}$. The gravitational binding energy of the
accretion system can be efficiently released by neutrino or MHD
processes, and provides the energy of a relativistic jet and then
the GRB phenomena. Later MacFadyen et al. (2001) suggested a new
type of collapsar (Type II collaspar), wherein the center black hole
forms after some delay, owing to the fallback of material that
initially moves outwards, but fails to achieve escape velocity. On
the other hand, The standard collapse, whose black hole forms
promptly due to the unsuccessful outgoing shock, is called as Type I
collapsar. The accretion rate in Type II collapsars is lower than
that in Type I collapsars, and the jet should be produced via MHD
processes only for such low accretion rate. Fryer et al. (2001)
discussed a third type of collapsar, which occurs for extremely
massive metal-deficient stars ($\sim300M_{\odot}$) existed in the
first generation stars. These collapses could produce a
$10^{54}$ergs jet and become the possible sources of some
high-redshift GRBs. Furthermore, Zhang, Woosley et al. (2003, 2004)
examined the propagation and breakout of relativistic jets produced
in the collapsar accretion process thought the massive Wolf-Rayet
stars using 2-D and 3-D simulations. The jet propagation and
breakout produces a variety of high-energy transients, ranging from
X-ray flares to classical GRBs.

However, collapsar model also faces some problems. For example, it
requires an unusually larger amount of angular momentum in stellar
inner regions compared to the common pulsars. Woosley \& Heger
(2006) suggested the very rapid rotating massive stars might result
from mergers or massive transfer in a binary, and single stars that
rotate unusually rapidly on the main sequence. However, such stars
might still retain a a massive hydrogen envelope and produce Type
II, not Type I supernovae. Yoon \& Langer (2005) considered the
evolution of massive magnetic stars where rapid rotation induces
almost chemically homogeneous evolution. Fryer \& Heger (2005) and
Fryer et al. (2007) discussed that single stars cannot be the only
progenitor for long bursts, but binary progenitors can match the
GRB-SNe association observational constraints better. I will discuss
the collapsar model more detailed in \S 1.3.

\textbf{$\blacksquare$ Magnetar Model}

Magnetized neutron stars and magnetars as the sources of
cosmological GRBs was fist discussed by Usov (1992), who considered
that such NSs are formed by accretion-induced-collapse of highly
magnetized WDs with surface magnetic fields $\sim10^{9}$ G. These
new formed rapidly rotating NSs or magnetars would lose their
rotational kinetic energy catastrophically on a timescale of seconds
or less via electromagnetic or gravitational radiation. An
electron-positron ($e^{-}e^{+}$) pairs plasma would be created by
the unstable strong electric fields which is generated by the
rotation of the magnetic fields. This $e^{-}e^{+}$ plasma flows away
from the NS at relativistic speeds and may produce GRBs. Duncan \&
Thompson (1992) also suggested that GRBs are powered by AIC
magnetars with their vast reservoirs of magnetic energy. They
discussed that the strong dipole magnetic fields in magnetars, which
could reach to $\sim3\times10^{17}$ G in principle, can be generated
by vigorous convection and the $\alpha-\Omega$ dynamo mechanism
during the first few tens of seconds after the NS formation.
Klu\'{z}niak \& Ruderman (1998) discussed a "transient magnetar"
model. The energy stored in the differential rotation is extracted
by the processes of wounding up and amplification of toroidal
magnetic fields. Magnetic buoyancy drives magnetic fields across the
NS surface, making NS to be a transient magnetar and release the
magnetic energy via reconnection. Such process can repeat several
times in the NS. Ruderman et al. (2000) estimated the buoyancy
timescale of each time is of the order $10^{-2}$ s, which may
associate with the temporal lightcurve structure of GRB prompt
emissions. The surface magnetic multipoles ($\sim10^{17}$ G) in the
"transient magnetar" phase is suppressed by surface shearing from
differential rotation and provide the energy of sub-bursts. Spruit
(1999) considered the similar buoyancy mechanism to produce GRBs,
while he suggested the NS in an X-ray binary environment, spin up by
accretion, loss angular momentum via gravitational wave radiation,
and generate strong differential rotation. Dai et al. (2006, see
also Gao \& Fan 2006) suggested the similar magnetic activity to
produce the early afterglow phenomena such as X-ray flares. Recently
Metzger et al. (2008) argued that extended emission of short-hard
GRBs (e.g. GRB 060614) is form the spin-down of magnetars.

A neutrino-driven thermal wind is dominated during the
Kelvin-Helmholtz (KH) cooling epoch after the NS formation, lasting
from few second to tens of seconds depending on the strength of
surface fields. After that, magnetic or Poynting wind becomes
dominated (Wheeler et al. 2000; Thompson 2003; Thompson et al.
2004). While a Poynting flux dominated flow may be dissipated in a
regular internal shocks. Usov (1994) and Thompson (1994) discuss a
scheme in which the energy is dissipated from the magnetic field to
the plasma and then via plasma instability to the observed
$\gamma$-rays outside the $\gamma$-rays photosphere at $\sim10^{13}$
cm. At this distance the MHD approximation of the pulsar wind breaks
down and intense electromagnetic waves are generated. The particles
are accelerated by these electromagnetic waves to Lorentz factors of
$10^{6}$ and produce the non thermal spectrum. Wheeler et al. (2000)
considered the magnetar scenario with the generation of
ultra-relativistic MHD waves (UMHDW), not the traditional Alfv\'{e}n
waves. If a UMHDW jet is formed it can drive shocks propagating
along the axis of the initial matter jet formed promptly during the
proto-NS phase. The shocks associated with the UMHDW jet could
generate GRBs by a process similar in the collapsar scenario. The
origin of collimated relativistic jets form magnetars was considered
by K\"{o}nigl \& Granot (2000) by analogy to pulsar wind nebulae
(Begelman \& Li 1992), that the interaction of wind from the
spinning-down magnetar with the surrounding star could facilitate
collimation. Bucciantini et al. (2007, 2008, 2009) presented
semi-analytic calculations and relativistic MHD simulations to show
the magnetized relativistic jets formation and propagation. However,
they only carried out low Lorentz factor less than 15 in their
simulations. Jets from magnetars may be accelerated to higher
Lorentz factor $\sim100-10^{3}$ at large radius several tens of
seconds after core bounce (Thompson et al. 2004; Metzger et al.
2007). Therefore, long-term simulations of magnetized jets
propagation is still needed in the future.

\textbf{$\blacksquare$ Quark Stars and Strange Stars}

A quark star or strange star is a hypothetical type of exotic star
composed of quark matter, or strange matter. These are ultra-dense
phases of degenerate matter theorized to form inside particularly
massive neutron stars. However, the existence of quark stars has not
been confirmed by astrophysical observations. The quark stars or
strange stars have been considered as the GRB progenitors since
Schramm \& Olinto (1992), who first briefly discussed the possibly
of conversion of neutron stars to strange stars, i.e., hadrons to
quarks as the origin energy source of GRBs. Cheng \& Dai (1996)
discussed the phase transition in the low-mass X-ray binaries
environment, while Schramm and Olinto (1992) close to an AGN. The
thermal energy released in the phase $\sim 10^{52}$ ergs transition
is mainly cooled by neutrino emission, and fireball is produced by
the process of $\gamma\gamma\rightarrow e^{-}e^{+}$ (Cheng \& Dai
1996). On the other hand, Ma \& Xie (1996) discussed the phase
transition process to produce the quark mass core and a super-giant
glitch of the order $\Delta\Omega/\Omega\sim 0.3$, which could
provide sufficient energy for GRBs in cosmological distance and SGRs
in Galactic distance. Following the neutron model of Kluzniak and
Ruderman (1998), Dai \& Lu (1998b) discussed the differentially
rotating strange stars with the buoyancy instability as the sources
of GRBs. Ouyed et al. (2002a, 2002b) suggested a model (quark-nova)
in which the engine both for short and long bursts is activated by
the accretion onto a quark star, which is formed in the core of a
neutron star. Later the process of "quark-nova explosion" (e.g.
Ouyed et al. 2007, 2009), i.e., the ejection of the outer layer of
the neutron star was studied, while the interaction between the
quark-nova eject and the collapsar enveloped is showed to be
possible to produce both GRBs. However, it is questionable today
whether a single model could explain both short bursts and long
bursts, which have quiet different proprieties and located in
different host galaxies. Paczy\'{n}ski and Haensel (2005) showed
that the surface of quark stars could acts as a membrane and allows
only ultrarelativistic matters to escape, and generate outflows with
large bulk Lorentz factors ($>300$).

The outcome of most quark star and strange star models is the
release of a large amount of energy within a short time, which can
provide the GRBs energy in cosmological distance. However, the main
problem today is that the existence of quark stars has not been
proofed. Most models shows the properties and activities of quark
stars, but seldom predicts the "key phenomena" for GRBs, which could
distinguish between the quark star activity and other compact star
activity. It is probably we need to determine the existence of quark
stars in other astrophysics fields and then go back to see the
observation effects of them in GRBs.

\textbf{$\blacksquare$ Non-standard Physical Models}

Astronomers usually like to choose theoretical models which are
based on confirmed physics laws and theories. Models that
incorporate astronomical objects which are not know to exist are not
encouraged at least today. Few astronomer will believe these models
until the related speculation physics theories are widely accepted
and the very observational features predicted by the models can be
confirmed. Recent developed physics theories could no doubt prompt
the development of theoretical astrophysics, which is actually part
of physics. However, astrophysics should always based on the
observation events first, not on physics theories, otherwise we will
be probably ``get lost". In addition to the models of collapsar,
compact objects mergers, magnetic fields activities and other
commonly accepted astrophysical processes, there are also several
speculative GRB models which are treated less seriously in current
state.

A white hole is the theoretical time reversal of a black hole.
Narlikar et al. (1974) first proposed that X-ray and $\gamma$-ray
transient may from while holes. The spectrum of while hole emission,
which satisfies a power law of index -3, should soft with time and
change to be the spectrum of X-ray and $\gamma$-ray bursts.
Trofimenko (1989) studied the structure of Kerr-Newman white hole in
detail, and showed the application in astrophysics, which include
explaining GRBs. Moreover, a cosmic string is a hypothetical
1-dimensional (spatially) topological defect in various fields
predicted in theoretical physics. Babul et al. (1987, revised by
Paczy\'{n}ski in 1988) showed that the cusps of superconducting
cosmic strings may be possible sources of very intense and highly
collimated bursts of energy. Cline \& Hong (1992), on the other
hand, considered the possibility of detecting Galactic "primordial
black holes" (PBH) by short-hard GRBs, as PBHs evaporating (Hawking
1974) could account for short-hard bursts with duration of several
milliseconds.

\section{Progenitor I: the Collapsar Model}

\subsection{Basic Collapsars Scenario}

In this section \S 1.3, we investigate the collapsar scenario in
detail. We classify collapsars into two types based on their
different ways to form the center black holes. Moreover, we also
discuss a third variety of collapsar occurs in the first generation
massive stars with high redshift and very low metallicity. The
classification is based on Woosley, Zhang, \& Herger (2002).

\textbf{\textit{Type I Collapsar}}

A standard (Type I) collapsar is one where the black hole forms
promptly in a Wolf-Rayet star, a blue supergiant, or a red
supergiant. There never is a successful outgoing shock after the
iron core first collapses. A star of 30 $M_{\odot}$ on the main
sequence evolved without mass loss would have a helium core of the
size 10-15 $M_{\odot}$. Larger stars that continued to lose mass
after exposing their helium core might also converge on this
configuration, depending on metallicity and the mass-loss rate
chosen. The iron core in such a star would be between 1.5 and 2.3
$M_{\odot}$, depending upon how convection and critical reaction
rate are treated. As the core collapses, mass begins to accrete from
the mantle. If neutrino energy deposition is unable to turn the
accretion around, the hot protoneutron star grows. Typically the
accretion rate is $\sim 0.5 M_{\odot}$s$^{-1}$. In a few seconds the
core has lost enough neutrinos, and grown to sufficient mass that it
collapses to a black hole. Material from the mantle and helium core
continue to accrete at a rate that declines slowly with time,
roughly as $t^{-1}$, and depends on the uncertain distribution of
angular momentum in the star. The disk forms on a free-fall time
scale, 446$\rho^{1/2}\sim 1$s for the stellar mantle, but will
continue to be fed as the rest of the star comes in. The polar
region, unhindered by rotation, will collapse first on a dynamic
time scale ($\sim 5$s), while the equatorial regions evolve on a
viscous time scale that is longer. The disk is geometrically thick
and typically has a mass inside 100 km of several tenths of a solar
mass.

MacFayden \& Woosley (1999) explored in particular the fate of a 14
$M_{\odot}$ helium core from a 35 $M_{\odot}$ main sequence star
using a two-dimensional hydrodynamics code. The evolution of the
massive star can be considered in three stages. First is transient
stage lasting roughly 2 s, during which low angular momentum
material in the equator and most of the material within a free fall
time along the axes falls through the inner boundary. A
centrifugally supported disk forms interior to roughly
200$(j_{16}/10)^{2}$ km, where $j_{16}\equiv j/(10^{16})$ cm$^{2}$
s$^{-1}$. The density near the hole and along its rotational axis
drops by an order of magnitude. The second stage is characterized by
a quasi-steady state in which the accretion delivers matter to the
hole at approximately the same rate at which it is fed at its outer
edge by the collapsing star. For $3\leq j_{16}\leq20$, the disk
forms at a radius at which the gravitational binding energy can be
efficiently radiated as neutrinos or converted to beamed outflows by
MHD processes (i.e., BZ process), and deposit energy in the polar
regions of the center black hole. The third stage is the explosion
of the star. This occurs on a longer timescale. Energy deposited
near the black hole along the rotation axes makes jets that blow
aside what remains of the star within about $10^{\circ}-20^{\circ}$
of the poles, typically $\sim 0.1 M_{\odot}$. The kinetic energy of
this material pushed aside is quite high, a few 10$^{51}$ ergs,
enough to blow up the star in an axially driven supernova, or
so-called hypernova. The properties of the relativistic jet which
produce a GRB and other high energy transients was studied in detail
by Aloy et al. (2000), Zhang et al. (2003) and Zhang \& Woosley
(2004). In supergiant stars, collapsar model predicts the jet
breakout also produces X-ray transients instead of GRBs with Type II
supernova (MacFadyen, Woosley \& Heger, 2001).

\textbf{\textit{Type II Collaspar}}

A variation on this theme is the "Type II collapsar", wherein the
black hole forms after some delay- typically a minute to an hour,
owing to the fallback of material that initially moves outwards, but
fails to achieve escape velocity. The time scale for such an event
is set by the interval between the first outgoing shock and the GRB
event. The fallback material onto the collapsed remnant is about
$0.1-5 M_{\odot}$, and turn the collapsed remnant into a black hole
surrounded by a accretion disk. The accretion rate, $\sim 0.001-0.01
M_{\odot}$ s$^{-1}$, is inadequate to produce a jet mediated by
neutrino annihilation but is similar to what has been invoked in MHD
models for GRBs. Type II collapsars should be common events,
probably more frequent than Type I. They are also capable of
producing powerful jets similar as Type I ones that might make GRBs.
Unfortunately their time scale may be, on the average, too long for
the typical long, soft bursts. If the GRB-producing jets is launched
within the first 100 s or so of the initial supernova shock, it
still emerges from the star before the supernova shock has gotten to
the surface, i.e, when the star is still dense enough to provide
collimation. Their accretion disks are also not hot enough to be
neutrino dominated and this may affect the accretion efficiency.
Table \ref{tab131} (i.e., Table 3 in MacFadyen, Woosley \& Heger
2001) outlines the diverse phenomena resulting from the two
collapsar types occurring in stars with varying radii, where $t_{\rm
engine}$ is the duration of GRB central engine, and $t_{\rm bo}$ is
the time of jet breakout.

\begin{table}
\begin{center} {\scriptsize
\begin{tabular}{lcc}
\hline
Parameters & Type I & Type II \\
\hline
Black hole formation timescale (s) & Prompt, $\tau\leq 1$ & Delayed, 30-3000  \\
$\dot{M}$($M_{\odot}$ s$^{-1}$) & 0.1 & 0.0001-0.01  \\
$\tau_{\rm accretion}$(s) & $\approx 10$ & 30-3000 \\
No H envelope (Wolf-Rayet star) & $t_{\rm engine}>t_{\rm bo}$, GRB+Type Ib/c SN & $t_{\rm engine}>t_{\rm bo}$, long GRB+Type Ib/c SN \\
Small H envelope (Blue supergiant) & $t_{\rm engine}<t_{\rm bo}$, Type II SN; XRT & $t_{\rm engine}\geq t_{\rm bo}$, long GRB+Type II SN \\
Large H envelope (Red supergiant) & $t_{\rm engine}\ll t_{\rm bo}$, Type II SN; XRT & $t_{\rm engine}\leq t_{\rm bo}$, Type II SN; XRT \\
\hline \hline
\end{tabular} }
\end{center}
\caption{The diverse phenomena resulting from Type I and Type II
collapsar occurring in stars with with varying radii. All collapsar
SNe are predominantly jet-powered and therefore asymmetric
(MacFayden, Woosley, \& Heger 2001).}\label{tab131}
\end{table}


\textbf{\textit{Type III Collapsar}}

A third variety of collapsar occurs for extremely massive
metal-deficient star (above $\sim 300 M_{\odot}$) that may have
existed in the early universe. For non-rotating stars with helium
core masses above 137 $M_{\odot}$ (main sequence mass 280
$M_{\odot}$), it is known that a black hole forms after the pair
instability is encountered. It is widely suspected that such massive
stars existed in abundance in the first generation after the Big
Bang at red shifts $\sim 5-20$. Fryer, Woosley \& Heger (2001)
considered the complete evolution of two zero-metallicity stars of
250 and 300 $M_{\odot}$. Despite their large masses, we argue that
the low metallicities of these stars imply negligible mass loss.
Evolving the stars with no mass loss, but including angular momentum
transport and rotationally induced mixing, the two stars produce
helium cores of 130 and 180 $M_{\odot}$. Explosive oxygen and
silicon burning is unable to drive an explosion in the 180
$M_{\odot}$ helium core, and it collapses to a black hole. For this
star, the calculated angular momentum in the preosupernova model is
sufficient to delay black hole formation. After the black hole
forms, accretion continues through a disk. The mass of the disk
depends on the adopted viscosity but may be quite large, up to 30
$M_{\odot}$ when the black hole mass is 140 $M_{\odot}$. The
accretion rate through the disk can be as large as $1-10 M_{\odot}$
s$^{-1}$. The interaction of the $10^{54}$ erg jet with surrounding
circumstellar gas may produce an energetic $\gamma$-ray transient,
but given the probable redshift and the consequent timescale and
spectrum, this model may have difficulty explaining typical GRBs.
The long, hard X-ray flashes rather than classical GRBs are expected
for these Type III collapsars.

According to Woosley, Zhang \& Heger (2002), the Type I and II
collapsar model which produce GRBs requires the star loses its
hydrogen envelope before death, because no relativistic jet can
penetrate the envelope in less than the light crossing time which is
typically 100 s for a blue supergiant and 1000 s for a red one.
Therefore, the Wolf-Rayet stars lost their hydrogen envelope are
most likely to be the candidates of GRBs sources. Fryer, Woosley \&
Hartmann (1999) discussed the possible stellar progenitors to form a
black hole with an accretion disk which satisfies the GRB production
environment. The hydrogen envelopes can be removed both from single
stellar wind or binary companions. After common envelope evolution,
which uncovers the helium core, the primary star collapses into a
black hole. As a result, binary systems should also be added into
the collapsar model. There are three possible scenarios leading to
collapsar formation.

\textbf{Scenarios A:}

Single-star collapsar formation scenario. Wolf-Rayet winds blow off
the hydrogen mantle of a rotation massive star, leaving behind a
massive helium core, which then collapses to a black hole. The
helium core must have a mass $\geq 10M_{\odot}$ to insure that the
core immediately forms a black hole without any supernova explosion.

\textbf{Scenarios B:}

Dominant collapsar formation scenario. Common envelope drives off
the hydrogen mantle of a rotation massive star. The helium core
collapses to form a GRB. The system may then go on to form a WD/BH
or BH/NS binary.

\textbf{Scenarios C:}

Collapsar scenario from the merger of a double helium binary. In
this phase, the two stars have nearly the same mass and the
secondary evolves off the main sequence before the primary
collapses, forming a double helium star binary. Then the helium
merge, producing a flattened helium core, which then powers the
collapsar.

Moreover, several additional scenarios were introduced later. If the
common-envelope phase in He-He binary merger occurs after helium
burning (case C mass transfer in binary theory), we also it "helium
case C" scenario. Other models include:

\textbf{Scenarios D:}

"Brown merger" scenario. Two stars in the binary have nearly equal
masses and hence the companion evolves off the main sequence before
the more massive star collapse. This scenario is termed due to the
work by Bethe \& Brown (1998), who discussed equal-mass stellar
binaries.

\textbf{Scenarios E:}

The progenitor interacts in a cluster. The cluster environment,
which many be more common in low-metallicity systems, enhances
mergers, and well produce massive, rapidly spinning cores.

In the next section we will discuss the requirements of collapsar
models based on the reasonable physics processes to produce GRBs as
well as recent observations.

\begin{figure}
\centering\resizebox{1.0\textwidth}{!} {\includegraphics{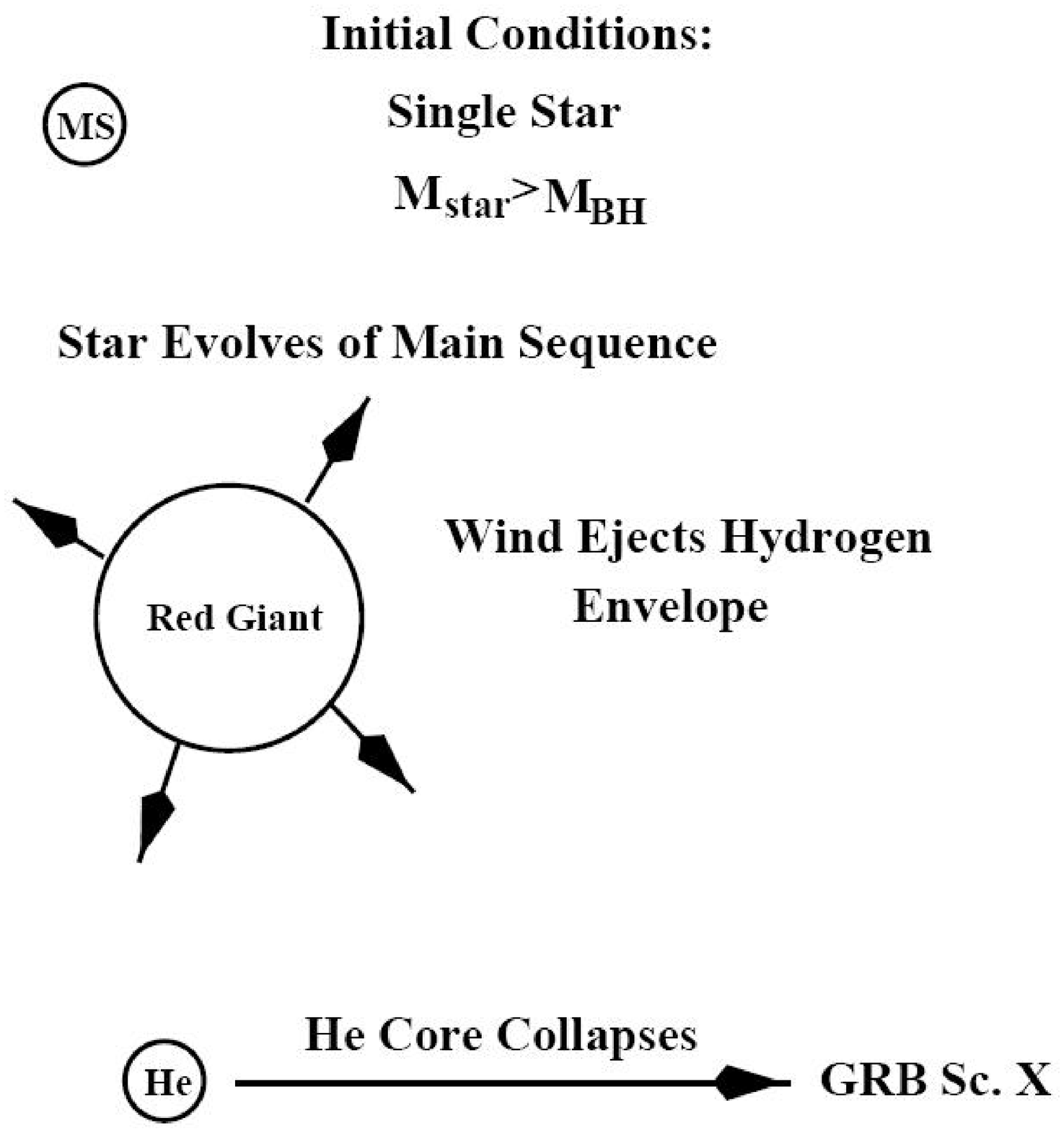}
\includegraphics{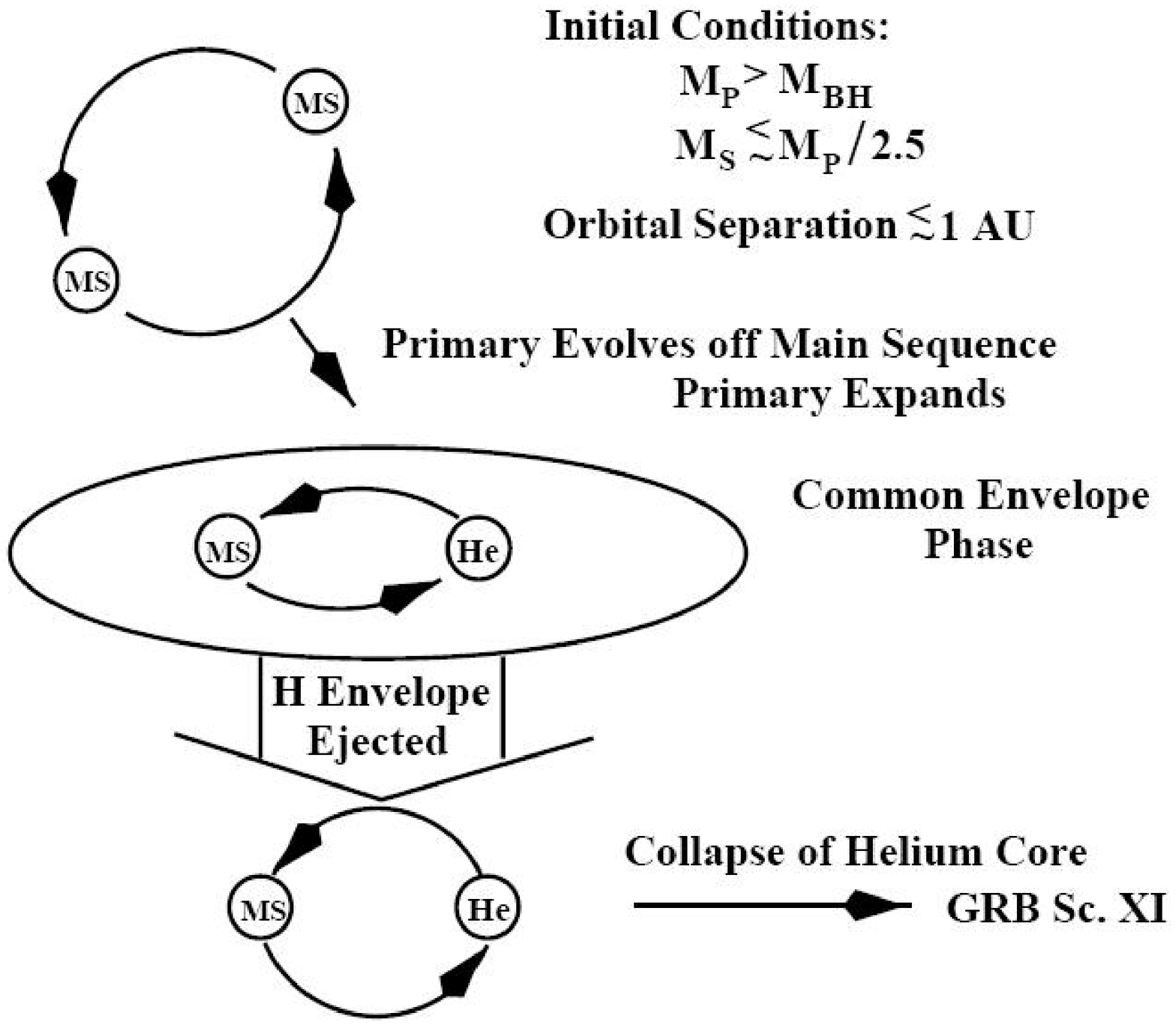}}
\\\resizebox{0.6\textwidth}{!} {\includegraphics{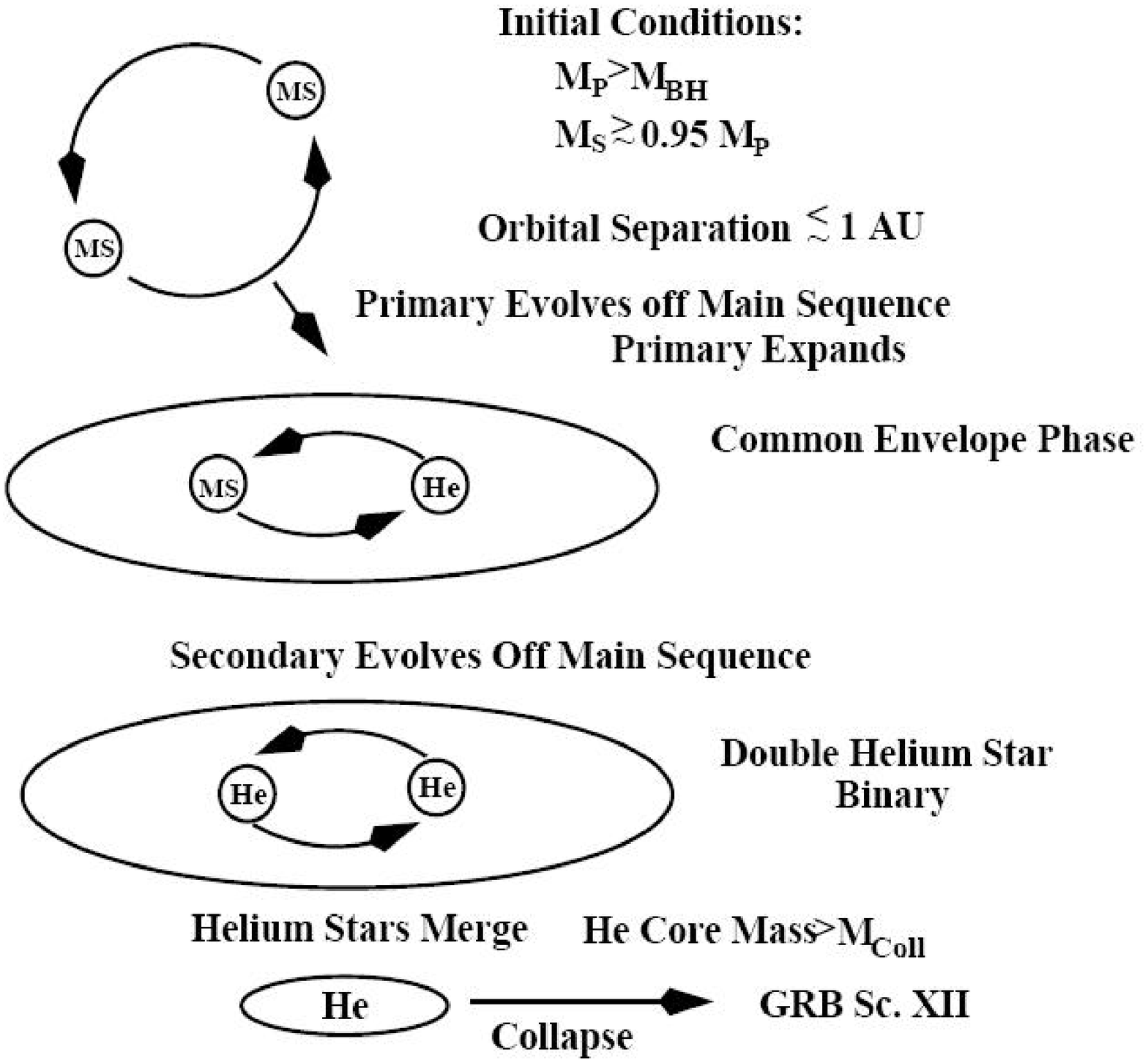}}
\caption{Three Scenarios A (\textit{upper left}), B (\textit{upper
right}), C (\textit{lower figure}) of collapsar formation proposed
by Fryer, Woosley and Hartmann (1999).}\label{fig131}
\end{figure}

Figure \ref{fig131} shows the first three scenarios to form
collapsars.

\subsection{Angular Momentum and Metallicity}

The collapsar model requires an unusably large angular momentum in
the inner regions of the massive progenitor stars lost their
hydrogen envelope to from a disk around a black hole of several
solar mass. The specific angular momentum
\begin{equation}
j=\frac{2GM}{3^{3/2}c}[1+2(3r_{lso}/r_{g}-2)^{1/2}]\simeq
10^{16}-10^{17} \rm cm^{2} \rm s^{-1}
\end{equation}
where $M$ is the gravitational mass of the black hole, $r_{lso}$ is
the radius of the last stable circular orbit, and $r_{g}=GM/c^{2}$.
Wolf-Rayet (WR) stars with a very rapidly rotating core are good
candidates for the progenitor. On the other hand, millisecond
pulsars, which are more than one order of magnitude faster than he
fastest observed pulsars, have a angular momentum $j\sim
10^{15}R_{10}^{-2}P_{1}^{-1} \rm cm^{2} \rm s^{-1}$ with
$R_{10}=R/10^{7}$cm as the pulsar radius and $P_{1}=P/10^{-3}$s as
the spin period. This angular momentum is still 1-2 orders of
magnitude lower than the ones that make collapsars. Recent stellar
evolution models which include angular momentum transport from the
core to the hydrogen envelope by magnetic torques (Spruit 2002) lead
naturally to rotation rates of pulsars in the range 10-15 ms, just
what is needed for common pulsars result from the deaths of read
supergiants (RSGs). However, the magnetic toques mechanism indicates
the most single stars end up with cores rotate much slower than the
ones in rapidly collapsars that lost their hydrogen envelopes. The
problem is, if typical massive star death produces common pulsars,
what special circumstance produce a GRB?

Magnetic torques are significant even when the envelope is removed.
It coming from fields generated by the differential rotation have a
tendency to enforce rigid rotation. Angular momentum is extracted
from the inner core when it contracts and is transported to the
outer layers which are spun up. Moreover, these magnetic torques
brake the core when extensive mass loss slows down the rotation of
the outer layers. Herger, Woosley \& Spruit (2005) have found that
magnetic torques decrease the final rotation rate of a collapsing
iron core by about a factor of 30 ro 50. Furthermore, even without
magnetic torques and even if the envelope of the star has been lost,
the vigorous mass loss of typical WR stars still carries away lots
of angular momentum. It has been known for some time that if the
magnetic torques are negligible, which is to say much weaker than
estimated by Spruit (2002), it is easy to give GRB progenitors the
necessary angular momentum; but then one must invoke another
mechanism to slow down from $10^{16}-10^{17}$ cm$^{2}$ s$^{-1}$ to
typical pulsars with rotation period $\geq 15$ms.

Woolsey \& Heger (2006) considered two ways to solve the problem.
The first assumes that a fraction of WR stars have a mass loss which
is not standard, for instance with a decrease of up to a long GRBs
are produced at high redshift where the metallicity Z is much lower
than in the solar system. Under these low metallicity conditions it
is known that WR stars have a lower mass loss rate. The mass-loss
rate of metal-deficient WR stars scales closer to $Z^{0.86}$.
Another positive effect expected in WR stars with low metallicity is
that they should rotate faster (Meynet \& Maeder 2005). Finally, the
mass loss anisotropies by stellar winds influence the loss of
angular momentum. These anisotropies may result from mass-losses
occurring predominantly at the poles, and help in reducing the loss
of angular momentum because the loss of polar mass takes away less
angular momentum than isotropic mass-loss. As a result, the GRB
progenitors might be the few O and B stars with high rotational
velocity ($\sim 400$km s$^{-1})$ representing a few percent of this
population. Moreover, because of the likely dependence of mass loss
on metallicity, GRBs will be favored in regions of low metallicity.
On the other hand, Yoon \& Langer (2005) showed that the hydrogen
envelopes retained due to the weakness of the stellar winds at low
metallicity may not be a problem, because such low Z stars could
become helium stars not by ejecting their hydrogen envelopes but
burning the hydrogen into helium under through chemically
homogeneous evolution. For low enough metallicity, this type of
evolution can lead to retention of sufficient angular momentum in CO
cores in the range from 10 $M_{\odot}$ to 40 $M_{\odot}$ to produce
GRBs according to the collapsar scenario.

\begin{figure}
\centering\resizebox{0.5\textwidth}{!} {\includegraphics{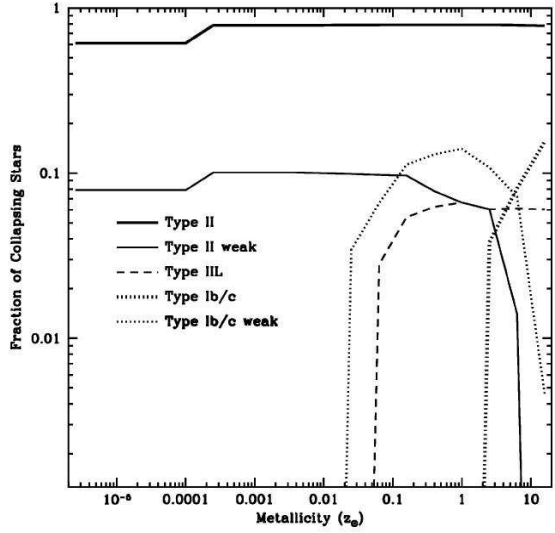}}
\caption{Single-star supernova rate as a fraction of total number of
collapsing stars as a function of metallicity calculated using the
stellar models (Heger et al. 2003; Fryer et al. 2007). Three classes
of Type II SNe in this figure include normal Type II SNe (Type II),
Type II SNe with weak SN explosions (Type II weak), and Type II SNe
that have lost most of their hydrogen envelope (Type IIL); and two
classes of Type Ib/c SNe are normal Ib/c SNe and weak Ib/c SNe. If
single stars dominate the Ib/c rate, these models predict only weak
Type Ib/c SNe below solar metallicity.}\label{fig1321}
\end{figure}

Both single-star models by Yoon \& Langer (2005) and Woosley \&
Heger (2006) have made strong predictions about the low metallicity
requirement of the progenitors. However, observations show that GRBs
occur in environments with a range of metallicities from 1/100th
solar to solar. The mean metallicity may be a high as 1/3-1/2 solar.
Although we are unlikely to develop an observation technique that
will permit a direct metallicity measurement of a GRB progenitor
today, we are able to infer the metallicity of gas near the GRB
progenitor by a few complementary approaches, for examples, by
observing absorption line spectroscopy of GRB afterglows, or
emission-line spectroscopy of H II regions within the GRB host
galaxy (Fryer et al. 2007). The present set of metallicity
measurements from $\approx$ 10 GRB afterglow spectra exhibit a large
dispersion of values from $\approx 1/100$ solar (Chen et al. 2005)
to nearly solar metallicity (Castro et al. 2003) with an average
metallicity of 1/3-1/2 solar. These observations strong constraint
on single-star models. In addition to the metallicity observation,
the supernovae associated with GRBs that are bright enough to be
studied in detail are almost Type Ic supernovae until
now\footnote{Thus far, the only five well-observed cases of SNe
associated with GRBs or XRFs: SN 1998bw/GRB 980425, SN 2003dh/GRB
030329, SN 2003w/GRB 031203, SN 2006aj/XRF 060218, and  SN 2008D/XRF
080109. All of these SNe expect for SN 2008D are of Type Ic.
However, XRF 080109 is much less energetic with $E\approx
2\times10^{46}$ ergs. }. If this result is universal, most
progenitor model must lose not only its hydrogen envelope but most
of its helium envelope as well. The single-star models, which
produce He-rich (Type II) supernovae, do not fit the existing data
of GRB-SN association as well. Furthermore, studying the progenitors
of Type Ib/c sypernovae may also provide some insight into the
progenitors of long GRBs. Evidence is growing that Type Ib/Ic
supernovae preferentially occur in high-metallicity systems. Figure
\ref{fig1321} shows the fraction of collapsing stars that form Type
II and Tpe Ib/c supernovae as a function of metallicity (Heger et
al. 2003). Note that single nonrotating stars produce Ib/c
supernovae only at metallicities above 0.02 solar. But we expect
these supernovae to have weak shocks and hence every very little
nickel. Without the high shock temperature and the radioactive
nickel to power the light curve, these supernovae will be dim.
Strong Ib/Ic supernovae are not produced at all until the
metallicity rises above solar! This conclusion could also be
maintained for rapidly rotating stars. As a result, the above
mentioned argues strongly show that the single-star models at low
metallicity, cannot explain all the long bursts. As the date get
better, the limitations on single-star models will become more
strict.

\begin{table}
\begin{center} {\scriptsize
\begin{tabular}{lcccc}
\hline
Scenario & Angular Momentum & Metallicity Trend & Surrounding & Associated Supernovae \\
 &  &  & Environment &  \\
\hline
Classic single &  Low? & Rate peaks $\sim 0.1Z_{\odot}$ & High wind, Low wind & H-rich to He-rich \\
Mixing single &  Good & $Z< 0.1 Z_{\odot}$ & Low wind & All He-rich \\
Classic binary & Low? & Rate$\uparrow$ $Z\downarrow$ & Tends to low wind & He-rich, He-poor \\
Tidal binary  & Good? & Rate$\uparrow$ $Z\downarrow$ & Tends to low wind & He-rich, He-poor \\
Brown merger  & Good? & Rate$\uparrow$ $Z\downarrow$ & Tends to low wind & He-rich, He-poor \\
Explosive ejection & Good? & Rate$\uparrow$ $Z\downarrow$ & Shell within 1 pc & He-poor \\
He merger & High & Rate$\uparrow$ $Z\downarrow$ & Tends to low wind & He-rich, He-poor \\
He case C & Good? & Rate$\uparrow$ $Z\downarrow$ & Tends to low wind & He-rich, more He-poor \\
Cluster & Good? & Rate$\uparrow$ $Z\downarrow$ & Tends to low wind & He-rich? \\
\hline \hline
\end{tabular} }
\end{center}
\caption{Different scenarios of collapsar progenitors (Fryer et al.
2007)}\label{tab132}
\end{table}

Therefore, we should also consider binary systems as the long GRB
progenitors, as suggested by Fryer et al. (1999). For many binary
progenitor models, the binary component is used to remove the
hydrogen envelope without the angular momentum loss that occurs in
wind mass loss. Also, the binary models tend to fit the metallicity
constraints well. Several binary progenitors can match the solid
observational constraints and also have the potential to match the
trends that we are currently seeing the observations mentioned
above. Table \ref{tab132} is the list of all the progenitors studied
and their basic predictions from Fryer et al. (2007).

\subsection{Jet Formation, Propagation and Breakout}

In collapsing massive stars, as a consequence of the in-fall into
the black hole of the matter which was initially situated along the
rotational axis and of the stagnation of matter in the equatorial
disk, a favorable geometry for jet outflow develops. Energy is
dissipated in the disk around a black hole by neutron annihilation
or MHD processes such as Blandford-Znajek (BZ) process, which can
power polar outflows, relativistically expanding bubbles of
radiation, and pairs and baryons focused by density and pressure
gradients into jets.

McFadyen \& Woosley (1999) showed that for a reasonable but
optimistic values, the total neutrino energy emitted by the disk
during 20s of accretion process is $3\times10^{53}$ ergs, and the
total energy deposited by neutrino annihilation was
$1.4\times10^{52}$ ergs. For less optimistic assumptions regarding,
the initial Kerr parameter and the neutrino transport, the emitted
energy was as low as $1.4\times10^{53}$ ergs and the deposited
energy $\leq1\times10^{51}$ ergs. But these numbers are obtained for
the particular model considered by the authors. A typical
factor-of-10 uncertainty can be due to the sensitivity of the model
to the accretion rate and to the Kerr parameter characterizing the
initial angular momentum of the black hole, but also to the
uncertainty in the neutrino efficiencies. The neutrino annihilation
efficiencies will be discussed in \S2.7. more detailed. Anyway,
after a fraction of a second, the energy to mass ration in these
jets became very large $\leq10^{22}$ ergs s$^{-1}$, corresponding to
a large asymptotic relativistic Lorentz factor. The problem of
"baryonic contamination" is circumvented because the energy
deposition blows a bubble of low density. Momentum and energy from
the annihilating neutrinos continues to be deposited in this bubble
even as the baryon fraction becomes small. The pressure gradient in
the bubble has a component pointing away from the polar axis that
tends to exclude gas from the polar region. This energy is naturally
directed outward along the axis. However, lacking a full special
relativistic calculation of the entire event, the jet propagation
cannot be accurately determined in MacFadyen \& Woosley (1999).

The propagation of relativistic jet through the collapsing rotating
massive star were studied by Aloy et al. (2000), Zhang et al. (2003,
2004). They did not discuss how jet form during the collapse
process, but focus on its propagation. Aloy et al. (1999) have
recalculated the 14 $M_{\odot}$ collapsar model of MacFadyen \&
Woosley, using a 3-D and fully relativistic code. The relativistic
jet forms as a consequence of an assumed energy deposition
($10^{50}-10^{51}$ ergs s$^{-1}$) within a $30^{\circ}$ cone around
the rotation axis. The jet flow is strongly beamed (few degrees)
spatially inhomogeneous and time-dependent. The jet is able to reach
the surface of the stellar progenitor (typically $R\sim
3\times10^{10}$ cm) intact. At breakout, the Lorentz factor of the
jet reaches $\Gamma\sim 33$. After breakout the jet accelerates into
the circumstellar medium, whose density is assumed to decrease
exponentially and then become constant, $\rho_{\rm ext}=10^{-5}$ g
cm$^{-3}$. Outside the star, the flow begins to expand laterally
also ($v\sim c$), but the beam remains very well collimated. At a
distance of 2.54 $R_{*}$, where the simulation ends, about 2s after
shock breakout, the Lorentz factor has increased to $\Gamma$=44 in
the core of the jet which is now highly collimated ($\sim
1^{\circ}$). At that time the jet has reached $7.5\times10^{10}$ cm,
which is $10^{2}$ to $10^{4}$ smaller than the distance at which the
fireball becomes optically thin. and the values of Lorentz factor
$\Gamma$ are still far from those required for the fireball model.

\begin{figure}
\centering\resizebox{1.0\textwidth}{!} {\includegraphics{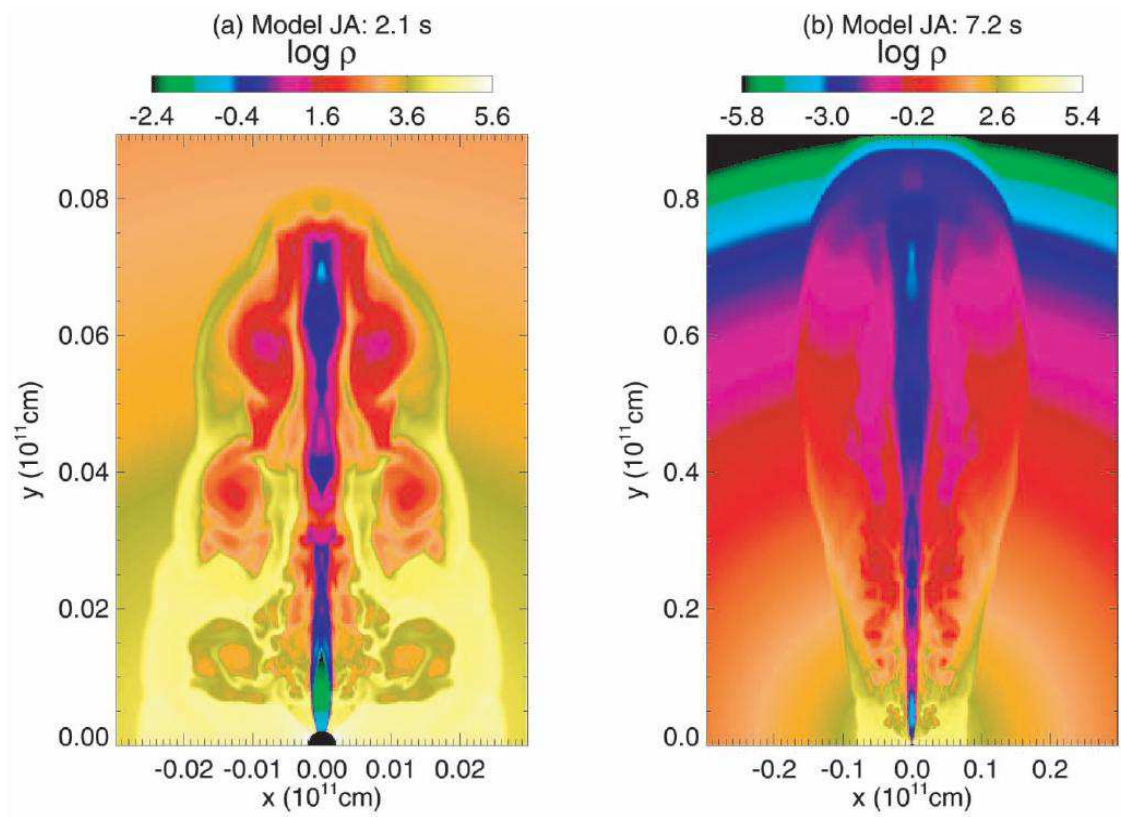}}
\caption{Density structure in the local rest frame for model JA at
(a) t=2.1 s and (b) 7.2 s. In (a), only the central region of the
star is shown. The radius of the star is $8\times10^{10}$ cm (Zhang,
Woosley, MacFadyen 2003).}\label{fig1331}
\end{figure}

\begin{figure}
\centering\resizebox{1.0\textwidth}{!} {\includegraphics{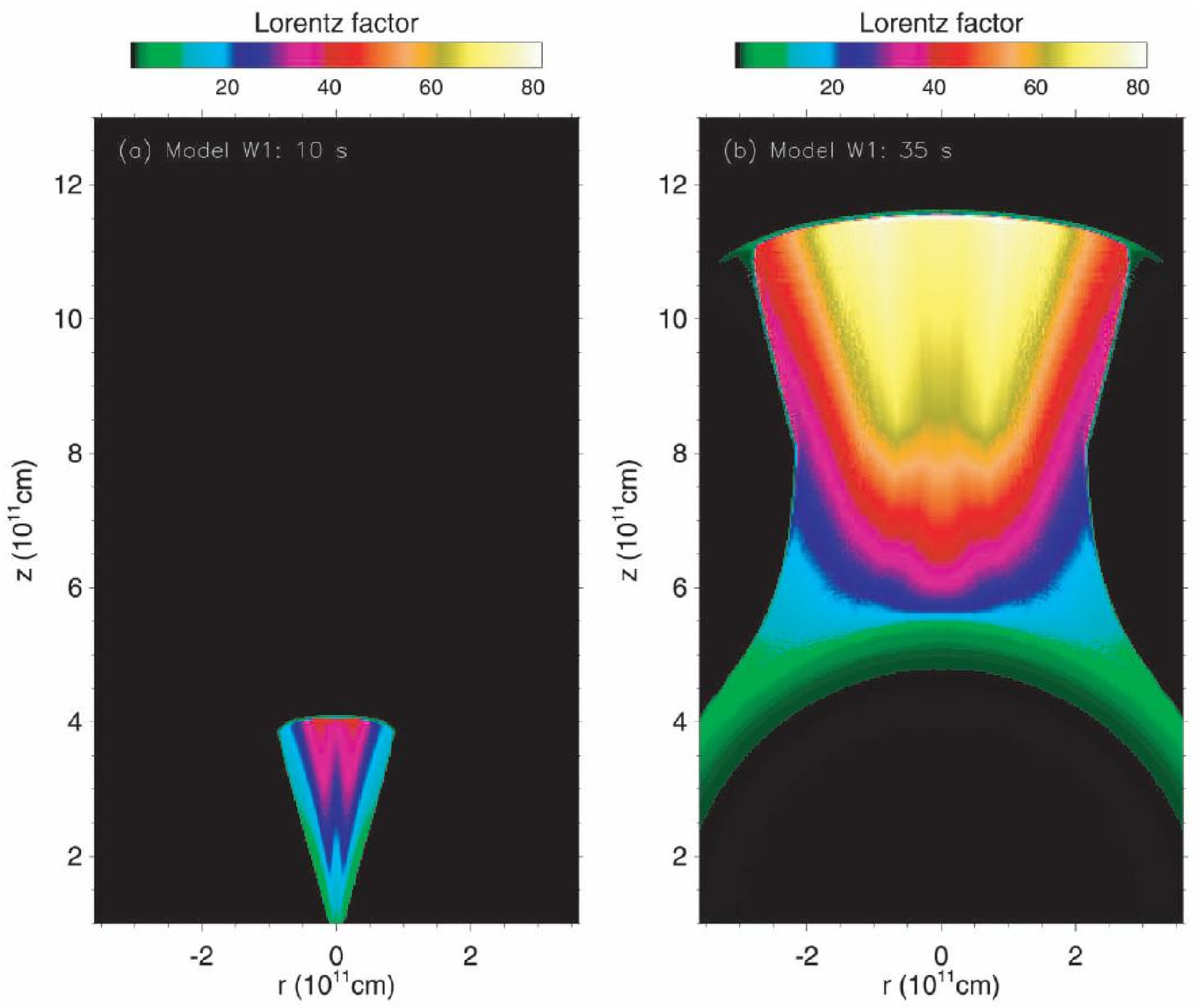}}
\caption{Bulk Lorentz factor for model W1 at (a) t=10 s and (b) t=35
s after the jets start to propagate in the stellar wind. The initial
opening angle is 3$^{\circ}$. At t=35 s, the opening angle is
$\sim15^{\circ}$. Because the power and Lorentz factor decrease
gradually after 10 s, the tail of the jet has much more lateral
expansion (Zhang, Woosley, MacFadyen 2003).}\label{fig1332}
\end{figure}

Zhang, Woosley \& MacFadyen (2003) and Zhang, Woosley \& Heger
(2004) go farther in the jet-stellar envelope interaction
simulations. While the simulation stops when the bulk Lorentz factor
$\Gamma$ reaches 44 in the work of Aloy et al. (2000), Zhang,
Woosley \& MacFadyen (2003) picked up the calculation when the jet,
which presumably began in a region $\sim 30$ km in size, has already
reaches a radius of 2000 km and do not consider what has gone on
inside. Each jet is parameterized by a radius where it is introduced
and by its initial Lorentz factor, opening angle, power, and
internal energy. In agreement with Aloy et al. (2000), Zhang,
Woosley \& MacFadyen (2003) found that relativistic jets are
collimated by their passage through the stellar mantle. Starting
with an initial half-angle of up to $20^{\circ}$, they emerge with
half-angles that, though variable with time, are around $5^{\circ}$.
Interaction of these jets with the star and their own cocoons also
causes mixing that sporadically decelerates the flow. The mixing
instabilities along the beams surface is chiefly responsible for the
variable Lorentz factor needed in the internal shock model and for
the complex light arrives in many GRBs. Moreover, the jet is shocked
deep inside the star following a brief period of adiabatic
expansion; and this shock converts most of the jet's kinetic energy
into internal energy. Eventually, the jet accelerates and breaks
free of the star with very large internal energy. Conversion of the
remaining internal energy gives terminal Lorentz factors along the
axis of $\sim 100-200$. Table \ref{tab133} list the jet models of
Zhang, Woosley \& MacFayden (2003) with different initial
parameters. Jet propagations is studied in two steps. inside the
star and in the stellar wind environment after breakout. Both steps
the jet is specified by six parameters as shown in Table
\ref{tab133}. Figure \ref{fig1331} and Figure \ref{fig1332}show the
selected calculation results (JA \& W1) of bulk Lorentz factor from
these models.
\begin{table}
\begin{center} {
\begin{tabular}{lcccc}
\hline
Model & $\dot{E}^{a}$ & $\theta_{0}^{b}$ & $\Gamma_{0}^{c}$ & $f_{0}^{d}$ \\
 & ($\times10^{51}$ ergs s$^{-1}$) & (deg) & &  \\
\hline
JA &  1.0 & 20 & 50 & 0.33 \\
JB &  1.0 & 5 & 50 & 0.33 \\
JC &  0.3 & 10 & 5 & 0.025 \\
\hline
W1 &  0.8 & 3 & 10 & 0.06 \\
W2 &  0.8 & 3 & 50 & 0.33 \\
\hline \hline
\end{tabular}}
\end{center}
\caption{Parameterized initial conditions for jet propagation inside
a star from radius $\approx 2000$ km (JA, JB, JC) and in the stellar
wind after breakout (W1, W2) from Zhang, Woosley \& MacFayden
(1999): $^{a}$ Energy deposition rate for each jet; $^{b}$ Initial
half-angle; $^{c}$ Initial Lorentz factor; $^{d}$ Initial ratio of
kinetic energy to total energy, which excludes the rest mass. They
did not consider the process of jet formation}\label{tab133}
\end{table}
Because of the large radio of internal to kinetic energy in both the
jet ($\geq80\%$) and its cocoon, the opening angle of the final jets
after breakout is significant greater than at breakout. Figure
\ref{fig1332} is an example of Lorentz factor at t=10 and t=35 after
the jets break out of the star and propagate in the star wind. A
small amount of material emerges at large angles, with a Lorentz
factor still sufficiently large to make a weak GRB. This suggests a
unified scenario which can explain various types of high-energy
transients from X-ray flashes to classic GRBs, depending on the
angle at which at standard collapsar is observed. Later Zhang,
Woosley \& Heger (2004) used new 2-D and 3-D simulations to further
studied the relativistic jet propagation and breakout in massive WR
stars and improved their former results. The highly relativistic
jets are ($3^{\circ}-5^{\circ}$, $\Gamma\geq 100$) is surrounded by
a cocoon with moderately relativistic ejecta ($\Gamma\sim 15$) that
expands and become visible at large polar angles ($\sim
10^{\circ}$). As a results, XRFs and GRBs should be different
expressions of the sam basic phenomenon from collapsars.

Another way to produced $\gamma$-ray and X-ray transients related
with cocoon was proposed by M\'{e}sz\'{a}ros \& Rees (2001) and
Ramirez-Ruiz, Celotti \& Rees (2002). Most of the energy output
during the period that the jet through the stellar core goes in a
cocoon of relativistic plasma surrounding the jet. This cocoon
material subsequently forms a bubble of magnetized plasma that takes
several hours to expand, subrelativistically, through the envelope
of a massive star. M\'{e}sz\'{a}ros \& Rees (2001) showed that the
shock waves and magnetic dissipation in the escaping bubble can
contribute a nonthermal UV/X-ray afterglow, and also excite Fe line
emission from thermal gas, in additional to the standard jet
deceleration power-law afterglow. Ramirez-Ruiz et al. (2002) also
discussed that a rebrightening caused by the cocoon photospheric
emission might appear with energy greater than $10^{50}$ ergs. If
the relativistic jet cames less energy and inertia than the cocoon
plasma itself, it will start to decelerate at a smaller radius than
the collimated cocoon fireball, and the afterglow would be dominated
by the emission of the cocoon material. On the other hand, if the
jet produced by the accretion maintains its energy for much longer
than it takes the jet head to reach the surface of the helium
envelope, the relativistic jet is likely to contain substantially
more energy than the off-axis cocoon material.

\begin{figure}
\centering\resizebox{0.6\textwidth}{!} {\includegraphics{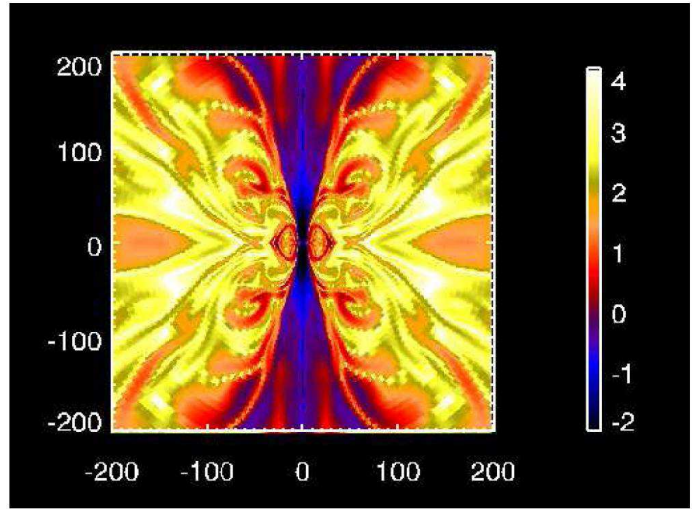}}
\caption{Contour of the plasma $\beta$ ($p_{gas}/p_{mag}$) at t =
1.773 s in logarithmic scale (Nagataki 2009).}\label{fig1333}
\end{figure}

Besides jet and cocoon's propagation and breakout, in Nagataki et
al. (2007) and Nagataki (2009), jet formation process in collapsars
is studied. Nagataki et al. (2007) investigated the dynamics of
collapsar using 2-D MHD simulations (Newtonian ZEUS-2D code) that
generates a jet, taking account of realistic equation of state
(contribution of electrons, positrons, radiation and ideal gas of
nuclei), neutrino cooling and heating processes, magnetic fields,
and gravitational force from the central black hole and
self-gravity. It is founded that neutrino heating processes
($\nu\bar{\nu}$ pair annihilation and $\nu_{e}\bar{\nu}_{e}$
captures on free nucleons) are not so efficient to launch a jet in
the Newtonian scenario. On the other hand, a jet could be launched
by $B_{\phi}$ fields that are amplified by the winding-up effect.
However, Tagataki et al. (2007) showed that the radio of total
energy relative to the rest mass energy in the jet at the final
stage of simulations suggest the bulk Lorentz factor of the jet
$\Gamma$ will not reach to as several hundreds, therefore the jet
will not be a GRB jet. Later Tagataki (2009) developed a 2-D general
relativistic MHD (GRMHD) code to further investigate the formation
of relativistic jets from a collapsar. He adopted a model from
Woosley \& Heger (2006), which corresponds to a 12 $M_{\odot}$ star
initially with 1\% of solar metallicity and relatively large iron
core of 1.82 $M_{\odot}$. Effects of magnetic fields are taken into
account with the vector potential $A_{\phi}\propto {\rm
max}(\rho/\rho_{\rm max}-0.2, 0)\textrm{sin}^{4}\theta$, where
$\rho_{\rm max}$ is the peak density in the progenitor. This
magnetic structure is initially introduced in Fishbone-Moncrief's
solution with magnetic fields (Gammie et al. 2003). On the other
hand, he did not consider the neutrino emission and heating
processes. Based on these considerations, it is shown that a jet is
launched from the center of the progenitor. Figure \ref{fig1333}
shows the contours of the plasma $\beta$($=p_{gas}/p_{mag}$) in
logarithms scale at $t=1.773$ s. The plasma $\beta$ is low in the
jet region but high in the accretion disk region. The jet is a
Poynting flux jet surrounded by a funnel-wall jet.
However, even at the final stage of the simulation with a time
evolution of 1.773 s, the bulk Lorentz factor of the jet is still
low, and the total energy of the jet is as small as $\sim 10^{48}$
ergs. Since the energy flux per unit rest-mass flux is $\sim 100$ at
the bottom of the jet, Tagataki (2009) suggested it is still
possible the jet Lorentz factor to be high enough when it propagates
outwards. Also, a long duration of the collapsar activity could
still make the total energy of jet large enough to $\sim
10^{50}-10^{51}$ ergs. Moreover, in this GRMHD simulations, jet is
launched mainly by the magnetic field amplified by the gravitational
collapse and differential rotation around the black hole, rather
than the Blandford-Znajek mechanism. Therefore, more detailed
simulations with a time evolution $\sim 10$ s should be operated for
determining the jet formation scenario. On the other hand, it is
interesting to describe the collapsar model as an entire consistent
scenario from the progenitor to the jet and cocoon breakout. What is
the role played by magnetic fields in angular momentum
transportation and jet formation and propagation? As Tagataki (2009)
studied Poynting flux jet rather than baryonic jet, it is still
interesting to simulate the propagation of such magnetized jets.

\subsection{GRB-SNe Association}

GRB-SN (XRF-SN) association greatly show the linking between the
GRBs and the collapse of massive stars in star formation regions.
Woosley \& Bloom (2006) reviewed the evidence for and theoretical
implications of this assocation detailed\footnote{However, their
review was written before the discovery of XRF 060218/SN 2006aj and
XRF 080109/SN 2008D.}. As GRB-SN association is one of the key
evidence to support the collapsar model, I also give a brief review
in this section.

There are at least five examples of GRB-SN or XRF-SN association had
been confirmed spectroscopically. Table \ref{tab134} gives the
properties of these GRB(XRF)-SN events. The first case of GRB-SN
association was GRB 980425, which has been associated with SN 1998
bw.
\begin{table}
\begin{center} {\scriptsize
\begin{tabular}{lcccccl}
\hline
GRB/SN & $z$ & E$_{\rm iso}$($\gamma$) & E$_{SN}$ &  M($^{56}$Ni) & SN type & References\\
 & & ($10^{49}$ ergs) & ($10^{52}$ ergs) & ($M_{\odot}$)& & \\
\hline
980425/1998bw &  0.0085 & 0.06-0.08 &  2-3& 0.5-0.7 & Ic-BL & Iwamoto et al. (1998)\\
030329/2003dh &  0.1685 & 1070 & 2-5 & 0.3-0.55 & Ic-BL & Stanek et al. (2003), Mazzali et al. (2003), etc \\
031203/2003lw &  0.1005 & 2.94$\pm$0.11 & 2-3 & 0.5-0.7 & Ibc-BL & Malesani et al. (2004), Gal-Yam et al. (2004), etc.\\
060218/2006aj &  0.0335 & 5.9$\pm$0.3 & 0.025 & 0.2$\pm$0.04 & Ic & Soderberg et al. (2006b), Mazzali et al. (2006)\\
080109/2008D  &  0.00649 & $\sim0.002$ & 0.3-0.6 & 0.05-0.1 & Ib & Soderberg et al. (2008), Mazzali et al. (2008)\\
\hline \hline
\end{tabular} }
\end{center}
\caption{Physical Properties of confirmed GRB-SNe association events
up to 2008.}\label{tab134}
\end{table}
GRB 980425 triggered detectors on broad both BeppoSAX and BATSE
(Kippen, 1998). At high energies, it was seemingly unremarkable
(Kippen 1998; Galama et al. 1998) with a typical soft spectrum
($E_{\rm peak}\approx 150$keV) and moderate duration ($\triangle
T\approx$23 s). GRB 980425 arose from the redshift $z=0.085$ galaxy,
and has $R_{\gamma, \rm iso}=8.5\pm0.1\times10^{47}$ ergs if
assuming isotropic emission. This energy is more than three orders
of magnitude fainter than the majority of long-duration GRBs. Any
collimation would imply an even smaller energy release in GRBs. The
evolution of SN 1998bw was unusually at all wavelengths. The
discovery of prompt radio emission just a few days after the GRB
(Kulkarni et al. 1998, see Figure ** in detail) was novel. Almost
irrespective of modeling assumptions, the rapid rise of radio
emission from SN 1998bw showed that the time of the SN explosion was
the same as the GRB to about one day. The brightness temperature
several days after the GRB suggested the radio photosphere moved
relativistically with $\Gamma\geq3$. The total supernova kinetic
energy is as large as $\sim 2-5\times10^{52}$ ergs more than ten
times the previous know energy of supernova. So the explosion is
called a 'hypernova'. Such a C+O star is the stripped core of a very
massive star that has lost both of its hydrogen and helium
envelopes.

The first truly solid evidence for a association between ordinary
GRBs and SNe came with the detection of the low-redshift
($z=0.1685$, Greiner et al. 2003) GRB 030329 and its associated
supernova, SN 2003dh. Shortly after its discovery (the brightest
burst HETE-2 ever saw), the afterglow of the GRB was very bright
($R\sim 13$mag). If faded slowly, undergoing several major
rebrightening events in the first few days (Burenin et al., 2003;
Greiner et al. 2003, etc.). Given the low redshift, several
spectroscopic campaigns were initiated. Spectra of the afterglow,
6.6 and 7.7 days after the GRB, showed a deviation from a pure
power-law and the mergence of broad SN spectral feature.
\begin{figure}
\centering\resizebox{0.6\textwidth}{!} {\includegraphics{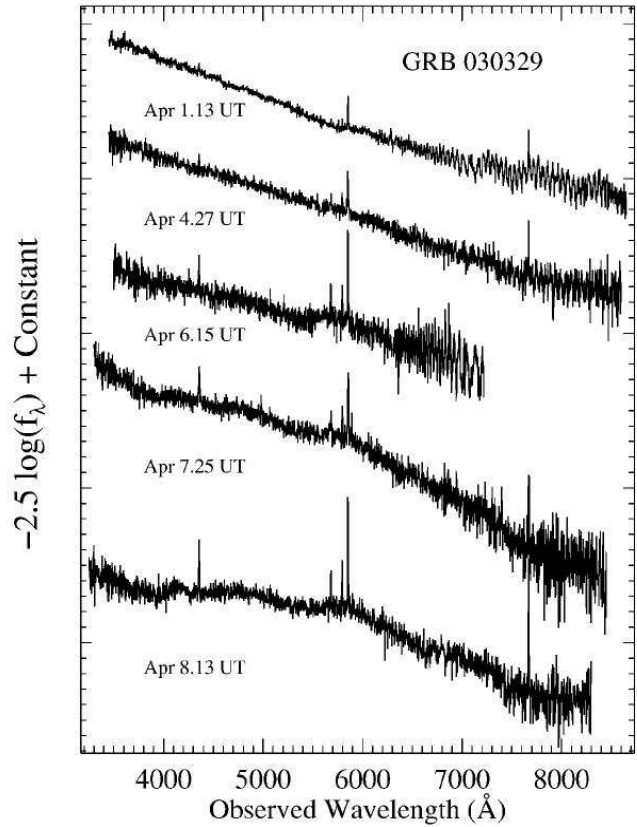}}
\caption{1.ˇŞEvolution of the GRB 03029/SN 2003 spectrum, from April
1.13 UT (2.64 days after the burst) to April 8.13 UT (9.64 days
after the burst). The early spectra consist of a power-law continuum
with narrow emission lines originating from H II regions in the host
galaxy at a $z=0.168$. Spectra taken after April 5 show the
development of broad peaks in the spectra characteristic of a
supernova (Stanek et al. 2003).}\label{fig1341}
\end{figure}
Figure \ref{fig1341} shows the evolution of the GRB 030329/Sn 2003dh
spectrum from Stanek et al. (2003). Spectrapolarimetric observations
at later times showed that the SN light was somewhat polarized
($P<1\%$) indicating mild asymmetry in the subrelativistic ejecta.
Given the broad spectral features, indicating high velocities ($\geq
250000$km s$^{-1}$; Stanek et al. 2003), and apparent absence of
hydrogen, helium and strong Si II $\lambda$6355 absorption, a
classification as Type Ic-BL was natural.

The association of XRF 080109 with SN 2008D is the first GRB-SN
event which shows a Type Ib supernova. The unabsorbed peak X-ray
luminosity of XRF 080109 (or called X-ray outburst by Soderberg et
al. 2008) is $L_{X,p}\approx6.1\times10^{43}$ergs s$^{-1}$ which the
fluence is $f_{X}\approx2\times10^{46}$ergs (Soderberg et al. 2008).
The peak luminosity of the XRF(XRO) is about 2 orders of magnitude
brighter than ULX outbursts, and about 3 orders of magnitude
brighter than Type I X-ray bursts, which additionally have blackbody
spectra. In comparison to other GRBs/XRFs, the value of $E_{X}$ is
about three orders of magnitude lower. Therefore, the properties of
XRF 080109 are distinct from those of all know X-ray transients.
These properties can be explained as the breakout of the supernova
shock through a dense wind surrounding the compact ($\sim 10^{11}$
cm) progenitor star. Moreover, SN 2008D also showed several unusual
features including (i) an early narrow optical peak (ii)
disappearance of the broad lines typical of SN Ic hypernovae and
(iii) development of helium lines as in SN Ib (Mazzali et al. 2008).
Detailed analysis shows that SN 2008D was not a normal supernova.
Its explosion energy $E\approx6\times10^{51}$ergs and ejected mass
$\sim 7M_{\odot}$ are intermediate between normal SNe Ibc and
hypernovae. SN 2008D is probably among the weakest explosions that
produce relativistic jets.

\begin{figure}
\centering\resizebox{0.6\textwidth}{!} {\includegraphics{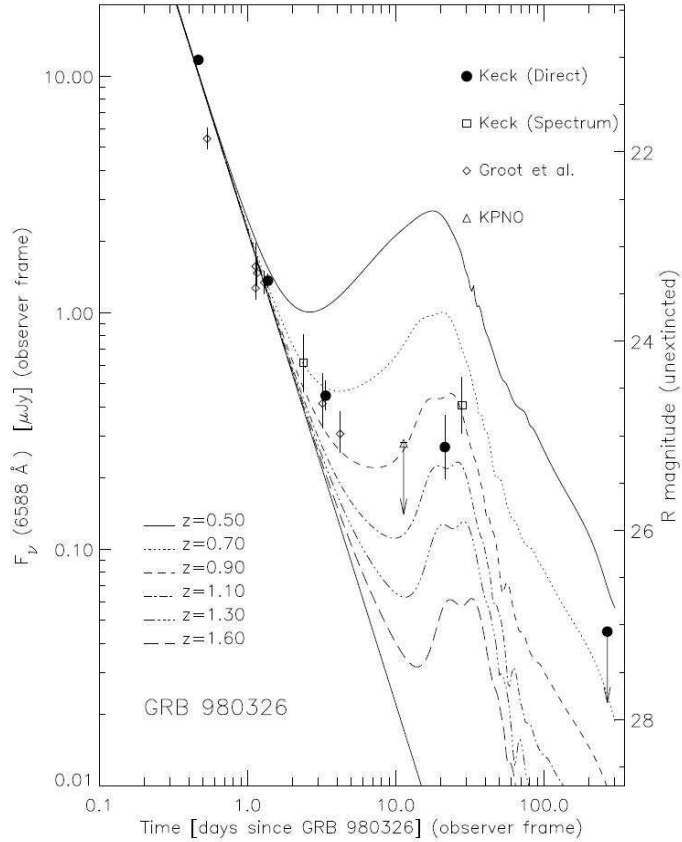}}
\caption{An "optical bump" on the afterglow of GRB 980326 (Bloom et
al. 1999). Models of SN 1998 bw at different redshifts are
shown.}\label{fig1342}
\end{figure}

Besides the above examples showed in Table \ref{tab134}, there are
other GRBs whose late-time deviations from the power-law decline
typically observed for optical afterglows have been seen in a number
of cases, and these bumps in the light curves have been interpreted
as evidence of supernovae. A report of a red emission "bump"
following GRB 980326 (Bloom \& Kulkarni, 1999), was interpreted as
due to a coincident SN at about a redshift of unity (see Figure
\ref{fig1342}).
Without a spectroscopic redshift for GRB 980326 and multi-band
photometry around the peak of the bump, the absolute peak brightness
and type of the purported SN could not be known. The available date
were also consistent with a dust echo (Esin \& Blandford, 2000), or
dust re-radiation (Waxman \& Draine, 2000) from material surrounding
the GRB. A subsequent reanalysis of the afterglow of GRB 970228
revealed evidence for a bump which appeared to be the same absolute
magnitude as SN 1998bw with similar rise time (Reichart, 1999;
Galama et al. 2000). Similar reports of bumps were made (Sahu et
al., 2000; Frucher et al., 2000; Bj\"{o}rnsson et al., 2001; Lazzati
et al., 2001; Castro-Tirado et al., 2001; Sokolov, 2001; Berger et
al., 2001; Gorosabel et al., 2005; Masetti et al., 2005), but mone
as significant and with as clear cut association to SNe as GRB
980326 and GRB 970228.

Concerted multi-epoch ground-based and space-based observing
campaigns following several GRBs strengthened the notation that
late-time bumps were indeed SNe. For example, the SN 0f GRB 011121
showed a spectral rollover during peak around 4000{\AA}, nominally
expected of core-collapse SNe in the photospheric phase. Garnavich
et al. (2003) showed evidence that the brightness and color
evolution resembled 1998S and Type IIn; on the other hand, Bloom et
al. (2002) showed consistency with a Ic-like curve interpolated
between the faint and fast 1994I and the bright and slow 1998bw. Zeh
et al. (2004) published an important photometric study of bumps in
GRBs, fitting 21 of the best sampled afterglows and finding evidence
of 9 bumps. Statistically significant evidence for bumps are found
in 4 GRB afterglows (990712, 991208, 011121, and 020405) while 5
have marginal significance (970228, 980703, 000911, 010921, and
021211). All of these GRBs had bumps claimed prior to the Zeh
analysis. Zeh et al. (2004) emphasize that all GRBs with $z\leq 0.7$
appear to have bumps, which of course, would be expected if all
long-duration GRBs have associated SNe. However, bump detection does
not necessarily imply a SN detection. There are, in fact, important
cases where multi-band photometry has curves of GRB990712 and XRF
030723, for example, do not appear consistent with a supernova.
Table \ref{tab1431} shows the properties of good candidate
supernovae associated with GRBs, XRFs and XRR (X-ray rich GRBs) up
to 2005.

\begin{table}
\begin{center} {\footnotesize
\begin{tabular}{lccccl}
\hline
GRB/SN & $z$ & Peak & $T^{a}_{\rm peak}$ &  SN likeness & References\\
 & & [mag] & [day] & designation)&  \\
\hline
XRF 020903 &  0.25 & $M_{V}=-18.6\pm0.5$ & $\sim15$ & Ic-BL & Soderberg et al. (2005) \\
GRB 011121/2001dk &  0.365 & $M_{V}$=-18.5 to -19.6 & 12-14 & I(IIn?) & Bloom et al. (2002), etc. \\
GRB 050525a &  0.606 & $M_{V}\approx-18.8$ & 12 & I & Della Valle et al. (2006) \\
GRB 021211/2002lt &  1.00 & $M_{V}=-18.4$ to -19.2 & $\sim14$ & Ic & Della Valle et al. (2004) \\
GRB 970228 & 0.695 & $M_{V}\sim-19.2$ & $\sim17$ & I & Galama et al. (2000) \\
XRR 041006 & 0.716 & $M_{V}=$-18.8 to -19.5 & 16-20 & I & Stanek et al. (2005), etc. \\
XRR 040924 & 0.859 & $M_{V}=-17.6$ & $\sim11$ & I & Soderberg et al. (2006a) \\
GRB 020405 & 0.695 & $M_{V}\sim-18.7$ & $\sim17$ & I & Price et al. (2003) \\
\hline \hline
\end{tabular} }
\end{center}
\caption{Properties of good candidate SNe associated with GRBs,
X-Ray Flashes, and X-Ray Rich GRBs, up to mid 2005 (Woosley \& Bloom
2006).}\label{tab1431}
\end{table}

The collapsar model suggested the supernova and the GRB derive their
energies from different sources. The supernova, and the $^{56}$Ni to
make it bright, are produced by a disk "wind" (MacFadyen \& Woosley,
1999; Kohri et al. 2005). This wind is subrelativistic with a speed
comparable to the escape velocity of the inner disk, or about 0.1c.
On the other hand, Nagataki et al. (2006) studied that $^{56}$Ni
produced in the jet of the collapsar is not sufficiently to explain
the observed amount of a hypernova when the duration of the
explosion is $\sim$10 s, which is considered to be the typical
timescale of explosion in the collapsar model. This result also
suggest that the sufficient $^{56}$Ni is provided from the accretion
disk outflow.

We should keep in mind there are at least some GRBs do not have
associated supernovae. Woosley \& Bloom (2006) discussed some
extrinsic biases which against detecting the associated SNe: for
example, poor localizations of bursts from their afterglows
dramatically hamper the ability for large aperture telescope to
discover emerging SNe; fainter SNe are expected from bursts that
occur near the line-of-sight through the Galaxy, or in especially
extinction-riddled regions of their host, which diminished the
sensitivity of the supernova observations; the host galaxies of GRBs
may still contaminate the light of GRB SNe at late times. However,
Fynbo et al. (2006) and Watson et al. (2007) discussed that no
supernova emission associated the two nearby long-duration bursts
060505 at $z$=0.089 and 060614 at $z$=0.125, otherwise the
associated supernovae should down to limits hundered of times
fainter than the archetypal SN 1998bw, and fainter than any type Ic
SN ever observed. Since the luminosity of a SN is roughly
proportional to the total amount of $^{56}$Ni produced in the
explosion it would mean that some GRBs produced very little
$^{56}$Ni ($<0.007M_{\odot}$). This would be explained as the case
for Type II collapsar (Fryer, Young, \& Hungerford 2006); or it is
possible that such GRBs without SNe are from low-mass, long-lived
progenitors which might be similar to those producing short GRBs.

\section{Progenitor II: Merges of Compact Stars}

\subsection{Binary Evolution and Mergers Rate}

Mergers of compact objects as the sources of short-hard GRBs have
been studied for years (Paczynski 1986, 1990; Eichler et al. 1989).
In the collapsar model, which is a good candidate for long-duration
GRBs, a spinning black hole with a accretion disk is formed. Mergers
of compact objects can also lead to the formation of a stellar-mass
black hole. The black hole would be unable to swallow the large
amount of angular momentum present in the binary system. The debris
from the tidally disrupted compact object forms a transient
accretion disk or tours, and ultimately falls into the black hole
and release a fraction of the gravitational binding energy. This
hyperaccreting disk scenario is similar to the central engine
suggested for long GRBs, and naturally explain the similarity
between these two phenomena.

Compact objects related to short-hard GRBs can be classified to
several types: neutron stars (NS-NS) mergers, black hole-neutron
star (BH-NS) mergers, black hole helium star (BH-He) mergers, black
hole white dwarf (BH-WD) mergers and merging white dwarfs binary
(WD-WD). The binaries formation scenario always begins with two
massive stars, which evolve to different types of binaries. Here we
provides various scenarios for forming NS/NS, BH/NS and BH/He based
on the review by Fryer, Woosley \& Hartmann (1999)\footnote{In my
paper I use NS-NS, BH-NS and BH-WD to mention the merger processes,
while NS/NS, BH/NS and BH/He to mention the binary systems before
mergers.}.

\textit{\textbf{NS/NS Scenarios}}
\begin{figure}
\centering\resizebox{1.0\textwidth}{!} {\includegraphics{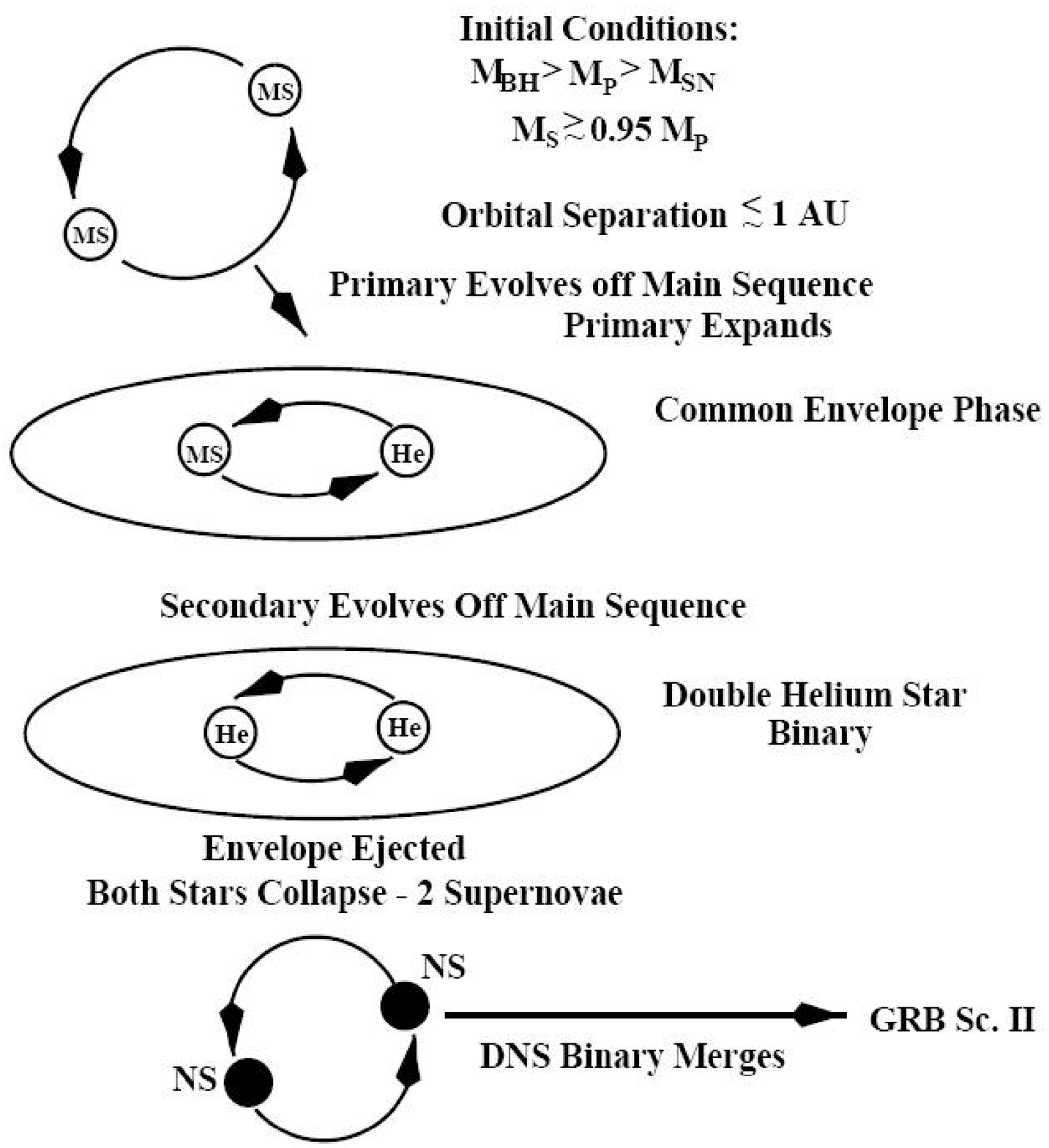}
\includegraphics{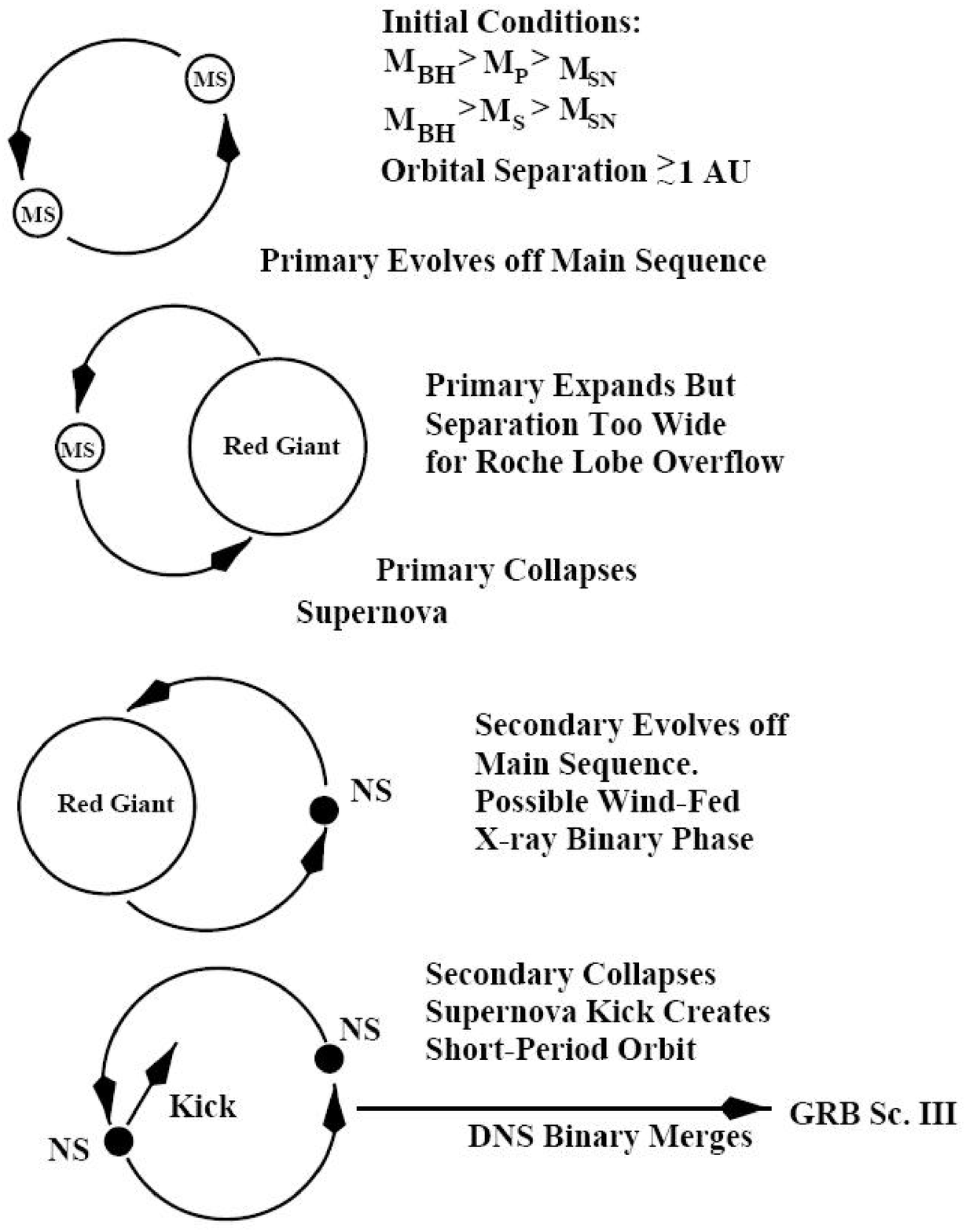}}
\caption{Two Scenarios NS/NS-b, NS/NS-c for neutron star binary
formation from Fryer, Woosley and Hartmann (1999). Scenario NS/NS-a,
which evolves through the common envelope phase, is unlike to form
NS/NS binary system (Bethe \& Brown 1998).}\label{fig141}
\end{figure}
\begin{itemize}
\item
\textit{Scenario NS/NS-a}: "Standard" double NS formation scenario.
The more massive star (primary) evolves off the main sequence,
overfills its Roche lobe, and transfers mass to its companions. The
primary then evolves to the end of its life, forms a NS in a SN
explosion. Next this system passes through an X-ray binary phase,
which then evolves through a common envelops phases in the standard
model. The neutron star spirals into the massive companion and forms
a NS/BH binary, which finally evolves to a close NS/NS binary, after
the explosion of the helium star.

However, farther calculations reveal that, during the common
envelope phase, the NS accretes at the Bondi-Hoyle rate and collapse
to form a black hole (Chevalier 1993, 1996; Brown 1995; Bethe \&
Brown 1998; etc.). As a result, this "standard" scenario for NS/Ns
systems may in fact form BH/NS binaries.

\item
\textit{Scenario NS/NS-b}: "Brown" mechanism for NS/NS binaries
(Brwon 1995). The initial binary system is comprised of two massive
stars of nearly equal mass. The secondary evolves off the main
sequence before the explosion of the primary as a supernova. The two
stars then enter a common envelope phase with two helium cores
orbiting within one combined hydrogen envelope. After the hydrogen
envelope is ejected, first one helium star, then the other,
explodes. if the kicks kicks imparted to the NSs during the
supernova explosion do not disrupt the binary, a NS/SN system can
form.

\item
\textit{Scenario NS/NS-c}: Kick scenario for double NS binaries. The
system avoids any common envelope phase but requires that the NS
formed in the explosion of the secondary receive a kick that places
it into an orbit that will allow the two NSs to merge within a
Hubble time. However, to aviod a common envelope phase, the
presupernova orbital separation must be $>1 AU$. The range of kick
magnitudes and directions that will produce a small orbit. This case
is probably rare compared to other scenarios.
\end{itemize}

Figure \ref{fig141} shows the scenarios to form NS/NS binary
systems.

\textit{\textbf{BH/NS Scenarios}}
\begin{itemize}
\item
\textit{Scenario BH/NS-a}: "Standard" BH/NS binary formation
scenario. The primary, with a mass greater than the limit for black
hole from fallback $M_{\rm BH}\approx25 M_{\odot}$, evolves off the
main sequence, overfill its Roche lobe, and transfers mass to its
companion. After the end of its life, the primary forms a BH, and
possibly, a supernova explosion. Then a binary consisting of a BH
and a massive star passes through an X-ray binary phase if this
system does not suffer from large kicks. Next the binary system
evolves through a common envelope, and the BH spiral into the
massive secondary and ejects the secondary's hydrogen envelope. The
BH/NS binary finally forms after the secondary's supernova
explosion.

\item
\textit{Scenario BH/NS-b}: This scenario is similar to the NS/NS-a
scenario. The neutron star gains too much matter during common
envelop phase and collapses into a BH. This formation scenario
produces a binary consisting of a low-mass BH ($\sim 3M_{\odot}$)
and a NS.

\item
\textit{Scenario BH/NS-c}: Kick scenario for BH/NS binaries. Just as
with the NS/NS system, a BH/NS binary can form under a third
mechanism, avoiding any common envelope evolution for appropriate
kick magnitudes and directions. However, as the NS/NS kick formation
scenario, the probability of this kick scenario is somewhat rare.
\end{itemize}

\textit{\textbf{BH/WD Scenarios}}

\begin{itemize}
\item
As WD/BH binary merge, the white dwarf is tidally disrupted and most
of its matter is converted into an accretion disk around the BH. The
accretion of this disk matter onto the BH may drive a GRB. The
energy from neutrino annihilation process is negligible for most
cases in WD-BH megers (Popham et al. 1999). The timescale of
accretion for disk formed in BH-WD mergers is in the range of
long-duration GRBs, but the rate of BH-WD merger is even lower than
NS-NS and BH-NS mergers. Therefore, BH/WD mergers are not likely to
be the leading causes of most GRBs. BH/WD binary may begin with a
binary consisting of a primary with mass great than $M_{\rm BH}$
mentioned above and a low-mass ($<10 M_{\odot}$) companion, or it
can also from neutron star system with low-mass ($<10 M_{\odot}$)
companions scenario.
\end{itemize}

Table \ref{tab141} provides a list of various rate estimates for
some of these possible short-hard GRBs progenitors based on the
review Lee \& Ramirez-Ruiz (2007), and compared them to the rate of
SNe events. However, there are many factors to cause the uncertain
of these rates, such as the supernova kicks, the mass ratio
distribution in binaries, mass limit for BH formation, stellar radii
and common envelope evolution and so on (e.g. Fryer et al. 1999;
Clark et al. 1979; Lipunov et al. 1987; Phinney 1991; Tutukov e\&
Yungelson 1993; Lipunov et al. 1995; Bloom et al. 1999; Belczynski
et al. 2001, 2002, 2006).

\begin{table}
\begin{center} {
\begin{tabular}{lllll}
\hline
Progenitor & Rate ($z=0$)\\
\hline
NS-NS   & 1-800\\
BH-NS  & 0.1-1000 \\
BH-WD  & 0.01-1000 \\
BH-He  & $\sim1000$\\
NS AIC  & 0.1-100\\
WD-WD & 3000 \\
SN Ib/c & 60000 \\
SN Ia &  150000 \\
SGRBs & 10(4$\pi/\Omega$)\\
\hline \hline
\end{tabular}
\caption{Estimated progenitors of SGRBs and their plausible rates in
yr$^{-1}$ Gpc$^{-3}$. The rates for these various occurrences are
plauged by a variety of uncertainties in many aspects. The observed
rates of Type (Ia+Ib/c) events given here provide a generous upper
limit (Lee \& Ramirez-Ruiz, 2007).\label{tab141}} }
\end{center}
\end{table}

The mergers rate and typical lifetime of NS/NS binaries are
estimated based on the observed system in our Galaxy or theoretical
population synthesis. The current detection of NS/NS binaries in our
Galaxy depends on at least one of the companions being a recycle
pulsar. The sample of observed NS/NS binaries is still small today.
We have only three NS/NS binaries examples: PRS B1913+16, PSR
B1934+12 and PSR J0737-3039. The first calculations of the Galactic
NS-NS merger rate, which were based on two systems (PRS B1913+16 \&
PSR B1934+12), showed the rate is between $10^{-6}$ to $10^{-5}$ per
yr per galaxy (Phinney 1991; Narayan, Piran \& Shemi 1991). The
discovery of the relativistic third binary pulsar PSR J0737-3039
makes the Galactic mergers rate increased to
$1.7\times10^{-5}-2.9\times10^{-4}$yr$^{-1}$ at 95\% confidence and
the local universe mergers rate to be 200-3000 Gpc$^{-3}$ yr$^{-1}$
(Kalogera et al. 2004). On the other hand, population synthesis
showed the NS-NS mergers rate span over two orders of magnitude,
which include the range inferred from observed Galactic binaries.
Bethe \& Brown (1998) suggested that the productions and mergers
rate of low-mass BH and NS should be an order of magnitudes greater
than NS-NS mergers. They carried a assumption that the semimajor
axes of binaries of heavy main sequence stars are distributed as
$da/a$, and that this distribution extends out to
$a=2\times10^{8}$km, or to orbital periods as long as 100 days.

The rate of BH-NS mergers could only be evaluated via population
synthesis. Bethe \& Brown (1998) concluded the BH-NS mergers rate is
$\sim 10^{-4}$ yr$^{-1}$ galaxy$^{-1}$, which was in a factor of
$\sim 10$ more merging than the commonly though before them
($\sim10^{-5}$ yr$^{-1}$ galaxy$^{-1}$). Evaluating the fraction of
binaries that survive both supernovae is important to an upper limit
on the local rate of compact objects mergers. However, this fraction
depends mostly on the distribution of the supernova kick velocity,
which is not well constrained. According to the review of short-hard
GRBs (SGRBs) by Nakar (2007), the upper limit on the local rate of
short-lived NS-NS and NS-BH binaries ($\tau\leq1$ Gyr) mergers is
$\sim$ 1000 Gpc$^{-3}$ yr$^{-1}$ (Cappellaro et al. 1999), and an
upper limit of $10^{4}$ Gpc$^{-3}$ yr$^{-1}$ on the local mergers
rate of long-lived binaries ($\tau\geq4$ Gyr) could be obtained by
relating it to the rate of core-collapse SNe at redshift 0.7 (Dahlen
et al. 2004). While the current observations point toward a lift
time of progenitors of SGRBs is dominated by several Gyr old, the
upper limit of the SGRBs must be $\sim 10^{4}$ Gpc$^{-3}$ yr$^{-1}$
if they are the results of mergers of compact binaries. On the other
hand, the observed BATSE local rate of SGRBs is $\sim10$ Gpc$^{-3}$
yr$^{-1}$ (Nakar, Gal-Yam \& Fox 2006), then the SGRBs rate due to
mergers must be in the range (Nakar 2007)
\begin{equation}
10\leq\Re_{SGRB=merger}\leq10^{4} \rm Gpc^{-3} yr^{-1},
\end{equation}
and the actual SGRBs local rate may be evaluated as (Nakar, Gal-Yam
\& Fox 2006)
\begin{equation}
\Re_{SGRB}\approx40f_{b}^{-1}\left(\frac{L_{\rm min}}{10^{49} \rm
ergs/s}\right)^{-1} \rm Gpc^{-3} yr^{-1},
\end{equation}
where $f_{b}^{-1}$ is the beaming factor (the fraction of the total
solid angle into which the prompt $\gamma$-rays are emitted) with
$1\ll f_{b}\leq100$, $L_{\rm min}$ is the lower cutoff of the
luminosity function ($\phi(L)\propto L^{-2}$, $L>L_{\rm min}$).

Several groups of researchers have estimated the progenitor lifetime
of SGRBs directly based on their host galaxy types, and showed that
their lifetime is about several Gyr. Zheng \& Ramirez-Ruiz (2007)
considered that the constraints of the SGRB progenitors lifetime can
be derived from separating the host galaxies into early and late
types. The convolution of the split star formation history and a
give lifetime distribution predicts the fraction of short-hard GRBs
in each host type. Zheng \& Ramirez-Ruiz (2007) suggested at least
half of the short GRB progenitors that can outburst within a Hubble
time have lifetimes greater than 7 Gyr. Gal-Yam et al. (2008)
analyzed new and archival observation of the fields of well-located
IPN SGRBs. They compared the distribution of SGRBs host types to
that of Type Ia SNe. SGRBs apparently occur in host galaxies of all
types, as do SNe Ia. However, compared to SNe Ia, SGRBs appear to
favor earlier type hosts. Even if the shorter values derived for the
typical Sn Ia delay time is adopted ($\sim1$ Gyr), the progenitors
appear to require a larger delay time, of order several Gry old.

On the other hand, the observed NS/NS systems in our Galaxy
represent a population with a typical lifetime of $\sim100$ Myr. A
binary with total mass 10$M_{1}M_{\odot}$ mass radio $Q_{M}$ and
period $P_{d}$ in a nearly circular orbit will merge in
\begin{equation}
\tau=1\times10^{9}P_{d}^{8/3}M_{1}^{-5/3}(1+Q_{M})(1+1/Q_{M})\rm yr
\end{equation}
Therefore for a merger to occur with a Hubble time, spiral-in most
have reduced to orbital period to $\leq1$ day. The merger rate
derived based on the observed Galactic sources is dominated by PSR
0737-3039 (Nakar 2007), which has a period of 2.4 hours and will
merge in about 85 Myr (Burgay et al. 2003). Therefore, this method
predict that the lifetime distribution of NS/NS binaries is
dominated by short-lived systems $\ll1$ Gyr. Since both the observed
SGRB sample and the Galactic NS.NS sample are small today, it is
somewhat early to make a determined conclusion about the lifetime of
short GRBs and Galactic NS/NS systems. However, if further
observations support the result that the observed Galactic binaries
are representative of the cosmological NS/NS population, and most
short GRBs are actual as old as several Gyr, then the candidates of
progenitors for most short GRBs should only be BH/NS rather than
NS/NS binaries.

\subsection{NS-NS Merger Simulations}

We followed Nakar (2006) and Lee \& Ramirez-Ruiz (2007) to review
the NS-NS and BH-NS merger simulations in \S 1.4.2 and \S 1.4.3.
Moreover, we also discuss the recent NS-NS and BH-NS simulation
results after 2007.

The numerical studies of merging binaries began in the 1980s and
1990s, with computations of the gravitational wave emission and
determinations of the stability and dynamics of the system at
separations comparable with the stellar radius under various
simplifying approximation (Oohara \& Nakamura 1989, 1990, 1992;
Nakamura \& Oohara 1989, 1991; Zhuge et al. 1994, 1996; Shibata et
al. 1992, 1993; Shibata 1997; Wilson et al. 1996; Marronett et al.
1998; Mathews et al. 2000; etc.). The thermodynamical evolution of
the fluid during mergers, with the potential for electromagnetic
energy release and GRBs was also considered by various groups
(Davies et al. 1994; Janka et al. 1996; Ruffert et al. 1996, 1997,
2001; Rosswog et al. 2003). These studies used Newtonian or
Post-Newtonian gravity with additional terms included in the
equation so motion to mimic gravitational radiation reaction by
various methods such as Lagrangian Smooth Particle Hydrodynamics
(SPH) and Eulerian Piecewise Parabolic Method (PPM).

For double NSs, synchronization during spiral-in is impossible
because the viscosity in neutron star matter is too small (Kochanek
1992) and so they merge with spin frequencies close to zero compared
with the orbital frequency (i.e., irrotational case). Contact occurs
at subsonic velocities and a shear larger then develops at the
interface. Modeling this numerically is extremely difficult, because
the larger is unstable at all wavelength s and vortices develop all
along it (Faber \& Rasio 2000), possibly amplifying the magnetic
field to extremely large values very quickly. The mass ratio in
double NS binaries is quite close to unity, so a fairly symmetrical
remnant would be a natural and expected outcome. However, in general
even small departures from unity can have important consequence in
the inner structure of the final object: the lighter star is
disturbed and spread over the surface of its massive companion,
which can remain largely undisturbed (Rasio \& Shapiro 1994; Rosswog
et al. 2000). The final configuration now consists of a
supra-massive neutron star surrounded again by a thick shock-heated
envelop and hot torus similar to that formed in BH-NS mergers, in
the range of 0.03-0.3$M_{\odot}$. The center of the remnant is
rapidly, and in many cases, differentially rotating, which can have
a profound impact on its subsequent evolution. A narrow funel along
the axis of symmetry is relatively baryonic free and may serve as a
potential site for launching a GRB jet (Ruffert \& Janka 1998,
1999).

\begin{figure}
\centering\resizebox{0.6\textwidth}{!} {\includegraphics{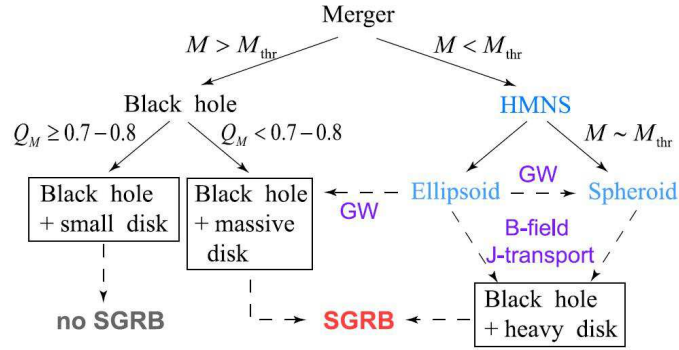}}
\caption{Summary about the outcome after the merger. "HMNS", "GW",
"B-field" and "J-transport" are hypermassive neutron star,
gravitational wave emission, magnetic field, and angular momentum
transport, respectively. "Small disk", "massive disk" and "heavy
disk" imply that the disk mass is $M_{\rm disk}\ll0.01M_{\odot}$,
$0.01M_{\odot}\leq M_{\rm disk}\leq0.03M_{\odot}$, and $M_{\rm
disk}\geq0.05M_{\odot}$, respectively (Shibata \& Taniguchi
2006).}\label{fig1421}
\end{figure}

In addition to Newtonian and Post-Newtonian simulations, General
Relativistic calculations have also been studied (Shibata \&
Ury\={u} 2000; Oechslin, Rosswog \& Thielemann 2002), and recently
progressed even further, going beyond the adoption of a simple
polytropic pressure-density relation (Shiabata et al. 2005, 2006;
Oechslin \& Janka 2006). The outcome is predictably complicated and
depends sensitively on the total mass of the system and the initial
mass ratio. For example, Shibata \& Taniguchi (2006) use a full
general simulations with a hybrid equation of state (EOS) that
mimics a realistic stiff EOS, for which the maximum allowed cold
mass of spherical neutron stars $M_{sph}$ is larger than
$2M_{\odot}$. ALso ,they focused on binary NSs of the ADM mass
(Arnowitt, Deser \& Misner 1962) $M\geq2.6M_{\odot}$. For an ADM
mass larger than a threshold mass $M_{thr}$, the merger results in
prompt formation of a black hole irrespective of the mass ratio
$Q_{M}$ with $0.7\leq Q_{M}\leq1$, but the disk mass steeply
increases with decreasing the value of $Q_{M}$ for given ADM mass
and EOS for $Q_{M}<0.7$. Here the value of $M_{ths}$ is about
$1.3\sim1.35 M_{sph}$, depending on EOS. For $M<M_{thr}$, the
outcome is a hypermassive neutron star (HMNS) of a large
ellipticity. Figure \ref{fig1421} is the results in Shibata \&
Taniguchi (2006). Kiuchi et al. (2009) recently studied longterm GR
simulation from NS/NS binary with a large initial orbital separation
to the formation of disk+BH system.

Besides studies on unmagnetized NS-NS mergers, full GR
magneto-hydrodynamics (GRMHD) simulations are indeed necessary to
model such NS-NS systems with strong magnetic fields. The first two
GRMHD codes capable of evolving the GRMHD equations in dynamical
spacetimes were developed by Duez et al. (2005) and Shibata \&
Sekiguchi (2005). Later they used these codes to study magnetic
fields in HMNSs, which are possible transient remnants of binary
NS-NS mergers (Duez et al. 2006a, 2006b; Shibata et al. 2006).
Secular angular momentum transport due to magnetic braking and the
magnetorotational instability results in the collapse of an HMNS to
a rotating black hole, accompanied by a gravitational wave burst.
The nascent black hole is surrounded by a hot, massive torus
undergoing quasistationary accretion and a collimated magnetic
field. Recently, Anderson et al. (2008) and Liu et al. (2008) have
sued GRMHD codes to study the magnetized NS-NS mergers directly.
Anderson et al. (2008) showed that the aligned magnetic fields delay
the full merger of the stars compared to the unmagnetized case. Liu
et al. (2008) showed that the effects of magnetic fields during and
short after the merger phase are significant but not dramatic. The
most important role of magnetic fields is long-term, secular
evolution of the newly formed disk+BH or HMNS system.

\subsection{BH-NS Merger Simulations}

\begin{figure}
\centering\resizebox{0.9\textwidth}{!} {\includegraphics{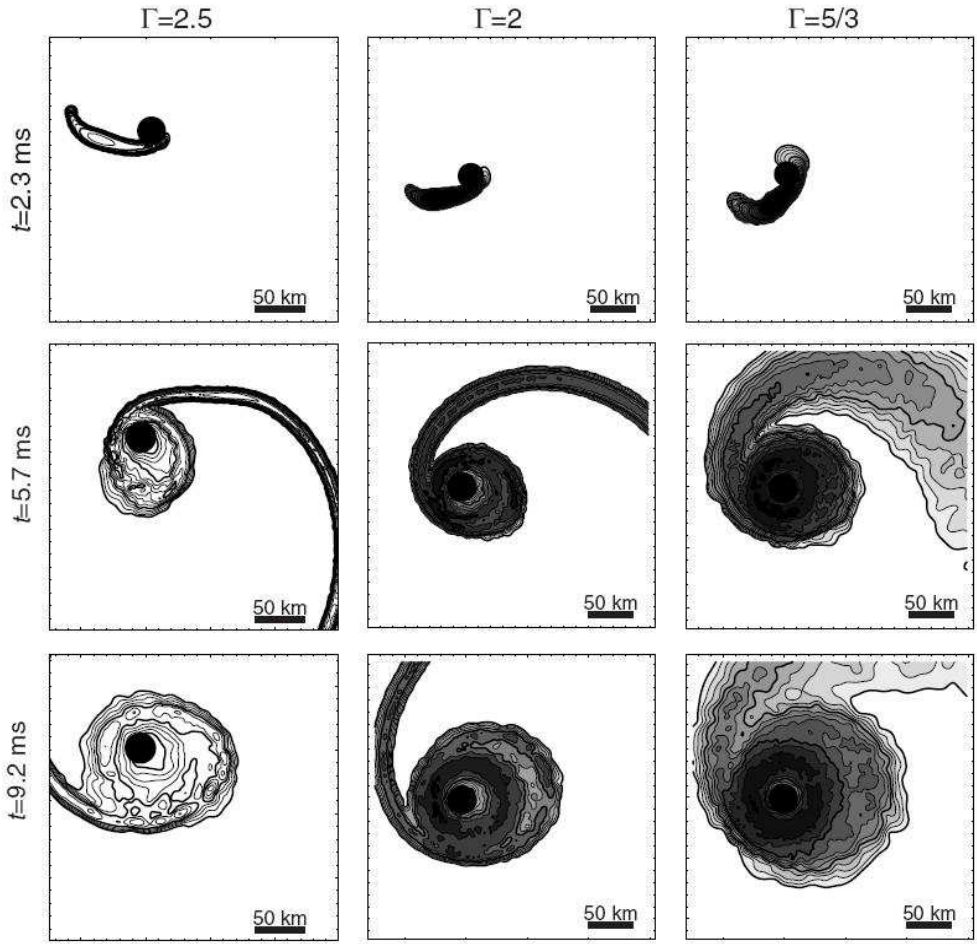}}
\caption{The tidal disruption of a NS by a BH in a binary (Lee \&
Ramirez-Ruiz, 2007) Each column depicts the interaction for a given
value of the adiabatic index $\Gamma$ in polytrope EOS. Logarithmic
density contours of density (with the lowest one at log ¦Ń(g
cm$^{-3}$) = 11), are shown in the orbital plane. All three cases
initially had identical mass ratios, $Q_{M}=0.31$.}\label{fig1431}
\end{figure}

Mergers of a black hole and a neutron star (BH-NS) is the second
class of mergers. Paczy\'{n}ski (1991) and Narayan et al. (1992)
also suggested that the merger of a neutron star with a preexisting
black hole of several solar masses might produce GRBs. In the
dominant formation scenario of binaries consisting of a BH and NS,
the BH is formed via hypercritical accretion during a common-envelop
phase (Bethe \& Brown 1998). The resulting BH has a very low mass
($\sim3M_{\odot}$) and a NS. Figure \ref{fig1431} shows the
Newtonian dynamical evolution of a merging BH-NS binary, during
which the NS (modeled as a polytrope) is tidally disrupted and an
accretion disk promptly forms (Lee \& Klu\'{z}niak 1999a, 1999b; Lee
2000, 2001).
The general conclusion of this body of work, which includes a mock
radiation reaction force to account for gravitational wave emission,
is that when the orbital separation is of the order of a few stellar
radius, the star becomes greatly deformed due to tidal effects, and
the system becomes dynamically unstable. for stellar component with
relatively stiff EOS ($\Gamma>5/2$), the instability is due to a
strong steepening of of the effective potential because of tidal
effects mass loss leads to an expansion and Roche lobe overflow,
further accelerating the process sand leading to a runaway in which
disruption occurs. All of the features anticipated in the work of
Lai et al. (1993, 1994) concerning the stability of the binary are
apparent, and are also seen in the Rasio \& Shapiro (1995). The main
additional result in terms of the dynamics is that configurations
that are stable in principle up to Roche lobe overflow are
de-stabilized by the mass transfer process itself, and fully merge
as well within a few orbital periods. It is also clear that in the
presence of gravitational radiation reaction, the system enters a
dynamical infall at small separations that no amount of mass
transfer can revert. Qualitatively, the result is largely
independent of the mass ratio, the assumed initial condition in
terms of NS spin, both tidally locked and irrotational cases, and
the assumed compressibility. Quantitatively, the details are
different, and are primarily reflected in the final disk mass and
the gravitational waves. Anyway, after a few initial orbital
periods, a BH of $M\sim3-5M_{\odot}$, surrounded by a thick and hot
debris disk (torus) with $M_{\rm disk}\sim0.01-0.1M_{\odot}$
approximately $4\times10^{7}$cm across is all that remains of the
initial couple (Lee \& Kluzniak 1999a, 1999b; Lee 2000, 2001; Duez
et al. 2002).

It is important to note that the debris disks are formed in only a
few dynamical timescale about a few millisecond, which is
practically instantaneous considering their later, more leisurely
evolution. The possibility that a fraction of the NS in a BH-NS
binary would survive the initial mass transfer episode and produce
cycles of accretion was explored at long timescales by Davies et al.
(2005). This was motivated by the fact that numerical simulations
employing relatively stiff equations of state at high binary mass
ratios Lee 2000; Rosswog et al. 2004) showed that this might
actually occur. However, this is most likely not the case, as the
required EOS was unrealistically stiff and gravity was essentially
computed in a Newtonian formalism. More recent calculations with
pseudo-Newtonian potentials and the use of General Relativity
consistently fail to reproduce this behavior for a range of
compressibleness in EOS (Rosswog 2005; Lee et al. 2005; Faber et al.
2006a, 2006b). Instead, the neutron star is promptly and fully
disrupted soon after the onset of mass transfer.

While Newtonian simulations seen to give a qualitatively correct
picture of the dynamics in at least some of the NS-NS mergers
scenarios, it is not clear at all that they are applicable also in
the case of BH-NS mergers. More importantly, it is also not even
clear how to estimate its effects with simple analytical
consideration. Computations of the location of the innermost stable
circular orbit (ISCO) in black hole systems indicate that tidal
disruption may be avoided completely, with the star plunging
directly beyond the horizon, essentially being accreted whole in a
matter of a millisecond (Miller 2005). This would preclude the
formation of a GRB lasting 10 to 100 times longer. But we stress
that, just as for stability considerations in the Newtonian case,
even if the location of the ISCO can be a useful guide in some
cases, it cannot accurately describe the dynamical behavior of the
system once mass transfer beings. In pseudo-Newtonian numerical
simulation (Rosswog 2005) and post-Newtonian (Prakash et al. 2004),
even when a near radial plunge is observed, the star is frequently
distorted enough by tidal forces that long tidal tails and disk-like
structures can form. The outcome is particularly sensitive to the
mass ratio $Q_{M}=N_{\rm NS}/M_{\rm BH}$. Moreover, the spin of the
BH is also very important, with rotating BHs favoring the creation
of disks. Dynamical calculations of BH-NS systems in
pseudo-Newtonian potential that mimic General Relativity effects
typically show that for mass ratios $Q_{M}\simeq0.25$ it is possible
to form a disk, although of lower mass than previously thought,
$M_{\rm disk}\approx10^{-2}M_{\odot}$. Faber et al. (2006a, 2006b)
presented their results for the dynamical merger phase of a BH-NS
binary for low mass ratio ($Q_{M}\simeq0.1$) using the conformal
flatness approximation for GR for a Schwarzschild BH. WHile a
polytropic relation with $\Gamma=2$ was assumed, the final outcome
is clearly dependent on the complex dynamics of mass transfer. Even
though at one point in the evolution most of the stellar material
lies within the analytically computed ISCO, tidal torques transfer
enough angular momentum to a large fraction of the fluid, producing
an accretion disk with $M_{\rm disk}\simeq0.1M_{\odot}$ by the end
of the simulation at $t\simeq70$ ms. Table \ref{tab143} gives a
summary of the disk properties for different progenitors based on
the calculations of various groups up to 2006 (Lee \& Ramirez-Ruiz
2007).

\begin{table}
\begin{center} {\footnotesize
\begin{tabular}{lllll}
\hline
Prog. & $M_{\rm disk}/M_{\odot}$ & Gravity, Method & EOS &  References\\
\hline
BH/NS &  0.1-0.3   & N, SPH &  Polytropes &  Lee \& Kluzniak (1999a, 1999b); Lee 2000, 2001\\
BH/NS &  0.03-0.04 & PW, SPH & Polytropes &  Lee et al. (2005) \\
BH/NS &  0.26-0.67 & N, Grid & LS &  Janka et al. (1999) \\
BH/NS &  0.001-0.1 & PW, SPH & Shen & Rosswog et al. (2004), Rosswog (2005)\\
BH/NS &  0.001-0.1 & GR, SPH & Polytropes & Faber et al. (2006a, 2006b)\\
\hline
NS/NS &  0.2-0.5  & N, SPH &  Polytropes &  Rasio \& Shapiro (1992, 1994) \\
NS/NS &  0.4 & N, SPH & Polytropes &  Davies et al. (1994) \\
NS/NS &  0.01-0.25 & N, Grid & LS &  Ruffert et al. (1996, 1997, 2001) \\
NS/NS &  0.25-0.55 & N, SPH & LS & Rosswog et al. (1999, 2000, 2002, 2003), etc.\\
NS/NS &  0.05-0.26 & GR, SPH & Shen & Oechislin \& Janka (2006)\\
NS/NS &  0.0001-0.01 & GR, Grid & APR & Shibata \& Taniguchi (2006)\\
\hline \hline
\end{tabular} }
\end{center}
\caption{Remnant disk masses in compact mergers of a variety of
equations of state: polytropes, Lattimer \& Swesty (LS), Shen et al.
(Shen); and various gravity methods: Newtonian (N), Paczynski \&
WIita (PW), General Relativity (GR). (from Lee \& Ramirez-Ruiz ,
2007)}\label{tab143}
\end{table}

Recently, more advanced full relativistic BH-NS simulations were
carried out by Shibata \& Ury\={u} (2007), Shibata \& Taniguchi
(2008), Duez et al. (2008), Etienne et al. (2009) and Shibata et al.
(2009). Shibata \& Taniguchi (2008) studied BH-NS binaries in a
quasi-circular orbit as the initial condition and adopted EOS with
$\Gamma=2$ and the irrotational velocity fields. They found the
resulting torus mass surrounding the BH depends significant on the
initial NS mass and radius: $M_{\rm disk}\leq0.05M_{\odot}$ for
$R_{\rm NS}=12$km and $M_{\rm disk}\sim0.15M_{\odot}$ for $R_{\rm
NS}=14.7$km. The total released energy is $\sim 10^{49}$ergs for NS
compactness $\leq0.145$. Etienne et al. (2009) explained the effects
of BH spins aligned and antialigned with the orbital angular
momentum. Only a small disk $M_{\rm disk}<0.01M_{\odot}$ forms for
the antialigned spin core  $a/M_{\rm BH}=-0.5$ and for the most
extreme-mass -ratio case ($Q_{M}=0.5$). On the other hand, a massive
hot disk $M_{disk}\approx0.7M_{\odot}$ forms in the rapidly spinning
aligned ($a/M_{BH}=0.75$). Such a disk could drive a short-hard GRB.

\subsection{WD-WD Merger Simulations}

WD-WD merger scenario (the double-degenerate scenario) has been
widely studied as a possible candidate for Type Ia supernova
progenitor since Iben \& Tutukov (1984) and Webbink (1984). On the
other hand, it is not likely that WD-WD mergers could directly lead
to GRB phenomena. However, researchers have showed that WD-WD
mergers could also form a neutron star remnant in some cases (Saio
\& Nomoto 1985, 1998; Kawai, Saio \& Nomoto 1987; Mochkovitch \&
Livio 1990; Timmes, Woosley \& Taam 1994; Mochkovitch, Guerrero \&
Segretain 1997; King, Pringle \& Wickramasinghe 2001). The newly
formed NS, particularly the magnetar or highly magnetized NS, could
be the central engine of GRBs (Levan et al. 2006; Chapman et al.
2006). We will discuss the magnetar and neutron star models in \S
1.5.1.

\begin{figure}
\centering\resizebox{1.0\textwidth}{!} {\includegraphics{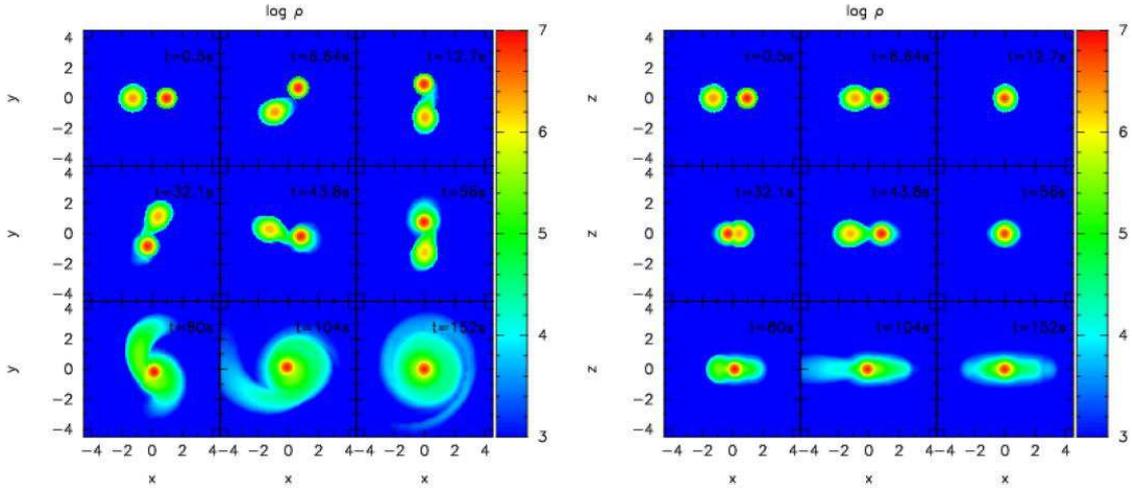}}
\caption{Temporal evolution of the density for the coalescence of
the $0.6+0.8M_{\odot}$ double WD binary system. The positions of the
particles have been projected onto the $xy$ plane (left panels) and
in the $xz$ plane (right panels). The disk and central object
evolution should be carried out in future study (Lor\'{e}n-Aguilar
et al. 2009).}\label{fig1441}
\end{figure}

The process of WD-WD merger itself was first studied by Mochkovitch
\& Livio (1989, 1990), using an approximate method. The full SPH
simulation have been carried out by Benz, Thielemann \& Hills
(1989), Benz, Cameron \& Bowers (1989), Benz, Hill \& Thielemann
(1989), Benz et al. (1990), Rasio \& Shapiro (1995), Segretain et
al. (1997), Guerrero et al. (2004); Lor\'{e}n-Aguilar et al. (2005)
Yoon et al. (2007), Shioya et al. (2007) and Lor\'{e}n-Aguilar et
al. (2009). For example, Benz, Hills \& Thielemann (1989) studied
the mergers of two WDs $0.6+0.6M_{\odot}$ and $0.7+0.9M_{\odot}$
respectively, using 5000 SPH particles. Segretain et al. (1997) used
$\sim6\times10^{4}$ SPH particles to simulate the coalescence of a
$0.9+0.6M_{\odot}$ system. Guerrero et al. (2004) studied a large
range of masses and chemical components of the merging binaries,
paying special attention to the issue of whether or not
thermonuclear runaway occurs during the process of merging.
Lor\'{e}n-Aguilar et al. (2004) studied gravitational wave (GW)
radiation from a $0.6+0.6M_{\odot}$ WDs binary, and recently
Lor\'{e}n-Aguilar et al. (2009) adopted a high-resolution SPH code
with an improved treatment of the artificial viscosity in this
algorithm to study the merging process again. Figure \ref{fig1441}
shows the temporal evolution of the density for the merger of the
$0.6+0.8M_{\odot}$ WD binary system.
The first step of WDs binary system evolution in these simulations
is more or less the same. The WDs are brought together by GW
radiation, until the less massive, and thus the less dense one fils
its Roche lobe and begins to transfer mass. However, depending on
the initial condition, mass transfer (accretion) could proceed
either stably or dynamically unstably, which lead to different
outcomes. If the mass transfer process is stable, mass will flow at
relatively low accretion rates with a period of order $\sim 10^{6}$
yr. The destruction of the accreting WD in a thermonuclear explosion
could produce a Type Ia supernova. On the contrary, unstable
accretion make the whole merging process to finish within a few
minutes. At such high accretion rate, the primary WD is believed to
collapse to form a NS, not to produce a SN Ia. The difference
between the two cases relies on the ability of the binary system to
return enough angular momentum back to the orbit. In fact, there are
two competing processes. On the one hand, the donor star is
supported by the pressure of degenerate electrons and, hence, it
will expand as it loses mass, thus enhancing the mass-transfer rate.
On the other, if orbital angular momentum is conserved the orbit
will expand as the donor star loses mass thus reducing the
mass-transfer rate. The precise trade-off between both physical
processes determines the stability of mass transfer.

Up to this date, almost all the WD-WD merger simulations only
adopted SPH code and showed that a compact central object surrounded
by a hot corona or disk as the merged configuration. More detailed
simulation should be carry out to calculate the evolution of the
disk with its angular momentum transfer and the final fate of the
central object have been done. And of course, GR or GRMHD methods
will instead SPH techniques for the future generation of WD-WD
merger simulations.

\subsection{Gravitational Waves from Merges}

\begin{figure}
\centering\resizebox{1.0\textwidth}{!} {\includegraphics{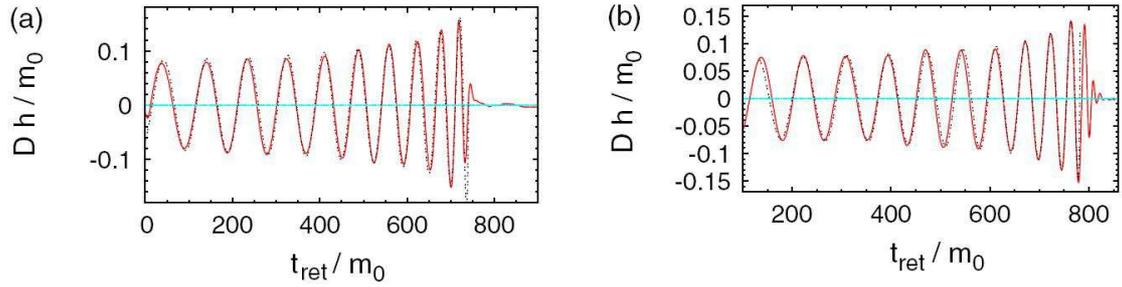}}
\caption{Gravitational waveforms observed along the z axis (solid
curve) for NS-BH merger models in Shibata et al. (2009). $m_{0}$ is
the total mass defined by $M_{\rm BH}$+$M_{\rm NS}$. $h$ denotes a
gravitational-wave amplitude and $D$ is a distance between the
source and an observer.}\label{fig1451}
\end{figure}

Gravitational radiation plays an observable role in the dynamics of
many known astronomical systems. Mergers of compact binaries (NS-NS,
BH-NS, BH-BH) are expected to be the first cosmological sources of
GW detected by ground-based observations. If short-hard GRBs are
generated by NS-NS or BH-NS mergers, then they are the
electromagnetic counterpart of the most accessible GW sources. In
principle the mergers are expected to have three different phases of
GW emission: the chirp inspiral signal, the merger signal and the
signal from the ringdown of the remaining BH. Figure \ref{fig1451}
shows an example of GW radiation from Shibata et al. (2009), who
studied NS with a polytropic $\Gamma=2$ and the irrotational
velocity field tidally disrupted and shallowed by the stellar-mass
BH.
\begin{figure}
\centering\resizebox{0.6\textwidth}{!} {\includegraphics{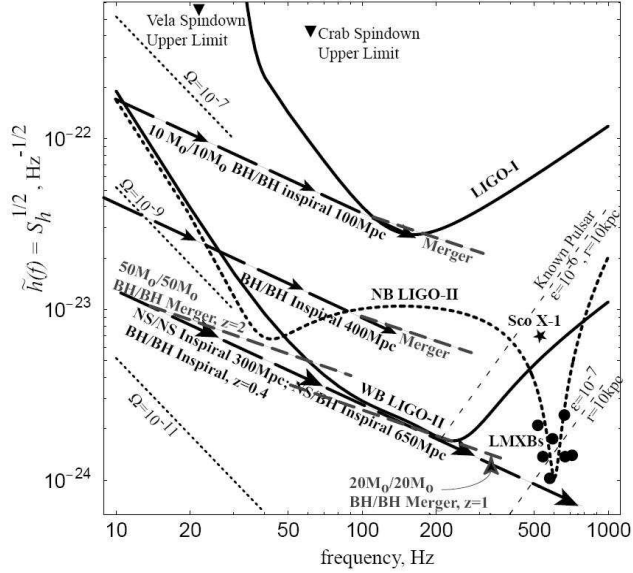}}
\caption{The noise $\tilde{h}(f)$ in several planned LIGO
interferometers plotted as a function of gravity-wave frequency $f$,
and compared with the estimated signal strengths $\tilde{h}_{s}(f)$
from various sources (Cutler \& Thorne 2002).}\label{fig1452}
\end{figure}
Figure \ref{fig1452} shows the three noise curves (LIGO-I,
"wide-band" LIGO-II and "narrow-band" LIGO-II), and the possible
signal from NS and BH binaries in the last few minutes of their
inspiral (Cutler \& Thorne 2002). At its current LIGO-I can detect a
NS-NS merger up to an average distance of about 15 Mpc, and a merger
of a NS with $10M_{\odot}$ BH up to about $30M_{\odot}$. Next
generation observatories, such as LIGO-II (advanced-LIGO), are
expected to be ten times more sensitive. LIGO-II would detect
mergers between NS binary up to $\sim300$ Mpc, and NS and 10
$M_{\odot}$ BH to 650 Mpc.

\section{Other Progenitor Models}

\subsection{Magnetized Neutron Star and Magnetar}

GRBs were once considered from pulsar-like activities in our Galaxy
before 1990s. The "cyclotron features" in the spectra  (20-40keV) of
several GRBs observed by KONUS and Ginga seemed to support the
conclusion that GRBs are produced by strongly magnetized
($\sim1.7\times10^{12}$ G) neutron stars with energy releasing rate
of $10^{31}$ ergs s$^{-1}$ (see \S 1.1 in detail). Later, however,
BATSE and BeppoSAX confirmed that GRBs actually come from
cosmological distances, which indicate that most Galactic pulsar
models were incorrect. New cosmological pulsar models involving much
more energetic pulsar-like activities ($\sim10^{50}$ ergs s$^{-1}$)
require the pulsar to be a millisecond-rotation-period NS with very
high surface magnetic fields up to $B_{s}>10^{14}$ G.

Neutron star might be formed via the accretion-induced collapse
(AIC) of WDs (Nomoto et al. 1979; Canal et al. 1980; Nomoto \& Kondo
1991; Usov 1992; Woosley \& Baron 1992; Thompson \& Duncan 1995;
Klu\'{z}niak and M. Ruderman 1998; Ruderman et al. 2000; Dessart et
al. 2006), the WD-WD binary merges (see \S 1.4.4) and NS-NS binary
mergers (see \S 1.4.3 and Figure \ref{tab1431}, or Dai et al. 2006;
Gao \& Fan 2006) as well as core collapse of massive stars.

\textit{\textbf{AIC of Magnetized WDs to NSs}}

\begin{figure}
\centering\resizebox{1.0\textwidth}{!} {\includegraphics{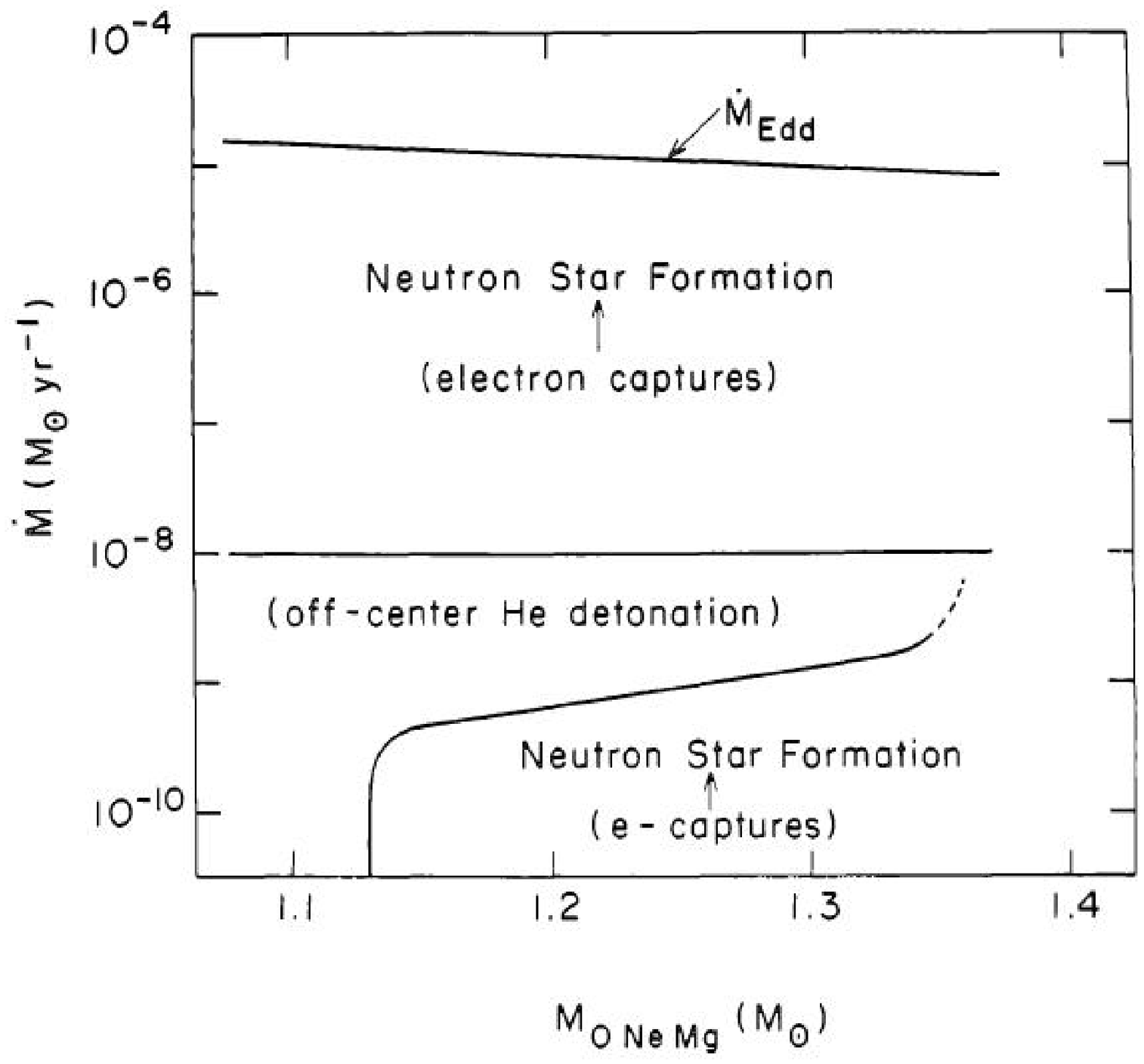}
\includegraphics{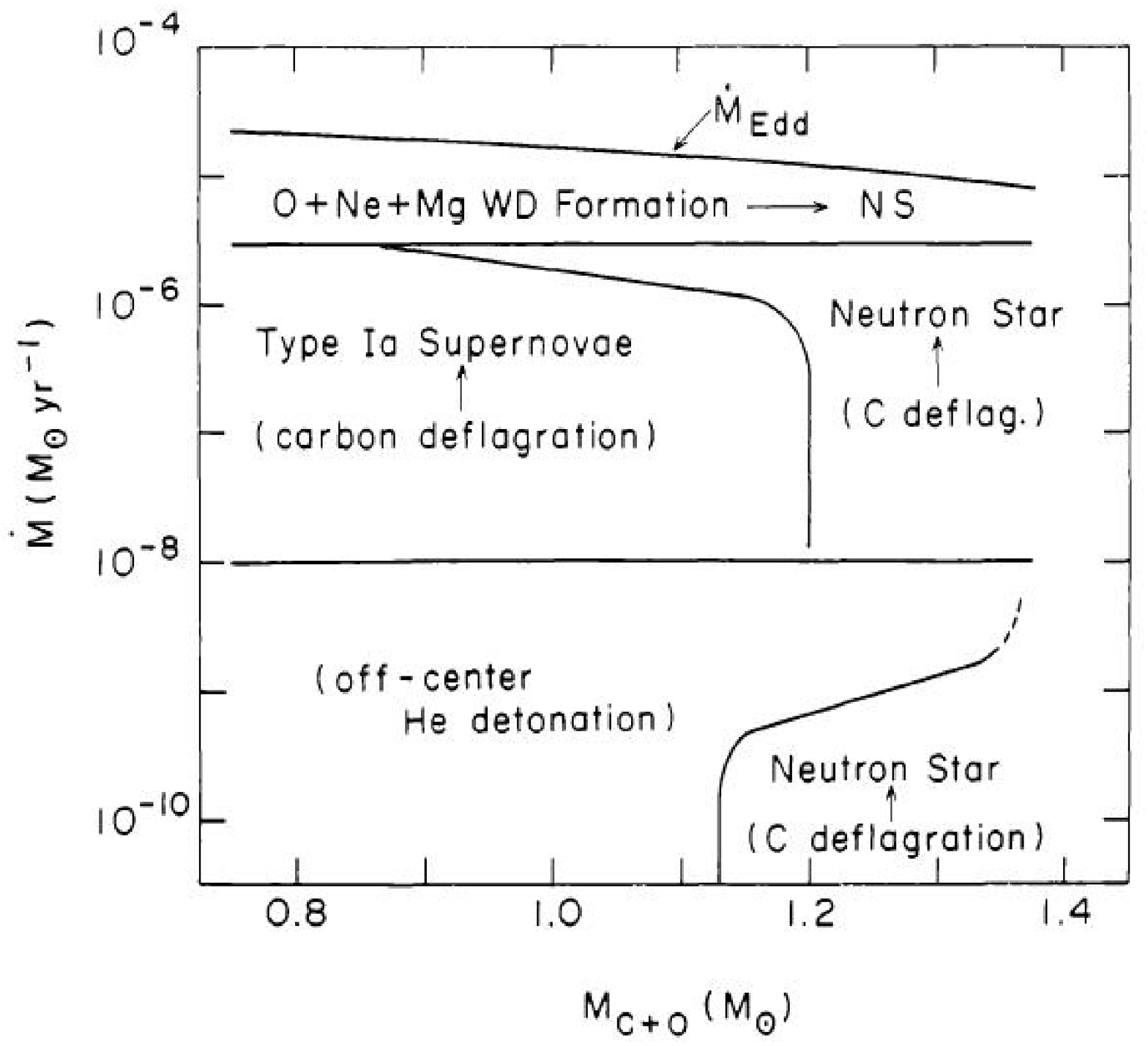}}
\caption{The final fate of accreting O+Me+Mg WD (\textit{left
panel}) and C+O WD (\textit{right panel}) expected for their initial
mass and accretion rate $\dot{M}$ (Nomoto \& Kondo
1991).}\label{fig1512}
\end{figure}

Some WDs in tight binaries can accrete enough mass from their
companions to initiate implosions as they approach the Chandrasekhar
limit. After such an implosion begins, there is a competition
between energy release from nuclear fusion reactions, which act to
explode the star, and the growing rate of electron captures, which
removes pressure support and accelerates collapse. The outcome of
such WD AIC is Type Ia SNe or NSs. Particularly, NSs could form when
the core evolves to sufficiently high density prior to carbon or
neon "ignition" (typically $\rho\geq10^{10}$g cm$^{-3}$), that
electron captures lead to an underpressure and consequent collapse
of the WD (Timmes \& Woosley 1992). Figure \ref{fig1512} shows the
final fate of accreting O+Ne+Mg and C+O WDs, which mainly depends on
the mass accretion rate. Also, the outcome for the case WD mass
$M\geq1.1M_{\odot}$ and that of $M\geq1.1M_{\odot}$ are different.
(Nomoto 1986, 1987; Nomoto \& Kondo 1991). Next we focus on the case
of successful NS formation by WD AIC.

\begin{figure}
\centering\resizebox{0.5\textwidth}{!} {\includegraphics{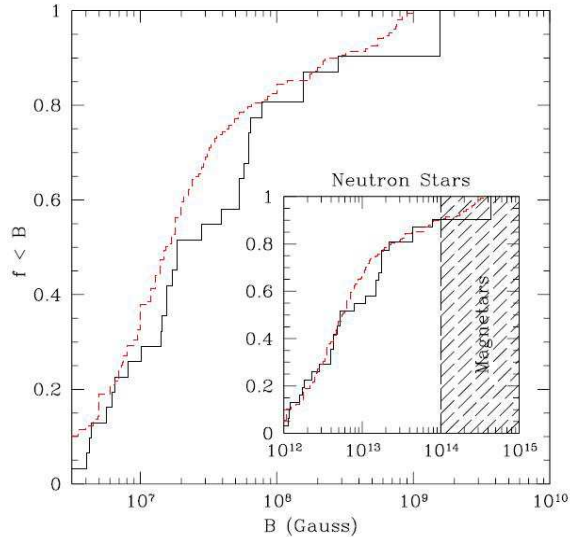}}
\caption{The distribution of WD magnetic fields seen in mCVs (black
line) and isolated WDs (red dashed) from Levan et al. (2006) and
Chapman et al. (2006).}\label{fig1513}
\end{figure}

Usov (1992) considered a WD with a surface magnetic fields $B_{s}$
to be a few times of $10^{9}$ G, which might exist, for instance, in
magnetic cataclysmic binaries. In the process of collapse, the
stellar radius changes from $R_{WD}\approx10^{9}$ cm for typical WD
to $R_{NS}\approx10^{6}$ cm for Ns, and the surface magnetic fields
on the newly formed NS can reach
${B}_{NS}=B_{WD}(R_{WD}/R_{NS})^{2}\sim 10^{15}$ G. Moreover, from
angular momentum conservation during the collapse process, the
expected angular velocity of the created NS can reach the maximum
possible value of NS $\Omega_{NS}\approx10^{4}$ s$^{-1}$ for the
angular velocity of WD $\Omega_{WD}\approx10^{-2}$ s$^{-1}$. As a
result, the highly magnetized WD could form a millisecond magnetar,
which would lose its rotational kinetic energy on a timescale of
second or less by electromagnetic or gravitational radiation. Levan
et al. (2006) discussed that the WDs with magnetic fields $10^{9}$ G
doa actual exist, although they are only in a small fraction
$(\sim10\%$) of all WDs (Figure \ref{fig1513}). Therefore, it is
plausible that the AIC of a WD could create a magnetar, whose
magnetic fields are organized mainly by collapse of the WD.
Klu\'{z}anik \& Ruderman (1998) and Ruderman et al. (2000), on the
other hand, requires a relatively much lower magnetic fields from
WDs $B_{WD}\approx10^{6}$ G and $B_{NS}\approx10^{12}$ G for a
millisecond-period NS formed by AIC of a WD. The magnetic fields can
farther amplified by differential rotating as well as by the AIC
process.

\textit{\textbf{WD-WD and NS-NS Binary Mergers}}

Strictly speaking, WD-WD merger to form a NS is another process of
AIC of a WD. As mentioned in \S 1.4.4., the less massive star in the
WD binary can be tidally disrupted and begin to transfer mass to the
primary. A high accretion rate induced by dynamical instability make
the accreting primary to become a accretion-induced NS. Levan et al.
(2006), Chapman et al. (2006), Chapman, Priddey \& Tanvir (2008,
2009) suggested that the scenario of WD-WD merger with a magnetar
remnant could be the progenitor of SGRs, and even be the progenitor
of a population of nearby short GRBs (Tanvir et al. 2005, Salvaterra
et al. 2008), which can be explained as the extra-Galactic SGR giant
flares (see also Nakar 2007).

NS-NS merger simulations show that HMNS could be formed in some
cases (see \S 1.4.3 and Figure \ref{fig1431}). Dai et al. (2006)
first suggested that, the new-born magnetic HMNSs, which is formed
by NS-NS binary mergers, could be a long-activity GRBs central
engine and produce the X-ray flares after the prompt emission (see
also Gao \& Fan 2006).

\textit{\textbf{Formation of Strong Magnetic Fields in NS}}

Besides AIC of WDs and compact object mergers, the more commonly way
to produce NSs is by core collapse of massive stars, which has been
studied for decades since Baade \& Zwicky (1934). The problem for
GRB progenitors in magnetar mor magnetized NS scenario is that, how
could the new-born NS to generate a strong magnetic fields, and
extract the rotating energy of NS via magnetic activities to produce
a GRB? (Another way to produce a GRB is via thermal hyperaccreting
onto the NS, as studied by Zhang \& Dai 2008, 2009a and Metzger,
Quataert \& Thompson 2008. However, the NS with its magnetic fields
is still considered the central engine to produce the prompt
extension emission or the early afterglow such as X-ray flares.)

The magnetic fields in a magnetar can be as strong as
$10^{15}-10^{16}$ G. Such strong fields are generated via three
processes: collapse of highly magnetized WDs, amplification from
differentially rotation and Paker instability inside the NS, and
convective dynamo action during a period of gradient-driven
convection after NS formation. Collapse-induced magnetic fields, as
mentioned before, satisfied the relation
$B_{NS}=B_{WD}(R_{WD}/R_{NS})^{2}$. In this AIC scenario, strong
magnetic fields is generated as soon as the NS forms. Differential
rotation and convection, on the other hand, could amplify the
magnetic fields by extracting the rotating energy to electromagnetic
energy for a period of time after NS formation (i.e., post-collapse
evolution of magnetic fields). Differential rotation may be caused
for two reasons (Thompson \& Duncan 1993; Ruderman et al. 2000).
First, NSs are less centrally condensed than the degenerate electron
pressure-supported cores out of which they form. For a polytropic
EOS, the central density $\rho_{c}$ of a WD is
$\rho_{c}(WD)\approx55\bar{\rho}_{WD}$, where $\bar{\rho}_{WD}$ is
the initial average WD density. The central density of a NS
satisfies $\rho_{c}(NS)\approx5\bar{\rho}_{NS}$ (Shapiro \&
Teukolsky 1983). Second, the angular momentum of various mass shells
are conserved to a first approximation in the collapse. As a result,
the central regions of the newly born NS should initially be
spinning much less rapidly than most of the mass in that star by a a
factor of about 0.2 (Ruderman et al. 2000).
\begin{figure}
\centering\resizebox{0.5\textwidth}{!} {\includegraphics{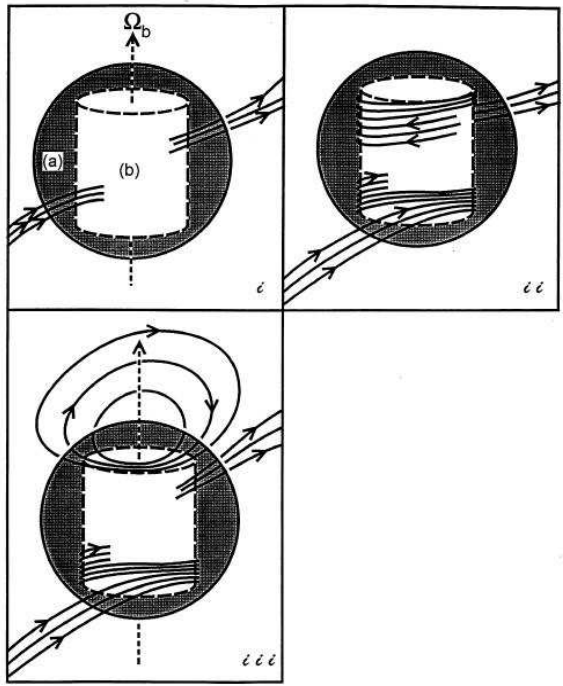}}
\caption{Magnetic field evolution in a differential rotation NS for
three stages (Klu\'{z}anik \& Ruderman 1998): (i) poloidal field
passing through the NS; (ii) toroidal field amplification; (iii)
effect of packer buoyancy instability. See \S 1.5.1 for detailed
discussion.}\label{fig1514}
\end{figure}
On the other hand, according to Thompson \& Duncan (1993),
convection in the NS may be driven by a radial gradient in the
lepton number per baryon (Epstein 1979) or by an entropy gradient
with the convective velocity $\sim1.3\times10^{8}$ cm$^{-1}$, after
the NS neutrinosphere has shrunk to a radius of $10-20$ km. NS
convection is a transient phenomenon and has an extremely high
magnetic Reynold number $\sim10^{17}$. The convective motions are
only mildly turbulent on scales larger than the $\sim10^{2}$ cm
neutrino mean free path, but the turbulence is well developed on
smaller scales. If the ratio of mean free path, but the turbulence
is well developed on smaller scale. If the ratio of mean magnetic
pressure to turbulent pressure equals that observed in the upper
convection zone of the Sun, and the initial rotation period exceeds
$\sim30$ ms, the energy for field amplification comes mostly from
convection, not different rotation, and fields as strong as
$3\times10^{15}$ G can be generated by convection during the first
few seconds after NS formation. On the other hand, large-scale
$\alpha-\Omega$ dynamo action is possible NS $P\leq30$ ms to
generate strong magnetic fields. For NSs with rapid rotation period
$P\sim 1$ ms, the fields generated by differential rotation may be
strong enough to suppress convection in some regions, the wrapping
of fields around the star by the shear motion allows the formation
of large scale magnetic structure. Duncan \& Thompson (1992) argued
that the fields could reach to $3\times10^{17}(P/1\textrm{ms})^{-1}$
G as the differential rotation is smoothed by growing magnetic
stresses (see also Klu\'{z}niak \& Ruderman 1998). When the strength
of magnetic fields has been sufficiently amplified, the Parker
buoyancy tends to change the topology of the fields, drive flux
ropes across the surface of the star, where they reconnect, diffuse
and eject into the surround medium, and directly provide the energy
of GRBs (e.g. Duncan \& Thompson 1992; Thompson \& Duncan 1993;
Klu\'{z}anik \& Ruderman 1998; Dai \& Lu 1998b; Ruderman et al.
2000; Dai et al. 2000).

\subsection{Supranova Model}
Vietri \& Stellar (1998, 1999) presented a scenario (they called
"supranova model") for the formation of GRBs occuring when a
hypermassive neutron star (HMNS) lose so much angular momentum that
centrifugal support against self-gravity becomes impossible and the
implodes to a BH about 2-3 $M_{\odot}$ surrounded by a transient
disk $\sim 0.1M_{\odot}$. The HMNS could be formed directly in the
SN explosion of a core with too much mass and angular momentum to
end up in a normal NS (Vietri \& Stellar 1998), or by accreting
matter onto a normal NS from a close companion during a low-mass
binary phase (LMXB), while the transport of mass and angular
momentum from the secondary makes the NS to become a HMNS (Vietri \&
Stellar 1999). The maximum mass for a differentially rotating HMNS
(Cook, Shapiro \& Teukolsky 1994a, 1994b) can reach to
$\approx2-3.5M_{\odot}$. Then the HMNS lose angular momentum via the
usual magnetic dipole radiation or gravitational waves radiation.
Slowly the HMNS collapses into a BH. Since the neutrinos inside the
HMNS have short mean free paths, the implosion from HMNS to BH is
almost adiabatic. A mass of $\sim 0.1M_{\odot}$ in the equatorial
belt can easily reach centrifugal equilibrium and forms a accretion
tours or disk. The new formed BH+disk system is similar to the final
results of collapsar and compact object mergers evolution. However,
the mechanism of extraction of energy from the system in the
supranova scenario is mainly via the conversion of Poynting flux
into a magnetized relativistic wind, not the baryonic or neutrino
annihilation process.

The advantage of the supranova model is that, it keeps the
environment baryon-free. The internal shock scenario requires a
small baryon contamination $\sim10^{-4} M_{\odot}$ to produce a
relativistic jet with the bulk Lorentz factor $\Gamma$ to be several
hundred (Rees \& M\'{e}sz\'{a}ros 1992). The powerful wind from a
collapsar may still contains lots of baryons, but his is not the
case for the supranova model. The GRB jet produced via the collapse
of a HMNS does not have to punch a whole through the stellar
envelope, both because the former SN explosion has swept away the
medium surrounding the remnant, and because of the clean and silent
collapse of the HMNS once centrifugal support weakens critically.
The small baryon contamination is a natural consequence of the
supranova scenario, with typical values of $\sim10^{-5} M_{\odot}$
well below the $\sim10^{-4} M_{\odot}$ upper limit.

The main two difficulties of the supranova model are that, the delay
between supernova and the following GRB is quite long in this model;
and the uncertainty of the torus mass and the lacking of an envelope
may not produce long GRBs. Vietri \& Stellar (1998) considered the
major source of HMNS angular momentum loss is through the usual
magnetic dipole radiation. The timescale between a normal SN where a
HMNS formed and the implosion to a BH is
\begin{equation}
t_{\rm sd}=\frac{J}{\dot{J}}=10{\rm
yr}\frac{j}{0.6}\left(\frac{M}{3M_{\odot}}\right)^{2}\left(\frac{15{\rm
km}}{R_{eq}}\right)^{6}\left(\frac{10^{4}{\rm
s}^{-1}}{\omega}\right)^{4}\left(\frac{10^{12}{\rm
G}}{B}\right)^{2},
\end{equation}
where the mass in the range of $M\approx2-3.5M_{\odot}$, equatorial
radius $R_{eq}\approx11-18$ km, angular velocity $\omega\approx
8000-12,000$ s$^{-1}$, and the angular momentum $j\approx0.6-0.78$.
However, it is clear in the case of events like GRB 980425 and GRB
030329 that the supernova and the GRB happened nearly
simultaneously-within a few days of each other at most (e.g. Hjorth
et al. 2003). Therefore, the supranova model, with the long delay
betwen SNs and GRBs, could probably only explain few GRB events.
Moreover, the main uncertainty in this model is the mass of the
formed disk or torus, although Vietri \& Stellar estimate a disk
$M_{\rm disk}\sim 0.1M_{\odot}$. The mass of the remaining disk
depends strongly on the EOS and magnetic fields strength (e.g.,
Cook, Shapiro \& Teukolsky, 1994a, 1994b; Shibata \& Shapiro 2002;
Shibata 2003; Shapiro 2004; Duez et al. 2006a, 2006b, Stephens et
al. 2008). A HMNS with a polytropic and soft EOS $\gamma-4/3\ll1$, a
massive disk may form after its collapse into a BH. On the other
hand, Shibata (2003) showed that a HMNS with $\gamma>3/2$ leaves a
disk $M_{\rm disk}<10^{-3}M_{\odot}$. However, strong magnetic
fields with the MHD angular momentum transfer make a massive disk
towards the equator of a HMNS even with a stiff EOS $\Gamma=2$ (Duez
et al. 2006a, 2006b; Stephens et al. 2008). Anyway, without an
remaining stellar envelope contained around the HMNS, the new formed
BH+disk system could only produce short GRBs at most. From this
point of view, whether the suprannova scenario allows a lifetime
$\geq 1$ Gyr, as mentioned in \S 1.4, is still questionable.

\chapter{Black Hole Accretion Disks, Accretion Flows and GRB Central Engines}
\label{chap2} \setlength{\parindent}{2em}

Most GRB progenitor scenarios lead to a formation of a stellar-mass
black hole and a hyperaccretion disk around it. The accreting black
hole system is commonly considered as the direct central engine of
GRBs. In this chapter, we first discuss the equations and the
classical solutions of the accreting black hole systems, and then
focus on the hyperaccreting black-hole disks, which could only be
cooled via neutrino emissions.

\section{Conservation Equations}

\textit{\textbf{Basic Equations}}

The general mass conservation or continuity equation is
\begin{equation}
\frac{\partial\rho}{\partial t}+\nabla\cdot(\rho
\textbf{v})=0\label{mass1}.
\end{equation}
Moreover, the general momentum conservation equation with viscosity,
i.e., the \textit{Navier-Stokes} equation is (Landau \& Lifshitz,
1959)
\begin{equation}
\frac{\partial \textbf{v}}{\partial
t}+(\textbf{v}\cdot\nabla)\textbf{v}=-\frac{1}{\rho}\nabla
p-\nabla\Phi+\frac{\eta}{\rho}\triangle
\textbf{v}+(\xi+\frac{\eta}{3})\nabla(\nabla\cdot
\textbf{v})\label{momen1},
\end{equation}
where $\Phi$ is the gravitational potential, the constant $\eta$ and
$\xi$ are called coefficients of viscosity. For incompressible
fluid, the term $(\xi+\eta/3)\nabla(\nabla\cdot \textbf{v})$ in the
right hand of equation {\ref{momen1}} is vanished.

For accretion disks under axial symmetry assumption, we adopt
cylindrical coordinates ($r,\varphi,z$), and take
$\partial/\partial\varphi=0$. Three velocity components of accretion
flow at a radius $r$ from the central compact object
($v_{r},v_{\varphi},v_{z}$) can be divided into two parts: an
toroidal component $v_{\phi}=\Omega r$ with $\Omega$ as the disk
angular velocity, and a small poloidal component
$\textbf{v}_{p}=(v_{r},v_{z})$. If we also consider the disk is in a
steady state, i.e., $\partial/\partial t=0$. Then the mass
conservation equation can be written as
\begin{equation}
\frac{1}{r}\frac{\partial}{\partial r}(\rho
rv_{r})+\frac{\partial}{\partial z}(\rho v_{z})=0\label{mass2}.
\end{equation}
The three components of the momentum equation are (Frank, King \&
Raine 2002)
\begin{equation}
v_{r}\frac{\partial v_{r}}{\partial r}+v_{z}\frac{\partial
v_{r}}{\partial
z}=(\Omega^{2}-\Omega_{K}^{2})r-\frac{1}{\rho}\frac{\partial
p}{\partial
r}+\frac{1}{\rho}\left[\frac{\partial\sigma_{rr}}{\partial
r}+\frac{\sigma_{rr}-\sigma_{\varphi\varphi}}{r}+\frac{\partial\sigma
rz}{\partial z}\right]\label{momen21},
\end{equation}
\begin{equation}
v_{r}\frac{\partial v_{z}}{\partial r}+v_{z}\frac{\partial
v_{z}}{\partial z}=-\Omega_{K}^{2}z-\frac{1}{\rho}\frac{\partial
p}{\partial
z}+\frac{1}{\rho}\left[\frac{1}{r}\frac{\partial(r\sigma_{rz})}{\partial
r}+\frac{\partial\sigma_{zz}}{\partial z}\right]\label{momen22},
\end{equation}
\begin{equation}
\frac{v_{r}}{r}\frac{\partial(r^{2}\Omega)}{\partial
r}+v_{z}\frac{\partial(r\Omega)}{\partial
z}=\frac{1}{\rho}\left[\frac{1}{r^{2}}\frac{\partial(r^{2}\sigma_{r\varphi})}{\partial
r}+\frac{\partial\sigma_{\varphi z}}{\partial
z}\right]\label{momen23},
\end{equation}
where $\Omega_{K}$ is the Kepler angular velocity
$\Omega_{K}=\sqrt{(GM)/r^{3}}$ with $M$ being the mass of central
star, and the nine components of stress tensor $\sigma$ for
incompressible fluid are
\begin{eqnarray}
&&\sigma_{r\varphi}=\sigma_{\varphi r}=\mu r\frac{\partial\Omega}{\partial r},\nonumber\\
&&\sigma_{rz}=\sigma_{zr}=\mu\left(\frac{\partial v_{z}}{\partial r}+\frac{\partial v_{r}}{\partial z}\right),\nonumber\\
&&\sigma_{\varphi z}=\sigma_{z\varphi}=\mu\frac{\partial(r\Omega)}{\partial z},\nonumber\\
&&\sigma_{rr}=2\mu\frac{\partial v_{r}}{\partial
r},\sigma_{zz}=2\mu\frac{\partial v_{z}}{\partial
z},\sigma_{\varphi\varphi}=2\mu\frac{v_{r}}{r}.
\end{eqnarray}\label{momen3}
The axial symmetry assumption and the disk geometry half-height
$H\ll$ radius $r$ allows us to take the approximation that the terms
$v_{z}$ and $\partial/\partial z$ are much smaller compared to other
terms in equations (\ref{momen21}) to (\ref{momen23}). The the
radial component of the momentum equation (\ref{momen21}) can be
simplified as
\begin{equation}
v_{r}\frac{\partial v_{r}}{\partial
r}=(\Omega^{2}-\Omega_{K}^{2})r-\frac{1}{\rho}\frac{\partial
p}{\partial r}+\frac{2}{\rho
r}\left[\frac{\partial}{\partial}\left(\mu\frac{\partial
v_{r}}{\partial r}\right)+\mu\frac{\partial}{\partial
r}\left(\frac{v_{r}}{r}\right)\right]\label{momen41}.
\end{equation}
The angular momentum equation (\ref{momen23}) to be
\begin{equation}
\frac{v_{r}}{r}\frac{\partial(r^{2}\Omega)}{\partial
r}=\frac{1}{\rho r^{2}}\frac{\partial}{\partial r}\left(\mu
r^{3}\frac{\partial\Omega}{\partial r}\right)\label{momen42}.
\end{equation}
The z-direction momentum equation (\ref{momen22}) be rewritten as
\begin{equation}
\Omega_{K}^{2}z+\frac{1}{\rho}\frac{\partial p}{\partial
z}\approx0\label{momen43}.
\end{equation}
For 'thin' disk with $z\ll H$, we can set $|\partial p/\partial
z|\sim p/H$ and $z\sim H$, the half-height of the disk $H$ satisfies
\begin{equation}
H\cong \frac{c_{s}}{\Omega_{K}}\label{h1}.
\end{equation}
where $c_{s}=\sqrt{p/\rho}$ is the isothermal sound speed.
Furthermore, the energy balance in a steady-state disk is
established among the heating rate $q^{+}$, the cooling rate $q^{-}$
and the advection rate $q_{\rm adv}$ per unit volume. The heating
rate term $q^{+}$ is usually contributed by local viscous
dissipation (Thompson 1972)
\begin{equation}
q^{+}=\mu r^{2}\left(\frac{d\Omega}{dr}\right)^{2}\label{ener1}.
\end{equation}
Cooling is mainly via phonon or high-energy particle emissions.  The
advection term $q_{\rm adv}$ is energy flux driven by entropy
gradient in the accretion process. Thus the local energy
conservation energy is written
\begin{equation}
q_{\rm adv}=\rho v_{r}T\frac{ds}{dr}=q^{+}-q^{-}\label{ener2}.
\end{equation}
where $s$ is the specify entropy and $T$ is the temperature of the
disk at radius $r$.

Conversation equations (\ref{mass2}), (\ref{momen41}),
(\ref{momen42}), (\ref{ener2}) and relation (\ref{h1}) are commonly
used for disks under the assumptions of axisymmetry and 'thin' disk
geometry (e.g., Abramowicz et al. 1988; Narayan \& Popham 1993;
Narayan \& Yi 1994).

\textbf{\textit{$\alpha$-Prescription}}

The radio of dynamic viscosity $\eta$ and the density $\rho$ ,i. e.,
$\nu=\eta/\rho$, is called the kinematic viscosity. We adopt the
relation that the kinematic viscosity $\nu$ is proportional to the
isothermal sound speed $c_{s}$ and the disk half-thickness $H$, and
write
\begin{equation}
\nu=\alpha c_{s}H\label{vis1}.
\end{equation}
with the viscosity parameter $\alpha\leq1$ in most cases. This
famous $\alpha$-prescription was first adopted by Shakura \& Sunyave
(1973) from another point of view. More detailed description based
on microphysics scenario can be seen in many classical references
(e.g., Landau \& Lifshitz 1959; Frank et al. 2002). Although MHD
simulations beyond $\alpha$-prescription for understanding the
physical mechanisms which may generate a viscosity in disks have
been carried out for years (see Balbus \& Hawley 1998 for a review),
the $\alpha$-prescription, which is a much simpler treatment with
clear physical scenario, is still very useful and will continue to
be used for analytic and semi-analytic calculations. Now we use the
relation $\mu=\nu\rho=\alpha c_{s}\rho H$ in all of our equations.

\textbf{\textit{Vertically Integrated Equations}}

A farther reasonable approximation treatment which could simplify
the group of partial differential equations and make them to be
differential equations, is the "vertical integration" treatment
along the $z$-direction. The disk may be described by vertically
average physical quantities as a function of radius $r$. The surface
density is $\sum=2\rho H$. Thus the equations of continuity, radial
and angular momentum (\ref{mass2}, \ref{momen41}, \ref{momen42}) are
\begin{equation}
\dot{M}=-2\pi rv_{r}\Sigma\label{mass3},
\end{equation}
\begin{equation}
v_{r}\frac{dv_{r}}{dr}=(\Omega^{2}-\Omega_{K}^{2})r-\frac{1}{\sum}\frac{d}{dr}\left(\Sigma
c_{s}^{2}\right)\label{momen5},
\end{equation}
\begin{equation}
\frac{v_{r}}{r}\frac{d(r^{2}\Omega)}{dr}=\frac{1}{\Sigma
r^{2}}\frac{d}{dr}\left(\Sigma\nu
r^{3}\frac{d\Omega}{dr}\right)\label{momen6},
\end{equation}
The angular equation (\ref{momen6}) can be further integrated. If we
take the angular velocity of the accretion flow
$\Omega\approx\Omega_{K}$, then we obtain the angular momentum
equation as
\begin{equation}
\nu\Sigma=\frac{\dot{M}}{3\pi}f\label{momen7},
\end{equation}
where $f=1-l_{0}/l$ with $l=r^{2}\Omega$ is the specific angular
momentum and $l_{0}$ is the special angular momentum constant at the
inner edge of the disk. Usually we can write $f=1-\sqrt{r_{*}/r}$
with $r_{*}$ being the inner edge of the disk (Frank et al. 2002).
Furthermore, combining equations (\ref{ener1}) and (\ref{momen7})
the vertically intergraded local viscous dissipation rate of a
nearly Keplerian disk is
\begin{equation}
Q^{+}=\frac{3}{8\pi}\frac{GM\dot{M}}{r^{3}}f\label{ener3},
\end{equation}
and the integrated energy equation is
\begin{equation}
\frac{1}{2}\Sigma
Tv_{r}\frac{ds}{dr}=\frac{3}{8}\frac{GM\dot{M}}{8\pi
r^{3}}f-Q^{-}\label{ener3}.
\end{equation}
The factor 1/2 in the left side of equation (\ref{ener3}) is added
because we only study the vertically-integrated energy equation over
a half-thickness $H$.

\section{Equations for Magnetized Flows}

For magnetized accretion flows, we need to consider the effects of
magnetic fields on momentum transfer and energy dissipation. The
momentum equation of accretion flows (\ref{momen1}) is modified as
\begin{equation}
\frac{\partial \textbf{v}}{\partial
t}+(\textbf{v}\cdot\nabla)\textbf{v}=-\frac{1}{\rho}\nabla
p-\nabla\Phi+\frac{\eta}{\rho}\triangle \textbf{v}+\frac{1}{\rho
c}\textbf{J}\times\textbf{B}\label{momen8},
\end{equation}
with the added term of Lorentz force in the right side of the
equation. The current density $J$ satisfies
\begin{equation}
\textbf{J}=\frac{4\pi}{c}(\nabla\times \textbf{B})\label{Curr1},
\end{equation}
In cylindrical coordinates ($r,\varphi,z$), the three components of
Lorentz force are
\begin{equation}
\frac{1}{\rho
c}(\textbf{J}\times\textbf{B})_{r}=-\frac{1}{2\rho}\frac{\partial}{\partial
r}\left(\frac{B_{\varphi}^{2}+B_{z}^{2}}{4\pi}\right)+\frac{B_{z}}{4\pi
\rho}\frac{\partial B_{z}}{\partial
z}-\frac{B_{\varphi}^{2}}{4\pi\rho r}\label{Lorenz1},
\end{equation}
\begin{equation}
\frac{1}{\rho
c}(\textbf{J}\times\textbf{B})_{\varphi}=\frac{B_{r}B_{\varphi}}{4\pi\rho}+\frac{B_{r}}{4\pi
\rho}\frac{\partial B_{\varphi}}{\partial r}+\frac{B_{z}}{4\pi
\rho}\frac{\partial B_{\varphi}}{\partial z}\label{Lorenz2},
\end{equation}
\begin{equation}
\frac{1}{\rho
c}(\textbf{J}\times\textbf{B})_{r}=-\frac{1}{2\rho}\frac{\partial}{\partial
z}\left(\frac{B_{\varphi}^{2}+B_{z}^{2}}{4\pi}\right)+\frac{B_{r}}{4\pi
\rho}\frac{\partial B_{z}}{\partial r}\label{Lorenz3}.
\end{equation}
The right side of equations (\ref{Lorenz1}), (\ref{Lorenz1}) and
(\ref{Lorenz1}) need to be added to the right side of equations
(\ref{momen21}), (\ref{momen22}) and (\ref{momen23}) respectively.

The energy equation should include Joule dissipation (or Ohmic
dissipation) as well as viscous dissipation,
\begin{equation}
\rho v_{r}T\frac{ds}{dr}=q^{+}_{\rm vis}+q^{+}_{\rm
Joule}-q^{-}\label{ener4},
\end{equation}
where
\begin{eqnarray}
&&q^{+}_{\rm Joule}=\frac{4\pi}{c^{2}}\eta_{m}\textbf{J}^{2}\nonumber\\
&&=\frac{\eta_{m}}{4\pi}\left(\frac{\partial B_{\varphi}}{\partial
z}\right)^{2}+\frac{\eta_{m}}{4\pi}\left(\frac{\partial
B_{r}}{\partial z}-\frac{\partial B_{z}}{\partial
r}\right)^{2}+\frac{\eta_{m}}{4\pi
r^{2}}\left[\frac{\partial}{\partial r}(r
B_{\varphi})\right]^{2}\label{jol1}
\end{eqnarray}
with $\eta_{m}$ being the magnetic diffusivity of the disk. If we
define the Alfv\'{e}n sound speeds along the three directions
$c_{r,\varphi,z}^{2}=B_{r,\varphi,z}^{2}/(4\pi\rho)$, then the
vertically integrated momentum equations read (Zhang \& Dai 2008b)
\begin{eqnarray}
v_{r}\frac{\partial v_{r}}{\partial
r}&=&(\Omega^{2}-\Omega_{K}^{2})r-\frac{1}{\Sigma}\frac{\partial}{\partial
r}(\Sigma c_{s}^{2})\nonumber\\
&&-\frac{1}{2\Sigma}\frac{\partial}{\partial
r}[\Sigma(c_{z}^{2}+c_{\varphi}^{2})]+\frac{1}{\sqrt{\Sigma}}c_{z}\frac{\partial}{\partial
z}(\sqrt{\Sigma}c_{r})-\frac{c_{\varphi}^{2}}{r}\label{momen91},
\end{eqnarray}
\begin{eqnarray}
\frac{v_{r}}{r}\frac{\partial(rv_{\varphi})}{\partial
r}+v_{z}\frac{\partial v_{\varphi}}{\partial
z}&=&\frac{1}{\Sigma r^{2}}\frac{\partial}{\partial r}\left(\Sigma\nu r^{3}\frac{\partial\Omega}{\partial r}\right)\nonumber\\
&&+\frac{c_{r}c_{\varphi}}{r}+\frac{c_{r}}{\sqrt{\Sigma}}\frac{\partial}{\partial
r}(\sqrt{\Sigma}c_{\varphi})+\frac{c_{z}}{\sqrt{\Sigma}}\frac{\partial}{\partial
z}(\sqrt{\Sigma}c_{\varphi})\label{momen92},
\end{eqnarray}
\begin{eqnarray}
\frac{v_{r}}{r}\frac{\partial v_{z}}{\partial r}&=&-\Omega_{K}^{2}z-\frac{1}{\Sigma}\frac{\partial}{\partial z}(\Sigma c_{s}^{2})\nonumber\\
&&-\frac{1}{2\Sigma}\frac{\partial}{\partial
z}[\Sigma(c_{r}^{2}+c_{\varphi}^{2})]+\frac{1}{\sqrt{\Sigma}}c_{r}\frac{\partial}{\partial
r}(\sqrt{\Sigma}c_{z})\label{momen93}.
\end{eqnarray}
And the energy conservation equation becomes
\begin{eqnarray}
\frac{1}{2}\Sigma v_{r}T\frac{ds}{dr}&=&\frac{3}{8\pi}\frac{GM\dot{M}}{r^{3}}f+\eta_{m}\left[\frac{\partial}{\partial z}\left(\sqrt{\Sigma}c_{\varphi}\right)\right]^{2}\nonumber\\
&&\eta_{m}\left[\frac{\partial}{\partial
z}\left(\sqrt{\Sigma}c_{r}\right)-\frac{\partial}{\partial
r}\left(\sqrt{\Sigma}c_{\varphi}\right)\right]^{2}+\frac{\eta_{m}}{r^{2}}\left[\frac{\partial}{\partial
r}(r\sqrt{\Sigma}c_{\varphi})\right]^{2}\label{momen93}.
\end{eqnarray}

Lovelace, Wang \& Sulkanen (1987) derived similar momentum equations
as (\ref{momen91}) to (\ref{momen93}). The magnetic fields symmetry
requires
\begin{eqnarray}
B_{r}(r,z)=+B_{r}(r,-z),\nonumber\\
B_{\varphi}(r,z)=+B_{\varphi}(r,-z),\nonumber\\
B_{z}(r,z)=-B_{z}(r,-z).\label{sym1}.
\end{eqnarray}
The $r,\varphi,z$-component equations are
\begin{equation}
\Sigma v_{r}\frac{\partial v_{r}}{\partial
r}=\Sigma\frac{v_{\varphi}^{2}}{r}-\Sigma\frac{GM}{r^{2}}-\frac{1}{8\pi
r^{2}}\frac{\partial}{\partial
r}[2Hr^{2}<(B_{\varphi}^{2}-B_{r}^{2})>]+\frac{1}{4\pi}B_{r}B_{z}|_{z=-H}^{z=H}\label{momen101},
\end{equation}
\begin{equation}
\frac{\dot{M}}{3\pi}f=\Sigma\nu+\frac{1}{3\pi\Omega}\left[\left(\frac{r_{*}}{r}\right)^{2}<B_{r}B_{\varphi}>_{r=r_{*}}-<B_{r}B_{\varphi}>\right]\label{momen102},
\end{equation}
\begin{equation}
\left(\frac{GM\Sigma}{2r^{3}}\right)\left(\int_{0}^{H}z\rho
dz/\int_{0}^{H}\rho
dz\right)=p|_{z=0}-\frac{B_{r}^{2}+B_{\varphi}^{2}}{8\pi}\label{momen103},
\end{equation}
respectively, where $<...>$ is defined as
\begin{equation}
<B_{\xi}>=\frac{1}{2H}\int_{-H}^{H}dzB_{\xi}(r,z)\label{nota1}.
\end{equation}

\section{Photon Radiation and Shakura-Sunyaev Solutions}
In accretion disks, the vertical energy transfer mechanism may be
either radiative or convective, depending on the temperature
gradient along the $z$-direction. We will discuss the convection
effect later in \S 1.1.4. In this section we first consider the
energy transfer is radiative. The specific flux is
\begin{equation}
F_{\nu}(z)=-\frac{4\pi}{3\kappa_{\nu}\rho}\frac{\partial
B_{\nu}(T)}{\partial z}\label{rad1},
\end{equation}
where $\kappa_{\nu}$ is the opacity coefficient for a particular
frequency $\nu$. Here we assume the radiation field in the disk has
reached thermal equilibrium in the layer $z$ to $z+dz$, and
$B_{\nu}(T)$ is the Planck blackbody function. Intergrading
(\ref{rad1}) over frequency we obtain
\begin{equation}
F(z)=\int_{0}^{\infty}F_{\nu}(z)dz=-\frac{c}{3\kappa_{R}\rho}\frac{\partial}{\partial
z}(a_{B}T^{4})=-\frac{4c\sigma_{B}}{3\kappa_{R}\rho}\frac{\partial}{\partial
z}(T^{4})\label{rad2},
\end{equation}
where $a_{B}$ is the radiation constant while $\sigma_{B}$ the
Stefan-Boltzmann constant, $\kappa_{R}$ is the Rosseland mean
opacity. Hubeny (1990, see also Popham \& Narayan 1995) derived an
more elaborate expression for the vertical flux
\begin{equation}
F(z)=\frac{\sigma_{B}T^{4}}{3(\tau/2+1/\sqrt{3}+1/3\tau_{a})}\label{rad3},
\end{equation}
where $\tau=\kappa_{R}\rho H$ and $\tau_{a}=\kappa_{a}\rho H$ are
the total and absorption  optical depths.

The generated dissipation energy in the disk is mainly cooled by
thermal photon emission. Thus we have the cooling rate $Q^{-}_{\rm
rad}$ over a half-thickness disk
\begin{equation}
Q^{-}_{\rm rad}=\int_{0}^{H}\frac{\partial F(z)}{\partial
z}dz=F(H)-F(0)\label{rad4}.
\end{equation}
We adopt the simplified energy equation, i.e., energy balance is
established between radiation and viscous dissipation
\begin{eqnarray}
Q^{-}_{\rm rad}&&=F(H)-F(0)=\frac{4\sigma_{B}T_{c}^{4}}{3\tau}\nonumber\\
&&=\int_{0}^{H}q^{+}_{\rm vis}(z)=Q^{+}_{\rm
vis}=\frac{3}{8\pi}\frac{GM\dot{M}}{r^{3}}f\label{rad4}.
\end{eqnarray}
$T_{c}$ is the central temperature at $z$=0, and we assume $T_{c}\gg
T(H)$. The main contribution to the opacity for a normal thermal
disk comes from Thompson scattering on free electrons, and free-free
absorption (Shakura \& Sunyave 1973)
\begin{equation}
\sigma_{T}=6.65\times10^{-25}\textrm{cm}^{2},
\sigma_{ff}=0.11T^{-7/2}n
\textrm{cm}^{3}\textrm{g}^{-1}\label{rad5},
\end{equation}
For a complete set of equations we also need thermodynamical
equation, or the expression of pressure, which include both gas
pressure and radiation pressure
\begin{equation}
p=p_{g}+p_{r}=\frac{\rho T\Re}{\mu_{\rm
acc}}+\frac{4\sigma_{B}}{3c}T_{c}^{4}\label{rad6},
\end{equation}
where $\Re$ is the gas constant, $\mu_{\rm acc}$ is the mean
molecular weight.

Combining equations (\ref{h1}), (\ref{vis1}), (\ref{momen7}),
(\ref{rad4}), (\ref{rad5}) and (\ref{rad6}), we are able to
determine the physical properties of a thin photon radiative disk as
functions of accretion rate $\dot{M}$, central object mass $M$,
radius $r$ and inner edge radius $r_{*}$. The analytical solutions
were first derived by Shakura \& Sunyaev (1973), who divided the
disk into three regions, and gave different solution in different
region:
\begin{figure}
\centering\resizebox{0.6\textwidth}{!} {\includegraphics{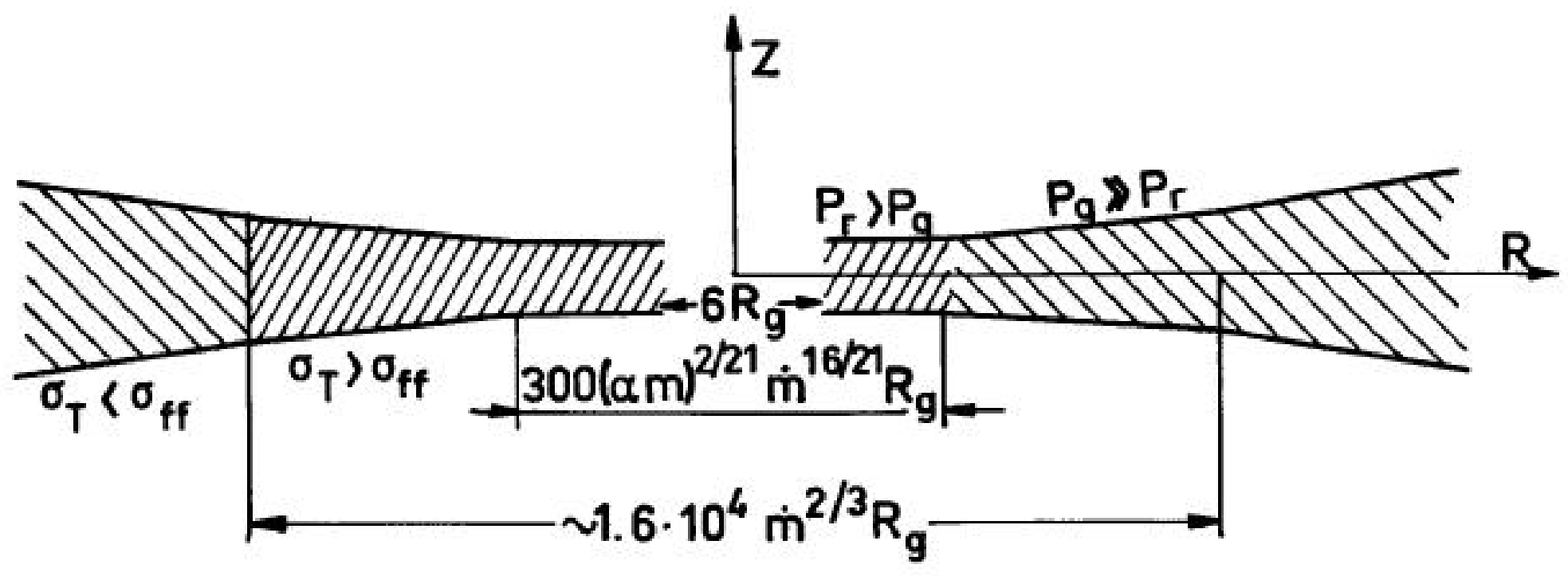}}
\caption{Physical scenario of Shakura \& Sunyavae
disk.}\label{fig213}
\end{figure}
a) $p_{r}\gg p_{g}$, $\sigma_{T}\gg\sigma_{ff}$
\begin{eqnarray}
&&\Sigma=4.6\alpha^{-1}m^{-1}\hat{r}^{3/2}f^{-1}\nonumber\\
&&T=2.3\times10^{7}(\alpha m)^{-1/4}\hat{r}^{-3/4}\nonumber\\
&&n=\Sigma/(2Hm_{p})=4.3\times10^{17}\alpha^{-1}\dot{m}^{-2}m^{-1}\hat{r}^{3/2}f^{-2}\nonumber\\
&&H=3.2\times10^{6}\dot{m}mf\nonumber\\
&&v_{r}=7.7\times10^{7}\alpha \dot{m}^{2}\hat{r}^{-5/2}f\label{SS1},
\end{eqnarray}
b) $p_{g}\gg p_{r}$, $\sigma_{T}\gg\sigma_{ff}$
\begin{eqnarray}
&&\Sigma=1.7\times10^{5}\alpha^{-4/5}\dot{m}^{3/5}m^{1/5}\hat{r}^{-3/5}f^{3/5}\nonumber\\
&&T=3.1\times10^{8}\alpha^{-1/5}\dot{m}^{2/5}m^{-1/5}\hat{r}^{-9/10}f^{2/5}\nonumber\\
&&n=4.2\times10^{24}\alpha^{-7/10}\dot{m}^{2/5}m^{-7/10}\hat{r}^{-33/20}f^{2/5}\nonumber\\
&&H=1.2\times10^{4}\alpha^{-1/10}\dot{m}^{1/5}m^{9/10}r^{21/20}f^{1/5}\nonumber\\
&&v_{r}=2\times10^{6}\alpha^{4/5}\dot{m}^{2/5}m^{-1/5}\hat{r}^{-2/5}f^{-3/5}\label{SS2},
\end{eqnarray}
b) $p_{g}\gg p_{r}$, $\sigma_{ff}\gg\sigma_{T}$
\begin{eqnarray}
&&\Sigma=6.1\times10^{5}\alpha^{-4/5}\dot{m}^{7/10}m^{1/5}\hat{r}^{-3/4}f^{7/10}\nonumber\\
&&T=8.6\times10^{7}\alpha^{-1/5}\dot{m}^{3/10}m^{-1/5}\hat{r}^{-3/4}f^{3/10}\nonumber\\
&&n=3\times10^{25}\alpha^{-7/10}\dot{m}^{11/20}m^{-7/10}\hat{r}^{-15/8}f^{11/20}\nonumber\\
&&H=6.1\times10^{3}\alpha^{-1/10}\dot{m}^{3/20}m^{9/10}\hat{r}^{9/8}f^{3/20}\nonumber\\
&&v_{r}=5.8\times10^{5}\alpha^{4/5}\dot{m}^{3/10}m^{-1/5}\hat{r}^{-1/4}f^{-7/10}\label{SS3},
\end{eqnarray}
where
\begin{equation}
m=\frac{M}{M_{\odot}},
\dot{m}=\frac{\dot{M}}{\dot{M}_{cr}}=\frac{\dot{M}}{3\times10^{-8}M_{\odot}yr^{-1}}\times\left(\frac{M_{\odot}}{M}\right),
\hat{r}=\frac{r}{3r_{g}}=\frac{1}{m}\left(\frac{r}{9\textrm{km}}\right)\label{nota2},
\end{equation}
and the units of surface density $\Sigma$, temperature $T$, number
density $n$, half-thickness $H$ and radial velocity $v_{r}$ are g
cm$^{-2}$, K, cm$^{-3}$, cm and cm s$^{-1}$. The general scenario of
the disk structure is summarized in Figure \ref{fig213}, which shows
the three regions having different physical properties.

\section{Accretion Flows with Self-Similar Structure}
We switch back to the basic energy equation (\ref{ener2})
\begin{equation}
q_{\rm adv}=q^{+}-q^{-}\nonumber
\end{equation}
where $q_{\rm adv}$ is the advection transport of energy driven by
the entropy gradient. In last section \S 2.1.3, we approximately set
$q_{\rm adv}\simeq0$. In fact, depending on the relative magnitudes
of the terms in equation (\ref{ener2}), we may identify three cases
of accretion flows (Narayan, Mahadevan \& Quataert, 1998)

\begin{itemize}
\item
$q^{+}\simeq qq^{-}\gg q_{\rm adv}$, the cooling-dominated flow
where all the energy released by viscous stresses is radiated. The
S-S thin disk solutions correspond to this case.

\item
$q_{\rm adv}\simeq q^{+}\gg qq^{-}$, an advection-dominated
accretion flow (ADAF) where almost all the viscous energy is stored
in the gas and is deposited into the central BH. For a given
$\dot{M}$, an ADAF is much less luminous than a cooling-dominated
flow. Moreover, the effect of convection in an ADAF is significant
for sufficiently low viscosity parameter $\alpha$. Convective ADAF
is called "convection-dominated accretion flow" (CDAF, Narayan et
al. 2000).

\item
$-q_{\rm adv}\simeq q^{-}\gg qq^{+}$, advection-cooling-balance
accretion flow, where energy generation by viscosity is negligible,
but the entropy of the inflowing gas is converted to radiation.

\end{itemize}

Self-similar treatment is very useful to describe the properties of
various fluid since Sedov (1969) and Taylor (1950). The analytical
ADAF solution based on self-similar treatment has been derived by
Narayan \& Yi (1994). The integrated  equations (\ref{mass3}),
(\ref{momen5}), (\ref{momen6}) and the energy equation
\begin{equation}
\Sigma
v_{r}T\frac{ds}{dr}=\frac{3+3\epsilon}{2}\Sigma\nu\frac{dc_{s}^{2}}{dr}-2c_{s}^{2}Hv_{r}\frac{d\rho}{dr}=f\frac{\alpha\Sigma
c_{s}^{2}r^{2}}{\Omega_{K}}\left(\frac{d\Omega}{dr}\right)^{2}\label{ener5}
\end{equation}
permit a self-similar solution of the form (Spruit et al. 1987):
\begin{equation}
\rho\propto r^{-3/2}, v_{r}\propto r^{-1/2}, \Omega\propto r^{-3/2},
c_{s}^{2}\propto r^{-1}\label{self1}
\end{equation}
Here $\epsilon=p_{r}/p$ is the fraction of radiation pressure, and
$f$ measures the degree to which the flow is advection-dominated.
Thus the self-similar solution of an ADAF is (Narayan \& Yi, 1994)
\begin{equation}
v_{r}(r)=-(5+2\epsilon')\frac{g(\alpha,\epsilon')}{3\alpha}v_{K}\label{self21}
\end{equation}
\begin{equation}
\Omega(r)=\left[\frac{2\epsilon'(5+2\epsilon')g(\alpha,\epsilon')}{9\alpha^{2}}\right]^{1/2}\Omega_{K}\label{self22}
\end{equation}
\begin{equation}
c_{s}^{2}(r)=\frac{2(5+2\epsilon')}{9}\frac{g(\alpha,\epsilon')}{\alpha^{2}}v_{K}^{2}\label{self23}
\end{equation}
where
\begin{equation}
v_{K}\equiv\left(\frac{GM}{r}\right)^{1/2},
\epsilon'\equiv\frac{\epsilon}{f}=\frac{1}{f}\left(\frac{5/3-\gamma}{\gamma-1}\right),
g(\alpha,\epsilon')\equiv\left[1+\frac{18\alpha^{2}}{(5+2\epsilon')}\right]^{1/2}-1\label{self3}
\end{equation}

Taking the assumption $\alpha^{2}\ll 1$, the solution (\ref{self21})
to (\ref{self23}) takes a simple form
\begin{equation}
\frac{v}{v_{K}}\approx-\frac{3\alpha}{5+2\epsilon'},
\frac{\Omega}{\Omega_{K}}\approx\left(\frac{2\epsilon'}{5+2\epsilon'}\right)^{1/2},
\left(\frac{c_{s}}{v_{K}}\right)^{2}\approx\frac{2}{5+2\epsilon'}\label{self4}
\end{equation}
In the limit of very efficient cooling, $f\rightarrow0$ and
$\epsilon'\rightarrow0$, we obtain $v_{r}, c_{s}\ll v_{K}$ and
$\Omega\rightarrow\Omega_{K}$, which correspond the Keplerian
accretion flows. In the opposite limit of an ADAF $f\rightarrow1$,
we have the extreme case $\Omega\rightarrow0$ for
$\gamma\rightarrow5/3$ and $c_{s}\sim v_{K}$, $H\sim c_{s}/v_{K}\sim
v_{K}/\Omega_{K}=r$, which correspond to the spherical Bondi
accretion limit.

Next we discuss the effect of convection in ADAFs. The $r$-direction
integrated form of angular momentum equation (\ref{momen42}) read
\begin{equation}
0=\dot{J}_{v}+\dot{J}_{\rm
adv}=-\alpha\frac{c_{s}^{2}}{\Omega_{K}}\rho
r^{3}\frac{d\Omega}{dr}+\rho rv_{r}\Omega r^{2}\label{conv1},
\end{equation}
where the viscosity next flux
\begin{equation}
\dot{J}_{v}=-\alpha\frac{c_{s^{2}}}{\Omega_{K}}\rho
r^{3}\frac{d\Omega}{dr}\label{conv2}.
\end{equation}
Analogous to the viscosity flux and viscosity parameter $\alpha$,
Narayan et al. (2000) introduced the "convection parameter"
$\alpha_{c}$ and the angular flux due to convection as
\begin{equation}
\dot{J}_{c}=-\alpha_{c}\frac{c_{s}^{2}}{\Omega_{K}}\rho
r^{3(1+g)/2}\frac{d}{dr}[\Omega r^{3(1-g)/2}],\label{conv3}
\end{equation}
with the index $g$ describes the physics of convective angular
momentum transport. When $g=1$, expression (\ref{conv3}) becomes
\begin{equation}
\dot{J}_{c}=-\alpha_{c}\frac{c_{s}^{2}}{\Omega_{K}}\rho
r^{3}\frac{d\Omega}{dr},\label{conv41}
\end{equation}
convection angular momentum flux is oriented down the angular
velocity gradient. While when $g=-1/3$,
\begin{equation}
\dot{J}_{c}=-\alpha_{c}\frac{c_{s}^{2}}{\Omega_{K}}\rho
r\frac{d\Omega r^{2}}{dr},\label{conv41}
\end{equation}
convection angular momentum flux is due to the specific angular
momentum gradient. The $r$-direction angular momentum equation reads
\begin{equation}
0=\dot{J}_{v}+\dot{J}_{c}+\dot{J}_{\rm
adv}=-\alpha\frac{c_{s}^{2}}{\Omega_{K}}\rho
r^{3}\frac{d\Omega}{dr}-\alpha_{c}\frac{c_{s}^{2}}{\Omega_{K}}\rho
r^{3(1+g)/2}\frac{d}{dr}[\Omega r^{3(1-g)/2}]+\rho rv_{r}\Omega
r^{2}.\label{conv5}
\end{equation}
We take the diffusion constant $K_{c}$ as
\begin{equation}
K_{c}=\alpha_{c}\frac{c_{s}^{2}}{\Omega_{K}},\label{conv6}
\end{equation}
and the energy flux via convection is
\begin{equation}
F_{c}=-K_{c}\rho\frac{de}{dt}=-\alpha_{c}\frac{c_{s}^{2}}{\Omega_{K}}\rho
T\frac{ds}{dr}.\label{conv7}
\end{equation}
Thus the energy transport by convection in cylindrical coordinates
is $r^{-1}d(rF_{c})/dr$. In this section we follow Narayan et al.
(2000) and adopt the expression $r^{-2}d(r^{2}F_{c})/dr$ in
spherical coordinates and obtain the energy equation including
convection as
\begin{equation}
\rho
v_{r}T\frac{ds}{dr}+\frac{1}{r^{2}}\frac{d}{dr}(r^{2}F_{c})=Q^{+}=(\alpha+\alpha_{c})\frac{c_{s}^{2}}{\Omega_{K}}\rho
r^{2}\left(\frac{d\Omega}{dr}\right)^{2}.\label{conv8}
\end{equation}

\begin{figure}
\centering\resizebox{0.5\textwidth}{!} {\includegraphics{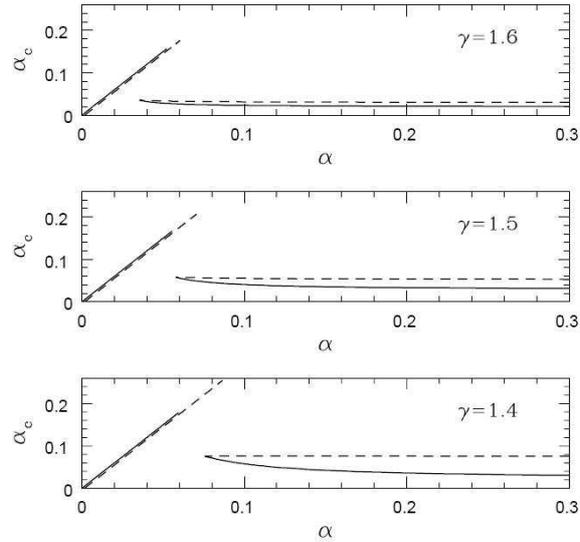}}
\caption{Variation of the convective parameter $\alpha_{c}$ as a
function of the viscosity parameter $\alpha$ for three values the
adiabatic index $\gamma$ (Narayan et al. 2000).}\label{fig214}
\end{figure}

The radial momentum equation (\ref{momen41}), and new angular
momentum and energy equations (\ref{conv5}) and (\ref{conv7}) allow
us to obtain the self-similar solutions for ADAF with convection. We
take the self-similar structure
\begin{equation}
\rho=\rho_{0}r^{-s}, v_{r}=v_{0}v_{K}, \Omega=\Omega_{0}\Omega_{K},
c_{s}^{2}=c_{0}^{2}v_{K}^{2}, H=c_{0}r,\label{conv9}
\end{equation}
where the index $s$ is equal to 3/2 in normal ADAF structure (e.g.,
Narayan \& Yi 1994), and $\rho_{0},v_{0},\Omega_{0},c_{0}$ is the
dimensionless constant to be determined. Thus we can derive the
relation
\begin{equation}
\Omega_{0}^{2}=1-(s+1)c_{0}^{2}\label{conv101}
\end{equation}
\begin{equation}
v_{0}=-\frac{3}{2}(\alpha+g\alpha_{c})c_{0}^{1}\label{conv102}
\end{equation}
\begin{equation}
\left(s-\frac{1}{\gamma-1}\right)v_{0}+\left(s-\frac{1}{2}\right)\left(s-\frac{1}{\gamma-1}\right)\alpha_{c}c_{0}^{1}
=\frac{9}{4}\Omega_{0}^{2}(\alpha+g\alpha_{c})\label{conv103}
\end{equation}
Moreover, another microphysics equation based on the mixing length
theory can be adopted (Narayan et al. 2000)
\begin{equation}
\alpha_{c}=\frac{l_{M}^{2}}{4\sqrt{2}(1+g)^{2}c_{0}^{2}}\left\{\frac{(1+s)[(\gamma+1)-(\gamma-1)s]}{\gamma}c_{0}^{2}-1\right\}^{1/2}\label{conv11}
\end{equation}
Combining equations (\ref{conv101}) to (\ref{conv103} and taking
$s=3/2$ (i.e., the case of normal ADAF), we have
\begin{equation}
\frac{5-3\gamma}{\gamma-1}\left[\frac{3}{2}\alpha+\left(\frac{3}{2}g-1\right)\alpha_{c}\right]c_{0}=\frac{9}{2}(\alpha+g\alpha_{c})\left(1-\frac{5}{2}c_{0}^{2}\right)\label{conv12}
\end{equation}
Equations (\ref{conv11}) and (\ref{conv12}) can be used to determine
$\alpha_{c}$ and $\alpha$ as the function of $c_{0}$. For $g<2/3$,
the minimum value of $\alpha$ can be reached when the sound speed
takes on its maximum value, $c_{0}^{2}=2/5$, and the minimum value
of $\alpha$ is
\begin{equation}
\alpha_{\rm
crit1}=\left(\frac{2}{3}-g\right)\frac{l_{M}^{2}}{20}\left(\frac{5-3\gamma}{\gamma}\right)^{1/2}>0.\label{conv13}
\end{equation}
Therefore, the normal ADAF self-similar solution can not exist when
$g<2/3$ and $\alpha$ satisfies $0<\alpha<\alpha_{\rm crit1}$ at the
same time. What happens if $\alpha<\alpha_{\rm crit1}$? If $\alpha$
is small and $g<2/3$, the viscous flux is unable to cope with the
with inward flux due to convection, and there is no consistent
accretion solution. However, a completely different solution is
possible if $g<0$. Equation (\ref{conv12}) could also be satisfied
if $s=1/2, \alpha=-g\alpha_{c}$ and $v_{0}=0$. The new self-similar
solution is referred as a "convective envelope" solution or
"convection-dominated accretion flow" (CDAF). We obtain
\begin{equation}
\alpha=-g\alpha_{c}=-\frac{gl_{M}^{2}}{9\sqrt{2}c_{0}^{2}}\left[\frac{3(\gamma+1)}{4\gamma}c_{0}^{2}-1\right]^{1/2},\label{conv14}
\end{equation}
and the maximum value of $\alpha$ for CDAF solution
\begin{equation}
\alpha_{\rm
crit2}=-\frac{gl_{M}^{2}}{12}\left(\frac{3-\gamma}{\gamma}\right)^{1/2}.\label{conv15}
\end{equation}
We list the different solutions with the range of parameters
$\alpha$ and $g$ following Narayan et al. 2000:

\begin{itemize}
\item
$g>2/3$, ADAF solution for $\alpha\leq1$

\item
$0<g<2/3$, ADAF solution for $\alpha\geq\alpha_{\rm crit1}$

\item
$g<0$, ADAF solution for $\alpha\geq\alpha_{\rm crit1}$; CDAF
solution for $\alpha>\alpha_{\rm crit2}$.
\end{itemize}
Figure \ref{fig214} shows the variation of $\alpha_{c}$ as a
function of $\alpha$ for various adiabatic index and $g=-1/3$. We
note that, since $\alpha_{\rm crit2}>\alpha_{\rm crit1}$, the
self-similar solution is not unique for $\alpha_{\rm
crit1}<\alpha<\alpha_{\rm crit2}$, and both ADAF and CDAF solutions
exist in this case.

Self-similar solutions for ADAF with large-scale magnetic fields
have also been studied (e.g.,Shadmehri 2004; Shadmehri \& Khajenabi
2005; Akizuki \& Fukue 2006; Shadmehri \& Khajenabi 2006; Ghanbari,
Salehi \& Abbassi 2007; Zhang \& Dai 2008b; Bu, Yuan \& Xie 2009).
The physical properties of magnetized flows are different to that
without magnetic fields. For example, Akizuki \& Fukue (2006)
examined the effects of toroidal magnetic fields and mass-loss from
the disk. They found that the disk becomes thick due to the magnetic
pressure, and the radial infalling velocity and rotation becomes
fast. Zhang \& Dai (2008b) also discussed the self-similar solutions
for CDAFs with large-scale fields. We will discuss these results in
\S 3 in details.

\section{NDAFs and Neutrino-Cooled Disks I: Formulae}

If the accreting matter has a sufficiently high temperature and
density, the photon optical depth is enormous and radiation cannot
escape. However, it can cooled by neutrino emission, leading to a
neutrino-dominated accretion flow (NDAF, Popham, Woosley \& Fryer
1999).

Neutrino-cooled accretion disk around BH has been considered as the
central engine of GRBs. Both processes of massive star collapses or
compact object mergers lead to the formation of BH + neutrino-cooled
disk systems. These disk are typically compact, with the bulk of the
mass residing with $4\times10^{7}$ cm of the BH ($\sim3-5M_{\odot}$)
and temperature ($10^{9}\textrm{K}\leq T\leq 10^{11}\textrm{K}$),
density ($10^{9}$ g cm$^{-3}$$\leq \rho\leq$ $10^{11}$ g cm$^{-3}$),
and a huge accretion rate up to $\sim1M_{\odot}$ s$^{-1}$. The
accretion timescale of these disks $t_{\rm acc}\sim M_{\rm
disk}/\dot{M}$ is short, from tens of milliseconds after mergers to
tens of seconds after collapses.

\textit{\textbf{Neutrino Cooling Rates and Optical Depths}}

The generated viscosity can be cooled via neutrino emission via
various cooling processes, including absorption processes such as
electron-positron pair annihilation, electron-positron capture rate
by nucleons, nucleon-nucleon bremsstrahlung, plasmon decay into
neutrinos, and scattering processes such as neutrino-nucleon
scattering and neutrino-electron scattering.

The rates of electron-positron pair annihilation into neutrinos are
(Burrows \& Thompson 2004)
\begin{equation}
q^{-}_{e^{-}e^{+}\rightarrow\nu_{e}\bar{\nu}_{e}}=2.56\times10^{33}T_{11}^{9}f(\eta_{e})
\,\textrm{ergs}\,\textrm{cm}^{-3}\,\textrm{s}^{-1},\label{pair1}
\end{equation}
\begin{equation}
q^{-}_{e^{-}e^{+}\rightarrow\nu_{\mu}\bar{\nu}_{\mu}}=q^{-}_{e^{-}e^{+}\rightarrow\nu_{\tau}\bar{\nu}_{\tau}}=1.09\times10^{33}T_{11}^{9}f(\eta_{e})
\,\textrm{ergs}\,\textrm{cm}^{-3}\,\textrm{s}^{-1},\label{pair2}
\end{equation}
where $T_{11}=T/10^{11}$ K,
\begin{equation}
f(\eta_{e})=\frac{F_{4}(\eta_{e})F_{3}(-\eta_{e})+F_{4}(-\eta_{e})F_{3}(\eta_{e})}{2F_{4}(0)F_{3}(0)}\label{pair3}
\end{equation}
and
\begin{equation}
F_{n}(\eta_{e})=\int_{0}^{\infty}\frac{x^{n}}{e^{x\mp\eta_{e}}+1}dx.\label{pair4}
\end{equation}

The electron-positron capture rate by nucleons (or, the URCA
reaction) is represented by the sum of three terms:
\begin{equation}
q_{eN}=q_{p+e^{-}\rightarrow n+\nu_{e}}+q_{n+e^{-}\rightarrow
p+\bar{\nu}_{e}}+q_{n\rightarrow
p+e^{-}+\bar{\nu}_{e}},\label{URCA1}
\end{equation}
with
\begin{equation}
q_{p+e^{-}\rightarrow
n+\nu_{e}}=\frac{G_{F}^{2}C_{V}^{2}(1+3g_{A}^{2})}{2\pi^{3}\hbar^{7}c^{6}}n_{p}\int_{Q}^{\infty}dE_{e}E_{e}\sqrt{E_{e}^{2}-m_{e}^{2}c^{4}}
(E_{e}-Q)^{3}\frac{1}{e^{(E_{e}-\mu_{e})/k_{B}T}+1},\label{URCA2}
\end{equation}
\begin{equation}
q_{n+e^{-}\rightarrow
p+\bar{\nu}_{e}}=\frac{G_{F}^{2}C_{V}^{2}(1+3g_{A}^{2})}{2\pi^{3}\hbar^{7}c^{6}}n_{n}\int_{m_{e}c^{2}}^{\infty}dE_{e}E_{e}\sqrt{E_{e}^{2}-m_{e}^{2}c^{4}}
(E_{e}+Q)^{3}\frac{1}{e^{(E_{e}+\mu_{e})/k_{B}T}+1},\label{URCA3}
\end{equation}
\begin{equation}
q_{n\rightarrow
p+e^{-}+\bar{\nu}_{e}}=\frac{G_{F}^{2}C_{V}^{2}(1+3g_{A}^{2})}{2\pi^{3}\hbar^{7}c^{6}}n_{n}\int_{m_{e}c^{2}}^{Q}dE_{e}E_{e}\sqrt{E_{e}^{2}-m_{e}^{2}c^{4}}
(Q-E_{e})^{3}\left(1-\frac{1}{e^{(E_{e}-\mu_{e})/k_{B}T}+1}\right),\label{URCA4}
\end{equation}
where the constants $G_{F}=1.463\times10^{-49}$ ergs cm$^{-3}$,
$C_{V}^{2}=0.947$, axial vector $g_{A}=1.26$, $Q$ is defined as
$(m_{n}-m_{p})c^{2}$, and $n_{p}$ and $n_{n}$ are the number density
of protons and neutrons in the flows.

The nucleon-nucleon bremsstrahlung rate through $n+n\rightarrow
n+n+\nu+\bar{\nu}$ is represented by
\begin{equation}
q^{-}_{\rm
brems}=3.4\times10^{33}T_{11}^{8}\rho_{13}^{1/3}\,\textrm{ergs}\,\textrm{cm}^{-3}\,\textrm{s}^{-1},\label{brems1}
\end{equation}
in the degeneracy regime of nucleons, and
\begin{equation}
q^{-}_{\rm
brems}=1.5\times10^{33}T_{11}^{5.5}\rho_{13}^{2}\,\textrm{ergs}\,\textrm{cm}^{-3}\,\textrm{s}^{-1},\label{brems2}
\end{equation}
in the nondegeneracy regime of nucleons (Hannestad \& Raffelt 1998).
Here $\rho_{13}=\rho/10^{13}$ g cm$^{-3}$.

Plasmon decay into neutrino through
$\tilde{\gamma}\rightarrow\nu_{e}+\bar{\nu}_{e}$ is estimated as
(Ruffert et al. 1996)
\begin{equation}
q^{-}_{\rm
plasmon}=1.5\times10^{32}T_{11}^{9}\gamma_{p}^{6}e^{-\gamma_{p}}(1+\gamma_{p})\left(2+\frac{\gamma_{p}^{2}}{1+\gamma_{p}}\right)\,\textrm{ergs}\,\textrm{cm}^{-3}\,\textrm{s}^{-1},\label{brems2}
\end{equation}
where
$\gamma_{p}=5.656\times10^{-2}[(\pi^{2}+3\eta_{e}^{2})/3]^{1/2}$.
However, in most cases, both the $n-n$ bremsstrahlung and plasmon
decay are less important than URCA and $e^{-}e^{+}$ pair
annihilation in a NDAF. Thus, the absorption optical depths for
three different types of neutrino can be calculated by (e. g.,
Popham et al. 1999; Kohri et al. 2005)
\begin{equation}
\tau_{a,\nu_{e}}=\frac{(q^{-}_{eN}+q^{-}_{e^{-}e^{+}\rightarrow\nu_{e}\bar{\nu}_{e}}+q^{-}_{\rm
brems}+q^{-}_{\rm
plasmon})H}{4(7/8\sigma_{B}T^{4})}\approx\frac{(q^{-}_{eN}+q^{-}_{e^{-}e^{+}\rightarrow\nu_{e}\bar{\nu}_{e}})H}{4(7/8\sigma_{B}T^{4})}\label{depth1}
\end{equation}
\begin{equation}
\tau_{a,\nu_{\mu,\tau}}=\frac{(q^{-}_{e^{-}e^{+}\rightarrow\nu_{\mu}\bar{\nu}_{\mu}}+q^{-}_{\rm
brems})H}{4(7/8\sigma_{B}T^{4})}\approx\frac{q^{-}_{e^{-}e^{+}\rightarrow\nu_{\mu}\bar{\nu}_{\mu}}H}{4(7/8\sigma_{B}T^{4})}\label{depth2}
\end{equation}

Furthermore, the neutrino scattering processes include elastic
scattering off background nucleons,
$\nu_{i=e,\mu,\tau}+\{p,\:n\}\rightarrow\nu_{i=e,\mu,\tau}+\{p,\:n\}$.
The total $\nu_{i}-p$ scattering cross section (Burrows \& Thompson
2004)
\begin{equation}
\sigma_{p}=\frac{\sigma_{0}}{4}\left(\frac{\varepsilon_{\nu}}{m_{e}c^{2}}\right)^{2}[(C_{V}-1)^{2}+3g_{A}^{2}(C_{A}-1)^{2}],\label{scatt1}
\end{equation}
and the transport cross section is
\begin{equation}
\sigma_{p}^{tr}=\frac{\sigma_{0}}{6}\left(\frac{\varepsilon_{\nu}}{m_{e}c^{2}}\right)^{2}[(C_{V}-1)^{2}+5g_{A}^{2}(C_{A}-1)^{2}],\label{scatt2}
\end{equation}
where $\sigma_{0}=1.705\times10^{-44}$ cm$^{2}$, $\varepsilon_{\nu}$
is the average neutrino energy. The total $\nu_{i}-n$ scattering
cross section is
\begin{equation}
\sigma_{n}=\frac{\sigma_{0}}{4}\left(\frac{1+3g_{A}^{2}}{4}\right),\label{scatt3}
\end{equation}
and the transport cross section is
\begin{equation}
\sigma_{n}^{tr}=\frac{\sigma_{0}}{4}\left(\frac{1+5g_{A}^{2}}{6}\right),\label{scatt3}
\end{equation}

Neutrino-electron scattering rate is given by Tubbs \& Schramm
(1975) for electrons with different physical properties: (a)
relativistic, nondegenerate electrons:
\begin{equation}
\sigma_{e\nu_{e}}=\frac{3}{8}\sigma_{0}\frac{\varepsilon_{\nu}}{m_{e}c^{2}}k_{B}T[(C_{V}+C_{A})^{2}+(C_{V}-C_{A})^{2}/3],\label{scatt4}
\end{equation}
(b) relativistic degenerate electrons, and high energy neutrinos
$\varepsilon_{\nu}\gg\varepsilon_{F}$ (Fermi energy of degenerate
electrons):
\begin{equation}
\sigma_{e\nu_{e}}=\frac{3}{32}\sigma_{0}\frac{\varepsilon_{\nu}\varepsilon_{F}}{m_{e}c^{2}}[(C_{V}+C_{A})^{2}+(C_{V}-C_{A})^{2}/3],\label{scatt5}
\end{equation}
(c) relativistic degenerate electrons, and low energy neutrinos
$\varepsilon_{\nu}\ll\varepsilon_{F}$
\begin{equation}
\sigma_{e\nu_{e}}=\frac{1}{32}\sigma_{0}\left(\frac{\varepsilon_{\nu}}{m_{e}c^{2}}\right)^{2}\left(\frac{\varepsilon_{\nu}}{\varepsilon_{F}}\right)[(C_{V}+C_{A})^{2}+(C_{V}-C_{A})^{2}/3],\label{scatt6}
\end{equation}
As a result, the neutrino scattering optical depth, which is
contributed by neutrino-nucleon scattering and neutrino-electron, is
given by
\begin{equation}
\tau_{s,\nu_{i}(e,\mu,\tau)}=H[\sigma_{p,\nu_{i}}n_{p}+\sigma_{n,\nu_{i}}n_{n}+\sigma_{e,\nu_{i}}(n_{e^{-}}+n_{e^{+}})]\label{depth3}
\end{equation}

Using the formulae of neutrino optical depths (\ref{depth1}),
(\ref{depth2}) and (\ref{depth3}) above, the total neutrino cooling
rate per unit area is (Popham \& Narayan 1995; Di Matteo et al.
2002)
\begin{equation}
Q^{-}_{\nu}=\sum_{i=e,\mu,\tau}\frac{(7/8)\sigma_{B}T^{4}}{(3/4)[\tau_{\nu_{i}}/2+1/\sqrt{3}+1/(3\tau_{a,\nu_{i}})]}\label{cooling1}
\end{equation}
where the total optical depth of the neutrino is
$\tau_{\nu_{i}}=\tau_{a,\nu_{i}}+\tau_{s,\nu_{i}}$. Formula
(\ref{cooling1}) is designed to operate in both the optically very
thin and optically very thick limits.

\textit{\textbf{Thermodynamics}}

The cooling term in energy equation (\ref{ener3}) has two
contributions: neutrino emission (see formula \ref{cooling1}) and
photodisintegration of $\alpha$-particles
\begin{equation}
Q^{-}=Q^{-}_{\nu}+Q^{-}_{\rm photodis},\label{cooling2}
\end{equation}
where
\begin{equation}
Q_{\rm
photodis}=6.8\times10^{38}\,\textrm{ergs}\,\textrm{cm}^{-3}\,\textrm{s}^{-1}
\rho\left(\frac{A}{4}\right)^{-1}\left(\frac{B}{28.3\textrm{Mev}}\right)\left(\frac{v_{r}}{\textrm{cm}\,\textrm{s}^{-1}}\right)
\left(\frac{dX_{nuc}/dr}{\textrm{cm}^{-1}}\right)\label{photodis1}
\end{equation}
with $A$ being the mass number of the nucleons, $B$ being the
binding energy of the nucleons (=28.3 Mev for $^{4}$He) and
$X_{nuc}$ being the mass fraction of free nucleons
\begin{equation}
X_{nuc}=\textrm{min}\{1,\quad295.5\rho_{10}^{-3/4}T_{11}^{9/8}\textrm{exp}(-0.8209/T_{11})\}.\label{photodis2}
\end{equation}
The effect of photodisintegration can be neglected in the region
with sufficiently high temperature and rare $\alpha$-particles
$X_{nuc}\approx1$.

The advection term in energy equation (\ref{ener3}) is (Di Matteo et
al. 2002; Liu et al. 2007)
\begin{eqnarray}
Q_{\rm adv}&=&\rho v_{r}HT\frac{ds}{dr}\nonumber\\
&&\approx\xi
v_{r}\frac{H}{r}T\left(\frac{4}{3}a_{B}T^{4}+\frac{3}{2}\rho\Re\frac{1+3X_{nuc}}{4}+\frac{4}{3}\frac{u_{\nu}}{T}\right),\label{adv1}
\end{eqnarray}
where $\xi\propto -d\textrm{ln}s/d\textrm{ln}r$ can be approximately
taken as equal to 1. Another expression of $Q_{\rm adv}$ can be seen
in Zhang \& Dai (2008a) that
\begin{equation}
Q_{\rm adv}\approx\frac{\dot{M}T}{4\pi
r^{2}}\left[\frac{\Re}{2}(1+Y_{e})\left(\frac{1+3X_{nuc}}{4}\right)+\frac{4}{3}g_{*}\frac{aT^{3}}{\rho}+\frac{4}{3}\frac{u_{\nu}}{T}\right],\label{adv2}
\end{equation}
where $Y_{e}$ is the radio of electron to nucleon number density,
the parameter $g_{*}=2$ for photons and $g_{*}=11/2$ for a plasma of
photons together with relativistic $e^{-}-e^{+}$ pairs.

The total pressure in the disk is
\begin{equation}
p=p_{e}+p_{\rm rad}+p_{\rm gas}+p_{v}.\label{pre1}
\end{equation}
The pressure $p_{e}$ is contributed by electrons and positrons
$p_{e}=p_{e^{-}}+p_{e^{+}}$ with
\begin{equation}
p_{e^{\pm}}=\frac{1}{3}\frac{m_{e}^{4}c^{5}}{\pi^{2}\hbar^{3}}
\int^{\infty}_{0}\frac{x^{4}}{\sqrt{x^{2}+1}}\frac{dx}{e^{(m_{e}c^{2}\sqrt{x^{2}+1}\mp\mu_{e})/k_{B}T}+1},\label{pre2}
\end{equation}
where $\mu_{e}$ is the chemical potential of the electron gas. The
radiation pressure is
\begin{equation}
p_{\rm rad}=a_{B}T^{4}/3,\label{pre3}
\end{equation}
while Popham et al. (1999), Narayan et al. (2001), Di Matteo et al.
(2002) took $p_{\rm rad}=11a_{B}T^{4}/12$, which include the
contribution of relativistic $e^{-}-e^{+}$ pairs. We calculate the
pressure of $e^{-}-e^{+}$ pairs in the term of $P_{e}$, not $P_{\rm
rad}$. Moreover, The gas pressure, can be written as
\begin{equation}
p_{\rm gas}=\rho\Re T\left(\frac{1+3X_{nuc}}{4}\right),\label{pre3}
\end{equation}
as ideal gas pressure is still a good approximation in the
neutrino-cooled disks. A more elaborate expression considered the
series correction is (Janiuk et al. 2007)
\begin{equation}
p_{i(p,n)}=\frac{2\sqrt{2}}{3\pi^{2}}\frac{(m_{i}c^{2})^{4}}{(\hbar
c)^{3}}\beta_{i}^{5/2}[F_{3/2}(\eta_{i},\beta_{i})+\beta_{i}F_{5/2}(\eta_{i},\beta_{i})/2],\label{pre4}
\end{equation}
where $\beta_{i}=kT/m_{e}c^{2}$, $\eta_{i}=\mu_{i}/k_{B}T$ with
$\mu_{i}$ being the chemical potential, and
$F_{k}(\eta_{i},\beta_{i})$ is the incomplete Fermi-Dirac integral
for an index $k$. The pressure of neutrinos $p_{\nu}=u_{\nu}/3$, on
the other hand, can be neglected in neutrino-cooled disks for most
cases.

\textit{\textbf{Charge Equilibrium and Chemical Equilibrium}}

The charge equilibrium requires
\begin{equation}
n_{p}=\frac{\rho Y_{e}}{m_{B}}=n_{e^{-}}-n_{e^{+}},\label{charge1}
\end{equation}
where
\begin{equation}
n_{e^{\mp}}=8\pi\left(\frac{m_{e}c}{h}\right)^{3}\int_{0}^{\infty}x^{2}dx\frac{dx}{e^{(m_{e}c^{2}\sqrt{x^{2}+1}\mp\mu_{e})/k_{B}T}+1},\label{charge2}
\end{equation}
The chemical equilibrium requires
\begin{eqnarray}
n_{p}(\Gamma_{p+e^{-}\rightarrow
n+\nu_{e}}+\Gamma_{p+\bar{\nu}_{e}\rightarrow
n+e^{+}}+\Gamma_{p+e^{-}+\bar{\nu}_{e}\rightarrow n}) \nonumber\\=
n_{n}(\Gamma_{n+e^{+}\rightarrow
p+\bar{\nu}_{e}}+\Gamma_{n\rightarrow
p+e^{-}+\nu_{e}}+\Gamma_{n+\nu_{e}\rightarrow
p+e^{-}})\label{chemical1}
\end{eqnarray}
We define a coefficient $K$ equal to
$G_{F}^{2}C_{V}^{2}(1+3g_{A}^{2})/2\pi^{3}\hbar^{7}c^{6}$, then the
transition reaction rates from neutrons to protons and from protons
to neutrons are
\begin{equation}
\Gamma_{p+e^{-}\rightarrow
n+\nu_{e}}=K\int_{Q}^{\infty}dE_{e}E_{e}\sqrt{E_{e}^{2}-m_{e}^{2}c^{4}}
(E_{e}-Q)^{2}f_{e^{-}},\label{chemical21}
\end{equation}
\begin{equation}
\Gamma_{p+e^{-}\leftarrow
n+\nu_{e}}=K\int_{Q}^{\infty}dE_{e}E_{e}\sqrt{E_{e}^{2}-m_{e}^{2}c^{4}}
(E_{e}-Q)^{2}(1-f_{e^{-}}),\label{chemical22}
\end{equation}
\begin{equation}
\Gamma_{n+e^{+}\rightarrow
p+\bar{\nu}_{e}}=K\int_{m_{e}c^{2}}^{\infty}dE_{e}E_{e}\sqrt{E_{e}^{2}-m_{e}^{2}c^{4}}
(E_{e}+Q)^{2}f_{e^{+}},\label{chemical23}
\end{equation}
\begin{equation}
\Gamma_{n+e^{+}\leftarrow
p+\bar{\nu}_{e}}=K\int_{m_{e}c^{2}}^{\infty}dE_{e}E_{e}\sqrt{E_{e}^{2}-m_{e}^{2}c^{4}}
(E_{e}+Q)^{2}(1-f_{e^{+}}),\label{chemical24}
\end{equation}
\begin{equation}
\Gamma_{n\rightarrow
p+e^{-}+\nu_{e}}=K\int_{m_{e}c^{2}}^{Q}dE_{e}E_{e}\sqrt{E_{e}^{2}-m_{e}^{2}c^{4}}
(Q-E_{e})^{2}(1-f_{e^{-}}),\label{chemical25}
\end{equation}
\begin{equation}
\Gamma_{n\leftarrow
p+e^{-}+\nu_{e}}=K\int_{m_{e}c^{2}}^{Q}dE_{e}E_{e}\sqrt{E_{e}^{2}-m_{e}^{2}c^{4}}
(Q-E_{e})^{2}f_{e^{-}},\label{chemical26}
\end{equation}
where $f_{e^{\mp}}=\{\textrm{exp}[(E_{e}-\mu_{e})/k_{B}T]+1\}^{-1}$.
If the neutrinos are perfectly thermalized and have an ideal
Fermi-Dirac distribution, we can simplify equation (\ref{chemical1})
as
\begin{equation}
\left(\frac{n_{n}}{n_{p}}\right)_{\rm
Eq}=\textrm{exp}\left\{\frac{\mu_{e}-Q}{k_{B}T}\right\}\label{chemical31}
\end{equation}
for the neutrino-opaque limit and (Yuan 2005)
\begin{equation}
\left(\frac{n_{n}}{n_{p}}\right)_{\rm
Eq}=\textrm{exp}\left\{\frac{2\mu_{e}-Q}{k_{B}T}\right\}\label{chemical32}
\end{equation}
for the neutrino-transparent limit\footnote{Beloborodov 2003 derived
the relation
$(n_{n}-n_{p})/(n_{n}+n_{p})=0.487(2\mu_{e}-Q)/(k_{B}T)$ for the
neutrino-transparent limit.}. A combined form for both
neutrino-opaque and transparent cases can be given by (Lee et al.
2005; Liu et al. 2007)
\begin{equation}
\textrm{ln}\left(\frac{n_{n}}{n_{p}}\right)_{\rm
Eq}=f(\tau_{\nu})\left\{\frac{2\mu_{e}-Q}{k_{B}T}\right\}+[1-f(\tau_{\nu})]\left\{\frac{\mu_{e}-Q}{k_{B}T}\right\}\label{chemical33}
\end{equation}

Let us give a brief summary in this section. We have list all the
formulae and equations for determining the structure of
neutrino-cooled disks as functions of accretion rate $\dot{M}$,
central BH mass $\dot{M}$, and the radius $r$ completely. The
accretion disks satisfies the conservation equations (\ref{mass3}),
(\ref{momen5}), (\ref{momen7}) and (\ref{ener3}). The total neutrino
cooling rate is expressed as formula (\ref{cooling1}), while
formulae (\ref{pair1}) to (\ref{depth3}) show how to calculate
various cooling processes and neutrino optical depths.
Photodisintegration is showed as formula (\ref{photodis1}), while
advection flux can be estimated using formula (\ref{adv1}) or
(\ref{adv2}). The total pressure is contributed by four terms as
showed in equation (\ref{pre1}), with formulae (\ref{pre2}) to
(\ref{pre4}) discuss each type of pressure. Finally, we need the
charge and chemical equilibrium equations (\ref{charge1}) and
(\ref{chemical1}), with density equation (\ref{charge2}) and
transition rates formulae (\ref{chemical21}) to (\ref{chemical26}).

\textit{\textbf{Effects of Highly Strong Magnetic Fields}}

Now we discuss the neutrino-cooled disk with highly strong magnetic
fields $\geq10^{14}$ G. Such ultrahigh magnetic fields may be
generated by amplification processes in disks such as differential
rotation, or directly form the central BH. Strong magnetic fields
could change the disk physical properties significantly. The effect
of magnetic fields on macrophysical conservation equations have been
discussed in \S 2.1.2 (see Xie et al. 2007 and Lei et al. 2008 for
more discussion). Here er mainly focus on microphysics. The main
modification in the case of strong magnetized disk comes from the
available phase space for the electrons. The phase space for $B=0$
should be replaced as
\begin{equation}
\frac{2}{h^{3}}\int d^{3}p\longrightarrow
\sum_{n=0}^{\infty}g_{nL}\int
\frac{eB_{m}}{h^{2}c}dp_{z},\label{magdisk1}
\end{equation}
where the free degree $g_{nL}=1$ for $n=1$ and $g_{nL}=2$ for
$n\geq1$. Thus the energy of electron is
\begin{equation}
E=\sqrt{p_{z}^{2}c^{2}+m^{2}c^{4}+2n_{L}eB\hbar c},\label{magdisk2}
\end{equation}
which is the form of "Landau $n_{L}$-level of energy". The electron
number density reads
\begin{equation}
n_{e^{\mp}}=\sum_{n=0}^{\infty}g_{nL}\int_{-\infty}^{\infty}dp\frac{eB_{m}}{h^{2}c}f_{e^{\mp}},\label{magdisk3}
\end{equation}
and the charge equilibrium condition satisfies
\begin{equation}
\frac{\rho
Y_{e}}{m_{B}}=\sum_{n=0}^{\infty}g_{nL}\int_{-\infty}^{\infty}dp\frac{eB_{m}}{h^{2}c}(f_{e^{-}}-f_{e^{+}}).\label{magdisk4}
\end{equation}
The pressure of electrons and positrons is modified as
\begin{equation}
P_{e_{\mp}}=\sum_{n=0}^{\infty}g_{nL}\int_{-\infty}^{\infty}dp\frac{eB_{m}}{3h^{2}}
\frac{p^{2}c}{\sqrt{p^{2}c^{2}+m^{2}c^{4}+2n_{L}eB_{m}\hbar
c}}f_{e^{\mp}}.\label{magdisk5}
\end{equation}
The various neutrino cooling rates should also be modified under in
strong magnetic fields environment. We give the examples of the
$e^{-}-e^{+}$ capture rate, which is the most important cooling term
in neutrino-cooled disks:
\begin{equation}
q_{p+e^{-}\rightarrow
n+\nu_{e}}=\frac{G_{F}^{2}C_{V}^{2}(1+3a^{2})}{8
\pi^{3}\hbar^{6}c^{5}}eB_{z}(m_{e}c^{2})^{4}\sum_{n=0}^{\infty}g_{nL}n_{p}\int
\frac{\varepsilon(\varepsilon-q)^{3}d\varepsilon}{\sqrt{\varepsilon^{2}-1-2n_{L}eB_{m}\hbar
c}}f_{e^{-}},\label{magdisk61}
\end{equation}
\begin{equation}
q_{n+e^{-}\rightarrow
p+\bar{\nu}_{e}}=\frac{G_{F}^{2}C_{V}^{2}(1+3a^{2})}{8
\pi^{3}\hbar^{6}c^{5}}eB_{z}(m_{e}c^{2})^{4}\sum_{n=0}^{\infty}g_{nL}n_{p}\int
\frac{\varepsilon(\varepsilon+q)^{3}d\varepsilon}{\sqrt{\varepsilon^{2}-1-2n_{L}eB_{m}\hbar
c}}f_{e^{+}},\label{magdisk62}
\end{equation}
\begin{equation}
q_{n\rightarrow
p+e^{-}+\bar{\nu}_{e}}=\frac{G_{F}^{2}C_{V}^{2}(1+3a^{2})}{8
\pi^{3}\hbar^{6}c^{5}}eB_{z}(m_{e}c^{2})^{4}\sum_{n=0}^{\infty}g_{nL}n_{p}\int
\frac{\varepsilon(q-\varepsilon)^{3}d\varepsilon}{\sqrt{\varepsilon^{2}-1-2n_{L}eB_{m}\hbar
c}}(1-f_{e^{-}}),\label{magdisk63}
\end{equation}

The above equations allows us to solve the structure of
neutrino-cooled disks with ultrahighly magnetic fields.

\section{NDAFs and Neutrino-Cooled Disks II: Solutions}

\textit{\textbf{Analytical Solutions}}

Popham, Woosley \& Fryer (1999) first studied the structure of
neutrino-cooled disks with huge accretion rate from $0.01M_{\odot}$
s$^{-1}$ to $10.0M_{\odot}$ s$^{-1}$. They simplified the formulae
of neutrino cooling emission (\ref{pair1}), (\ref{URCA1}) in last
section as
\begin{equation}
q_{\nu\bar{\nu}}\approx5.0\times10^{33}T_{11}^{9}\,\textrm{ergs}\,\textrm{cm}^{-3}\,\textrm{s}^{-1}\label{Popham1}
\end{equation}
for pair annihilation, and
\begin{equation}
q_{eN}\approx9.0\times10^{33}\rho_{10}T_{11}^{6}X_{nuc}\,\textrm{ergs}\,\textrm{cm}^{-3}\,\textrm{s}^{-1}\label{Popham2}
\end{equation}
for the capture of pairs on nucleons. Thus the neutrino cooling rate
over a half-thickness disk can be written as
\begin{equation}
Q_{\nu}^{-}=(q_{\nu\bar{\nu}}+q_{eN})H.\label{Popham3}
\end{equation}
Moreover, the pressure of electrons is assumed mainly due to
relativistic degenerate electrons; $P_{e}\propto(\rho _{e}Y)^{4/3}$
with $Y_{e}$ being fixed as 0.5. Based on these simplifications,
Narayan, Piran \& Kumar (2001) derived the analytical solutions of
NDAFs. They took the disk mass ($m_{d}M_{\odot}$) and the outer
radius ($r_{out}R_{S}$, $R_{S}$ is the Schwarzschild radius of the
central BH). GRBs could only be produced when the disk is small
enough, i.e., $r_{out}<118\alpha_{-1}^{-2/7}m_{3}^{-1}m_{d}^{3/7}$
with $\alpha_{-1}=\alpha/0.1$, $m_{3}=M/(3M_{\odot})$ and
$\dot{m}=\dot{M}/M_{\odot}\,\textrm{s}^{-1}$, because small disk can
be sufficiently cooled via neutrino emission and accreted most of
its mass into the central BH. On the other hand, large accretion
disk with its CDAF structure are unlikely to produce GRBs since very
little cooling mass reaches the BH. When the outer radius satisfies
$26.2\alpha_{-1}^{-2/7}m_{3}^{-46/49}m_{d}^{20/49}<r_{out}<118\alpha_{-1}^{-2/7}m_{3}^{-1}m_{d}^{3/7}$,
gas pressure dominates in the NDAF. As a result, the temperature,
accretion timescale and accretion rate of the disk are
\begin{eqnarray}
&&T_{11}=0.548\alpha_{-1}^{1/5}m_{3}^{-1/5}r^{-3/10},\nonumber\\
&&t_{acc}=2.76\times10^{-2}\alpha_{-1}^{-6/5}m_{3}^{6/5}r_{out}^{4/5},\nonumber\\
&&\dot{m}=36.2\alpha_{-1}^{6/5}m_{3}^{-6/5}m_{d}r_{out}^{-4/5}.\label{NPK01}
\end{eqnarray}
When $r_{out}<26.2\alpha_{-1}^{-2/7}m_{3}^{-46/49}m_{d}^{20/49}$,
electron degeneracy pressure takes over from gas pressure and the
disk quantities are
\begin{eqnarray}
&&T_{11}=0.800\alpha_{-1}^{1/6}m_{3}^{-13/42}m_{d}^{1/21}r^{-5/12},\nonumber\\
&&t_{acc}=2.82\times10^{-3}\alpha_{-1}^{-1}m_{3}^{13/7}m_{d}^{-2/7}r_{out}^{3/2},\nonumber\\
&&\dot{m}=355\alpha_{-1}m_{3}^{-13/7}m_{d}^{9/7}r_{out}^{-3/2}.\label{NPK02}
\end{eqnarray}
However, these solutions are derived based on the assumption that
the accreting gas is neutrino optically thin. Since the NDAF may
become neutrino optically thick when
$r<20.5\alpha_{-1}^{-4/17}m_{3}^{-110/119}m_{d}^{46/119}$. Inside
the radius, neutrino transport should be considered.

The accretion timescale $t_{acc}$ is sensitive dependence on the
viscosity parameter $\alpha$. However, the accretion timescale is
unlikely to be sufficiently long to explain long GRBs, because the
above discussions are based on the scenario that the disk or debris
torus is formed after mergers immediately without supplementary
feeding material. If the accretion disk can be fed by fallback of
material after the collapse of a massive star, as in the collapsar
model, then the timescale of the burst id determined by fallback,
not accretion. Such a disk can produce long GRBs. On the other hand,
as showed in numerical calculations, a disk with a fixed huge
hyperaccretion rate may contains different regions such as
convection-dominated region, neutrino-dominated and
gas-pressure-dominated or degeneracy-pressure-dominated region at
the same time. This scenario is different to that in Narayan et al.
(2001) that a disk is either CDAF or NDAF, depending on its size and
total mass.

Kohri \& Mineshige (2002) used semi-analytical method to study NDAFs
with high surface density $\Sigma\geq10^{20}$ g cm$^{-2}$, which is
realized when about a solar mass of accreting matter is contained
within a disk of size $\sim5\times10^{6}$ cm. Electron degeneracy
pressure dominates over gas and radiation pressure. This result is
consistent with that in Narayan et al. (2001). On the other hand,
Kohri \& Mineshige (2002) also showed the discrepancies between
their results and those in Narayan et al. (2001). For example, the
relation $p_{\rm rad}\gg p_{\rm gas}$ is always realized in the
nondegenerate electron region at $T\geq m_{e}c^{2}/k_{B}$, while
Narayan et al. (2001) showed $p_{\rm gas}>p_{\rm rad}$ in this
region. More elaborate numerical calculations (e.g., Kohri et al.
2005) support the analytical result of Kohri \& Mineshige (2002).
Table \ref{tab216} gives the five cases for $T-\Sigma$ and
$\dot{M}-\Sigma$ relations from Kohri \& Mineshige (2002).
\begin{table}
\begin{center} {\scriptsize
\begin{tabular}{lcccccc}
\hline\hline
Region & $p$ & $Q^{+}_{\rm vis}$ & $Q^{-}_{\nu}$ & $T-\Sigma$ relation & $\dot{M}-\Sigma$ relation \\
\hline
I   & $p_{\rm gas}$ & $\alpha\Sigma Tr^{-3/2}M^{1/2}$  & $Q_{\rm rad}^{-}\propto \Sigma^{-1}T^{4}$ & $\alpha^{1/3}\Sigma^{2/3}r^{-1/2}M^{1/6}$ & $\alpha^{4/3}\Sigma^{5/3}rM^{-1/3}$\\
II  & $p_{\rm rad}$ & $\alpha\Sigma^{-1}T^{8}r^{3/2}M^{-1/2}$ & $Q_{\rm rad}^{-}\propto \Sigma^{-1}T^{4}$ & $\alpha^{-1/4}\Sigma^{0}r^{-3/8}M^{1/6}$ & $\alpha^{-1}\Sigma^{-1}r^{3/2}M^{-1/2}$\\
III & $p_{\rm rad}$ & $\alpha\Sigma^{-1}T^{8}r^{-3/2}M^{-1/2}$ & $Q_{\rm adv}^{-}\propto \alpha\Sigma^{-3}T^{16}r^{11/2}M^{-5/2}$ & $\alpha^{0}\Sigma^{1/4}r^{-1/2}M^{1/4}$ & $\alpha\Sigma r^{1/2}M^{1/2}$\\
IV & $p_{\rm d,rel}$& $\alpha\Sigma^{9/7}Tr^{-27/14}M^{9/14}$ & $Q_{\rm adv}^{-}\propto \alpha\Sigma T^{2}r^{-1/2}M^{-1/2}$ & $\alpha^{0}\Sigma^{1/7}r^{-5/7}M^{4/7}$ & $\alpha\Sigma^{9/7}r^{15/14}M^{-5/14}$\\
V  & $p_{\rm d,rel}$& $\alpha\Sigma^{9/7}Tr^{-27/14}M^{9/14}$ & $Q_{\nu}^{-}\propto \eta_{e}^{9}T^{9}\Sigma^{1/7}r^{9/7}M^{-3/7}$ & $\Sigma^{4/7}r^{-6/7}M^{2/7}$ & $\alpha\Sigma^{9/7}r^{15/14}M^{-5/14}$\\
\hline \hline
\end{tabular} }
\end{center}
\caption{$T-\Sigma$ and $\dot{M}-\Sigma$ relations in various
regions. Here $p$ shows the dominant type of pressure. $p_{\rm
d,rel}$ is the electron relativistic degeneracy pressure (Kohri \&
Mineshige 2002).}\label{tab216}
\end{table}

Zhang \& Dai (2008a) also obtained a group of analytical solutions
both for ADAF we do not consider the effect of convection) and NDAF
region in a disk with fixed accretion rate. The disk region for a
radiation-pressure-dominated ADAF is
\begin{eqnarray}
&&r_{6}f^{1/5}>2.28m^{-3/5}\dot{m}_{d}^{6/5}\alpha_{-1}^{-2},\nonumber\\
&&r_{6}f^{-7/3}<74.6\left(1+Y_{e}\right)^{-8/3}m^{7/3}\dot{m}_{d}^{-2/3}\alpha_{-1}^{2/3},\nonumber\\
&&r_{6}f^{-7/3}<74.6\left(1+Y_{e}\right)^{-8/3}m^{7/3}\dot{m}_{d}^{-2/3}\alpha_{-1}^{2/3},\label{ZD1}
\end{eqnarray}
where the notations in these relations are slightly different to
those in Narayan et al. (2001): $\dot{M}=0.01\dot{m}_{d}M_{\odot}$
s$^{-1}$, $M=1.4mM_{\odot}$, $r=10^{6}r_{6}$ cm. Also, the
gas-pressure-dominated ADAF region satisfies
\begin{eqnarray}
&&r_{6}^{47/22}f^{-20/11}>128\left(1+Y_{e}
\right)^{-26/11}m^{29/22}\dot{m}_{d}^{10/11}\alpha_{-1}^{-21/11},\nonumber\\
&&r_{6}f^{-7/3}>74.6\left(1+Y_{e}\right)^{-8/3}
m^{7/3}\dot{m}_{d}^{-2/3}\alpha_{-1}^{2/3}.\label{ZD2}
\end{eqnarray}
The neutrino-cooled dominated region, where satisfies $Q_{\rm
adv}<Q_{\nu}^{-}$ is\footnote{In this thesis, neutrino-cooled
accretion flows include both ADAFs and NDAFs, because both ADAFs and
NDAFs are cooled by neutrino emissions. In analytical solutions we
consider $Q_{\rm adv}>Q_{\nu}^{-}$ for ADAF case and $Q_{\rm
adv}<Q_{\nu}^{-}$ for NDAF case respectively.}
\begin{equation}
r_{6}^{47/22}f^{-20/11}<128\left(1+Y_{e}
\right)^{-26/11}m^{29/22}\dot{m}_{d}^{10/11}\alpha_{-1}^{-21/11},\label{ZD3}
\end{equation}
and the gas-pressure-dominated NDAF is the region where
\begin{eqnarray}
&&r_{6}f^{-20/33}>3.18Y_{e}^{80/33}\left(1+Y_{e}
\right)^{-36/11}m^{1/3}\dot{m}_{d}^{20/33}\alpha_{-1}^{-38/33},\nonumber\\
&&r_{6}f^{-20/33}<6.90\alpha_{-1}^{-38/33}f^{20/33}m^{1/3}\dot{m}_{d}^{20/33}(1+Y_{e})^{-28/33}.\label{ZD4}
\end{eqnarray}
Finally, we derived the region where is
degeneracy-pressure-dominated NDAF
\begin{eqnarray}
&&r_{6}f^{-7/3}>8.07\times19^{3}m^{7/3}\dot{m}_{d}^{-2/3}\alpha_{-1}^{2/3}Y_{e}^{-44/3}\left(1+Y_{e}
\right)^{12}\alpha_{-1}^{-38/33},\nonumber\\
&&r_{6}f^{-7/3}>174m^{7/3}\alpha_{-1}^{2.3}\dot{m}_{d}^{-2/3}Y_{e}^{-8/3},\label{ZD5}
\end{eqnarray}
which could only be satisfied for very large radius.

On the other hand, for a fixed radius $r$, we could obtain the range
of $\dot{m}_{d}$ for two types of accretion flows with various
dominated pressure. More details about the analytical solutions of
neutrino-cooled disk will be discussed in \S 3.3.1.

\textit{\textbf{Numerical Solutions}}

Numerical methods have been carried out to study the neutrino-cooled
disks first by Popham et al. (1999), then by Di Matteo et al.
(2002), Lee \& Ramirez-Ruiz (2002), Lee et al. (2004, 2005), Janiuk
et al. (2004, 2007), Kohri et al. (2005), Gu et al. (2006), Chen \&
Beloborodov (2007), Liu et al. (2007), Shibata et al. (2007), etc.
Their numerical outcomes are slightly different with each other, due
to the different simplifications and energy transport forms.

Popham, Woosley \& Fryer (1999) adopted the conservation equations
based on Kerr geometry, but used simplified equations of neutrino
emission and pressure mentioned above. They did not consider the
effect of neutrino opacity. The disk is advection-dominated outside
about $10^{8}$ cm, and neutrino-dominated inside. Di Matteo et al.
(2002) followed Kohri \& Mineshige (2002) to discuss various
neutrino cooling processes, and focused on the effect of neutrino
opacity at high $\dot{M}$. They found that, for
$\dot{M}\geq1M_{\odot}$ s$^{-1}$, neutrinos are sufficiently trapped
that energy advection becomes the dominant cooling mechanism in the
flow. Thus the neutrino luminosity would decrease for high accretion
rate, and $\nu\bar{\nu}$ annihilation above the neutrino-cooled disk
is an inefficient mechanism. Similar to Popham et al. (1999), they
mainly considered the contribution of relativistic degeneracy
pressure $p_{e}\propto (\rho Y_{e})^{4/3}$ in the term of electron
pressure, and fix $Y_{e}$ to be 0.5 in their calculations. Gu et al.
(2006), on the other hand, argued that in the optically thick region
advection does not necessary dominate over neutrino cooling because
the advection factor $f_{\rm adv}=Q_{\rm adv}/Q_{\rm vis}^{+}$ is
relevant to the geometrical depth rather tan the optical depth of
the flow (Abramowicz et al. 1986)
\begin{equation}
f_{\rm adv}\propto(H/r)^{2}.
\end{equation}
This conclusion is consistent with the more elaborate numerical
calculations by Kohri et al. (2005).

\begin{figure}
\centering\resizebox{0.7\textwidth}{!} {\includegraphics{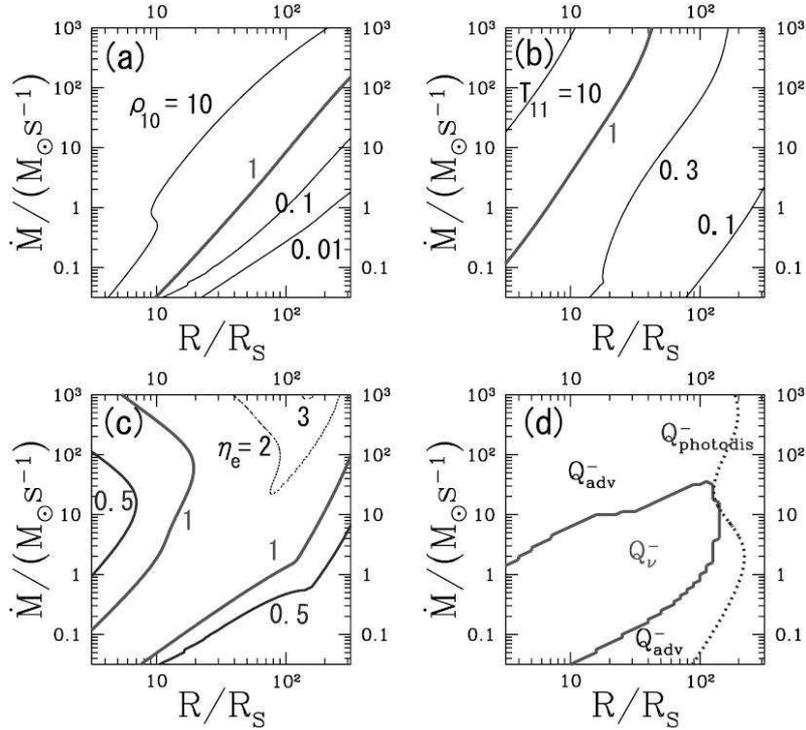}}
\caption{contours of matter density $\rho_{10}=\rho/10^{10}$ g
cm$^{-3}$, temperature $T_{11}=T/10^{11}$ K, the degeneracy
parameter $\eta_{e}$ and the dominant cooling process in the
radius-accretion rate parameter plane (kohri et al.
2005).}\label{fig2161}
\end{figure}

\begin{figure}
\centering\resizebox{0.7\textwidth}{!} {\includegraphics{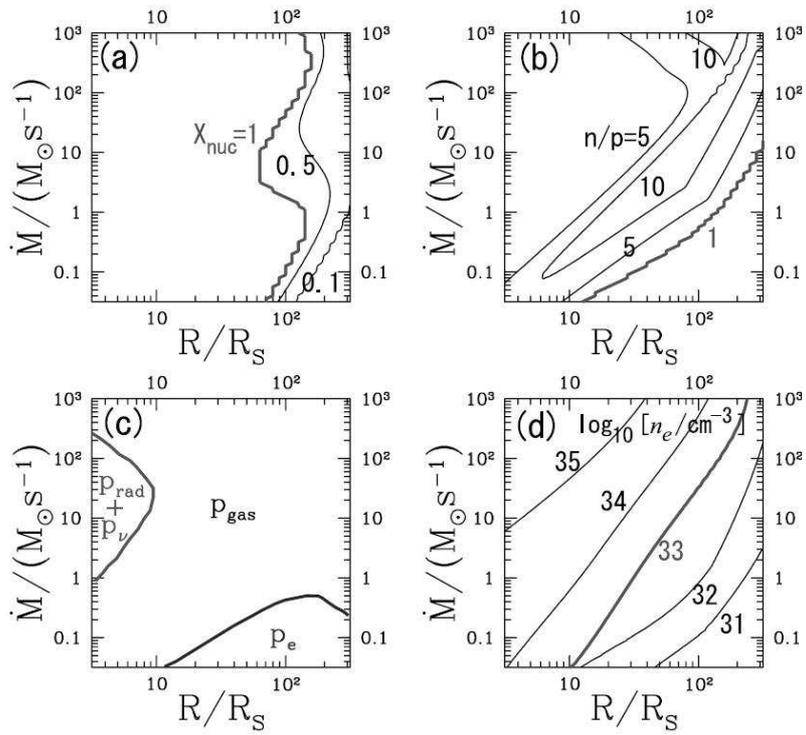}}
\caption{Contour of the nucleon fraction $X_{nuc}$,
neutron-to-proton ratio, the dominant source of pressure and number
density of electrons (kohri et al. 2005).}\label{fig2162}
\end{figure}

Kohri, Narayan \& Piran (2005) calculated the electron pressure and
density using the Fermi-Dirac distribution (\ref{pre2}) and
(\ref{charge2}), and adopt the full equation (\ref{cooling1}) for
neutrino cooling rate, thermal equation (\ref{chemical31}) for
chemical equilibrium calculation. Figure \ref{fig2161} shows the
contours of matter density $\rho_{10}=\rho/10^{10}$ g cm$^{-3}$,
temperature $T_{11}=T/10^{11}$ K, the degeneracy parameter
$\eta_{e}$ and the dominant cooling process in the radius-accretion
rate parameter plane. Moreover, Figure \ref{fig2162} is the contour
of the nucleon fraction $X_{nuc}$, neutrino-to-proton ratio, the
dominant source of pressure and number density of electrons. These
two figures well describe the physical properties of the Newtonian
Keplerian neutrino-cooled disk from accretion rate less than
$0.1M_{\odot}\,\textrm{s}^{-1}$ to
$10^{3}M_{\odot}\,\textrm{s}^{-1}$ in the disk of size
$\sim6\times10^{7}(M/M_{\odot})$ cm.

\begin{figure}
\centering\resizebox{0.8\textwidth}{!} {\includegraphics{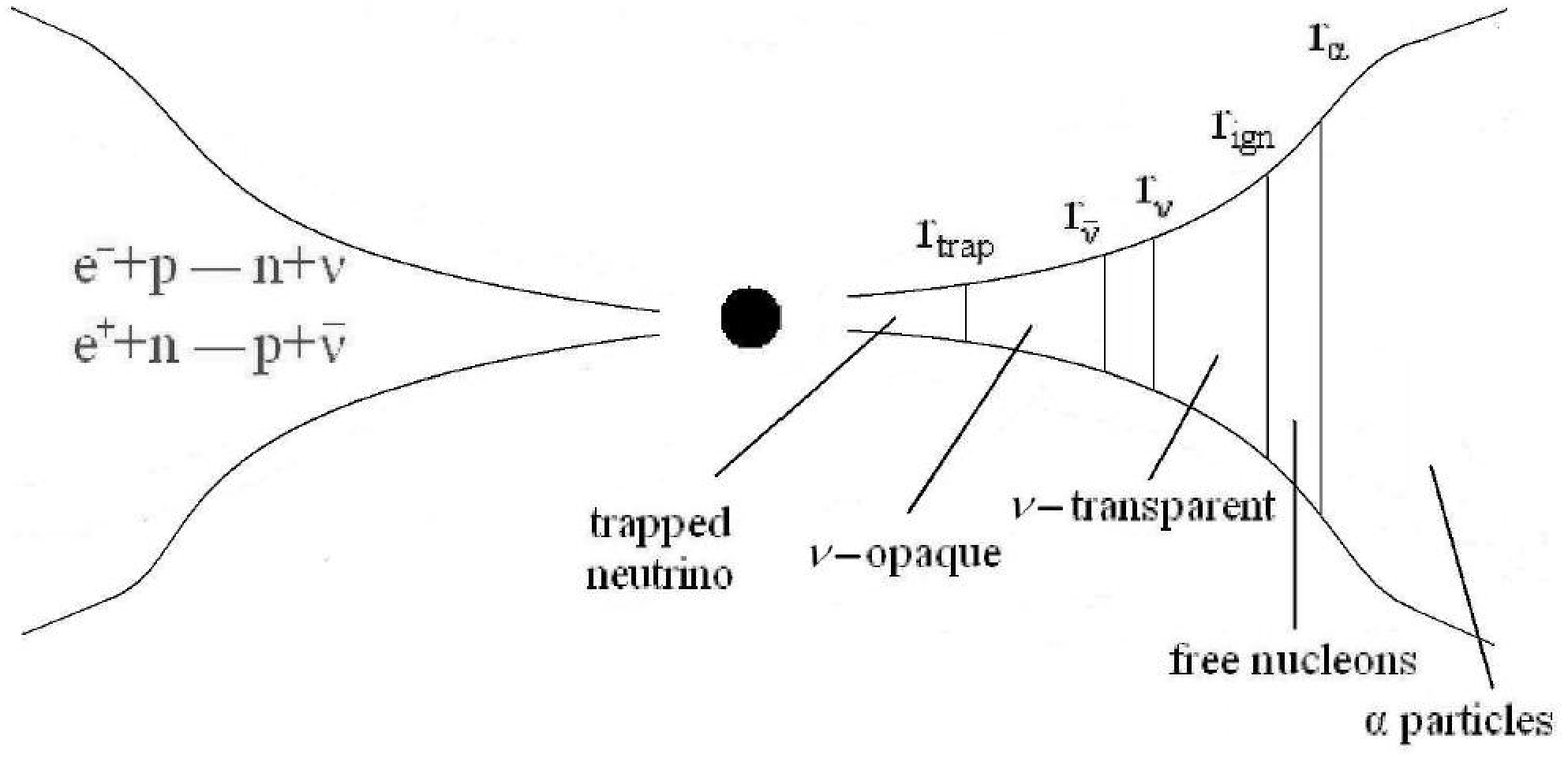}}
\caption{}\label{fig2163}
\end{figure}

Similar to Popham et al. (1999), Chen \& Beloborodov (2007)
presented general relativistic calculation of macrophysical
structure of neutrino-cooled disks around Kerr BHs. Also, they
adopted Fermi-Dirac distributions to calculate the electron density
and pressure. However, they only considered the $e^{-}-e^{+}$ pair
capture onto nucleons among various neutrino cooling processes, and
simplified the equations of neutrino transfer and total cooling
rate. Figure \ref{fig2163} gives the scenario of an accretion disk
around a Kerr BH with a huge accretion rate, where radius
$r_{\alpha}$, $r_{\rm ign}$, $r_{\nu}$, $r_{\bar{\nu}}$, $r_{\rm
tr}$ are five characteristic radii:
\begin{itemize}
\item
Radius $r_{\alpha}$, where 50\% of the $\alpha$-particles are
decomposed into three nucleons via photodisintegration process.

\item
"Ignition" radius $r_{\rm ign}$, where the neutrino emission
switches on, i.e., the accretion flow becomes neutrino-cooled. The
ignition radius exists only if $\dot{M}>\dot{M}_{\rm ign}$, where
the critical accretion rate $\dot{M}_{\rm ign}$ depends on the
viscosity parameter $\alpha$ and spin parameter $a$.

\item
Radius $r_{\nu}$, where the disk becomes opaque for neutrinos and
they relax to a thermal distribution.

\item
Radius $r_{\bar{\nu}}$, where the disk becomes opaque for
antineutrinos. Then both neutrinos and antineutrinos are now in
thermal equilibrium with the matter.

\item
Radius $r_{\rm tr}$, where the timescale of neutrino diffusion
becomes longer than the accretion timescale, and neutrinos get
rapped and advected into the black hole.
\end{itemize}

2D and 3D time-dependent simulations were given by Lee \&
Ramirez-Ruiz (2002), Lee et al. (2004, 2005) and Setiawan et al.
(2004). Lee \& Ramirez-Ruiz (2002) took the disk initial condition
from 3D BH-NS simulations and studied the time dependence of the
disk structure for 0.2 s in 2D cylindrical coordinates ($r,z$). They
showed the dynamical evolution of the accretion disks  formed after
BH-NS mergers. A neutrino-cooled disk could produce a short
impulsive energy input $L_{\nu\bar{\nu}}\propto t^{-5/2}$ via
$\nu\bar{\nu}$-annihilation, while strong fields $B\approx10^{16}$ G
being anchored in the dense matter could also power a short GRB.
However, the equation of state was simplified in Lee \& Ramirez-Ruiz
(2002) with a ideal gas and adiabatic index $\Gamma=4/3$. Also, they
assumed that all the energy dissipated by viscosity is radiated
away. Next Lee et al. (2004) improved the equations of state
following Popham et al. (1999) and Di Matteo et al. (2002), and
considered pair capture on nucleons as the main neutrino cooling
process. The improved simulations last approximately 1 s. The
inclusion of neutrino optical depth effects produces important
qualitative temporal and spatial transitions in the evolution and
structure of the disk, which may directly reflect on the duration
and variability of short GRBs. Later, more elaborate 2D simulations
were carried out, also by Lee et al. (2005), who used much more
details detailed equation of state, neutrino emission processes and
optical depths. They construct the bridging formula for chemical
equilibrium in both optical thin and thick cases (i.e., equation
[\ref{chemical32}]). However, the effect of neutrino absorption was
still neglected, since the total neutrino cooling luminosity is
computed according
\begin{equation}
L_{\nu}=\int\rho^{-1}(q_{eN}+q_{e^{-}e^{+}}+q_{\rm
brem}+q_{plasmon}).\label{Disksim1}
\end{equation}
Also, as conservation equations were not included, the initial
conditions were also taken from the results of BH-NS merger
simulations. Setiawan, Ruffert, \& Janka (2004) presented 3D
simulations of a time-dependent hyperaccretion disk around a BH,
while the initial data is obtained from Janka et al. (1999) with a
BH mass of $4.017M_{\odot}$ and torus mass changing from
$0.0120M_{\odot}$ to $0.1912M_{\odot}$. The accretion duration in
their model is about tens of milliseconds.

The above 2D and 3D simulations stared with initial disk conditions
obtained from the early-phase simulations of compact stars merger,
and focused on disk dynamical evolution less than 1 s. Janiuk et al.
(2004) self-consistently obtained the initial disk structure based
on the equations as showed in Popham et al. (1999), Di Matteo et al.
(2002), Kohri et al. (2005); and extended the study of disk
evolution to much longer timescale. The time-dependent evolution is
mainly cased by density and temperature variation, which can be
determined by time-dependent equations of mass and angular momentum
conservation
\begin{equation}
\frac{\partial\Sigma}{\partial
t}=\frac{1}{r}\frac{\partial}{\partial
r}\left[3r^{1/2}\frac{\partial}{\partial
r}(r^{1/2}\nu\Sigma)\right],\label{Janiuk1}
\end{equation}
\begin{equation}
\frac{\partial T}{\partial t}+v_{r}\frac{\partial T}{\partial
r}=\frac{T}{\Sigma}\frac{4-3\chi}{12-9\chi}\left(\frac{\partial\Sigma}{\partial
t}+v_{r}\frac{\partial\Sigma}{\partial
t}+v_{r}\frac{\partial\Sigma}{\partial
r}\right)+\frac{T}{pH}\frac{1}{12-9\chi}(Q^{+}_{\rm
vis}-Q^{-}_{\nu}),\label{Janiuk2}
\end{equation}
where $\chi=(p-p_{\rm rad})/p$. Moreover, the contribution of
photodisintegration varies with time as
\begin{equation}
Q_{\rm photo}^{-}\propto v_{r}\frac{\partial X_{nuc}}{\partial
r}+\frac{\partial X_{nuc}}{\partial t}.\label{Janiuk3}
\end{equation}
\begin{figure}
\centering\resizebox{0.5\textwidth}{!} {\includegraphics{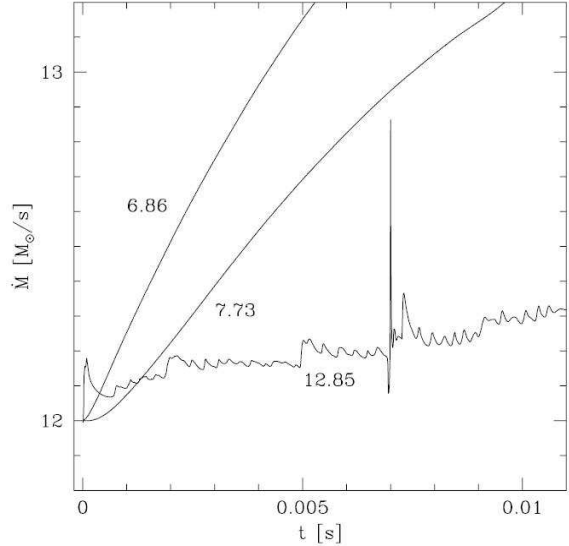}}
\caption{The evolution of local accretion, for several chosen radial
locations in the disk: 6.87, 7.73 and 12.85 $R_{s}$. The starting
accretion rate is $\dot{M}=12M_{\odot}$ s$^{-1}$ (Janiuk et al.
2007).}\label{fig2164}
\end{figure}
Janiuk et al. (2007) found that, for sufficiently large accretion
rates ($\dot{M}\geq10 M_{\odot}$ s$^{-1}$, the inner regions of the
disk develop a viscous and thermal instability, which might be
relevant for GRB observations. For example, Figure \ref{fig2164}
showed the behavior of the local accretion rate in the unstable disk
at several chosen locations with the instability strip. Furthermore,
Metzger, Piro \& Quataert (2008) presented the dynamical evolution
of the neutrino-cooled disks in a timescale of $\sim100$ s based on
their "ring model", which treats the disk as a single annulus that
is evolved forward in time. They also obtained analytic self-similar
time-depend solution that the disk mass $M_{d}\propto t^{-5/8}$, the
accretion rate $\dot{M}\propto t^{-13/8}$ for neutrino-cooled thin
disk, and $M_{d}\propto t^{-1/3}$, $\dot{M}\propto t^{-4/3}$ for
late-time advective case.

\section{Neutrino Annihilation and Jet Production}

In neutrino-cooled disks and other compact high-energy objects, a
significant fraction of the gravitational energy is converted during
the accretion into neutrino flux. This energy is available to
extracted via $\nu\bar{\nu}$-annihilation and subsequent pair
production. The problem is that, could the process of
$\nu\bar{\nu}$-annihilation launch a relativistic jet, and provide
sufficient energy for a GRB? We have mentioned some results of
$\nu\bar{\nu}$-annihilation above disks, now we discuss this process
more detailed.

Motivated by the delayed explosion of Type II supernovae, the energy
deposition rate due to the $\nu\bar{\nu}$-annihilation above the
spherical neutrinosphere has been calculated by Cooperstein et al.
(1986), Goodman et al. (1987), Berezinsky \& Prilutsky (1987) and
later by Salmonson \& Wilson (1999). The $\nu\bar{\nu}\rightarrow
e^{-}e^{+}$ deposition per unit time per volume in Newtonian case is
(Goodman et al. 1987)
\begin{equation}
q_{\nu\bar{\nu}}(\textbf{r})=\int\int
f_{\nu}(\textbf{p}_{\nu},\textbf{r})f_{\bar{\nu}}(\textbf{p}_{\bar{\nu}},\textbf{r})
\{\sigma|\textbf{v}_{\nu}-\textbf{v}_{\bar{\nu}}|\varepsilon_{\nu}\varepsilon_{\bar{\nu}}\}
\frac{\varepsilon_{\nu}+\varepsilon_{\bar{\nu}}}{\varepsilon_{\nu}\varepsilon_{\bar{\nu}}}
\textrm{d}^{3}\textbf{p}_{\nu}\textrm{d}^{3}\textbf{p}_{\bar{\nu}},\label{anni1}
\end{equation}
where $f_{\nu}$ and $f_{\bar{\nu}}$ are the numbers of neutrinos and
antineutrinos per unit volume per momentum state, $\textbf{v}_{\nu}$
is the velocity of the neutrino, and $\sigma$ is the appropriate
rest-frame cross section, which can be calculated as
\begin{equation}
\{\sigma|\textbf{v}_{\nu}-\textbf{v}_{\bar{\nu}}|\varepsilon_{\nu}\varepsilon_{\bar{\nu}}\}
=2\mathscr{K}G_{F}^{2}(\varepsilon_{\nu}\varepsilon_{\bar{\nu}}-\textbf{p}_{\nu}\cdot\textbf{p}_{\bar{\nu}}c^{2})^{2},\label{anni2}
\end{equation}
where $G_{F}^{2}=5.29\times10^{-44}$ cm$^{2}$ Mev$^{-2}$ and
\begin{equation}
\mathscr{K}=\frac{1\pm4\textrm{sin}^{2}\theta_{W}+8\textrm{sin}^{4}\theta_{W}}{6\pi},\label{anni3}
\end{equation}
with the plus sign being for $\nu_{e}\bar{\nu}_{e}$ pairs and minus
sign being for $\nu_{\mu}\bar{\nu}_{\mu}$ and
$\nu_{\tau}\bar{\nu}_{\tau}$ pairs, and
$\textrm{sin}^{2}\theta_{W}\approx0.23$. Taking
$\textrm{d}^{3}\textbf{p}_{\nu}=(\varepsilon_{\nu}^{2}/c^{3})\textrm{d}\varepsilon_{\nu}
\textrm{d}\varepsilon_{\nu}$, we obtain
\begin{equation}
q_{\nu\bar{\nu}}(\textbf{r})=\frac{2\mathscr{K}G_{F}^{2}}{c^{5}}\Theta(\textbf{r})
\int\int
f_{\nu}f_{\bar{\nu}}(\varepsilon_{\nu}+\varepsilon_{\bar{\nu}})\varepsilon_{\nu}^{3}\textrm{d}\varepsilon_{\nu}\varepsilon_{\bar{\nu}}^{3}\textrm{d}\varepsilon_{\bar{\nu}},\label{anni3}
\end{equation}
where the angular integration factor
\begin{equation}
\Theta(\textbf{r})=\int\int(1-\Omega_{\nu}\cdot\Omega_{\bar{\nu}})^{2}\textrm{d}\Omega_{\nu}\textrm{d}\Omega_{\nu}.\label{anni4}
\end{equation}
The phase factor integrations are
\begin{equation}
\int_{0}^{\infty}f_{\nu}\varepsilon_{\nu}^{3}\textrm{d}\varepsilon_{\nu}=\frac{2(k_{B}T)^{4}}{h^{3}}\frac{7\pi^{4}}{120},\label{anni51}
\end{equation}
\begin{equation}
\int_{0}^{\infty}f_{\nu}\varepsilon_{\nu}^{4}\textrm{d}\varepsilon_{\nu}=\frac{2(k_{B}T)^{5}}{h^{3}}\frac{45\zeta(5)}{2}.\label{anni51}
\end{equation}
Thus the $\nu\bar{\nu}$-annihilation energy deposition rate is
\begin{equation}
q_{\nu\bar{\nu}}(\textbf{r})=\frac{14\mathscr{K}G_{F}^{2}\pi^{4}\zeta(5)}{c^{5}h^{6}}(k_{B}T)^{9}\Theta(\textbf{r})
\propto T^{9}\Theta(\textbf{r}),\label{anni6}
\end{equation}
where the angular integration can be analytically derived as
\begin{equation}
\Theta(\textbf{r})=\frac{2\pi^{3}}{3}(1-x)^{4}(x^{2}+4x+5)\label{anni7}
\end{equation}
with $x=\sqrt{1-(R_{\nu}/r)^{2}}$. Here $R_{\nu}$ is the radius of
neutrinosphere. Goodman et al. (1987) estimated the total rate of
annihilation energy input
\begin{eqnarray}
&&Q(\nu_{e}\bar{\nu}_{e})=\int_{R_{\nu}}^{\infty}q(\nu_{e}\bar{\nu}_{e})4\pi
r^{2}dr=0.682\times10^{51}\,\textrm{ergs}\,\textrm{s}^{-1}\nonumber\\
&&Q(\nu_{\mu}\bar{\nu}_{\mu})=Q(\nu_{\tau}\bar{\nu}_{\tau})=0.585\times10^{51}\,\textrm{ergs}\,\textrm{s}^{-1}\nonumber\\
&&Q({\rm
total})=1.9\times10^{51}\,\textrm{ergs}\,\textrm{s}^{-1}.\label{anni8}
\end{eqnarray}
Salmonson \& Wilson (1999) calculated the annihilation rate in
Schwarzchild coordinates
\begin{equation}
Q_{\nu\bar{\nu}}=\int_{R_{\nu}}^{\infty}q_{\nu\bar{\nu}}\frac{4\pi
r^{2}dr}{\sqrt{1-(2GM/c^{2}r)}}\propto
L_{\infty}^{9/4}R_{\nu}^{-3/2},\label{anni8}
\end{equation}
where $L_{\infty}$ is the luminosity at $r\rightarrow\infty$. It is
found that the efficiency of $\nu\bar{\nu}$-annihilation is enhanced
over the Newtonian values up to a factor of more than 4 times for a
spherical neutrinosphere.

Asano \& Fukuyama (2000), on the contrary , argued that the
gravitational effects do not substantially change the annihilation
energy deposition rate. They studied the gravitational effects on
the annihilation rate for neutrinosphere with either spherical or
disk geometry, assuming the neutrinosphere is isothermal and the
gravitational fields by the Schwarzchild BH. The main discrepancy
between Salmonson \& Wilson (1999) and Asano \& Fukuyama (2000), is
probably that, the former considered the proper energy deposition
rate per unit proper time, but the latter computed the deposition
rate per unit world time under the isothermal assumption (Birkl et
al. 2007). Later Asano \& Fukuyama (2001) investigated the effects
of Kerr BH on the neutrino-cooled accretion disks. They found that
the energy deposition rate can be either decreased or increased up
to a factor about 2, depending on the Kerr parameter $a$ and
temperature distribution of the neutrinosphere.

Miller et al. (2003) considered the full 3D calculations of the
$\nu\bar{\nu}$-annihilation above a thin disk around a Kerr BH using
full geodesic equations.Formula (\ref{anni1}) is still used to
calculate the energy deposition rate at a point $\bar{x}$. The
momentum and velocity vector in (\ref{anni1}) is measured in the
local frame at $\bar{x}$ with an orthonormal base
\begin{eqnarray}
&&(\textbf{e}_{t}^{\alpha})=\frac{1}{\sqrt{g_{tt}}}(1,0,0,0),\quad(\textbf{e}_{r}^{\alpha})=\frac{1}{\sqrt{-g_{rr}}}(1,0,0,0),\nonumber\\
&&(\textbf{e}_{\theta}^{\alpha})=\frac{1}{\sqrt{-g_{\theta\theta}}}(0,0,1,0),\nonumber\\
&&(\textbf{e}_{\phi}^{\alpha})=\sqrt{\frac{g_{tt}}{g_{t\phi}^{2}-g_{\phi\phi}g_{tt}}}(0,0,1,0)
\end{eqnarray}
where ($t,r,\theta,\phi$) is measured in the global Boyer-Lindquist
spacetime metric. The steps for evaluating $f_{\nu}(\textbf{p})$
with a chosen direction $\textbf{p}$ in the local frame in
(\ref{anni1}) are as follows:
\begin{itemize}
\item
Transform $\textbf{p}$ to global Boyer-Linquist frame by
$p_{G}^{\alpha}=p^{\beta}e_{\beta}^{\alpha}$.
\item
Trace the geodesics to the disk starting at the observer point back
to the point at the disk ($t_{D},x_{D}$) and 4-momentum
$p_{D}=(E_{D},p_{D})$. The geodesic equation is
\begin{equation}
\frac{dx^{\alpha}}{d\lambda}=p_{G}^{\alpha},\quad\frac{dp^{\alpha}_{G}}{d\lambda}=-\Gamma_{\beta\gamma}^{\alpha}p^{\alpha}_{G}p^{\gamma}_{G}
\end{equation}

\item
Compute
$f_{\nu}(t,\bar{x},p_{G})=f_{\nu}(t_{D},x_{D},p_{D})=[1+\textrm{exp}(E_{D}/k_{B}T_{D})]^{-1}$,
where $E_{D}=p_{D}\cdot U_{D}$ with $u_{D}(t_{D},x_{D})$ being the
4-velocity at the disk.
\end{itemize}

In many cases, $f_{\nu}(\textbf{p})=0$ because the geodesics cannot
be traced back to the surface of the disk. Birkl et al. (2007) used
the similar geodesics trace method to investigate the effects of
general relativity (GR) on the $\nu\bar{\nu}$-annihilation, and they
studied different shapes of the neutrinospheres, such as spheres,
thin disks and thick accretion torus. It was found that, GR effects
increase the total annihilation rate measured by an observer at
infinity by a factor of two when the neutrinosphere is a thin disk,
but the increase is only $\approx25\%$  for toroidal and spherical
neutrinospheres.

\begin{figure}
\centering\resizebox{1.0\textwidth}{!} {\includegraphics{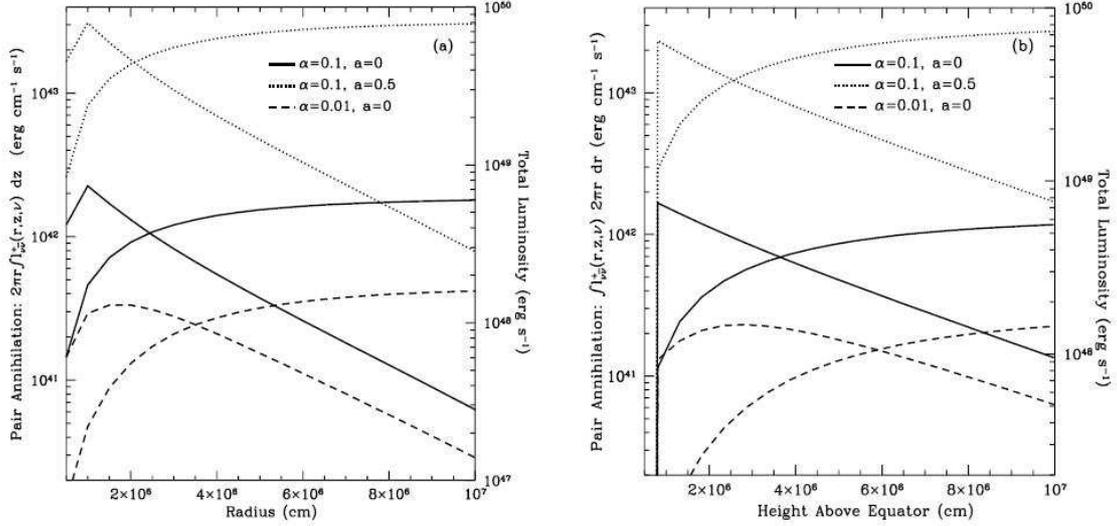}}
\caption{$\nu\bar{\nu}$ annihilation luminosity distribution along
the disk radius (left panel) and the z-direction (right panel) for
$\dot{M}=0.1M_{\odot}$ s$^{-1}$ and the mass of BH $M=3.0M_{\odot}$.
The total luminosity out to each radius is also plotted (Popham et
al. 1999).}\label{fig2171}
\end{figure}

\begin{figure}
\centering\resizebox{0.5\textwidth}{!} {\includegraphics{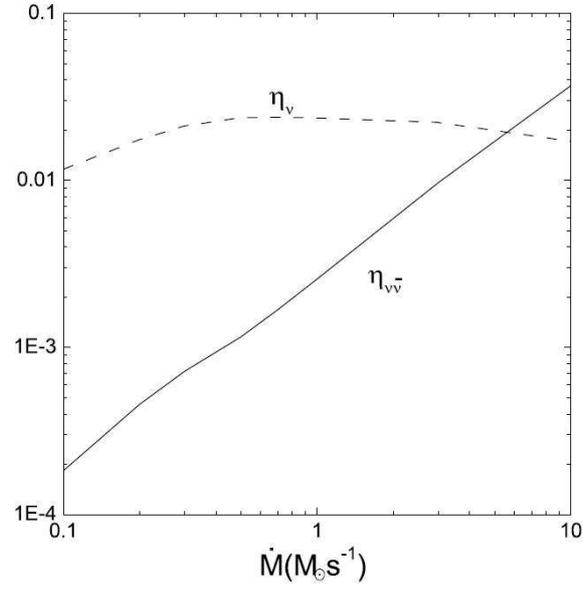}}
\caption{Neutrino emission efficiency $\eta_{\nu}$ and neutrino
annihilation efficiency $\eta_{\nu\bar{\nu}}$ with $M=3M_{\odot}$,
$\alpha=0.1$ (Liu et al. 2007).}\label{fig2172}
\end{figure}

\begin{figure}
\centering\resizebox{0.5\textwidth}{!} {\includegraphics{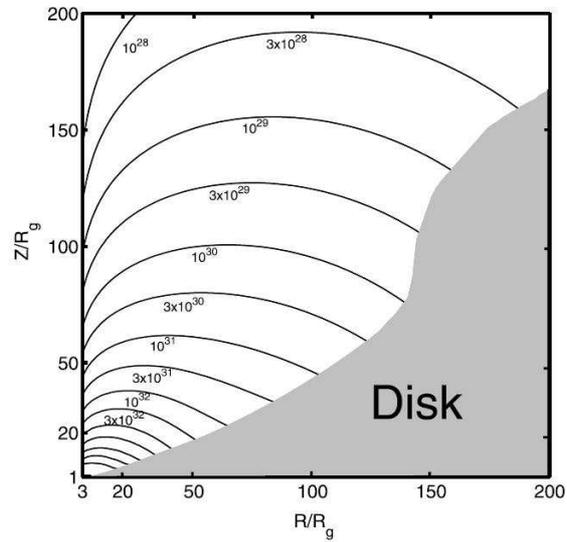}}
\caption{Contours of the $\nu\bar{\nu}$-annihilation luminosity
above the disk in units of ergs s$^{-1}$ cm$^{-2}$. The parameters
are the same as in Fig. \ref{fig2172} (Liu et al.
2007).}\label{fig2173}
\end{figure}

The above works mainly focus on the process of
$\nu\bar{\nu}$-annihilation and the effects of GR. The properties of
disk or torus neutrinosphere is simplified. On the other hand, some
researchers, as mentions in last section \S 2.1.6, derived the disk
structure and neutrino luminosity by solve the group of equations of
state, hydrodynamics, thermodynamics and microphysics in detail; but
they adopted a relatively simple method to calculate the
$\nu\bar{\nu}$-annihilation process. Since GR effect cannot change
the result dramatically, Newtonian scenario is a good approximation
for annihilation calculation. Popham, Woosley \& Fryer (1999)
calculated the neutrino annihilation at any point above a
neutrino-cooled disk by a simplified expression (see also Ruffert et
al. 1997; Rosswog et al. 2003)
\begin{eqnarray}
l_{\nu\bar{\nu}}^{+}(\nu_{i}\bar{\nu}_{i})&=&A_{1}\Sigma_{k}\frac{\Delta
L_{\nu_{i}}^{k}}{d_{k}^{2}}\Sigma_{k'}\frac{\Delta
L_{\nu_{i}}^{k'}}{d_{k'}^{2}}[\varepsilon_{\nu_{i}}+\varepsilon_{\bar{\nu}_{i}}](1-\textrm{cos}\theta_{kk'})^{2}\nonumber\\
&&+A_{2}\Sigma_{k}\frac{\Delta
L_{\nu_{i}}^{k}}{d_{k}^{2}}\Sigma_{k'}\frac{\Delta
L_{\nu_{i}}^{k'}}{d_{k'}^{2}}\frac{\varepsilon_{\nu_{i}}+\varepsilon_{\bar{\nu}_{i}}}{\varepsilon_{\nu_{i}}\varepsilon_{\bar{\nu}_{i}}}
(1-\textrm{cos}\theta_{kk'}),\label{anni9}
\end{eqnarray}
where $A_{1}\approx1.7\times10^{-44}$ cm ergs$^{-2}$ s$^{-1}$ and
$A_{2}\approx1.6\times10^{-56}$ cm ergs$^{-2}$ s$^{-1}$ are the
neutrino cross section constants for electron neutrinos. The disk is
modeled as a grid of cells in the plane with neutrino mean energy
$\varepsilon_{\nu_{i}}$ and luminosity $\Delta L_{\nu_{i}}^{k}$. For
each pair of cells and a given point above the scale half-thickness
of the disk $H(r)$, $d_{k}$ is the distance from each cell to that
point, and ($\theta$) is angle between the neutrinos and
antineutrinos from the pair of cells interact. The summation over
all pairs of cells gives the energy density from pair production
$l_{\nu\bar{\nu}}^{+}(\nu_{i}\bar{\nu}_{i})$ at that point.
Moreover, integrating over the distance above the plane and taking
advantage of the cylindrical symmetry of the disk,
\begin{equation}
2\pi
r\int_{H(r)}^{\infty}l_{\nu\bar{\nu}}^{+}(\nu_{i}\bar{\nu}_{i})dz,\label{anni9}
\end{equation}
demonstrates the distribution of annihilation energy deposition rate
along the disk radius $r$. Further integrating the luminosity over
the equatorial raids gives the total neutrino annihilation
luminosity. Figure \ref{fig2171} gives the $\nu\bar{\nu}$
annihilation luminosity distribution along the disk radius and the
z-direction for $\dot{M}=0.1M_{\odot}$ s$^{-1}$ and the mass of BH
$M=3.0M_{\odot}$. The total luminosity out to each radius is also
plotted in Figure \ref{fig2171} for the same models. In these
examples, a half of the annihilation energy is injected in a narrow
beam with radii$\leq2\times10^{6}$ cm.
\begin{table}
\begin{center} {\footnotesize
\begin{tabular}{lcccccc}
\hline\hline
$\dot{M}$ & $\alpha$ & $a$ & $M$ & $L_{\nu}$ & $L_{\nu\bar{\nu}}$ & Efficiency \\
($M_{\odot}$ s$^{-1}$) & & & ($M_{\odot}$) & (10$^{51}$ ergs s$^{-1}$) & (10$^{51}$ ergs s$^{-1}$) & \% \\
\hline
0.01 & 0.1 & 0 & 3 & 0.015 & $3.9\times10^{-8}$ & 0.0003 \\
0.01 & 0.03 & 0 & 3 & 0.089 & $2.9\times10^{-7}$ & 0.0003 \\
0.01 & 0.01 & 0 & 3 & 0.650 & $9.0\times10^{-6}$ & 0.001 \\
0.01 & 0.1 & 0.5 & 3 & 0.036 & $5.9\times10^{-7}$ & 0.002 \\
0.01 & 0.01 & 0 & 10 & 0.049 & $6.4\times10^{-9}$ & $10^{-5}$ \\
0.05 & 0.1 & 0.5 & 3 & 1.65 & $1.8\times10^{-3}$ & 0.11 \\
0.1 & 0.1 & 0 & 3 & 3.35 & $3.0\times10^{-3}$ & 0.09 \\
0.1 & 0.1 & 0 & 3 & 6.96 & $1.7\times10^{-3}$ & 0.02 \\
0.1 & 0.03 & 0 & 3 & 6.15 & $8.0\times10^{-4}$ & 0.01 \\
0.1 & 0.01 & 0.5 & 3 & 8.03 & 0.039 & 0.5 \\
0.1 & 0.1 & 0.95 & 3 & 46.4 & 2.0 & 4.2 \\
0.1 & 0.1 & 0.95 & 6 & 26.2 & 0.79 & 3.0 \\
1.0 & 0.1 & 0 & 3 & 86.3 & 0.56 & 0.6 \\
1.0 & 0.1 & 0.5 & 3 & 142 & 3.5 & 2.5 \\
10.0 & 0.1 & 0 & 3 & 781? & 200? & 26? \\
10.0 & 0.1 & 0.5 & 3 & 1280? & 820? & 64? \\
\hline \hline
\end{tabular} }
\end{center}
\caption{The neutrino annihilation efficiency for different cases.
(Popham, Woosley, \& Fryer 1999).}\label{tab2171}
\end{table}
Table \ref{tab2171} gives the neutrino annihilation energies and
efficiencies for all models studied by Popham et al. (1999). The
efficiency increases with the increasing of accretion rate and spin
parameter $a$. The energy conversion efficiencies become extremely
high for $\dot{M}=10M_{\odot}$ s$^{-1}$ . However, the results
showed in Table \ref{tab2171} is too optimistic, as they do not
consider neutrino opacity and overestimate the neutrino luminosity.
Di Matteo et al. (2002) showed that the effect of neutrino opacity
becomes significant for $\dot{M}>1M_{\odot}$ s$^{-1}$, and
recalculated the annihilation rate by
\begin{eqnarray}
L_{\nu\bar{\nu}}&\sim&6\times10^{-35}\frac{2\varepsilon_{\nu}L_{\nu}^{2}}{\pi
c^{2}[1-(r_{\rm min}/r_{rm surf})^{2}]^{2}}\nonumber\\
&&\times\frac{1}{r_{rm
surf}}\int_{0}^{\infty}d\epsilon\Phi(\textrm{cos}\theta_{\rm
min}-\textrm{cos}\theta_{\rm
surf})^{2}\,\textrm{ergs}\,\textrm{s}^{-1},\label{anni10}
\end{eqnarray}
where $L_{\nu}$ is the neutrino luminosity, $r_{\rm min}$ is the
inner edge of the disk, $r_{\rm surf}$ is the radius where
$\tau_{\nu}=2/3$, and $\varepsilon_{\nu}$ is the average energy of
the escaping neutrinos. Di Matteo et al. (2002) demonstrated that
the $L_{\nu\bar{\nu}}$ increases up to its maximum value of
$\sim10^{50}\,\textrm{ergs}\,\textrm{s}^{-1}$ at
$\dot{M}>1M_{\odot}$ s$^{-1}$ and decreases for large $\dot{M}$.
Thus Di Matteo et al. (2002) concluded that the annihilation in
hyperaccreting BHs is an inefficient mechanism for liberating large
amounts of energy. However, they neglected the neutrino emission
from the region with neutrino optical depth $\tau>2/3$. Gu et al.
(2006), on the other hand, showed that the energy released by
annihilation is still sufficient, if the contribution from the
optically thick region of the neutrino-cooled disks $\tau>2/3$ is
included. Liu et al. (2007) improved the results of Gu et al. (2007)
based on more elaborate calculations similar to Kohri et al. (2005).
The electron degeneracy and the lower $Y_{e}<0.5$ resulting from the
neutronization processes indeed suppress the neutrino emission
considerably, thus the resulting $L_{\nu}$ and $L_{\nu\bar{\nu}}$
are lower than the simplified calculations by Gu et al. (2006).
However, the corrected $L_{\nu\bar{\nu}}$ still reaches to
$\sim10^{52}$ ergs s$^{-1}$ for $\dot{M}\sim10M_{\odot}$ s$^{-1}$,
thus can provide the energy of GRBs. Figure \ref{fig2172} gives the
neutrino radiation efficiency $\eta_{\nu}(\equiv
L_{\nu}/\dot{M}c^{2})$ and neutrino annihilation efficiency
$\eta_{\nu\bar{\nu}}(\equiv L_{\nu\bar{\nu}}/L_{\nu}$) as a function
of accretion rate. $\eta_{\nu\bar{\nu}}$ increases rapidly with
increasing $\dot{M}$. Figure \ref{fig2173} is the contour of $2\pi
rl_{\nu\bar{\nu}}^{+}$ in units of ergs s$^{-1}$ cm$^{-2}$. Similar
to the results in Popham et al. (1999), the
$\nu\bar{\nu}$-annihilation strongly concentrates near the central
region of space. Nearly 60\% of the total annihilation luminosity is
ejected from the region $r<20r_{g}$. Also, Figure \ref{fig2173}
shows that the annihilation luminosity varies more rapidly along the
$z$-direction than along the $r$-direction.

Although the space distribution of the $\nu\bar{\nu}$-annihilation
energy deposition rate can be calculated, it is less clear how could
such process produce a narrowly collimated relativistic jet, which
is required in the GRB inner shock model. In the collapsar scenario,
as mentioned in \S 1.3.3, Aloy et al. (2000) and Zhang et al. (2003,
2004) showed that the jets from the central BH+disk systems are
collimated by their passage through passage through the stellar
mantle. But how can a jet, or called ``the seed of jet", forms
inside $10^{17}$ cm? MacFayden et al. (1999, 2001) discussed that it
is not possible to produce a strong jet very early when the momentum
of the infalling material along the axis is too high. At $\sim1$ s,
for example, the density of the infalling material is $\geq10^{7}$ g
cm$^{-3}$ and its velocity is $\sim10^{10}$ cm s$^{-1}$ for Type I
collapsar. Thus any energy deposited by neutrino will be advected
into the BH. As time passes, the density especially the density
along the polar direction declines, and a disk is formed with the
accretion rate up to $\sim0.1M_{\odot}$ s$^{-1}$ about 10 s after
core collapse. Thus MacFayden et al. (1999, 2001) discussed that the
jets could be initiated when the disk has fully formed, although the
initial jets are not well collimated. However, their work did not
includes the detailed microphysics of the
$\nu\bar{\nu}$-annihilation heating for jets formation. More
detailed Mewtonian simulations (e.g., Nagataki et al. 2007) and GR
simulations (e.g., Nagataki et al. 2009) showed that, the
$\nu\bar{\nu}$-annihilation processes are not efficient to launch a
relativistic jet.
\begin{figure}
\centering\resizebox{0.9\textwidth}{!} {\includegraphics{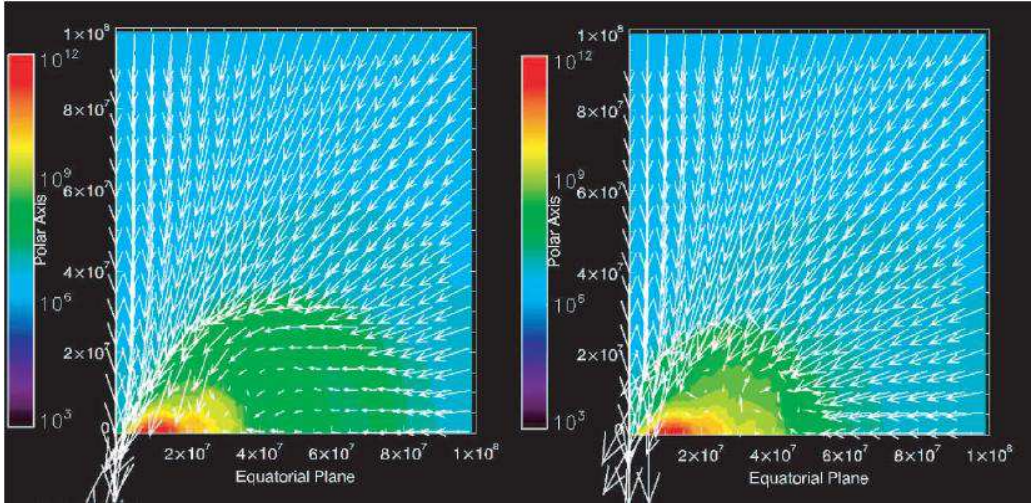}}
\caption{Density contour with velocity fields for BH-disk model
without magnetic fields in Nagataki et al. (2007): $t=2.1 s$ (left
panel) and $t=2.2 s$ (right panel). The central region is
$r\leq10^{8}$ cm in this figure. No jet is launched.}\label{fig2174}
\end{figure}
Figure \ref{fig2174} shows the density contour of the central region
of collapsar without magnetic fields at $t=2.1$ s and $t=2.2$ s. No
jet occurs in this case, partly because the accretion rate which
drops monotonically from $\sim0.1M_{\odot}$ s$^{-1}$ to
$\sim10^{-3}M_{\odot}$ s$^{-1}$ is not high enough to provide energy
along the polar axis. It is possible longterm simulations (at least
$\sim10$ s) is needed to see whether the jets could be formed.

Besides the collapsar scenario, jets formation after compact object
mergers has also been studied. As the environment outside the
BH+disk systems $>10^{7}$ cm does not collimate the initial jets as
showed in the stellar mantle of collapsars in this case, other
mechanism should be considered. For example, Rosswog et al. (2003)
found that the $\nu\bar{\nu}$-annihilation may fail to explain the
apparent isotropized energies implied for short GRBs, unless the jet
is beamed into less than 1 per cent of the solid angle. They argued
that the energetic, neutrino-driven wind may have enough pressure,
and thus provide adequate collimation to the
$\nu\bar{\nu}$-annihilation-driven jet, based on the mechanism
pointed out by Levinson \& Eichler (2000) that the pressure and
inertia of a baryonic wind can lead to the collimation of a
baryon-poor jet, However, such pressure-driven collimation was not
self-consistently obtained in the simulations of Rosswog et al.
(2003). As a result, it is still an open question that whether the
$\nu\bar{\nu}$-annihilation mechanism is efficient to launch a
relativistic collimated jet which can produce a GRB.

\section{MHD Processes and Jet Production}

Jets produced in accreting systems surrounded by magnetic fields via
MHD processes have been widely studied in various astrophysics
objects such as active galactic nuclei, galactic microquasars as
well as GRBs. Strong magnetic fields up to $10^{14}-10^{15}$ G is
required to launch relativistic Poynting-flux-jets from the
stellar-mass BH+ diks systems and rive the GRB phenomena. Even if
the initial magnetic fields are low in the progenitors such as weak
magnetic massive collapsar, it is still expected that magnetic
fields can be amplified by the magneto-rotational instability (MRI,
Balbus \& Hawley, 1991; or see Balbus \& Hawley 1998 for a review).
Although the mechanisms for the creation of MHD jets are still a
matter of much debate. there are two basic types of MHD mechanisms
that may account for the jet formation (Garofalo 2009). The first
involves a magnetocentrifugal or MHD wind similiar to the solar wind
phenomena (e.g., Blandford \& Payne 1982; Koide et al. 1998, 1999).
The second class of mechanism applied only to BH accretors, which
drive jets by their ergosphere (e.g., Pernose 1969; Blandford \&
Znajek 1977).

Blandford \& Payne (1982) illustrated that a centrifugally driven
wind from the disk is possible if the poloidal component of the
magnetic field makes an angle of less than $\sim60^{\circ}$ with the
disk surface. At large distances from the disk, the toroidal
component of the field becomes important and collimates the outflow
into a pair of anti-parallel jets moving perpendicular to the disk.
Close to the disk, the flow is probably driven by gas pressure in a
hot magnetically dominated corona. As a result, magnetic stresses
can extract the angular momentum from a thin disk and enable matter
to be acreted, independently of the disk viscosity. Although the
basic purpose of Blandford \& Payne (1982) was to explain the radio
jets which emerging from the AGNs, this mechanism does also work in
the stellar-mass BH+ disk system as the GRB central engine. However,
this mechanism tends to produce relatively cold jets, in the sense
that the thermal energy of the jet is not initially large compared
to either the rest mass of the jet or the jet kinetic energy.
MacFadyen et al. (2001) discussed that cold MHD jets may be
collimated both by magnetic fields and by geometry.

\begin{figure}
\centering\resizebox{0.7\textwidth}{!} {\includegraphics{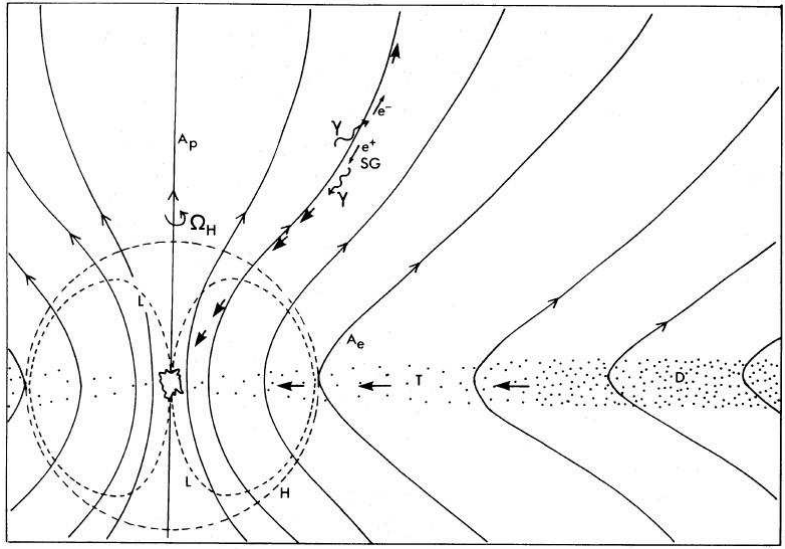}}
\caption{Schematic cross-section of BH and magnetosphere. A current
$I$ is flowing from the magnetosphere into the hole, and back out of
the hole into the disk D lying in the $\theta=\pi/2$ plane. Between
the disk and the hole there is a transition region $T$ in which the
matter is falling from the disk to the hole. More details can be
seen in Blandford \& Znajek (1977). }\label{fig2181}
\end{figure}

Christodolou (1970) first discussed that for a rotating hole
described by the Kerr metric, a portion of the rest mass can be
regarded as ``reducible" in the snes that it can be removed and
extracted to infinity. The possibility of extracting the reducible
mass was discussed by Pernose (1969) that the existence of negative
energy orbits within the ergosphere surrounding a Kerr BH permits
the mechanical extraction of energy via certain types of particle
collision (see Teukolsky \& Press 1974 for a similar process study).
However, neither process seems likely to operate effectively in any
astrophysical situation\footnote{Piran \& Shaham (1975, 1977), as
showed in \S 1,  suggested a Galactic GRB model that GRBs are
produced in the X-ray binary systems, are Compton scattered by
tangentially moving electrons deep in the ergosphere of the rotating
black hole via the Penrose mechanism.}. Blandford \& Znajek (1977)
studied a rotating BH surrounded by a stationary axisymmetric
magnetized plasma (see Figure \ref{fig2181}).
They showed an effective mechanism which can extracted the
rotational energy of the BH by ergosphere. This mechanism appears
when the charges around the BH are sufficient to provide the
force-free condition. The fundamental equations describing a
stationary axiymmetric magnetosphere were derived in Blandford \&
Znajek (1977), and the details of the energy and angular momentum
balance were discussed. The poloidal components of the conserved
electromagnetic energy flux $\mathscr{E}_{r}$ and angular momentum
flux $\mathscr{L}_{r}$ satisfy
\begin{equation}
\mathscr{E}_{r}=\omega(\Omega_{H}-\omega)\left(\frac{A_{\phi,\theta}}{r_{+}^{2}+a^{2}cos^{2}\theta}\right)
(r_{+}^{2}+a^{2})\epsilon_{0},\label{MHD1}
\end{equation}
\begin{equation}
\mathscr{L}_{r}=\mathscr{E}_{r}/\omega,\label{MHD2}
\end{equation}
where $\Omega_{H}\equiv a/(r_{+}^{2}+a^{2})$ is the angular velocity
of the hole with $r_{+}$ being the radius of the event horizon,
$\omega$ is the ``rotation frequency" of the electromagnetic fields,
and $A_{\phi}$ is the $\phi$-component of a vector potential
$A_{\mu}$. Here the Boyer-Lindqust coordinates are adopted. Thus
$\mathscr{E}_{r},\mathscr{L}_{r}\geq0$ implies
$0\leq\omega\leq\Omega_{H}$\footnote{According to the numerical
resutls of Komissarov (2001) and the argument of MacDonald \& Throne
(1982), $\omega$ adjusts to $\sim\Omega_{H}/2$ even at large $a$.}.
Furthermore, Blandford \& Znajek (1977) adopted a perturbation
technique to provide approximate solutions for slowly rotating holes
$a\ll1$. Particularly, for the force-free magnetosphere with a
paraboloidal magnetic force, the extracted energy appears to be
focused along the rotation axis, and the power radiated satisfies
\begin{equation}
L_{BZ}\approx\frac{1}{32}\frac{\omega(\Omega_{H}-\omega)}{\Omega_{H}^{2}}B^{2}r_{+}^{2}a^{2}c.\label{MHD3}
\end{equation}
Although the formula (\ref{MHD3}) was originally obtained for
$a\ll1$, Komissarov (2001) has performed a numerical study of the BZ
solution concluding that the formula is valid at least up to
$a=0.9$.

GRMHD simulations in the Kerr metric in 2D (McKinney \& Gammie 2004;
Mckiney 2005, 2006; Garofalo 2009) and in 3D (De Villiers et al.
2003, 2005; Hirose et al. 2004; Krolik et al. 2005; Hawley \& Krolik
2006) have been carried out to investigate the properties of the MHD
jets from the central BHs surrounded by the magnetized plasma. For
example, McKinney \& Gammie (2004) used a general relativistic MHD
code (HARM) to evolve a weakly magnetized thick disk around a Kerr
BH. They found an outward electromagnetic energy flux on the event
horizon as anticipated by Blandford \& Znajek (1977). The funnel
region near the polar axis of the BH is consistent with the BZ
model. The outward electromagnetic energy flux is, however,
overwhelmed by the inward flux of energy associated with the
rest-mass and internal energy of the accreting plasma. This result
that suggested confirms work by Ghosh \& Abramowicz (1997) that
suggested the BZ luminosity should be small or comparable to the
nominal accretion luminosity. Hawley \& Krolik (2006) examined the
unbound axial outflows (jets) that develop within 3D GRMHD
simulations. Different to most of other studies that magnetic field
boundary conditions are imposed on the simulations, they simulated
how accretion dynamics self-consistently creates magnetic field and
explore the outflows and jets. They jets have two major components:
a matter-dominated outflow that moves at a modest velocity
($v/c\sim0.3$) along the centrifugal barrier surrounding an
evacuated axial funnel, and a highly relativistic Poynting
flux-dominated jet within the funnel. The funnel-wall jet is
accelerated and collimated by magnetic and gas pressure forces in
the inner torus and the surrounding corona.
\begin{figure}
\centering\resizebox{0.5\textwidth}{!} {\includegraphics{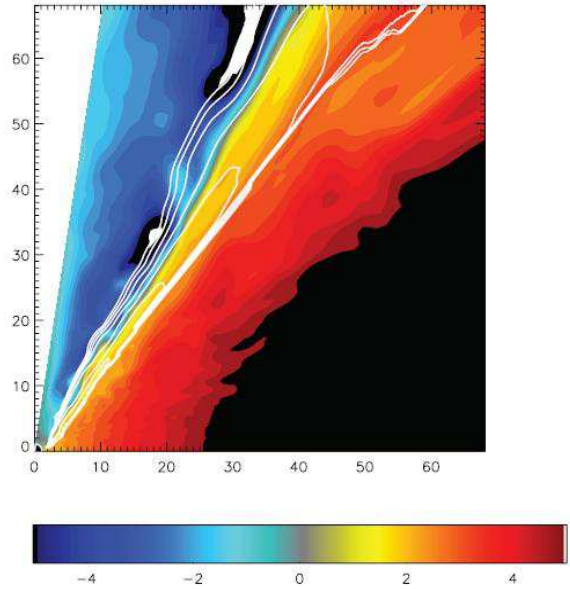}}
\caption{Specific angular momentum and mass flux at late time in the
KDPs simulation in Hawley \& Krolik (2006), Colar contours show the
mass-weighted mean angular momentum in the funnel-wall outflow. Th
radius is in unit of $M$ with $G=c=1$.}\label{fig2182}
\end{figure}
The Poynting flux jet results from the formation of a large-scale
radial magnetic field within the funnel. This field is spun by the
rotating spacetime of the BH. Figure \ref{fig2182} shows the
specific angular momentum and mass flux at late time of MHD outflows
and jets evolution in Hawley \& Krolik (2006). Their results are in
general agreement with those of Blandford \& Znajek (1977), but
there are also specific differences from the classical
Blandford-Znajek model. For instance, it is not in general in a
state of force-free equilibrium in the magnetically dominated funnel
region. Also, the electromangetic luminosity in the outflows that
places entirely as a by-produce of accretion in Hawley \& Krolik
(2006), whereas in the classical BZ model there was zero accretion.
Recently, Garofalo (2009) studies the spin-dependence of the BZ
effect. Large scale magnetic fields are enhanced on the BH compared
to the inner accretion flow and that the ease with which this occurs
for lower prograde BH spin, produces a spin dependence in the BZ
effect.

The GRMHD simulations mentioned above are not directly applicable to
the GRB case, as they do not consider the dense disks with huge
accretion rate with a fraction of solar mass per second and the
neutrino cooling and heating processes. Nagataki (2009) studied the
outgoing Poynting flux exists at the horizon of the central BH
around the polar region inside a collapsar. They concluded that the
jet is launched mainly by the magnetic field amplified by the
gravitational collapse and differential rotation around the BH,
rather than the BZ mechanism in this study. However, no microphysics
is included in the code such as nuclear reactions, neutrino
processes, and equation of state for dense matter. Further studied
should be done in subsequent work.

\chapter{Hyperaccretion Disks around Neutron Stars}
\label{chap3} \setlength{\parindent}{2em}

\section{Introduction}
Gamma-ray bursts (GRBs) are commonly divided into two classes:
short-duration, hard-spectrum bursts, and long-duration,
soft-spectrum bursts. The observations have provided growing
evidence that short bursts result from the mergers of compact star
binaries and long bursts originate from the collapses of massive
stars (for recent reviews see Zhang \& M\'esz\'aros 2004; Piran
2004; M\'esz\'aros 2006; Nakar 2007). It is usually assumed that
both a compact-star merger and a massive-star collapse give rise to
a central black hole and a debris torus around it. The torus has a
mass of about $0.01-1 M_{\odot}$ and a large angular momentum enough
to produce a transient accretion disk with a huge accretion rate up
to $\sim 1.0 M_{\odot}\,{\rm s}^{-1}$. The accretion timescale is
short, e.g., a fraction of one second after the merger of two
neutron stars, and tens of seconds if a disk forms due to fallback
of matter during the collapse process.

The hyperaccretion disk around a black hole is extremely hot and
dense. The optical depth of the accreting gas is so enormous that
radiation is trapped inside the disk and can only be advected
inward. However, in some cases, this hot and dense disk can be
cooled via neutrino emission (Narayan et al. 1992). According to the
disk structure and different cooling mechanisms, flows in the disk
fall into three types: advection-dominated accretion flows (ADAFs),
convection-dominated accretion flows (CDAFs), and neutrino-dominated
accretion flows (NDAFs). The first two types of flow are
radioactively inefficient (Narayan et al. 1998, 2000, 2001) and the
final type cools the disk efficiently via neutrino emission (Popham
et al. 1999; Di Matteo et al. 2002). In view of effects of these
three flows, hyperaccretion disks around black holes have been
studied both analytically and numerically (e.g., see Popham et al.
1999; Narayan et al. 2001; Kohri \& Mineshige 2002; Di Matteo et al.
2002; Gu et al. 2006; Chen \& Beloborodov 2007; Liu et al. 2007;
Janiuk et al. 2007).

However, newborn neutron stars have been invoked to be central
engines of GRBs in some origin/afterglow models. First, the
discovery of X-ray flares by Swift implies that the central engines
of some GRBs are in a long-living activity (at least hundreds of
seconds) after the bursts (Zhang 2007). This provides a challenge to
conventional hyperaccretion disk models of black holes. Recently,
Dai et al. (2006) argued that newborn central compact objects in the
GRBs could be young neutron stars, at least transiently-existing
neutron stars, rather than black holes (for alternative models see
Perna et al. 2006 and Proga \& Zhang 2006). These neutron stars may
have high angular momentum and their maximum mass may be close to or
slightly larger than the upper mass limit of nonrotating
Tolman-Oppenheimer-Volkoff neutron stars. According to this
argument, Dai et al. (2006) explained X-ray flares of short GRBs as
being due to magnetic reconnection-driven events from
highly-magnetized millisecond pulsars. It is thus reasonable to
assume a unified scenario: a prompt burst originates from a
highly-magnetized, millisecond-period neutron star surrounded by a
transient hyperaccretion disk, and subsequent X-ray flares are due
to a series of magnetic activities of the neutron star.

Second, the shallow decay phase of X-ray afterglows about several
hundreds of seconds after a sizable fraction of GRBs discovered by
Swift has been understood as arising from long-lasting energy
injection to relativistic forward shocks (Zhang 2007). It was
proposed before Swift observations that pulsars in the unified
scenario mentioned above can provide energy injection to a forward
shock through magnetic dipole radiation, leading to flattening of an
afterglow light curve (Dai \& Lu 1998a; Zhang \& M\'esz\'aros 2001;
Dai 2004). Recent model fitting (Fan \& Xu 2006; Yu \& Dai 2007) and
data analysis (Liang et al. 2007) indeed confirm this result. An
ultra-relativistic pulsar wind could be dominated by
electron/positron pairs and its interaction with a postburst
fireball gives rise to a reverse shock and forward shock (Dai 2004).
The high-energy emission due to inverse-Compton scattering in these
shocks is significant enough to be detectable with the upcoming {\em
Gamma-ray Large-Area Space Telescope} (Yu et al. 2007).

Third, we note that, in some origin models of GRBs (e.g., Klu\'zniak
\& Ruderman 1998; Dai \& Lu 1998b; Wheeler et al. 2000; Wang et al.
2000; Paczy\'nski \& Haensel 2005), highly-magnetized neutron stars
or strange quark stars surrounded by hyperaccretion disks resulting
from fallback of matter could occur during the collapses of massive
stars or the mergers of two neutron stars. A similar neutron star
was recently invoked in numerical simulations of Mazzali et al.
(2006) and data analysis of Soderberg et al. (2006) to understand
the properties of supernova SN 2006aj associated with GRB 060218. In
addition, hyperaccretion disks could also occur in type-II
supernovae if fall-back matter has angular momentum. It would be
expected that such disks play an important role in supernova
explosions via neutrino emission, similar to some effects reviewed
by Bethe (1990).

Based on these motivations, we here investigate a hyperaccretion
disk around a neutron star. To our knowledge, this paper is the
first to study hyperaccretion disks around neutron stars related
possibly with GRBs. Chevalier (1996) discussed the structure of
dense and neutrino-cooled disks around neutron stars. He considered
neutrino cooling due to electron-positron pair annihilation, which
is actually much less important than the cooling due to
electron-positron pair capture in the hyperaccreting case of our
interest, since the accretion rate assumed in Chevalier (1996)
($\sim M_{\odot}\,{\rm yr}^{-1}$) is much less than that of our
concern ($\sim 0.1M_{\odot}\,{\rm s}^{-1}$). In this chapter we
consider several types of neutrino cooling by using elaborate
formulae developed in recent years. In addition, we focus on some
differences between black hole and neutron star hyperaccretion. A
main difference is that, the internal energy in an accretion flow
may be advected inward into the event horizon without any energy
release if the central object is a black hole, but the internal
energy must be eventually released from the disk if the central
object is a neutron star, because the stellar surface prevents any
heat energy from being advected inward into the star. Since the
accretion rate is always very high, the effective cooling mechanism
in the disk is still neutrino emission, and as a result, the
efficiency of neutrino cooling of the entire disk around a neutron
star should be higher than that of a black hole disk.

There have been a lot of works to study accretion onto neutron stars
in binary systems (e.g., Shapiro \& Salpeter 1975; Klu\'zniak \&
Wilson 1991;  Medvedev \& Narayan 2001; Frank et al. 2002) and
supernova explosions (e.g., Chevalier 1989, 1996; Brown \&
Weingartner 1994; Kohri et al. 2005). In the supernova case,
spherically symmetric accretion onto neutron stars (the so-called
Bondi accretion) was investigated in detail. In particular, Kohri et
al. (2005) tried to use the hyperaccretion disk model with an
outflow wind to explain supernova explosions. In the binary systems,
as the accretion rate is not larger than the Eddington accretion
rate ($\sim 10^{-8}M_{\odot}\,{\rm yr}^{-1}$), the physical
properties of the disk must be very different from those of a
hyperaccretion disk discussed here.

This chapter is organized as follows: in \S 3.3.2, we describe a
scheme of our study of the structure of a quasi-steady disk. We
propose that the disk around a neutron star can be divided into two
regions: inner and outer disks. Table \ref{tab31} gives the notation
and definition of some quantities in this chapter. In order to give
clear physical properties, we first adopt a simple model in \S3.3,
and give both analytical and numerical results of the disk
properties. In \S3.4, we study the disk using a state-of-the-art
model with lots of elaborate considerations about the thermodynamics
and microphysics, and compare results from this elaborate model with
those from the simple model. \S3.5 presents conclusions and
discussions.
\begin{table}
\begin{center}\footnotesize{
\begin{tabular}{|l|l|r|r|r|r|}
\hline
notation & definition & \S/Eq. \\
\hline
 $m$ & mass of the central neutron star, $m=M/1.4M_{\odot}$ & \S3.3.1, (\ref{3s11})\\
 $\dot{m}_{d}$ & mass accretion rate, $\dot{m}_d=\dot{M}/0.01M_{\odot}\,{\rm s}^{-1}$ & \S3.3.1, (\ref{3s11})\\
 $Y_{e}$ & ratio of the electron to nucleon number density in the disk & \S3.2.2, (\ref{3a02})\\
 $r_{*}$ & radius of the neutron star & \S3.2.3, (\ref{3e231}) \\
 $r_{\rm out}$ & outer radius of the disk & \S3.2.3, (\ref{3e231}) \\
 $\Omega_{*}$ & angular velocity of the stellar surface & \S3.3.3, (\ref{3e331}) \\
 $\varepsilon$ & efficiency of energy release in the inner disk & \S3.2.3, (\ref{3e4}) \\
 $\tilde{r}$ & radius between the inner and outer disks & \S3.2.3, (\ref{3en02}) \\
 $\bar{r}$ & radius between the neutrino optically-thick \& -thin regions & \S3.3.2, (\ref{3e10}) \\
 $k=\bar{r}/\tilde{r}$ & parameter to measure the neutrino optically-thick region & --- \\
 $f=1-\sqrt{\frac{r_{*}}{r}}$ & useful factor as a function of $r$& \S3.2.2, (\ref{3e21}) \\
 $\tilde{f}=1-\sqrt{\frac{r_{*}}{\tilde{r}}}$ & value of $f$ at radius $\tilde{r}$ & \S3.3.2, (\ref{3e11}) \\
 $\bar{f}_{\nu}$ & average efficiency of neutrino cooling in the outer disk & \S3.2.3, (\ref{3e231}) \\
\hline
\end{tabular}
\caption{Notation and definition of some quantities in this
chapter.}}
\end{center}\label{tab31}
\end{table}

\section{Description of Our Study Scheme}  
\subsection{Motivations of a two-region disk}  
We study the quasi-steady structure of an accretion disk around a
neutron star with a weak outflow. We take an accretion rate to be a
parameter. For the accretion rates of interest to us, the disk flow
may be an ADAF or NDAF. In this chapter, we do not consider a CDAF.
Different from the disk around a black hole, the disk around a
neutron star should eventually release the gravitational binding
energy of accreted matter (which is converted to the internal energy
of the disk and the rotational kinetic energy) more efficiently.

The energy equation of the disk is (Frank et al. 2002)
\begin{equation}
\Sigma v_{r}T\frac{ds}{dr}=Q^{+}-Q^{-},\label{3a1}
\end{equation}
where $\Sigma$ is the surface density of the disk, $v_{r}$ is the
radial velocity, $T$ is the temperature, $s$ is the entropy per unit
mass, $r$ is the radius of a certain position in the disk, and
$Q^{+}$ and $Q^{-}$ are the energy input (heating) and the energy
loss (cooling) rate in the disk. From the point of view of
evolution, the structure of a hyperaccretion disk around a neutron
star should be initially similar to that of the disk around a black
hole, because the energy input is mainly due to the local viscous
dissipation, i.e., $Q^{+}=Q^{+}_{\rm vis}$. However, since the
stellar surface prevents the matter and heat energy in the disk from
advection inward any more, a region near the compact object should
be extremely dense and hot as accretion proceeds. Besides the local
viscous heating, this inner region can also be heated by the energy
($Q^{+}_{\rm adv}$) advected from the outer region of the disk.
Thus, the heat energy in this region may include both the energy
generated by itself and the energy advected from the outer region,
that is, we can write $Q^{+}=Q^{+}_{\rm vis}+Q^{+}_{\rm adv}$.
Initially, such a region is so small (i.e., very near the compact
star surface) that it cannot be cooled efficiently for a huge
accretion rate ($\sim 0.01-1M_{\odot}\,{\rm s}^{-1}$). As a result,
it has to expand its size until an energy balance between heating
and cooling is built in this inner region. Such an energy balance
can be expressed by $Q^{+}=Q^{-}$ in the inner region of the disk.

Once this energy balance is built, the disk is in a steady state as
long as the accretion rate is not significantly changed. For such a
steady disk, therefore, the structure of the outer region is still
similar to that of the disk around a black hole, but the inner
region has to be hotter and denser than the disk around a black
hole, and could have a different structure from both its initial
structure and the outer region that is not affected by the neutron
star surface.

Based on the above consideration, the hyperaccretion disk around a
neutron star could have two different regions. In order to discuss
their structure using a mathematical method clearly, as a reasonable
approximation, we here divide the steady accretion disk into an
inner region and an outer region, called inner and outer disks
respectively. The outer disk is similar to that of a black hole. The
inner disk, depending on its heating and cooling mechanisms
discussed above, should satisfy the entropy conservation condition
$Tds/dr\propto Q^{+}-Q^{-}= 0$, and thus we obtain $P\propto
\rho^{\gamma}$, where $P$ and $\rho$ are the pressure and the
density of the disk, and $\gamma$ is the adiabatic index of the disk
gas. Also, the radial momentum equation is
\begin{equation}
(\Omega^{2}-\Omega_{K}^{2})r-\frac{1}{\rho}\frac{d}{dr}(\rho
c_{s}^{2})=0,\label{3b01}
\end{equation}
where $\Omega$ and $\Omega_{K}$ are the angular velocity and
Keplerian angular velocity of the inner disk, and
$c_{s}=\sqrt{P/\rho}$ is the isothermal sound speed. We here neglect
the radial velocity term $v_{r}dv_{r}/dr$ since we consider the
situation $v_{r}dv_{r}/dr\ll |\Omega^{2}-\Omega_{K}^{2}|r$, which
can still be satisfied when $v_{r}\ll r\Omega_{K}$ with $\Omega\sim
\Omega_{K}$ but $|\Omega-\Omega_{K}|\geq v_{r}/r$. Equation
(\ref{3b01}) gives $\Omega\propto r^{-3/2}$ and $c_{s}\propto
r^{-1/2}$. Moreover, from the continuity equation, we have
$v_{r}\propto (\rho rH)^{-1}$ with the disk's half-thickness
$H=c_{s}/\Omega\propto r$. Thus we can derive a self-similar
structure in the inner region of the hyperaccretion disk of a
neutron star,
\begin{equation}
\rho \propto r^{-1/(\gamma-1)},\,\, P \propto
r^{-\gamma/(\gamma-1)},\,\, v_r \propto
r^{(3-2\gamma)/(\gamma-1)},\label{3b02}
\end{equation}
This self-similar structure has been given by Chevalier (1989) and
Brown \& Weingartner (1994) for Bondi accretion under the adiabatic
condition and by Medvedev \& Narayan (2001) and Medvedev (2004) for
disk accretion under the entropy conservation condition. In
addition, if the gas pressure is dominated in the disk, we have
$\gamma=5/3$ so that equation (\ref{3b02}) becomes $\rho \propto
r^{-3/2}$, $P \propto r^{-5/2}$, and $v_r \propto r^{-1/2}$, which
have been discussed by Spruit et al. (1987) and Narayan \& Yi
(1994).

An important problem we next solve is to determine the size of the
inner disk. Since the total luminosity of neutrinos emitted from the
whole disk significantly varies with the inner-region size, we can
estimate the inner region size by solving an energy balance between
neutrino cooling and heating in the entire disk. The details will be
discussed in \S 3.2.3.

Finally, we focus on two problems in this subsection. First, we have
to discuss a physical condition of the boundary layer between the
outer and inner disks. There are two possible boundary conditions.
One condition is to assume that a stalled shock exists at the
boundary layer. Under this assumption, the inner disk is a
post-shock region, and its pressure, temperature and density just
behind the shock are much higher than those of the outer disk in
front of the shock. The other condition is to assume that no shock
exists in the disk, and that all the physical variables of two sides
of this boundary change continuously. We now have to discuss which
condition is reasonable.

Let us assume the mass density, pressure, radial velocity, and
internal energy density of the outer disk along the boundary layer
to be $\rho_{1}$, $P_{1}$, $v_{1}$, and $u_1$, and the corresponding
physical variables of the inner disk to be $\rho_{2}$, $P_{2}$,
$v_{2}$, and $u_2$ at the same radius. Thus the conservation
equations are
\begin{equation}
\begin{array}{ll}\rho_{1}v_{1}=\rho_{2}v_{2}
\\P_{1}+\rho_{1}v_{1}^{2}=P_{2}+\rho_{2}v_{2}^{2}
\\(u_{1}+P_{1}+\rho_{1}v_{1}^{2}/2)/\rho_{1}=(u_{2}+P_{2}
+\rho_{2}v_{2}^{2}/2)/\rho_{2}.
\end{array}
\end{equation}
From the Rankine-Hugoniot relations, we know that if $P_{2}\gg
P_{1}$, the densities of two sides of the boundary layer are
discontinuous, which means that a strong shock exists between the
inner and outer disks. On the other hand, if $P_{1} \sim P_{2}$, we
can obtain $\rho_{1} \sim \rho_{2}$, which means that only a very
weak shock forms at this boundary layer, or we can say that no shock
exists. Therefore, we compare $P_{1}$ and $\rho_{1}v_{1}^{2}$ of the
outer disk. If $P_{1} \ll \rho_{1}v_{1}^{2}$ or $c_{s} \ll v_{1}$,
we can assume that a stalled shock exists at the boundary layer,
i.e., the first boundary condition is correct. If $P_{1} \gg
\rho_{1}v_{1}^{2}$ or $c_{s} \gg v_{1}$, otherwise, we consider
$P_{1} \sim P_{2}$, and thus no shock exists. In \S 3.3 and \S 3.4,
we will use this method to discuss which boundary condition is
reasonable.

Second, what we want to point out is that the effect of the magnetic
field of the central neutron star is not considered in this chapter.
We estimate the order of magnitude of the Alfv\'{e}n radius by using
the following expression (Frank et al 2002), $r_A\simeq
2.07\times10^{4}\dot{m}_{d}^{-2/7}m^{-1/7}\mu_{30}^{4/7}$, where
$\dot{m}_{d}=\dot{M}/0.01M_{\odot}\,{\rm s}^{-1}$ is the accretion
rate, $m=M/1.4M_\odot$ is the mass of the neutron star, and
$\mu_{30}$ is the magnetic moment of the neutron star in units of
$10^{30}{\rm G}\,{\rm cm}^{3}$. Let $r_{*}$ be the neutron star
radius. If the stellar surface magnetic field $B_{s}\leq B_{s,{\rm
cr}}=2.80\times10^{22} \dot{m}_{d}^{1/2}m^{1/4}r_{*}^{-5/4}\,$G or
$B_{s}\leq B_{s,{\rm cr}}=0.89\times10^{15}
\dot{m}_{d}^{1/2}m^{1/4}\,$G when $r_{*}=10^{6}$ cm, then the
Alfv\'{e}n radius $r_A$ is smaller than $r_{*}$. The critical value
$B_{s,{\rm cr}}$ increases with increasing the accretion rate. This
implies that the stellar surface magnetic field affects the
structure of the disk significantly if $B_{s}\geq B_{s,{\rm cr}}\sim
10^{15}-10^{16}\,$G for typical accretion rates. Therefore, we
assume a neutron star with surface magnetic field weaker than
$B_{s,{\rm cr}}$ in this chapter. This assumption is consistent with
some GRB-origin models such as Klu\'zniak \& Ruderman (1998), Dai \&
Lu (1998b), Wang et al. (2000), Paczy\'nski \& Haensel (2005), and
Dai et al. (2006), because these models require a neutron star or
strange quark star with surface magnetic field much weaker than
$B_{s,{\rm cr}}$.

\subsection{Structure of the outer disk}   
Here we discuss the structure of the outer disk based on the
Newtonian dynamics and the standard $\alpha$-viscosity disk model
for simplicity. The structure of the hyperaccretion disk around a
stellar-mass black hole has been discussed in many previous works.
The method and equations we use here are similar to those in the
previous works since the outer disk is very similar to the disk
around a black hole.

We approximately consider the angular velocity of the outer disk to
be the Keplerian value $\Omega_K=\sqrt{GM/r^{3}}$. The velocity
$\Omega_K$ should be modified in a relativistic model of accretion
disks (Popham et al. 1999; Chen \& Beloborodov 2007), but we do not
consider it in this chapter. We can write four equations to describe
the outer disk, i.e., the continuity equation, the energy equation,
the angular momentum equation and the equation of state.

The continuity equation is
\begin{equation}
\dot{M}=4\pi r\rho v_{r}H\equiv 2\pi r\Sigma v_r,\label{3e1}
\end{equation}
where the notations of the physical quantities have been introduced
in $\S$3.2.1.

In the outer disk, the heat energy could be advected inward and we
take $Q_{\rm adv}^{-}=(1/2)\Sigma Tv_{r}ds/dr$ to be the quantity of
the energy advection rate, where the factor $1/2$ is added because
we only study the vertically-integrated disk over a half-thickness
$H$. Thus the energy-conservation equation (\ref{3a1}) is rewritten
as
\begin{equation}
Q^{+}=Q_{\rm rad}^{-}+Q_{\rm adv}^{-}+Q_{\nu}^{-}.\label{3e2}
\end{equation}

The quantity $Q^{+}$ in equation (\ref{3e2}) is the viscous heat
energy generation rate per unit surface area. According to the
standard viscosity disk model, we have
\begin{equation}
Q^{+}=\frac{3GM\dot{M}}{8\pi r^{3}}f\label{3e21}
\end{equation}
where $f=1-(r_{*}/r)^{1/2}$ (Frank et al. 2002).

The quantity $Q_{\rm rad}^{-}$ in equation (\ref{3e2}) is the photon
cooling rate per unit surface area of the disk. Since the disk is
extremely dense and hot, the optical depth of photons is always very
large and thus we can take $Q_{\rm rad}^{-}= 0$ as a good
approximation.

The entropy per unit mass of the disk is (similar to Kohri \&
Mineshige 2002)
\begin{equation}
s=s_{\rm gas}+s_{\rm rad}=\frac{S_{\rm gas}}{\rho}+\frac{S_{\rm
rad}}{\rho}=\sum
_{i}n_{i}\left\{\frac{5}{2}\frac{k_{B}}{\rho}+\frac{k_{B}}{\rho}\textrm{ln}\left[\frac{(2\pi
k_{B}T)^{3/2}}{h^{3}n_{i}}\right]\right\}+S_{0}+\frac{2}{3}g_{*}\frac{aT^{3}}{\rho},\label{3a01}
\end{equation}
where the summation runs over nucleons and electrons, $k_{B}$ is the
Boltzmann constant, $h$ is the Planck constant, $a$ is the radiation
constant, $S_{0}$ is the integration constant of the gas entropy,
and the term $2g_{*}aT^{3}/3\rho$ is the entropy density of the
radiation with $g_{*}=2$ for photons and $g_{*}=11/2$ for a plasma
of photons and relativistic $e^{+}e^{-}$ pairs. We assume that
electrons and nucleons have the same temperature. Then we use
equation (\ref{3a01}) to calculate $ds/dr$ and approximately take
$dT/dr\approx T/r$ and $d\rho/dr\approx \rho/r$ to obtain the energy
advection rate,
\begin{equation}
Q_{\rm
adv}^{-}=v_{r}T\frac{\Sigma}{2r}\left[\frac{R}{2}\left(1+Y_{e}\right)+\frac{4}{3}g_{*}\frac{aT^{3}}{\rho}
\right],\label{3a02}
\end{equation}
where the gas constant $R=8.315\times10^{7}$ ergs mole$^{-1}$
K$^{-1}$ and $Y_{e}$ is the ratio of electron to nucleon number
density. The first term in the right-hand bracket of equation
(\ref{3a02}) comes from the contribution of the gas entropy and the
second term from the contribution of radiation.

The quantity $Q_{\nu}^{-}$ in equation (\ref{3e2}) is the neutrino
cooling rate per unit area. The expression of $Q_{\nu}^{-}$ will be
discussed in detail in $\S$3.3 and $\S$3.4.

In this chapter we ignore the cooling term of photodisintegration
$Q_{\rm photodis}^{-} $, and approximately take the free nucleon
fraction $X_{\rm nuc}\approx 1$. For the disks formed by the
collapses of massive stars, the photodisintegration process that
breaks down $\alpha$-particles into neutrons and protons is
important in a disk region at very large radius $r$. However, the
effect of photodisintegration becomes less significant for a region
at small radius, where contains less
$\alpha$-particles\footnote{Kohri et al. (2005), Chen \& Beloborodov
(2007), and Liu et al. (2007) discussed the value of nucleon
fraction $X_{\rm nuc}$ and the effect of photodisintegration as a
function of radius for particular parameters such as the accretion
rate and the viscosity parameter $\alpha$. The former two papers
show that $X_{\rm nuc}\approx 1$ and $Q_{\rm photodis}\approx 0$ for
$r\leq 10^{2}r_{g}$. Although the value of $X_{\rm nuc}$ from Liu et
al. (2007) is somewhat different from the previous works, the ratio
of $Q_{\rm photodis}^{-}/Q^{+}$ in their work also drops
dramatically for $r\leq 10^{2}r_{g}$. Therefore, it is convenient
for us to neglect the photodisintegration process for $r\leq
10^{2}r_{g}$ or $r\leq 400$ km for the central star mass
$M=1.4M_{\odot}$.}. On the other hand, for the disks formed by the
mergers of double compact stars, there will be rare
$\alpha$-particles in the entire disk. As a result, we reasonably
take all the nucleons to be free ($X_{\rm nuc}\approx 1$) and
neglect the photodisintegration process, since we mainly focus on
small disks or small regions of the disks (as we will mention in
\S3.3.4 with the outer boundary $r_{\rm out}$ to be 150 km.)

Furthermore, the angular momentum conservation and the equation of
sate can be written as
\begin{equation}
\nu \Sigma=\frac{\dot{M}}{3\pi}f,\label{3e3}
\end{equation}

\begin{equation}
P=P_{e}+P_{\rm nuc}+P_{\rm rad}+P_{\nu},\label{3e31}
\end{equation}
where $\nu=\alpha c_{s}H$ in equation (\ref{3e3}) is the kinematic
viscosity and $\alpha$ is the classical viscosity parameter, and
$P_{e}$, $P_{\rm nuc}$, $P_{\rm rad}$ and $P_{\nu}$ in equation
(\ref{3e31}) are the pressures of electrons, nucleons, radiation and
neutrinos. In $\S$3.3 we will consider the pressure of electrons in
extreme cases and in $\S$3.4 we will calculate the $e^{\pm}$
pressure using the exact Fermi-Dirac distribution function and the
condition of $\beta$-equilibrium.

Equations (\ref{3e1}), (\ref{3e2}), (\ref{3e3}) and (\ref{3e31}) are
the basic equations to solve the structure of the outer disk, which
is important for us to study the inner disk.

\subsection{Self-similar structure of the inner disk}  
In \S 3.2.1 we introduced a self-similar structure of the inner
disk, and described the method to determine the size of the inner
disk. Now we will establish the energy conservation equation in the
inner disk. We assume that $\tilde{r}$ is the radius of the boundary
layer between the inner and outer disks, and that $\tilde {\rho}$,
$\tilde {P}$, and $\tilde {v}_r$ are the density, pressure and
radial velocity of the inner disk just at the boundary layer
respectively. From equation (\ref{3b02}), thus, the variables in the
inner disk at any given radius $r$ can be written by
\begin{equation}
\rho=\tilde{\rho}(\tilde{r}/r)^{1/(\gamma-1)}, P=\tilde
{P}(\tilde{r}/r)^{\gamma/(\gamma-1)}, v_r=\tilde{v}_r
(\tilde{r}/r)^{(2\gamma-3)/(\gamma-1)}.\label{3en02}
\end{equation}

We take the outer radius of the accretion disk to be $r_{\rm out}$.
The total energy per unit time that the outer disk advects into the
inner disk is (Frank et al. 2002)
\begin{equation}
\dot{E}_{\rm
adv}=(1-\bar{f}_{\nu})\frac{3GM\dot{M}}{4}\left\{{\frac{1}
{\tilde{r}}\left[1-\frac{2}{3}\left(\frac{r_{*}}
{\tilde{r}}\right)^{1/2}\right]-\frac{1}{r_{\rm out}}
\left[1-\frac{2}{3}\left(\frac{r_{*}}{r_{\rm
out}}\right)^{1/2}\right]} \right\},\label{3e231}
\end{equation}
where $\bar{f}_{\nu}$ is the average neutrino cooling efficiency of
the outer disk,
\begin{equation}
\bar{f}_{\nu}=\frac{\int_{\tilde{r}}^{r_{\rm out}}Q_{\nu}^{-}2\pi r
dr}{\int_{\tilde{r}}^{r_{\rm out}}Q^{+}2\pi r dr}.\label{3ee02}
\end{equation}

If the outer disk flow is mainly an ADAF, $Q_{\nu}^{-}\ll Q^{+}$,
then $\bar{f}_{\nu} \sim 0$; if the outer disk flow is mainly an
NDAF, $Q_{\nu}^{-}\gg Q_{\rm rad}^{-}$ and $Q_{\nu}^{-}\gg Q_{\rm
adv}^{-}$, we have $\bar{f}_{\nu} \sim 1$. The heat energy in the
inner disk should be released more efficiently than the outer disk,
we can still approximately take $\Omega\simeq \Omega_{K}$ in the
inner disk, and the maximum power that the inner disk can release is
\begin{eqnarray}
L_{\nu,max}&\approx&\frac{3GM\dot{M}}{4}\left\{\frac{1}{3r_{*}}-\frac{1}{r_{\rm
out}}\left[1-\frac{2}{3}\left(\frac{r_{*}}{r_{\rm
out}}\right)^{1/2}\right]\right\}\nonumber\\
&&-\bar{f}_{\nu}\frac{3GM\dot{M}}{4}\left\{\frac{1}{\tilde{r}}
\left[1-\frac{2}{3}\left(\frac{r_{*}}{\tilde{r}}\right)^{1/2}\right]-
\frac{1}{r_{\rm out}}\left[1-\frac{2}{3}\left(\frac{r_{*}}{r_{\rm
out}}\right)^{1/2}\right]\right\},
\end{eqnarray}
where we have integrated vertically over the half-thickness. The
first term in the right-hand of this equation is the total heat
energy per unit time of the entire disk, and the second term is the
power taken away through neutrino cooling in the outer disk.
Considering $r_{\rm out}\gg r_*$, we have
\begin{equation}
L_{\nu,max}\approx \frac{3GM\dot{M}}
{4}\left\{\frac{1}{3r_{*}}-\frac{\bar{f}_{\nu}}{\tilde{r}}
\left[1-\frac{2}{3}\left(\frac{r_{*}}{\tilde{r}}\right)^{1/2}\right]\right\}.
\end{equation}

The maximum energy release rate of the inner disk is
$GM\dot{M}/4r_{*}$ if the outer disk flow is mainly an ADAF and
$\bar{f}_{\nu} \sim 0$. This value is just one half of the
gravitational binding energy and satisfies the Virial theorem. If
the outer disk flow is an NDAF, then the energy release of the inner
disk mainly results from the heat energy generated by its own.

Following the above consideration, most of the energy generated in
the disk around a neutron star is still released from the disk, so
we have an energy-conservation equation,
\begin{equation}
\int_{r_{*}}^{\tilde{r}}Q_{\nu}^{-}2\pi rdr=\varepsilon
\frac{3GM\dot{M}}{4} \left\{{\frac{1}{3
r_{*}}-\frac{\bar{f}_\nu}{\tilde{r}}\left[1-\frac{2}{3}
\left(\frac{r_{*}}{\tilde{r}}\right)^{1/2}\right]}\right\},\label{3e4}
\end{equation}
where $\varepsilon$ is a parameter that measures the efficiency of
the energy release. If the central compact object is a black hole,
we have $\varepsilon\approx 0$ and the inner disk cannot exist. If
the central compact object is a neutron star, we can take
$\varepsilon\approx 1$ and thus we are able to use equation
(\ref{3e4}) to determine the size of the inner disk.

\section{A Simple Model of the Disk} 
In $\S$3.2, we gave the equations of describing the structure of a
hyperaccretion disk. However, additional equations about
microphysics in the disk are needed. In order to see the physical
properties of the entire disk clearly, we first adopt a simple model
for an analytical purpose. Comparing with \S 3.4, we here adopt a
relatively simple treatment with the disk microphysics similar to
Popham et al. (1999) and Narayan et al. (2001), and discuss some
important physical properties, and then we compare analytical
results with numerical ones which are also based on the simple
model.

If the disk is optically thin to its own neutrino emission, the
neutrino cooling rate can be written as a summation of four terms
including the electron-positron pair capture rate, the
electron-positron pair annihilation rate, the nucleon bremsstrahlung
rate and the plasmon decay rate, that is, $Q_{\nu}^{-}=(\dot{q}_{\rm
eN}+\dot{q}_{e^{+}e^{-}}+\dot{q}_{\rm brems} +\dot{q}_{\rm
plasmon})H$ (Kohri \& Mineshige 2002). We take two major
contributions of these four terms and use the approximative
formulae:
$\dot{q}_{e^{+}e^{-}}=5\times10^{33}T_{11}^{9}\textrm{ergs\,cm}^{-3}
\,\textrm{s}^{-1}$, and $\dot{q}_{\rm eN}=9\times10^{23}\rho
T_{11}^{6}\textrm{ergs\,cm}^{-3}\,\textrm{s}^{-1}$. Thus equation
(\ref{3e2}) can be rewritten as
\begin{equation}
\frac{3GM\dot{M}}{8\pi r^{3}}f=\frac{\dot{M}T}{4\pi
r^{2}}\left[\frac{R}{2}\left(1+Y_{e}\right)+\frac{22}{3}\frac{aT^{3}}{\rho}\right]
+(5\times10^{33}T_{11}^{9}+9\times10^{23}\rho
T_{11}^{6})\frac{c_{s}}{\Omega_K}.\label{3e201}
\end{equation}
If neutrinos are trapped in the disk, we use the blackbody limit for
the neutrino emission luminosity: $Q_{\nu}^{-}\sim
(\frac{7}{8}\sigma_{B}T^{4})/\tau$, where $\tau$ is the neutrino
optical depth. We approximately estimate the neutrino optical depth
as $(\dot{q}_{e^{+}e^{-}}+\dot{q}_{\rm
eN})H/(4\times\frac{7}{8}\sigma_{B}T^{4})$.

Moreover, we take the total pressure in the disk to be a summation
of three terms $P_{e}$, $P_{\rm nuc}$ and $P_{\rm rad}$, and neglect
the pressure of neutrinos: $P_{e}=n_{e}kT+K_{1}\left(\rho
Y_{e}\right)^{4/3}$, $P_{\rm nuc}=n_{\rm nuc}kT$, and $P_{\rm
rad}=11aT^{4}/12$, where $K_{1}\left(\rho Y_{e}\right)^{4/3}$ is the
relativistic degeneracy pressure of electrons, $n_e$ and $n_{\rm
nuc}$ are the number densities of electrons and nucleons
respectively. Here we also neglect the non-relativistic degeneracy
pressure of nucleons.
We thus obtain
\begin{equation}
P=P_{e}+P_{\rm nuc}+P_{\rm rad}=\rho
\left(1+Y_{e}\right)RT+K_1\left(\rho Y_{e}
\right)^{4/3}+\frac{11}{12}aT^{4},\label{3e32}
\end{equation}
where $K_1=\frac{2\pi hc}{3}\left(\frac{3}{8\pi
m_{p}}\right)^{4/3}=1.24\times10^{15}\,{\rm cgs}$.

Equations (\ref{3e3}), (\ref{3e201}) and (\ref{3e32}) can be solved
for three unknowns (density, temperature and pressure) of the steady
outer disk as functions of radius $r$ for four given parameters
$\alpha$, $Y_{e}$, $\dot{M}$ and $M$ in the simple model. Once the
density, temperature and pressure profiles are determined, we can
present the structure of the outer disk and further establish the
size and the structure of the inner disk.

\subsection{The outer disk}   
We analytically solve equations (\ref{3e3}), (\ref{3e201}) and
(\ref{3e32}) in this subsection. Our method is similar to that of
Narayan et al. (2001). However, what is different from their work is
that we use the same equations to obtain both ADAF and NDAF
solutions in different conditions. Besides, we find that the factor
$f=1-(r_{*}/r)^{1/2}$ cannot be omitted because of its importance.
In this subsection, for convenience, we expand the solution range to
the entire disk (i.e., $r_{*}<r<r_{\rm out}$) rather than just
consider it in the outer region. The size of the inner disk, which
depends on the structure of the outer disk, will be solved in
$\S$3.3.2.

First we show a general picture. If the accretion rate is not very
high, most of the energy generated in the disk is advected inward
and we call the disk as an advection-dominated disk. As the
accretion rate increases, the density and temperature of the disk
also increase and the neutrino cooling in some region of the disk
becomes the dominant cooling mechanism. Thus we say that this region
becomes neutrino-dominated. When the accretion rate is sufficiently
large, the disk may totally become neutrino-dominated. Also, there
are some other factors such as the mass of the central neutron star,
$M$, and the electron-nucleon ratio, $Y_{e}$, are also able to
influence the disk structure.

In order to obtain analytic solutions, we have to make some
assumptions. First of all, we assume that the disk flow is an ADAF
with the radiation pressure to be dominated. Let the mass of neutron
star $M=1.4mM_{\odot}$, the accretion rate
$\dot{M}=\dot{m}_{d}\times 0.01M_{\odot}\,{\rm s}^{-1}$,
$\alpha=0.1\alpha_{-1}$, $r=10^{6}r_{6}\,$cm, $\rho=10^{11}\rho_{11}
\,\textrm{g}\,\textrm{cm}^{-3}$, $T=10^{11}T_{11}\,{\rm K}$, and
$P=10^{29}P_{29}\,\textrm{ergs\,cm}^{-3}$. Combining equations
(\ref{3e3}), (\ref{3e201}) and (\ref{3e32}) to eliminate pressure
$P$, we obtain
\begin{equation}
\begin{array}{ll}
m\dot{m}_{d}fr_{6}^{-3}=0.199\dot{m}_{d}T_{11}^{4}\rho_{11}^{-1}r_{6}^{-2}\\
m^{2/3}\rho_{11}^{1/3}\dot{m}_{d}^{2/3}f^{2/3}\alpha_{-1}^{-2/3}r_{6}^{-2}=
0.953 T_{11}^{4}.\label{3s110}
\end{array}
\end{equation}
Then we find that the density and temperature in the disk are
\begin{equation}
\begin{array}{ll} \rho_{11}=0.0953 \dot{m}_{d} m^{-1/2}f^{-1/2}\alpha_{-1}^{-1}r_{6}^{-3/2}\\
T_{11}=0.832
m^{1/8}\dot{m}_{d}^{1/4}f^{1/8}\alpha_{-1}^{-1/4}r_{6}^{-5/8}.
\end{array}\label{3s11}
\end{equation}
Also, the pressure of the disk from equation (\ref{3e32}) becomes
\begin{equation}
P_{29}=3.32m^{1/2} \dot{m}_{d}f^{1/2}\alpha_{-1}^{-1}r_{6}^{-5/2}.
\label{3s12}
\end{equation}
Equations (\ref{3s11}) and (\ref{3s12}) tell us that for an
advection-dominated disk with the radiation pressure to be
dominated, the density $\rho$, pressure $P$ and temperature $T$ all
decrease as radius $r$ increase. These quantities increase with
increasing the accretion rate $\dot{m}_{d}$, and are all independent
of the electron fraction $Y_{e}$. $\rho$ decreases but $P$ and $T$
increase with increasing the mass of the central compact star.
Besides, in the region near the neutron star surface where the
radius $r$ is relatively small, the term $f$ is able to decrease $P$
and $T$ but increase $\rho$ significantly.

We have to check the validity of the assumption made in deriving the
above solution. First, the assumption of ADAF is valid only if
$Q_{\rm adv}^{-}>Q_{\nu}^{-}$. Since the radiation entropy in the
quantity $Q_{\rm adv}^{-}$ is always much larger than the gas
entropy in an ADAF, and $Q_{\rm eN}^{-}$ is usually the dominant
term $Q_{\nu}^{-}$ (where we consider that the outer disk is
optically thin to neutrinos), $Q_{\rm adv}^{-}>Q_{\rm eN}^{-}$
requires
\begin{equation}
r_{6}f^{1/5}>2.28m^{-3/5}\dot{m}_{d}^{6/5}\alpha_{-1}^{-2}.\label{3iq1}
\end{equation}

Furthermore, we require that the gas pressure and the degeneracy
pressure are smaller than the radiation pressure ( $P_{\rm rad}>
P_{\rm gas}$ and $P_{\rm rad}> P_{\rm deg}$ ). From equations
(\ref{3s11}) and (\ref{3s12}), we obtain
\begin{equation}
r_{6}f^{-7/3}<74.6\left(1+Y_{e}\right)^{-8/3}m^{7/3}\dot{m}_{d}^{-2/3}\alpha_{-1}^{2/3},\label{3iq21}
\end{equation}
and
\begin{equation}
r_{6}f^{-7/3}<174m^{7/3}\dot{m}_{d}^{-2/3}\alpha_{-1}^{2/3}Y_{e}^{-8/3}.\label{3iq22}
\end{equation}

Inequation (\ref{3iq1}) gives the range of an advection-dominated
region in the disk with the radiation pressure to be dominated, and
this region decreases with decreasing the mass of the center compact
star $m$, $\alpha$ or increasing the accretion rate $\dot{m}_{d}$.
For a fixed radius $r$, we can rewrite inequation (\ref{3iq1}) as
\begin{equation}
\dot{m}_{d}<0.504m^{1/2}\alpha_{-1}^{5/3}r_{6}^{5/6}f^{1/6}.\label{3e301}
\end{equation}
which means that a larger radius allows a larger upper limit of the
accretion rate for the radiation-pressure-dominated region. On the
other hand, if the parameters $m$, $\alpha$ and $\dot{m}_{d}$ in
some region of the disk do not satisfy inequations (\ref{3iq1}) or
(\ref{3e301}), the other types of pressure can exceed the radiation
pressure but the region can still be advection-dominated. For an
analytical purpose, we want to discuss the range of different types
of pressure in two extreme cases where $Y_{e}\sim 1$ or $Y_{e}\ll
1$. From inequation (\ref{3iq21}), we can see that, since the
minimum value of $r_{6}f^{-7/3}$ is 19.9, when $\dot{m}_{d}>
0.453m^{7/2}\alpha_{-1}$ for $Y_{e}\sim 1$, or $\dot{m}_{d}>
7.25m^{7/2}\alpha_{-1}$ for $Y_{e}\ll 1$ , the gas pressure takes
over the radiation pressure in the disk and the entire disk becomes
gas-pressure-dominated. On the other hand, the degeneracy pressure
is larger than the radiation pressure at a very large radius if the
electron fraction $Y_{e}$ is not very small. However, we do not
consider this situation for an ADAF region of the disk, because the
degeneracy pressure, even if larger than the radiation pressure,
cannot exceed the gas pressure.

When the gas pressure is dominated and the outer disk flow is still
an ADAF, we consider another set of equations,
\begin{equation}
\begin{array}{ll}
m\dot{m}_{d}fr_{6}^{-3}=0.199\dot{m}_{d}T_{11}^{4}\rho_{11}^{-1}r_{6}^{-2}\\
m^{2/3}\rho_{11}^{1/3}\dot{m}_{d}^{2/3}f^{2/3}\alpha_{-1}^{-2/3}r_{6}^{-2}=1.14
\left(1+Y_{e}\right)\rho_{11}T_{11}.
\end{array}
\end{equation}
and obtain the density, temperature and pressure of the disk,
\begin{equation}
\begin{array}{ll} \rho_{11}=0.556\left(1+Y_{e}\right)^{-12/11}
m^{5/11}\dot{m}_{d}^{8/11}f^{5/11}\alpha_{-1}^{-8/11}r_{6}^{-21/11}\\
T_{11}=1.29\left(1+Y_{e}\right)^{-3/11}m^{4/11}
\dot{m}_{d}^{2/11}f^{4/11}\alpha_{-1}^{-2/11}r_{6}^{-8/11}\\
P_{29}=5.98\left(1+Y_{e}\right)^{-4/11}m^{9/11}
\dot{m}_{d}^{10/11}f^{9/11}\alpha_{-1}^{-10/11}r_{6}^{-29/11} .
\end{array}\label{3e302}
\end{equation}
Equation (\ref{3e302}) tells us that for an advection-dominated disk
with dominant gas pressure, $\rho$, $P$ and $T$ increase with
increasing parameters $\dot{m}_{d}$ and $m$ or decreasing $Y_{e}$.

The assumption of a gas-pressure-dominated ADAF disk requires
$Q^{-}_{\rm adv}>Q^{-}_{\nu}$, and thus we obtain
\begin{equation}
r_{6}^{47/22}f^{-20/11}>128\left(1+Y_{e}
\right)^{-26/11}m^{29/22}\dot{m}_{d}^{10/11}\alpha_{-1}^{-21/11}.\label{3e303}
\end{equation}
$P_{\rm gas}>P_{\rm rad}$ can be written by
\begin{equation}
r_{6}f^{-7/3}>74.6\left(1+Y_{e}\right)^{-8/3}
m^{7/3}\dot{m}_{d}^{-2/3}\alpha_{-1}^{2/3}.\label{3iq22}
\end{equation}
$P_{\rm gas}>P_{\rm deg}$ leads to
\begin{equation}
r_{6}f^{-7/3}<8.07\times10^{3}\left(1+Y_{e}\right)^{12}Y_{e}^{-44/3}m^{7/3}\dot{m}_{d}^{-2/3}\alpha_{-1}^{2/3},\label{3e304}
\end{equation}
which is satisfied for a large parameter space.

If $Y_{e}\sim 1$ and the parameters $m$ and $\alpha$ make
$0.453m^{7/2}\alpha_{-1}<\dot{m}_{d}<1.73m^{-29/20}\alpha_{-1}^{21/10}$
valid, the entire disk becomes an advection-dominated disk with the
gas pressure to be dominated. However, if $Y_{e}\ll 1$, such a disk
cannot exist, since inequation (\ref{3e303}) cannot be always
satisfied in the entire disk and some region of the disk would
become neutrino-dominated. Also, when the mass accretion rate
$\dot{m}_{d}$ becomes higher, most region of the disk becomes
neutrino-dominated.

In the region where neutrino cooling is efficient, using equations
(\ref{3e3}), (\ref{3e201}) and (\ref{3e32}) to eliminate $P$ and
$T$, we have an equation about the density,
\begin{eqnarray}
m^{2/3}\dot{m}_{d}^{2/3}f^{2/3}\alpha_{-1}^{-2/3}r_{6}^{-2}
&=&0.618\left(1+Y_{e}\right)\rho_{11}^{5/9}m^{7/36}
\dot{m}_{d}^{1/9}f^{1/9}r_{6}^{-7/12}\alpha_{-1}^{1/18}
\nonumber\\&&+0.794Y_{e}^{4/3}\rho_{11}+0.0810m^{7/9}\dot{m}_{d}^{4/9}
f^{4/9}r_{6}^{-7/3}\alpha_{-1}^{2/9}\rho_{11}^{-7/9}.
\end{eqnarray}
From this equation, we can obtain two solutions if $\dot{m}_{d}$ is
large enough to satisfy the assumption of NDAF. The first solution
is gas or relativistic degeneracy pressure-dominated, and the second
is radiation pressure-dominated. From the second solution, we can
find a hotter but thinner disk than the first solution. However,
using the second solution to calculate $Q_{\rm adv}^{-}$ and
$Q_{\nu}^{-}$, we find $Q_{\rm adv}^{-}> Q_{\nu}^{-}$, which
contradicts with the assumption of NDAF. Therefore, we have to
choose the first solution. We further assume that the gas pressure
dominates over the degeneracy pressure and obtain
\begin{equation}
\begin{array}{ll} \rho_{11}=2.38\left(1+Y_{e}\right)^{-9/5}m^{17/20}\dot{m}_{d}f\alpha_{-1}^{-13/10}r_{6}^{-51/20} \\
T_{11}=0.490\left(1+Y_{e}\right)^{1/5}m^{1/10}
\alpha_{-1}^{1/5}r_{6}^{-3/10} \\
P_{29}=9.71\left(1+Y_{e}\right)^{-3/5}m^{19/20}
\dot{m}_{d}f\alpha_{-1}^{-11/10}r_{6}^{-57/20}.
\end{array}\label{3s2}
\end{equation}
In this case, the temperature is independent of the accretion rate
$\dot{m}_{d}$. In addition, as $Y_{e}$ decreases (that is, the disk
contains more neutrons), the temperature decreases, which is quite
different from the situation of ADAF.

Finally we check the gas pressure-dominated assumption. Using
equation (\ref{3s2}) and assuming $P_{\rm gas}>P_{\rm deg}$, we
obtain
\begin{equation}
r_{6}f^{-20/33}>3.18Y_{e}^{80/33}\left(1+Y_{e}
\right)^{-36/11}m^{1/3}\dot{m}_{d}^{20/33}\alpha_{-1}^{-38/33}.
\label{3e5}
\end{equation}
Inequation (\ref{3e5}) is always satisfied if $Y_{e}\ll 1$, and thus
we can say that the gas pressure-dominated assumption is valid.
However, if $Y_{e}\sim 1$, a part of the disk becomes degeneracy
pressure-dominated if
$\dot{m}_{d}>64.5\alpha_{-1}^{19/10}m^{-11/20}$. In particular, in
the case of $Y_{e}\sim 1$ and large accretion rate $\dot{m}_{d}$, a
set of solutions on the part of the disk are
\begin{equation}
\begin{array}{ll} \rho_{11}=1.26Y_{e}^{-4/3}m^{2/3}\dot{m}_{d}^{2/3}
f^{2/3}\alpha_{-1}^{-2/3}r_{6}^{-2} \\
T_{11}=0.526Y_{e}^{4/27}m^{13/108}\dot{m}_{d}^{1/27}f^{1/27}\alpha_{-1}^{7/54}
r_{6}^{-13/36} \\
P_{29}=7.85Y_{e}^{-4/9}m^{8/9}\dot{m}_{d}^{8/9}f^{8/9}
\alpha_{-1}^{-8/9}r_{6}^{-8/3}.\label{3e3031}
\end{array}
\end{equation}
which describe an NDAF with the degeneracy pressure to be dominated.
However, we should remember that in deriving the above solutions we
have not considered the constraint of $r>\tilde{r}$.

Here we make a summary of $\S$ 3.3.1. We used an analytical method
to solve the density, pressure and temperature of the outer disk
based on a simple model discussed at the beginning of $\S$3.3. The
accretion flow may be ADAF or NDAF with the radiation, gas or
degeneracy pressure to be dominated. We fix radius $r$ and show the
possible cases in the disk with different accretion rate
$\dot{m}_{d}$ in Table \ref{tab32}. Moreover, for different electron
fraction $Y_{e}$ and fixing $m=1$ and $\alpha_{-1}=1$, we calculate
the upper limit of $\dot{m}_{d}$. If $Y_{e}\sim 1$, the
advection-dominated region in the disk can be radiation or gas
pressure-dominated, and the neutrino-cooled region can be gas or
degeneracy pressure-dominated. However, if $Y_{e}\ll 1$, the gas
pressure-dominated region in the ADAF case is very small, and the
degeneracy pressure-dominated region cannot exist. In \S 3.3.4 we
will obtain similar results by using a numerical method.
\begin{table}
\begin{center}
\begin{tabular}{rccccrc}
\hline \hline
dominant pressure & accretion flow & range of $\dot{m}_{d}$ & $Y_{e}\sim 1$ & $Y_{e}\ll 1$ \\
\hline
$P_{\rm rad}$....................... & ADAF & (a) \& (b) & 0.453 & 7.25\\
$P_{\rm gas}$....................... & ADAF & (c) \& (d) & 1.73 & ---\\
$P_{\rm gas}$....................... & NDAF & (e) \& (f) & 64.5 & $\infty$\\
$P_{\rm deg}$....................... & NDAF & (g) & $\infty$ & ---\\
\hline \hline
\end{tabular}
\caption{Range of the accretion rate in different regions as
follows:
$(\textrm{a}):\,\dot{m}_{d}<0.504m^{1/2}\alpha_{-1}^{5/3}r_{6}^{5/2}f^{1/6}$;
$(\textrm{b}):\,\dot{m}_{d}<644(1+Y_{e})^{-4}m^{7/2}\alpha_{-1}r_{6}^{-3/2}f^{7/2}$;
$(\textrm{c}):\,\dot{m}_{d}>644(1+Y_{e})^{-4}m^{7/2}\alpha_{-1}r_{6}^{-3/2}f^{7/2}$;
$(\textrm{d}):\,\dot{m}_{d}<4.81\times10^{-3}(1+Y_{e})^{13/5}m^{-29/20}\alpha_{-1}^{21/10}r_{6}^{47/20}f^{-2}$;
$(\textrm{e}):\,\dot{m}_{d}>4.81\times10^{-3}(1+Y_{e})^{13/5}m^{-29/20}\alpha_{-1}^{21/10}r_{6}^{47/20}f^{-2}$;
$(\textrm{f}):\,\dot{m}_{d}<0.148Y_{e}^{-4}(1+Y_{e})^{27/5}m^{-11/10}\alpha_{-1}^{19/10}r_{6}^{33/20}f^{-1}$;
$(\textrm{g}):\,\dot{m}_{d}>0.148Y_{e}^{-4}(1+Y_{e})^{27/5}m^{-11/10}\alpha_{-1}^{19/10}r_{6}^{33/20}f^{-1}$.}
\end{center}\label{tab32}
\end{table}

The solutions given in this subsection can also be used to discuss
the properties of the disk around a black hole. Our analytical
solutions of the outer disk are similar to those of Narayan et al.
(2001) and Di Matteo et al. (2002), who took $Y_{e}$ to be a
parameter. Narayan et al. (2001) found that advection-dominated
disks can be radiation or gas pressure-dominated, and
neutrino-dominated disks can be gas or degeneracy pressure-dominated
instead. This is consistent with our conclusion for $Y_{e} \sim 1$.
However, these authors did not consider the factor
$f=1-\sqrt{r_{*}/r}$, which is an important factor because a small
$r$, as we have mentioned above, can dramatically change the
parameter space of the outer disk. Di Matteo et al. (2002) discussed
the different pressure components (their Fig. 2), which is also
consistent with our conclusion. Chen \& Beloborodov (2007)
calculated the value of $Y_{e}$ and showed that $Y_{e}\ll 1$ when
$r$ is small. According to the above discussion, therefore, the
degeneracy-pressure-dominated region in the NDAF disk cannot exist.
This is consistent with Chen \& Beloborodov (2007) that the pressure
in a neutrino-cooled disk is dominated by baryons (gas).

On the other hand, our analytical results are partly different with
Kohri et al. (2005) and Chen \& Beloborodov (2007) which showed that
the electron pressure is dominant in some advection-dominated
regions of the disk. This difference is mainly because that we take
$P_{\rm rad}=11aT^{4}/12$ in our analytical model, which includes
the contribution of relativistic electron-positron pairs. However,
Kohri et al. (2005) and Chen \& Beloborodov (2007) took $P_{\rm
rad}=aT^{4}/3$ and calculated the pressure of $e^{+}e^{-}$ pairs in
the term of electron pressure $P_{e}$. As a result, the radiation
pressure we consider in this subsection is actually the pressure of
a $\gamma$-$e^{+}e^{-}$ plasma.

In the following subsection, we will solve the structure of the
inner disk and use the value of $\tilde{r}$ to further constrain the
solutions that we have obtained.

\subsection{The inner disk}  
\subsubsection{Boundary layer between the inner disk and outer disk}
We use the method discussed in \S 3.2.1 and compare the radial
velocity with the local speed of sound of the outer disk using the
results given in section \S 3.3.1.

The radial velocity of the outer disk at radius $r$ is
\begin{equation}
v_{1}=\frac{\dot{M}}{2\pi r\Sigma}=\frac{\dot{M}}{4\pi r\rho_{1}
H}=\frac{\dot{M}\sqrt{GM}}{4\pi r^{5/2}\sqrt{P_{1} \rho_{1}}}.
\end{equation}
Hence, we have
\begin{equation}
\frac{v_{1}^{2}}{c_{s}^{2}}\sim
\frac{\rho_{1}v_{1}^{2}}{P_{1}}=\frac{\dot{M}^{2}GM}{16\pi^{2}r^{5}P_{1}^{2}}
=0.0465\frac{\dot{m}_{d}^{2}m}{r_{6}^{5}P_{1,29}^{2}}.
\end{equation}

In the case where the outer disk is ADAF and the radiation pressure
is dominated, using solutions (\ref{3s11}) and (\ref{3s12}), we have
\begin{equation}
\frac{v_{1}^{2}}{c_{s}^{2}}\sim
4.22\times10^{-3}\alpha_{-1}^{2}f^{-1}.
\end{equation}
Therefore, we see $v_{1}\ll c_{s}$ for typical values of the
parameters.

In the case of ADAF with the gas pressure to be dominated, using
solution (\ref{3e302}), we have
\begin{equation}
\frac{v_{1}^{2}}{c_{s}^{2}}\sim
1.302\times10^{-3}\left(1+Y_{e}\right)^{8/11}
m^{-7/11}\dot{m}_{d}^{2/11}\alpha_{-1}^{20/11}f^{-18/11}r_{6}^{3/11}.\label{3e3211}
\end{equation}

For NDAF, using expression (\ref{3s2}) to compare the radial
velocity with the speed of sound, we have
\begin{equation}
\frac{v_{1}^{2}}{c_{s}^{2}}\sim
4.94\times10^{-4}\left(1+Y_{e}\right)^{6/5}
m^{-9/10}\alpha_{-1}^{11/5}f^{-2}r_{6}^{7/10}.\label{3e3212}
\end{equation}
Note that this ratio is independent of the accretion rate. From
(\ref{3e3211}) and (\ref{3e3212}), we still find that $v_{1}\ll
c_{s}$ is always satisfied except for the region very near the
surface of the neutron star, where the factor $f$ is very small.
This region, however, is so small that it belongs to the inner disk
where we have to use the self-similar structure, which we will
discuss later in details.

Therefore, for the hyperaccretion disk discussed in this chapter, as
the disk is extremely hot and dense, the radial velocity is always
subsonic. So there is no stalled shock between the inner and outer
disks. Thus, all physical variables between two sides of the
boundary layer between two regions of the disk change continuously.
Besides, the rotational velocity is assumed to be the Keplerian
value and has no jump at the boundary layer.

\subsubsection{Solution of the inner disk }  
We now study the inner disk analytically based on the results given
in $\S$3.1. The main problem that we should solve in this subsection
is to determine the size of the inner disk for a range of parameters
and to describe the structure of the inner disk. In the case where
the radiation pressure is dominated, by using the self-similar
structure (\ref{3en02}), we obtain the temperature of the inner
disk,
\begin{equation}
T_{11}=\left(\frac{P_{29}}{6.931}\right)^{1/4}=1.01m^{1/6}\tilde{\rho}_{11}^{1/2}\dot{m}_{d}^{1/6}
\tilde{f}^{1/6}\alpha_{-1}^{-1/6}\tilde{r}_{6}^{\frac{2-\gamma}{4(\gamma-1)}}r_{6}^{\frac{-\gamma}{4(\gamma-1)}}.\label{3e11}
\end{equation}
where $\tilde{r}$ and $\tilde{\rho}$ are the radius and density of
the boundary layer between the inner and outer disk, and
$\tilde{f}=1-(r_{*}/\tilde{r})^{1/2}$.

Using the self-similar condition and the above expression of
$T_{11}$, we find the total neutrino cooling rate,
\begin{equation}
\int_{r_{*}}^{\tilde{r}}Q_{\nu}^{-}2\pi
rdr=10^{51}\times\int120\tilde{\rho}_{11}^{7/6}
m^{5/6}\dot{m}_{d}^{4/3}\alpha_{-1}^{-4/3}\tilde{f}^{4/3}
\tilde{r}_{6}^{\frac{9-4\gamma}{2(\gamma-1)}}r_{6}^{\frac{\gamma-6}{2(\gamma-1)}}dr_{6}.
\end{equation}

The outer disk flow is mainly an ADAF, using the solution of a
radiation pressure-dominated ADAF (i.e., solutions \ref{3s11} and
\ref{3s12}), we have the total neutrino cooling rate,
\begin{eqnarray}
L_\nu  = 1.55\times10^{52}\,{\rm ergs}\,{\rm
s}^{-1}\left(\frac{\gamma-1}{8-3\gamma}\right)
m^{1/4}\dot{m}_{d}^{5/2}
 \tilde{f}^{3/4} \alpha_{-1}^{-5/2}\tilde{r}_{6}^{\frac{25-15\gamma}
{4(\gamma-1)}}\left(r_{*,6}^{\frac{3\gamma-8}{2(\gamma-1)}}-
\tilde{r}_{6}^{\frac{3\gamma-8}{2(\gamma-1)}}\right).
\end{eqnarray}
From the energy-conservation equation (\ref{3e4}), the position of
the boundary layer satisfies the following equation,
\begin{equation}
\tilde{r}_{6}^{\frac{5(5-3\gamma)}{4(\gamma-1)}}\left(1-\sqrt{\frac{r_{*,6}}{\tilde{r}_{6}}}\right)^{3/4}
\left(r_{*,6}^{\frac{3\gamma-8}{2(\gamma-1)}}-
\tilde{r}_{6}^{\frac{3\gamma-8}{2(\gamma-1)}}\right)
=0.0597\left(\frac{8-3\gamma}{\gamma-1}\right)m^{3/4}\dot{m}_{d}^{-3/2}\alpha_{-1}^{5/2}r_{*,6}^{-1}\label{3e6}
\end{equation}
where we take $\varepsilon\sim 1$ and $\bar{f}_{\nu}\sim 0$ in
equation (\ref{3e4}). The left-hand term of equation (\ref{3e6})
increases with increasing $\tilde{r}_{6}$, so $\tilde{r}_{6}$
decreases if $\dot{m_{d}}$ increases in the right-hand term of this
equation. In other words, the size of the inner disk decreases as
the accretion rate increases. From equation (\ref{3e6}), we can also
see that its solution, $\tilde{r}_{6}$, is independent of $Y_{e}$,
and increases with the mass of the central star. In addition, since
the gas pressure has its own contribution to the disk, the actual
adiabatic index $\gamma$ is larger than 4/3, which can makes the
solution of equation (\ref{3e6}) larger. For an analytical purpose,
we assume several different sets of parameters to solve equation
(\ref{3e6}). Table \ref{tab33} clearly shows that, in the
radiation-pressure-dominated disk with an advection-dominated outer
region, as the accretion rate increases, the value of $\tilde{r}$
decreases, and that $\tilde{r}$ increases with increasing $\gamma$
or decreasing $m$.
\begin{table}
\begin{center}
\begin{tabular}{rccccrc}
\hline \hline
$\dot{m}_{d}$ & 0.01 & 0.05 & 0.1 & 0.5 \\
\hline
Case 1 & 6.38 & 3.53 & 2.76 & 1.65 \\
Case 2 & 13.87 & 5.66 & 3.92 & 1.84 \\
Case 3 & 6.83 & 3.76 & 2.94 & 1.74 \\
\hline \hline
\end{tabular}
\caption{Equation (\ref{3e6}) gives $\tilde{r}_{6}$ in several
cases. Case 1: $m=1, \gamma=4/3$; case 2: $m=1, \gamma=1.4$; case 3:
$m=2.0/1.4, \gamma=4/3$.}\label{tab33}
\end{center}
\end{table}

In the case where the gas pressure is dominated and the outer disk
flow is still an ADAF, we obtain the temperature of the inner disk,
\begin{equation}
T_{11}=\left(1+Y_{e}\right)^{-1}\frac{1}{8.315}
\frac{P_{29}}{\rho_{11}} =\left(1+Y_{e}\right)^{-1}0.874m^{2/3}
\tilde{\rho}_{11}^{-2/3}\dot{m}_{d}^{2/3}\tilde{f}^{2/3}
\alpha_{-1}^{-2/3}\tilde{r}_{6}^{-1}r_{6}^{-1}.
\end{equation}
Similarly, from equations (\ref{3e4}) and (\ref{3e302}), we have
\begin{equation}
\int_{r_{*}}^{\tilde{r}}Q_{\nu}^{-}2\pi
rdr=10^{51}\times\int45.0\tilde{\rho}_{11}^{-10/3}\left(1+Y_{e}\right)^{-6}m^{23/6}\dot{m}_{d}^{13/3}\tilde{f}^{13/3}
\tilde{r}_{6}^{(\frac{1}{\gamma-1}-\frac{13}{2})}r_{6}^{-(4+\frac{1}
{\gamma-1})}dr_{6}.
\end{equation}
The neutrino luminosity of the inner disk reads
\begin{eqnarray}
L_\nu & = & 3.53\times10^{53}\,{\rm ergs}\,{\rm
s}^{-1}\left(\frac{\gamma-1}{3\gamma-2}\right)
\left(1+Y_{e}\right)^{-26/11}\nonumber \\ & & \times
m^{51/22}\dot{m}_{d}^{21/11}
\tilde{f}^{\frac{31}{11}}\alpha_{-1}^{-21/11}\tilde{r}_{6}^{(\frac{1}
{\gamma-1}-\frac{3}{22})}\left(r_{*,6}^{\frac{2-3\gamma}{\gamma-1}}-
\tilde{r}_{6}^{\frac{2-3\gamma}{\gamma-1}}\right).
\end{eqnarray}
The energy-conservation equation of the inner disk is
\begin{eqnarray}
\left(1-\sqrt{\frac{r_{*,6}}{\tilde{r}_{6}}}\right)^{-31/11}\tilde{r}_{6}^{(\frac{-1}
{\gamma-1}+\frac{3}{22})}\left(r_{*,6}^{\frac{2-3\gamma}{\gamma-1}}-
\tilde{r}_{6}^{\frac{2-3\gamma}{\gamma-1}}\right)^{-1}
\nonumber\\
=\frac{382(\gamma-1)}{3\gamma-2}\left(1+Y_{e}\right)^{-26/11}
m^{29/22}\dot{m}_{d}^{10/11}\alpha_{-1}^{-21/11}r_{*,6}.\label{3e321}
\end{eqnarray}

From equation (\ref{3e321}), we see that the size of the inner disk
($\tilde{r}$) also decreases with increasing the accretion rate
$\dot{m}_{d}$. Table \ref{tab34} gives solutions of equation
(\ref{3e321}) with different sets of parameters. We can see that
$\tilde{r}$ also decrease with increasing $\tilde{\gamma}$, $m$ or
decreasing $Y_{e}$.
\begin{table}
\begin{center}
\begin{tabular}{rccccrc}
\hline \hline
$\dot{m}_{d}$ & 0.5 & 1 & 3 & 5 & 10 \\
\hline
Case 1 & 2.17 & 1.91 & 1.62 & 1.52 & 1.42\\
Case 2 & 2.05 & 1.83 & 1.59 & 1.50 & 1.40\\
Case 3 & 1.68 & 1.54 & 1.38 & 1.33 & 1.27\\
Case 4 & 1.97 & 1.76 & 1.52 & 1.44 & 1.36\\
\hline \hline
\end{tabular}
\caption{Equation (\ref{3e321}) gives $\tilde{r}_{6}$ in several
cases. Case 1: $Y_{e}=1, \gamma=5/3, m=1$; case 2: $Y_{e}=1,
\gamma=3/2, m=1$; case 3: $Y_{e}=1/9, \gamma=5/3, m=1$; case 4:
$Y_{e}=1, \gamma=5/3, m=2.0/1.4$.}\label{tab34}
\end{center}
\end{table}

We above study the case where the outer disk is advection-dominated,
and find that the size of the inner disk always increase with the
accretion rate. In the case where the outer disk is mainly
neutrino-dominated, using expression (\ref{3s2}) and equation
(\ref{3e4}), we obtain an energy-conservation equation in the inner
disk,
\begin{equation}
2.77\left(\frac{\gamma-1}{3\gamma-2}\right)
\tilde{r}_{6}^{\frac{2\gamma-1}{\gamma-1}}
\tilde{f}\left(r_{*,6}^{\frac{2-3\gamma}{\gamma-1}}-
\tilde{r}_{6}^{\frac{2-3\gamma}{\gamma-1}}\right)
=0.924\left\{\frac{1}{r_{*,6}}-\frac{3\bar{f}_{\nu}}{\tilde{r}_{6}}\left[1-\frac{2}{3}
\left(\frac{r_{*,6}}{\tilde{r}_{6}}\right)^{1/2}\right]\right\}.\label{3e7}
\end{equation}

This equation shows us that $\tilde{r}_{6}$ in NDAF is independent
of $Y_{e}$, $m$, $\dot{m}_{d}$, and $\alpha$, but only dependent on
$\gamma$ and $\bar{f}_{\nu}$. The size of the inner disk is constant
no matter how much the components, the accretion rate of the disk,
and the mass of the central neutron star are. If the outer disk is
mainly an NDAF, we have $\bar{f}_{\nu}\sim 1$. We choose several
different sets of parameters to obtain the solution of equation
(\ref{3e7}) (see Table \ref{tab35}). As $\bar{f}_{\nu}$ increases,
the value of $\tilde{r}$ decreases slightly. Here we also consider
an intermediate case of $\gamma$ between $5/3$ and $4/3$, the
decline of $\gamma$ makes the size of the inner disk decrease.
\begin{table}
\begin{center}
\begin{tabular}{rccccrc}
\hline \hline
$\bar{f}_{\nu}$$\setminus$$\gamma$ & 5/3 & 3/2 & 4/3 \\
\hline
0.9 & 1.31 & 1.28 & 1.26 \\
0.7 & 1.46 & 1.43 & 1.39 \\
0.5 & 1.56 & 1.53 & 1.47 \\
\hline \hline
\end{tabular}
\caption{Equation (\ref{3e7}) gives $\tilde{r}_{6}$. We take
$\bar{f}_{\nu}$ and $\gamma$ as parameters to solve this
equation.}\label{tab35}
\end{center}
\end{table}

However, in the above discussion about an NDAF, we have not
considered the effect of neutrino opacity but simply assumed that
neutrinos escape freely. Actually, if the accretion rate is
sufficiently large and the disk flow is mainly an NDAF, and the
disk's region near the neutron star surface can be optically thick
to neutrino emission. With increasing the accretion rate, the area
of this optically thick region increases. We now estimate the effect
of neutrino opacity on the structure of the inner disk. Let the
region of $r_{*}<r<\bar{r}$ be optically thick to neutrino emission,
the region of $\bar{r}<r<\tilde{r}$ be optically thin, and the
electron/positron pair capture reactions be the dominant cooling
mechanism. Thus equation (\ref{3e4}) becomes
\begin{equation}
\int_{r_{*}}^{\bar{r}}\frac{4\left(\frac{7}{8}\sigma_{B}T^{4}\right)}{\tau}2\pi
rdr+\int_{\bar{r}}^{\tilde{r}}9\times10^{34}\rho_{11}T_{11}^{6}H2
\pi rdr =\frac{3GM\dot{M}}{4} \left\{{\frac{1}{3
r_{*}}-\frac{\bar{f}_{\nu}}{\tilde{r}}\left[1-\frac{2}{3}\left(\frac{r_{*}}
{\tilde{r}}\right)^{1/2}\right]}\right\}.\label{3e10}
\end{equation}
where we take $\varepsilon\approx 1$. Using the self-similar
relations and performing some derivations, we find
\begin{eqnarray}
18.7\left(1+Y_{e}\right)^{8/5}\left(\frac{\gamma-1}{2-\gamma}\right)
m^{-1/5}\dot{m}_{d}^{-1}\tilde{f}^{-1}\alpha_{-1}^{8/5}
\tilde{r}_{6}^{(\frac{18}{5}-\frac{1}{\gamma-1})}
\left[\bar{r}_{6}^{-1+1/(\gamma-1)}
-r_{*,6}^{-1+1/(\gamma-1)}\right]
\nonumber\\+2.77\left(\frac{\gamma-1}{3\gamma-2}\right)m\dot{m}_{d}\tilde{f}
\tilde{r}_{6}^{\frac{2\gamma-1}{\gamma-1}}\left(\bar{r}_{6}^{\frac{2-3\gamma}
{\gamma-1}}-\tilde{r}_{6}^{\frac{2-3\gamma}{\gamma-1}}\right)
=0.924m\dot{m}_{d}\left[\frac{1}{r_{*,6}}-\frac{3\bar{f}_{\nu}}
{\tilde{r}_{6}}+\frac{2\bar{f}_{\nu}}{\tilde{r}_{6}}\left(\frac{r_{*,6}}
{\tilde{r}_{6}}\right)^{1/2}\right]. \label{3e8}
\end{eqnarray}
The solution shown by equation (\ref{3e8}) gives $\tilde{r}$. We
take $\bar{f}_{\nu}\approx 1$ and $\alpha=0.1$. Moreover, we define
a new parameter $k=\bar{r}_{6}/\tilde{r}_{6}$, and assume several
sets of parameters to give the solution of equation (\ref{3e8}).
\begin{table}
\begin{center}
\begin{tabular}{rccccrc}

\hline \hline
$\dot{m}_{d}$ & 80 & 100 & 120 & 150 \\
\hline
Case 1 & 4.28 & 5.13 & 5.94 & 7.10 \\
Case 2 & 6.26 & 7.48 & 8.65 & 10.33 \\
Case 3 & 5.09 & 6.09 & 7.05 & 8.43 \\
Case 4 & 4.10 & 4.99 & 5.86 & 7.11 \\
Case 5 & 4.87 & 5.90 & 6.89 & 8.32 \\
\hline \hline
\end{tabular}
\caption{Equation (\ref{3e8}) gives $\tilde{r}_{6}$ in several
different cases. Case 1: $Y_{e}=1, m=1, k=1, \gamma=5/3$; case 2:
$Y_{e}=1/ 9, m=1, k=1, \gamma=5/3$; case 3: $Y_{e}=1, m=2.0/1.4,
k=1, \gamma=5/3$, Case 4: $Y_{e}=1, m=1, k=0.7, \gamma=5/3$; case 5:
$Y_{e}=1, m=1, k=1, \gamma=3/2$. }\label{tab36}
\end{center}
\end{table}

From Table \ref{tab36}, we can see that the size of the inner disk
increases with the accretion rate. In addition, an increase of $m$
or $k$, or an decrease of $Y_{e}$ also makes the inner-disk size
larger. We will compare the analytical results from Tables
\ref{tab33} to \ref{tab36} with numerical results in $\S$3.3.4 in
more details.

\subsection{Discussion about the stellar surface boundary}  
Now we want to discuss the physical condition near the neutron star
surface. We know that if the rotational velocity of the neutron star
surface is different from that of the inner boundary of the disk,
then the disk can act a torque on the star at the stellar radius.
Here we take the rotational velocity of the inner disk $\Omega\simeq
\Omega_{K}$ approximately as mentioned in \S 3.2.1 and \S 3.2.3.
Then we always have the stellar surface angular velocity
$\Omega_{*}$ to be slower than that of the inner disk $\Omega_{K}$
(i.e., $\Omega_{*}< \Omega_{K}$). As a result, the kinetic energy of
the accreted matter is released when the angular velocity of the
matter decreases to the angular velocity of the neutron star
surface. From the Newtonian dynamics, we obtain a differential
equation,
\begin{equation}
G_{*}r_{*}=\frac{d(I\Omega_{*})}{dt},\label{3e331}
\end{equation}
where $G_{*}r_{*}=\dot{M}r_{*}^{2}(\Omega_K-\Omega_{*})$ is the
torque acting on the star surface from the disk, $I$ is the moment
of inertia of the star, $I=\xi M r_{*}^{2}$, with $\xi$ being a
coefficient. The above equation can be further written as
\begin{equation}
[\Omega_K-\Omega_{*}(1+\xi)]dM=\xi Md\Omega_{*}. \label{3s3}
\end{equation}
From equation (\ref{3s3}) we can see that if
$\Omega_{*,0}<\Omega_K/(1+\xi)$ initially, we always have
$d\Omega_{*}/dM>0$, i.e., the neutron star is spun up as accretion
proceeds. On the other hand, if $\Omega_{*,0}>\Omega_K/(1+\xi)$
initially, then the star is always spun down by the disk. The limit
value of $\Omega_{*}$ is $\Omega_K/(1+\xi)$ in both cases. The
solution of equation (\ref{3s3}) is
\begin{equation}
\left|\Omega_{*}-\frac{\Omega_K}{1+\xi}\right|=\left(\frac{M_{0}}{M_{0}+\Delta
M}\right)^{\frac{\xi+1}{\xi}}\left|\Omega_{*,0}-\frac{\Omega_K}{1+\xi}\right|\approx
\left[1-\left(\frac{\xi+1}{\xi}\right)\frac{\Delta
M}{M_{0}}\right]\left|\Omega_{*,0}-\frac{\Omega_K}{1+\xi}\right|,\label{3e9}
\end{equation}
where $M_{0}$ is the initial mass of the neutron star, and $\Delta
M$ is the mass of the accreted matter from the disk by the star. The
change of $\Omega_{*}$ depends on the ratio $\Delta M/M_{0}$. For
example, if we assume $M_{0}=1.4M_{\odot}$, $\Delta M=0.01M_{\odot}$
and set $\xi=2/5$, then we have
$(\frac{\Omega_K}{1+\xi}-\Omega_{*})=0.98(\frac{\Omega_K}{1+\xi}-\Omega_{*,0})$.
Or if we take $\Delta M=0.1M_{\odot}$, $\xi=2/5$, and the central
stellar mass is unchanged, then we obtain
$(\frac{\Omega_K}{1+\xi}-\Omega_{*})=0.79(\frac{\Omega_K}{1+\xi}-\Omega_{*,0})$.

Here we consider that the energy released from the surface boundary
is also taken away by neutrino emission, and assume that neutrinos
emitted around the stellar surface are opaque. We can estimate the
temperature of the neutron star surface through
$\frac{7}{8}\sigma_{B}T^{4}4\pi r_{*}H\sim
\frac{GM\dot{M}}{2r_{*}}(2-\varepsilon)$, where $H$ is the half
thickness of the inner boundary of the disk, and the parameter
$\varepsilon$ has the same meaning as that in $\S$2.3. Then we can
estimate the surface temperature as $T_{11}\sim
0.415(2-\varepsilon)^{1/4}m^{1/4}\dot{m}_{d}^{1/4}H_{6}^{-1/4}r_{*,6}^{-1/2}$.
This estimation of the temperature is only valid if the inner disk
is optically thin for neutrinos and only the surface boundary is
optically thick. If the inner disk, due to a large accretion rate,
becomes optically thick to neutrino emission, the surface
temperature should be higher.

\subsection{Comparison with numerical results}
In order to give an analytical solution of the accretion disk around
a neutron star in the simple model, we can choose the dominant terms
in equations (\ref{3e201}) and (\ref{3e32}). We now consider one
type of pressure to be dominated, and assume extreme cases from ADAF
to NDAF. For example, we consider the neutrino-dominated region with
$\bar{f}_{\nu}\sim1$ and the advection-dominated region with
$\bar{f}_{\nu}=0$. In this subsection, we solve the
hyperaccretion-disk structure numerically based on the simple model.
In order to compare the numerical results with what we have obtained
analytically, we keep on with using the equations and definitions of
all the parameters in this chapter. However, we first need to point
out several approximations and some differences between numerical
and analytical methods based in the simple model.

First of all, we choose the range of $r_{6}$ from 1 to 15 in our
numerical calculations. In other words, we take the range of the
accretion disk to be from the surface of the neutron star to the
radius of 150 km as the outer boundary. During the compact-star
merger or massive-star collapse, the torus around a neutron star has
only some part that owns a large angular momentum to form a debris
disk, so the mass of the disk may be smaller than the total mass of
the torus. From view of simulations (Lee \& Ramirez-Ruiz 2007), if
the debris disk forms through the merger of two neutron stars, its
outer radius can be slightly smaller than the value we give above.
On the other hand, the disk size may be slightly larger than that we
assume above if the disk forms during the collapse of a massive
star. The changes of physical variables of the disk due to a change
of the outer radius may be insignificant, and we don't discuss the
effect of the outer radius in this chapter.

Second, we fix the viscosity parameter $\alpha$ to be 0.1. If
$\alpha$ decreases (or increases), the variables of the disk
increases (or decreases). More information can be seen in analytical
solutions in \S 3.3.1 and \S 3.3.2. In numerical calculation, we
take $\alpha$ to be a constant.

Third, we still set $Y_{e}$ as a parameter in numerical calculations
in \S 3.4. In order to show results clearly, we consider two
conditions: one is an extreme condition with $Y_{e}=1$, which means
that the disk is made mainly of electrons and protons but no
neutrons; the other is $Y_{e}=1/9$, which means that the number
ratio of electrons, protons and neutrons is $1:1:8$. An elaborate
work should consider the effect of $\beta$-equilibrium, and we will
discuss it in detail in $\S$3.4.

For numerical calculations, we consider all the terms of the
pressure and an intermediate case between ADAF and NDAF. In
analytical calculations we take the adiabatic index $\gamma$ of the
inner disk to be 5/3 if the disk is gas pressure-dominated or
$\gamma=4/3$ if the disk is radiation or degeneracy
pressure-dominated. In numerical calculations, however, it is
convenient to introduce an ``equivalent" adiabatic index $\gamma$
based on the original definition of $\gamma$ from the first law of
thermodynamics, $\gamma=1+P/u$, where $P$ is the pressure of the
disk at a given $r$, and $u$ is the internal energy density at the
same radius. Therefore $\gamma$ is a variable as a function of
radius. We obtain the self-similar structure,
\begin{equation}
\frac{\rho(r)}{\rho(r+dr)}=\left(\frac{r}{r+dr}\right)^{-1/(\gamma(r)-1)},
\,\,\,\,
\frac{P(r)}{P(r+dr)}=\left(\frac{r}{r+dr}\right)^{-\gamma(r)/(\gamma(r)-1)}.\label{3e34}
\end{equation}
If $\gamma$ does not vary significantly in the inner disk, the
difference between the approximate solution where $\gamma$ is a
constant and the numerical solution where we introduce an
``equivalent" $\gamma$ is not obvious.

In addition, some region of the accretion disk is optically thick to
neutrino emission when $\dot{m}_{d}$ is sufficiently large. We use
the same expressions of neutrino optical depth and emission at the
beginning of \S3. We also require that the neutrino emission
luminosity per unit area is continuous when the optical depth
crosses $\tau=1$.

\begin{figure}
\centering\resizebox{0.7\textwidth}{!} {\includegraphics{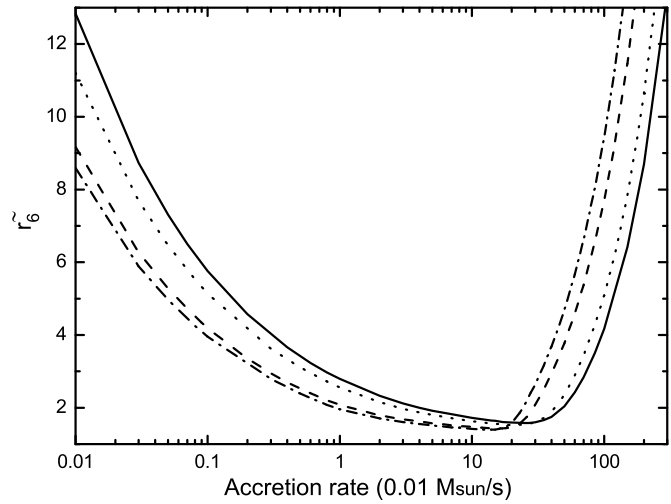}}
\caption{The radius $\tilde{r}_{6}$ of the boundary layer between
the inner and outer disks as a function of accretion rate
$\dot{m}_{d}$ in the simple model for several sets of parameters:
(1) $M=1.4M_{\odot}$ and $Y_{e}=1.0$ ({\em solid line}), (2)
$M=1.4M_{\odot}$ and $Y_{e}=1/9$ ({\em dashed line}), (3)
$M=2.0M_{\odot}$ and $Y_{e}=1.0$ ({\em dotted line}), and (4)
$M=2.0M_{\odot}$ and $Y_{e}=1/9$ ({\em dash-dotted
line}).}\label{fig31}
\end{figure}

We first calculate the value of $\tilde{r}$, which is the radius of
the boundary layer between the inner and outer disks. Figure
\ref{fig31} shows $\tilde{r}$ as a function of $\dot{m}_{d}$ for
different values of $m$. We can see that when the disk flow is an
ADAF at a low accretion rate, $\tilde{r}$ decreases monotonously as
the accretion rate increases until the value of $\tilde{r}$ reaches
a minimum. At this minimum, the outer disk flow is an NDAF and most
of the neutrinos generated from the outer disk can escape freely,
and the effect of neutrino opacity is not important. If the
accretion rate is higher and makes the effect of neutrino opacity
significant, the value of $\tilde{r}$ increases with increasing the
accretion rate. From Figure \ref{fig31}, we see that when the
accretion rate is either low or sufficiently high, $\tilde{r}$ is
very large and even reaches the value of the outer radius, which
means that the inner disk totally covers the outer disk. In this
situation, since the outer disk is covered, no part of the disk is
similar to that of the accretion disk around a black hole as we
discussed in $\S$3.1, and thus we say that the entire disk becomes a
self-similar structure and that the physical variables of the entire
disk are adjusted to build the energy balance between heating and
neutrino cooling. For this situation, we do not want to discuss in
details any more. We focus on the accretion rate which allows the
two-steady parts of disks to exist.

\begin{figure}
\centering\resizebox{1.0\textwidth}{!}{\includegraphics{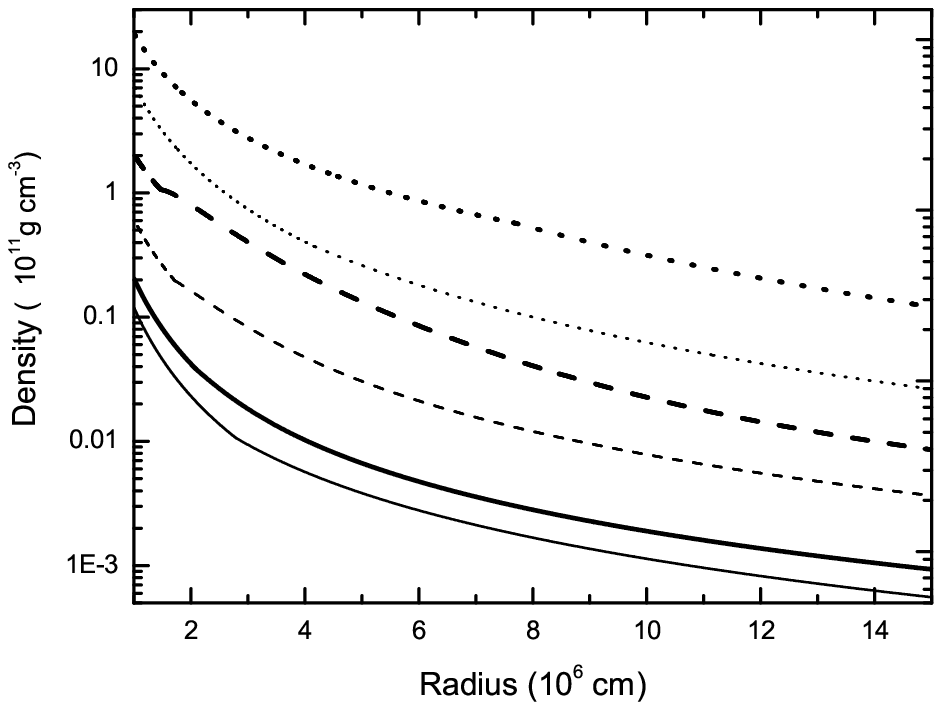}
\includegraphics{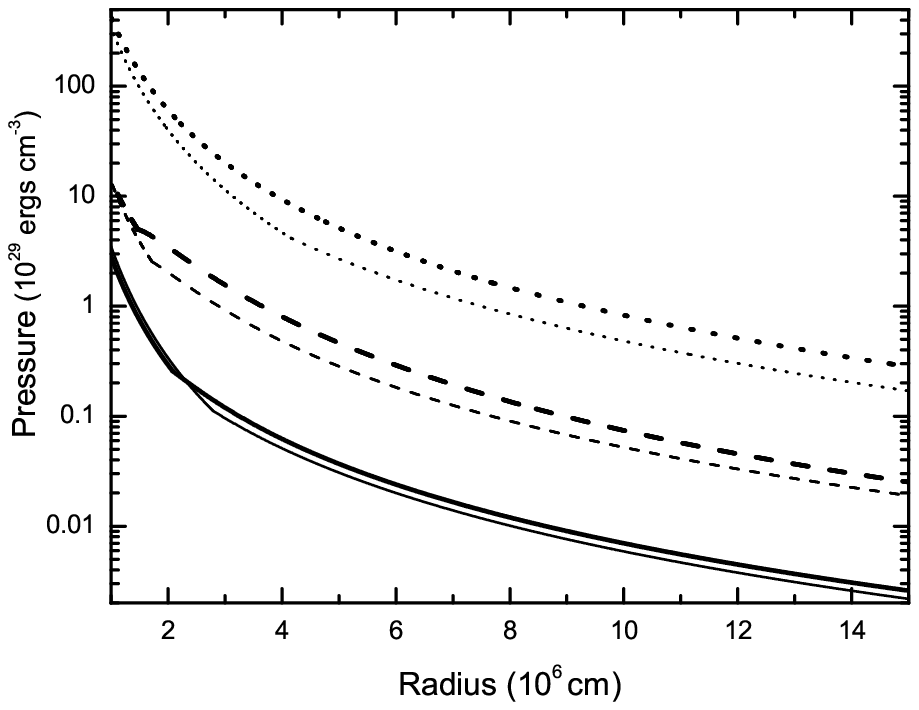}}\\
\centering\resizebox{1.0\textwidth}{!}{\includegraphics{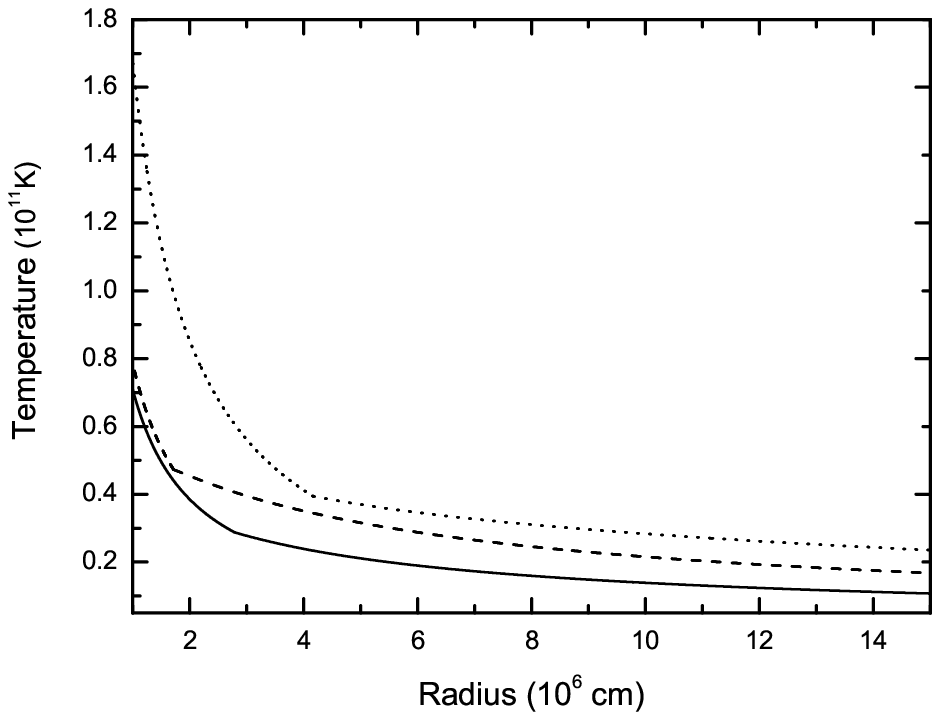}
\includegraphics{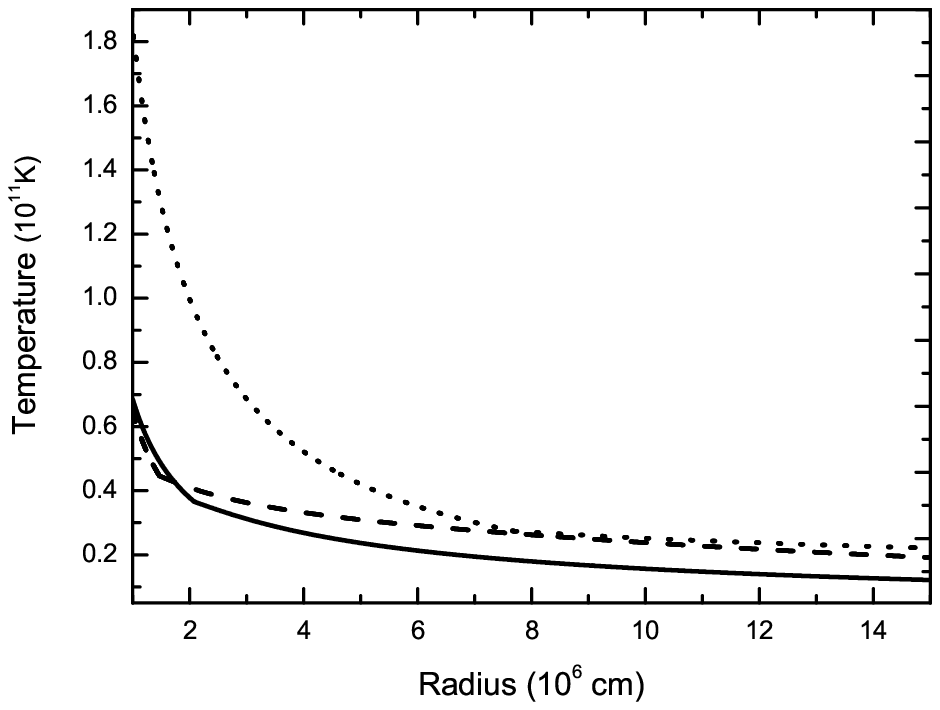}}
\caption{The density (in units of $10^{11}$g cm$^{-3}$), pressure
(in units of $10^{29}$ ergs cm$^{-3}$) and temperature (in units of
$10^{11}$K) of the disk as functions of radius $r$ (in units of
$10^{6}$ cm) in the simple model for several sets of parameters: (1)
$M=1.4M_{\odot}$, $Y_{e}=1.0$, and $\dot{M}=0.01M_{\odot}\,{\rm
s}^{-1}$ ({\em thin solid line}), (2) $M=1.4M_{\odot}$, $Y_{e}=1.0$,
and $\dot{M}=0.1M_{\odot}\,{\rm s}^{-1}$ ({\em thin dashed line}),
(3) $M=1.4M_{\odot}$, $Y_{e}=1.0$, and $\dot{M}=1.0M_{\odot}\,{\rm
s}^{-1}$ ({\em thin dotted line}), (4) $M=2.0M_{\odot}$,
$Y_{e}=1.0$, and $\dot{M}=0.01M_{\odot}\,{\rm s}^{-1}$ ({\em thick
solid line}), (5) $M=2.0M_{\odot}$, $Y_{e}=1.0$, and
$\dot{M}=0.1M_{\odot}\,{\rm s}^{-1}$ ({\em thick dashed line}), and
(6) $M=2.0M_{\odot}$, $Y_{e}=1.0$, and $\dot{M}=1.0M_{\odot}\,{\rm
s}^{-1}$ ({\em thick dotted line}). }\label{fig32}
\end{figure}

In Figure \ref{fig32}, we choose several special conditions to plot
the density, temperature and pressure of the whole disk as functions
of radius $r$. If the parameter $\dot{m}_{d}$ is larger, or the disk
contains more neutrons (i.e., $Y_{e}$ becomes smaller), then the
density, temperature and pressure of the disk are larger, and the
change of the density and pressure is more dramatic than that of the
temperature. The change of the temperature cannot be very large
because it greatly affects the neutrino cooling rate of the disk.

\begin{figure}
\centering\resizebox{0.7\textwidth}{!}{\includegraphics{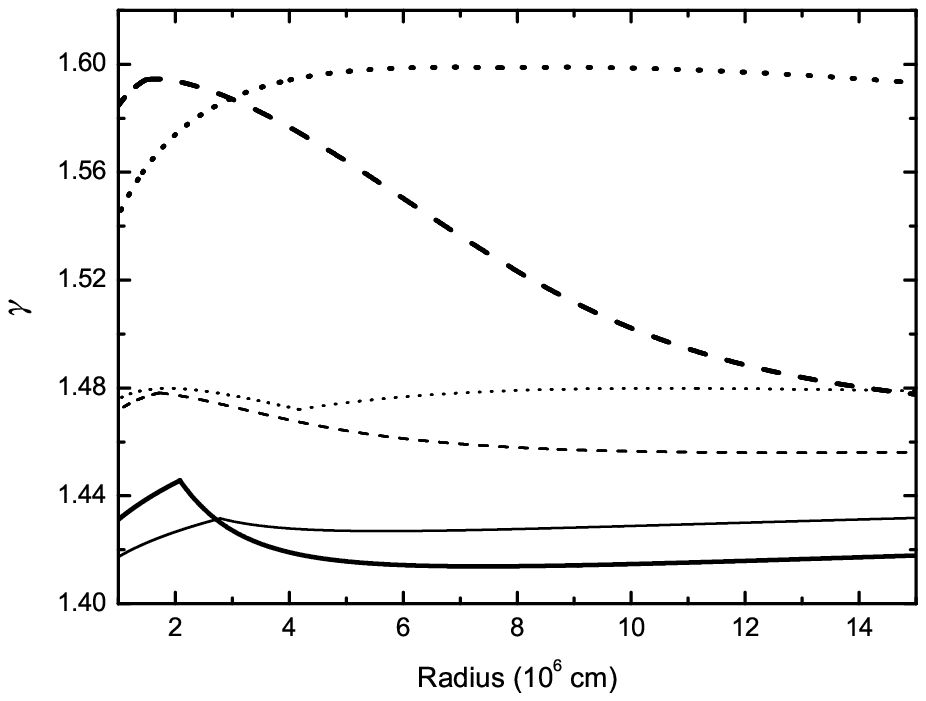}}
\caption{The ``equivalent" adiabatic index $\gamma$ of the disk as a
function of radius $r$ in the simple model. The meanings of
different lines are the same as those in Fig. 3.2.}\label{fig33}
\end{figure}

If the mass of the central neutron star ($m$) becomes larger or the
electron fraction $Y_{e}$ becomes smaller, then the value of
$\tilde{r}$ in the monotonous decreasing segment of
$\dot{m}_{d}-\tilde{r}$ becomes smaller, and the value of
$\tilde{r}$ in the monotonous increasing segment of
$\dot{m}_{d}-\tilde{r}$ is larger. And the minimum value of
$\tilde{r}$ is almost independent of $Y_{e}$ and $m$. All of these
conclusions are consistent with the analytical solutions, except for
the case of the advection-dominated outer disk with the radiation
pressure to be dominated.  In $\S$3.3.2.2, we found that the size of
the inner disk increases with increasing $m$ for our analytical
solutions. However, by calculating the ``equivalent" adiabatic index
$\gamma$, we find that $\gamma$ decreases slightly with increasing
$m$. This makes the value of $\tilde{r}_{6}$ decrease, which is also
consistent with the analytical results (see Table \ref{fig33}).
Figure \ref{fig33} shows the ``equivalent" adiabatic index $\gamma$
as a function of radius of the entire disk for several different
sets of parameters. In the case where the accretion rate is low, the
radiation pressure is important. As the accretion rate increases,
the gas pressure becomes more dominant and the value of $\gamma$ is
larger. On the other hand, the ratio of the degeneracy pressure to
the total pressure is larger in the case of a higher accretion rate
and $Y_{e}\sim 1$. However, the gas pressure is always dominant for
the accretion rate chosen here. We see from Figure 3 that the change
of $\gamma$ in the inner disk is insignificant.
\begin{figure}
\resizebox{\hsize}{!} {\includegraphics{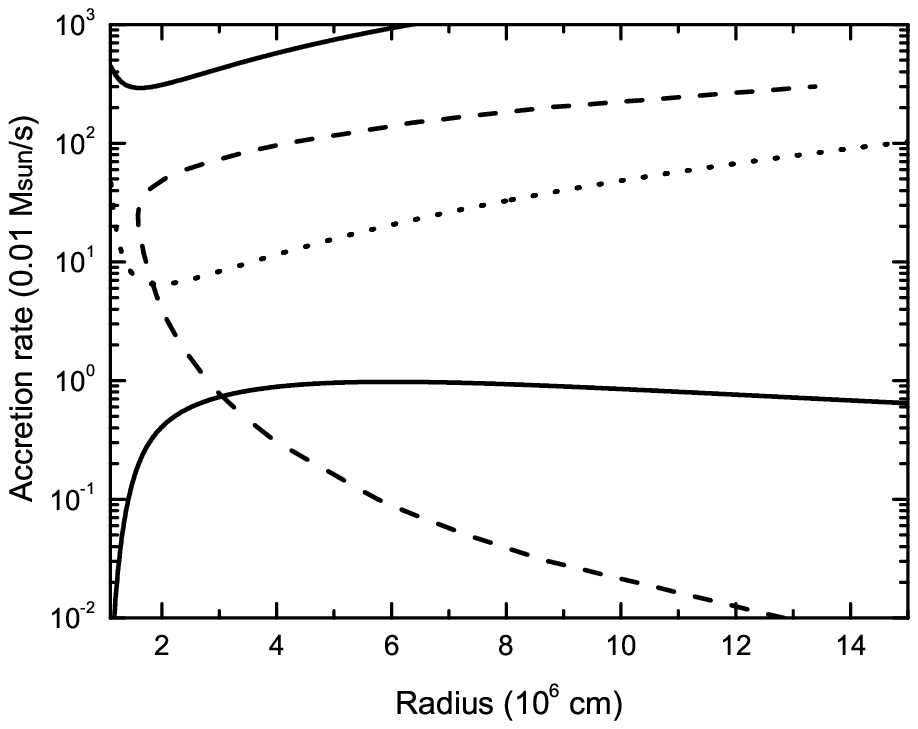}
\includegraphics{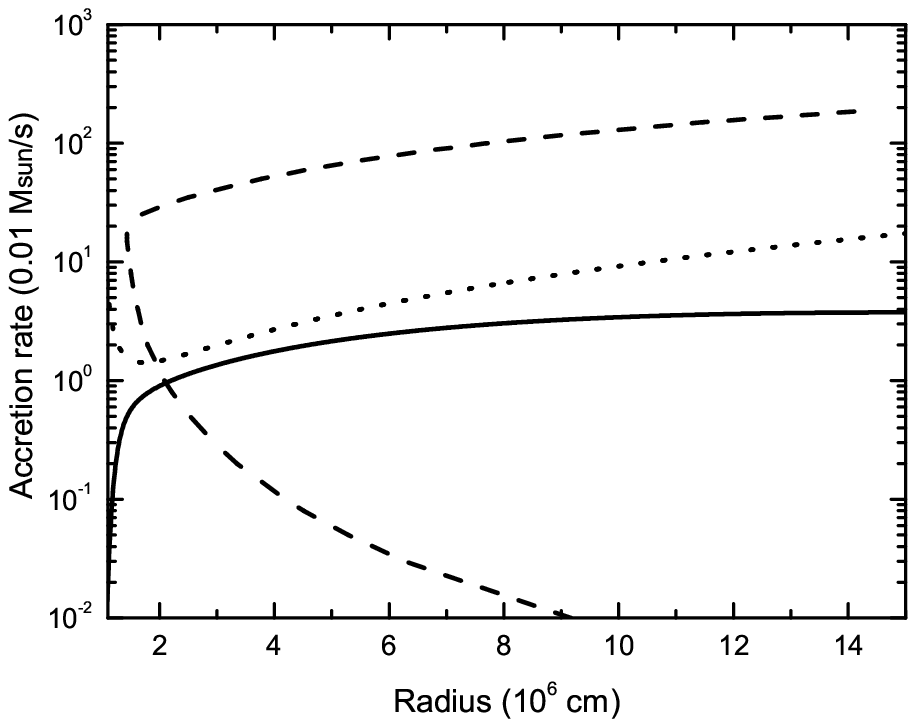}}
\caption{The distribution of a dominant pressure in the
$\dot{m}_{d}-r$ plane for two different values of $Y_{e}$. (a) {\em
Left panel}: $Y_{e}=1$. (b) {\em Right panel}: $Y_{e}$=1/9. The
neutron star $M=1.4M_{\odot}$. The dotted lines in two panels show
the boundary where NDAF becomes significant as a result of higher
accretion rate. The dashed lines show the inner disk size
$\tilde{r}$ as a function of $\dot{m}_{d}$ for different $Y_{e}$.
The three regions in Figure \ref{fig34}a divided by two solid lines from
bottom to top are radiation, gas and degeneracy pressure-dominated
regions. Two regions divided by one solid line in Figure 4b are
radiation and gas pressure-dominated regions.}\label{fig34}
\end{figure}

Figure \ref{fig34} shows the distribution of a dominant pressure in
the $\dot{m}_{d}-r$ plane for different $Y_{e}$. The results are
quite similar to what we have discussed in \S 3.3.1. At low
accretion rates, most of the disk is radiation pressure-dominated,
and some inner region of the disk can be degeneracy
pressure-dominated only if $Y_{e} \sim 1$ and the accretion rate is
sufficiently high. In addition, the gas pressure-dominated region in
the ADAF case is smaller if $Y_{e}$ is smaller, and the disk flow
can become an NDAF for a lower accretion rate if $Y_{e}$ is smaller.
We also plot $\tilde{r}$ in the $\dot{m}_{d}-r$ plane to obtain
final results. We note that in the case of $m=1$ and $Y_{e}=1$, the
region where the degeneracy pressure is dominant in the outer disk
is entirely covered by the region of the inner disk, which means
that the outer disk is hardly degeneracy pressure-dominated. In
order to see this result more clearly, we here take the value of the
accretion rate in Figure \ref{fig34} to be wider. We do not plot the
distribution of the dominant pressure of the inner disk in this
chapter. In most cases, the dominant pressure of the inner disk is
similar to that of the outer disk.

\begin{figure}
\resizebox{\hsize}{!} {\includegraphics{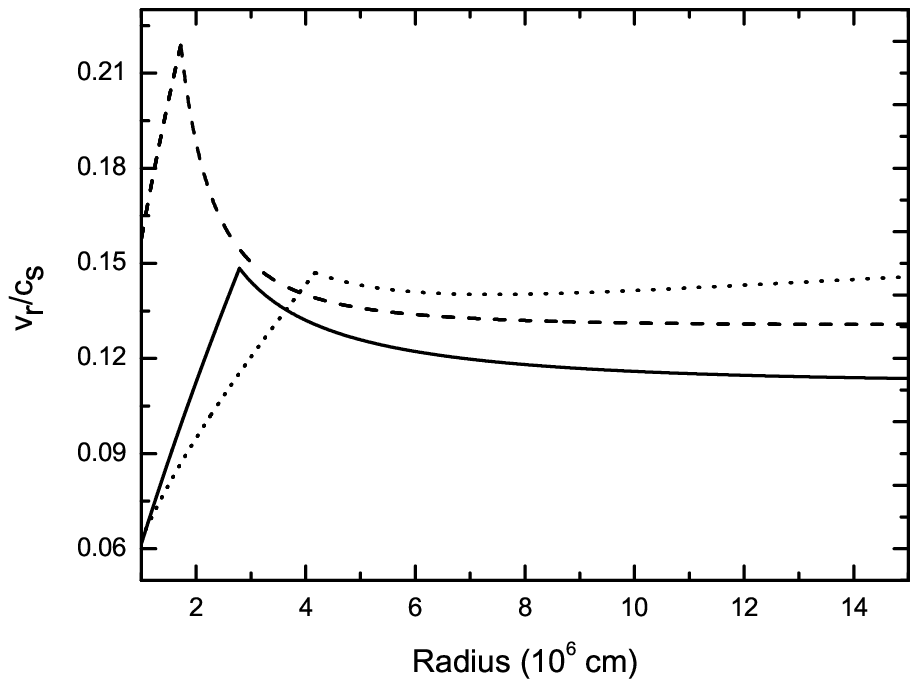}
\includegraphics{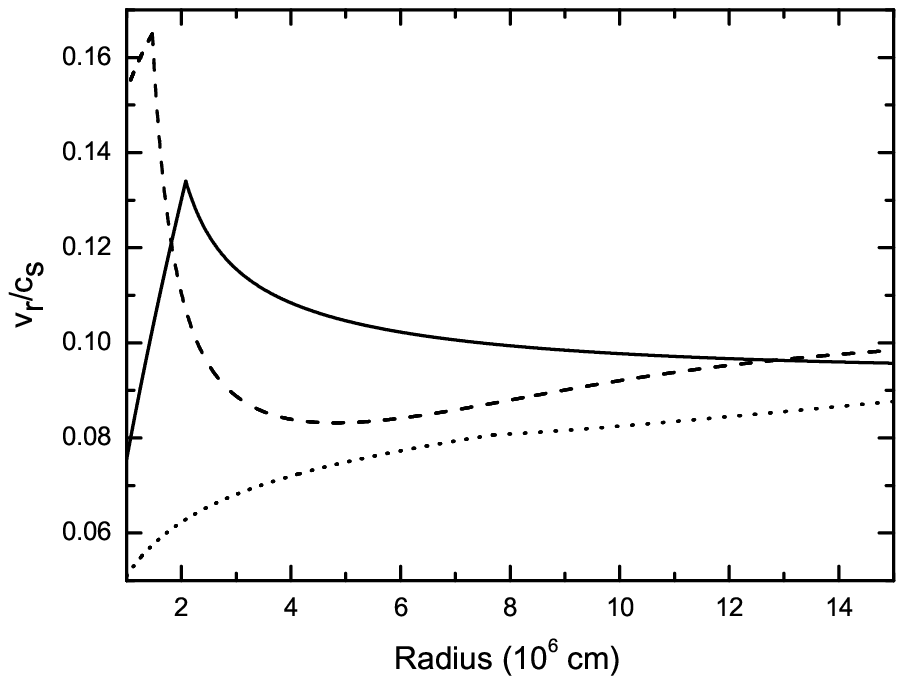}}
\caption{The ratio of the radial velocity $v_{r}$ and local speed of
sound $c_{s}$ as a function of $r$. (a) {\em Left panel}:
$M=1.4M_{\odot}$, $Y_{e}=1.0$; (b) {\em Right panel}:
$M=1.4M_{\odot}$, $Y_{e}=1/9$. The accretion rate:
$\dot{m}_{d}=0.01M_{\odot}\,{\rm s}^{-1}$ ({\em solid line}),
$\dot{m}_{d}=0.1M_{\odot}\,{\rm s}^{-1}$ ({\em dashed line}), and
$\dot{m}_{d}=1.0M_{\odot}\,{\rm s}^{-1}$ ({\em dotted
line})}\label{fig35}
\end{figure}

Figure \ref{fig35} shows the ratio of the radial velocity $v_{r}$
and local speed of sound $c_{s}$ as a function of $r$. The ratio is
always much smaller than unity, which means that the accretion flow
is always subsonic and no stalled shock exists in the disk. In many
cases the peak of this ratio is just at the boundary between the
inner and outer disks. The reason can be found in $\S$3.2.1, where
we gave the analytical expression of the ratio.
$f=1-(r_{*}/r)^{1/2}$ is the major factor that affects the value of
$v_{r}/c_{s}$ of the outer disk. However, in the inner disk,
$v_{r}/c_{s}$ always decreases since the isothermal sound speed can
be greater at smaller radii (i.e., $c_{s}\propto r^{-1/2}$), and the
radial velocity of the accreting gas, which satisfies the
self-similar solution (\ref{3en02}), cannot change dramatically for
the disk matter to strike the neutron star surface.

\begin{figure}
\centering\resizebox{0.7\textwidth}{!} {\includegraphics{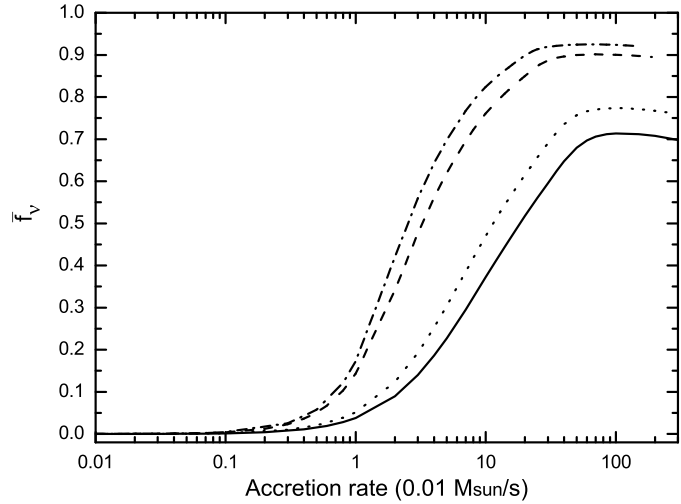}}
\caption{The average neutrino emission efficiency $\bar{f}_{\nu}$ in
the outer disk as a function of $\dot{m}_{d}$ (where the meanings of
different lines are the same as those in Fig. 3.1.)}\label{fig36}
\end{figure}

Figure \ref{fig36} gives the change of $\bar{f}_{\nu}$ of the outer
disk as a function of $\dot{m}_{d}$ for several different sets of
parameters (where $\bar{f}_{\nu}$ is defined by equation
\ref{3ee02}). In analytical calculations, we approximatively take
$\bar{f}_{\nu}= 0$ in an ADAF and  $\bar{f}_{\nu}\sim 1$ in an NDAF.
Here we give the numerical result. $\bar{f}_{\nu}$ increases
monotonously as the accretion rate increases, and reaches the value
$\sim 1$ as most of the disk region becomes NDAF. However, if the
accretion rate is so large that the neutrino opacity begins to play
a significant role, then $\bar{f}_{\nu}$ decreases with increasing
$\dot{m}_{d}$. Furthermore, an increase of $M$ or decrease of
$Y_{e}$ can make the value of $f_{\nu}$ and the average
$\bar{f}_{\nu}$ be larger. This means that the efficiency of
neutrino emission from the disk is higher if the disk contains more
neutrons or the central neutron star is more massive.
\begin{figure}
\resizebox{\hsize}{!} {\includegraphics{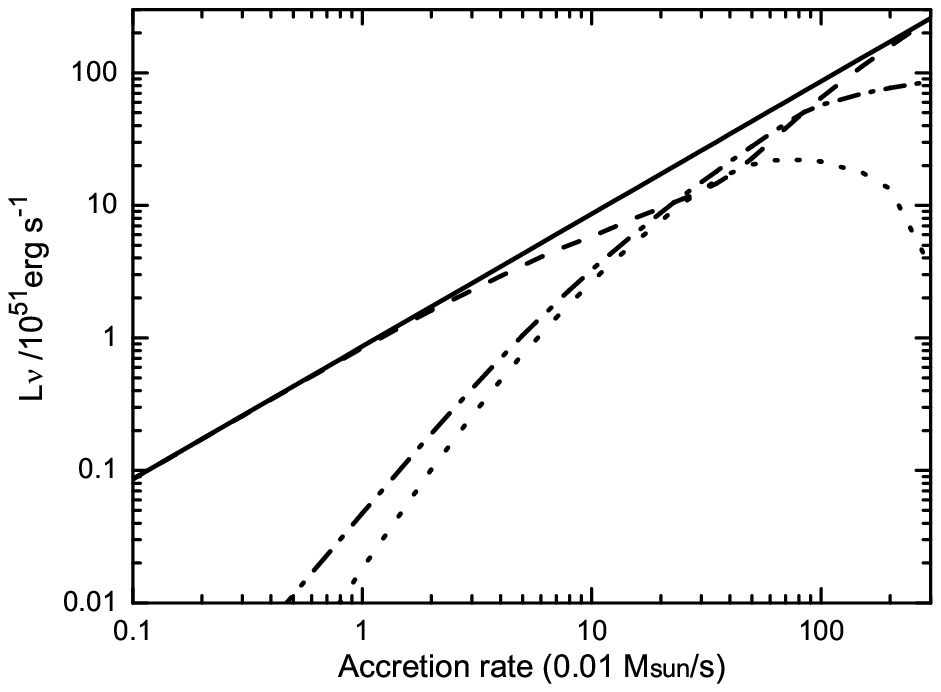}
\includegraphics{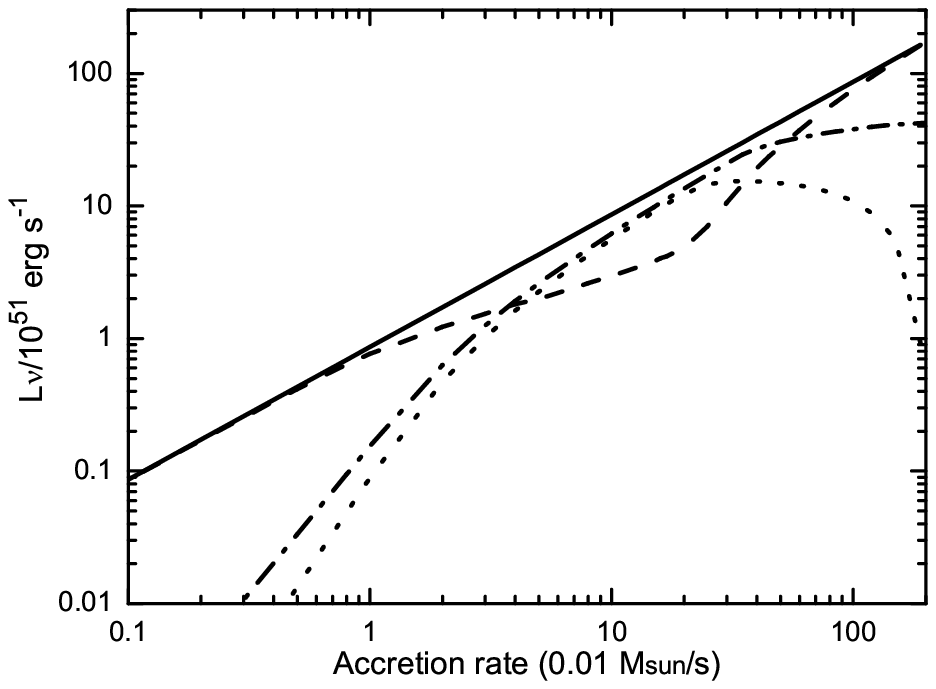}}
\caption{Neutrino luminosity from the disk around a neutron star in
the simple model. $L_\nu$ is in units of $10^{51}$ergs\,s$^{-1}$.
(a) {\em Left panel}: $M=1.4M_{\odot}$ and $Y_{e}=1.0$. (b) {\em
Right panel}: $M=1.4M_{\odot}$ and $Y_{e}=1/9$. The solid line
corresponds to the maximum energy release rate of the disk around a
neutron star, the dashed line to the neutrino luminosity from the
inner disk, the dotted line to the neutrino luminosity from the
outer disk, and the the dash-dotted line to the neutrino luminosity
from a black hole disk.}\label{fig37}
\end{figure}

Figure \ref{fig37} shows the total neutrino emission luminosity of
the entire disk around a neutron star as a function of accretion
rate for parameters $M$ and $Y_{e}$ (where we do not consider
neutrino emission from the stellar surface discussed in \S 3.3.3),
and we compare it with the neutrino luminosity from a black-hole
disk. Also, we calculate the total neutrino luminosity from the
inner and outer disks. Here we roughly take the mass of the black
hole to be the same as that of the neutron star, and the innermost
stable circular orbit of the disk has a radius which is equal to the
radius of the neutron star. We approximately use the Newtonian
dynamics for simplicity. In Fig. \ref{fig37}, we find that the
difference in neutrino luminosity between the neutron-star and
black-hole cases is a strong function of the accretion rate. When
the accretion rate is low ($\dot{m}_{d}\leq 10$), the total neutrino
luminosity of the black-hole disk $L_{\nu,\rm BH}$ is much smaller
than that of the neutron-star disk $L_{\nu,\rm NS}$, but $L_{\nu,\rm
BH}$ and $L_{\nu,\rm NS}$ are similar for a moderate accretion rate
($\dot{m}_{d}$ from 10 to 100). Actually, this result is consistent
with the general scenario introduced in \S3.2.1 and the basic result
shown in Fig. \ref{fig31}: for a low accretion rate, the black-hole
disk is mainly advection-dominated with most of the viscous
dissipation-driven energy to be advected into the event horizon of
the black hole, and we have $L_{\nu,\rm BH}\ll GM\dot{M}/(4r)$. On
the other hand, a large size of the inner disk of the neutron-star
disk for a low accretion rate makes the neutrino emission efficiency
be much higher than its black-hole counterpart. However, for a
moderate accretion rate, the black-hole disk is similar to the
neutron-star disk, which owns a quite small inner disk, and we have
$L_{\nu,\rm BH}\sim L_{\nu,\rm NS}$. Moreover, neutrino opacity
leads the value of $L_{\nu,\rm BH}$ to be less again compared with
$L_{\nu,\rm NS}$ for a high accretion rate, as this opacity
decreases the neutrino emission efficiency in the black-hole disk
but increases the size of the neutron-star disk again to balance the
heat energy release.

\section{An Elaborate Model of the Disk}
In the last section we first studied the disk structure
analytically. To do this, we used several approximations. First of
all, we took the pressure as a summation of several extreme
contributions such as the gas pressure of nucleons and electrons,
and the radiation pressure of a plasma of photons and $e^{+}e^{-}$
pairs. However, electrons may actually be degenerate or partially
degenerate, and the neutrino pressure should also be added to the
total pressure. Following Kohri et al. (2005), a lot of works about
hyperaccretion disks used the Fermi-Dirac distribution to calculate
the pressure of electrons and even the pressure of nucleons. Second,
neutrino cooling we used in the last section is simplified,
following Popham et al. (1999) and Narayan et al. (2001) and
neglecting the effect of electron degeneracy and the effect of
different types of neutrinos and their different optical depth. In
fact, these effects may be significant in some cases. Third, we took
the electron-nucleon-radio $Y_{e}$ as a constant parameter in our
analytic model in $\S$3.3. Realistically, $Y_{e}$ should be
calculated based on $\beta$-equilibrium and neutronization in
hyperaccretion disks. In this section, we still use the assumption
of outer and inner disks discussed in $\S$3.2, but consider a
state-of-the-art model with lots of elaborate (more physical)
considerations on the thermodynamics and microphysics in the disk,
which was recently developed in studying the neutrino-cooled disk of
a black hole. In addition, we compare results from this elaborate
model with those of the simple model discussed in $\S$3.3.

\subsection{Thermodynamics and microphysics}
The total pressure in the disk can be written as: $P=P_{\rm
nuc}+P_{\rm rad}+P_{e}+P_{\nu}$. We still consider all the nucleons
to be free ($X_{\rm nuc}\approx 1$) as mentioned in \S 3.2.2, and
ignore the photodisintegration process. Also, we replace the term of
radiation pressure $11aT^{4}/12$ in the simple model by $aT^{4}/3$
in this section, because the pressure of $e^{+}e^{-}$ pairs can be
calculated in the electron pressure $P_{e}$ with the Fermi-Dirac
distribution:
\begin{equation}
P_{e^{\pm}}=\frac{1}{3}\frac{m_{e}^{4}c^{5}}{\pi^{2}\hbar^{3}}
\int^{\infty}_{0}\frac{x^{4}}{\sqrt{x^{2}+1}}\frac{dx}{e^{(m_{e}c^{2}\sqrt{x^{2}+1}\mp\mu_{e})/k_{B}T}+1},\label{3z02}
\end{equation}
where $x=p/m_ec$ is the dimensionless momentum of an electron and
$\mu_{e}$ is the chemical potential of the electron gas. $P_{e}$ is
the summation of $P_{e^{-}}$ and $P_{e^{+}}$. In addition, we take
the neutrino pressure to be
\begin{equation}
P_{\nu}=u_{\nu}/3,\label{3z03}
\end{equation}
where $u_{\nu}$ is the energy density of neutrinos.

The ``equivalent" adiabatic index can be expressed by
\begin{equation}
\gamma=1+(P_{\rm nuc}+P_{\rm rad}+P_{e}+P_{\nu})/(u_{\rm nuc}+u_{\rm
rad}+u_{e}+u_{\nu}). \label{3z061}
\end{equation}
with the inner energy density to be
\begin{equation}
u_{\rm gas}=\frac{3}{2}P_{\rm gas},\label{3z04}
\end{equation}
\begin{equation}
u_{\rm rad}=3P_{\rm rad},\label{3z05}
\end{equation}
\begin{equation}
u_{e^{\pm}}=\frac{m_{e}^{4}c^{5}}{\pi^{2}\hbar^{3}}
\int^{\infty}_{0}\frac{x^{2}\sqrt{x^{2}+1}}{e^{(m_{e}c^{2}\sqrt{x^{2}+1}\mp\mu_{e})/k_{B}T}+1}dx.\label{3z06}
\end{equation}
We then use equation (\ref{3e34}) to obtain the self-similar inner
disk.

In addition, we add the equation of charge neutrality among protons,
electrons and positrons to establish the relation between $\rho$,
$Y_{e}$ and $\mu_{e}$.
\begin{equation}
n_{p}=\frac{\rho Y_{e}}{m_{B}}=n_{e^{-}}-n_{e^{+}},\label{3z062}
\end{equation}
where we use the Fermi-Dirac form to calculate $n_{e^{-}}$ and
$n_{e^{+}}$.

Moreover, in the elaborate model, we adopt the improved formula of
the neutrino cooling rate $Q_{\nu}^{-}$, the inner energy density of
neutrinos $u_{\nu}$ , as well as the absorption and scattering
optical depth for three types neutrinos
$\tau_{a,\nu_{i}(e,\mu,\tau)}$ and $\tau_{s,\nu_{i}(e,\mu,\tau)}$
following a series of previous work (e.g., Popham \& Narayan 1995,
Di Matteo et al. 2002, Kohri et al. 2005, Gu et al. 2006, Janiuk et
al. 2007 and Liu et al. 2007). The three types of neutrino cooling
rate per unit volume are
\begin{equation}
\dot{q}_{\nu_{e}}=\dot{q}_{\rm eN}+\dot{q}_{e^{-}e^{+}\rightarrow
\nu_{e}\bar{\nu}_{e}}+\dot{q}_{\rm brem}+\dot{q}_{\rm
plasmon},\label{3z09}
\end{equation}
\begin{equation}
\dot{q}_{\nu_{\mu}}=\dot{q}_{\nu_{\tau}}=\dot{q}_{e^{-}e^{+}\rightarrow
\nu_{\tau}\bar{\nu}_{\tau}}+\dot{q}_{\rm brem}, \label{3z091}
\end{equation}
where the meanings of four terms $\dot{q}_{\rm eN}$,
$\dot{q}_{e^{-}e^{+}\rightarrow \nu_{i}\bar{\nu}_{i}}$,
$\dot{q}_{\rm brem}$, and $\dot{q}_{\rm plasmon}$ have been shown at
the beginning of $\S$3.3. Here $\dot{q}_{\rm eN}$ and $\dot{q}_{\rm
plasmon}$ are only related to $\dot{q}_{\nu_{e}}$. Moreover,
$\dot{q}_{\rm eN}$ is a summation of three terms,
\begin{equation}
\dot{q}_{\rm eN}=\dot{q}_{p+e^{-}\rightarrow
n+\nu_{e}}+\dot{q}_{n+e^{+}\rightarrow
p+\bar{\nu}_{e}}+\dot{q}_{n\rightarrow
p+e^{-}+\bar{\nu}_{e}}.\label{3z092}
\end{equation}
The formulae of three terms in equation (\ref{3z092}) are the same
as Kohri et al. (2005), Janiuk et al. (2007) and Liu et al. (2007),
who considered the effect of electron degeneracy. In addition, we
use the same formulae of $\dot{q}_{e^{-}e^{+}\rightarrow
\nu_{i}\bar{\nu}_{i}}$, $\dot{q}_{\rm brem}$, and $\dot{q}_{\rm
plasmon}$ as the early works such as Kohri et al. (2002, 2005) and
Liu et al. (2007).

Finally, different from the simple model in $\S$3.3 in which we took
the electron fraction $Y_{e}$ as a free parameter, in this section
we calculate $Y_{e}$ by considering the $\beta$-equilibrium in the
disk among electrons and nucleons following Lee et al. (2005) and
Liu et al. (2007)
\begin{equation}
\textrm{ln}\left(\frac{n_{n}}{n_{p}}\right)
=f(\tau_{\nu})\frac{2\mu_{e}-Q}{k_{B}T}+[1-f(\tau_{\nu})]\frac{\mu_{e}-Q}{k_{B}T},\label{3z10}
\end{equation}
with the weight factor $f(\tau_{\nu})$=exp$(-\tau_{\nu_{e}})$ and
$Q=(m_{n}-m_{p})c^{2}$. Equation (\ref{3z10}) is a combined form to
allow the transition from the neutrino-transparent limit case to the
neutrino-opaque limit case of the $\beta$-equilibrium.

\subsection{Numerical results in the elaborate model}
Using equations (\ref{3e1}), (\ref{3e2}), (\ref{3e3}) and
(\ref{3e31}) in $\S$3.2.2 and the improved treatment in $\S$3.4.1,
we can solve the structure of the outer disk. Then using equations
(\ref{3en02}) and (\ref{3e4}) in $\S$3.2.3, we determine the size
and the structure of the inner disk, and calculate the neutrino
luminosity of the entire disk.

\begin{figure}
\resizebox{\hsize}{!} {\includegraphics{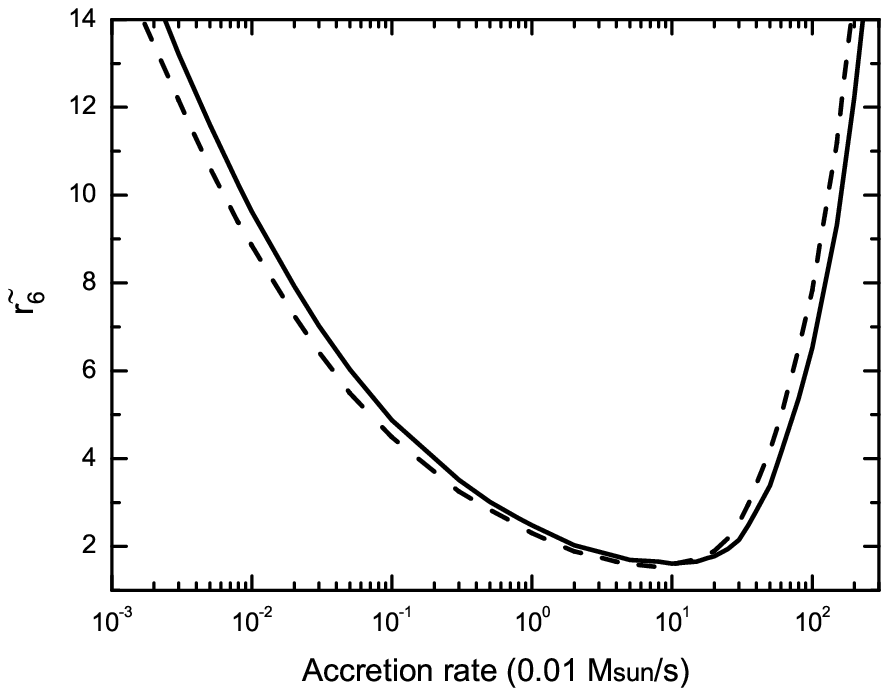}
\includegraphics{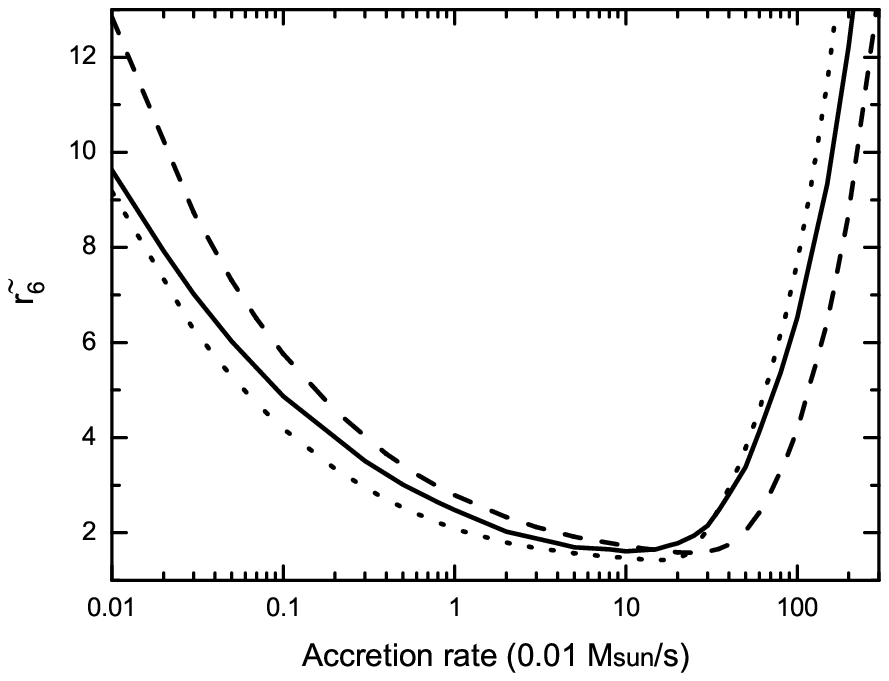}}
\caption{(a) {\em Left panel}: the radius $\tilde{r}_{6}$ between
the inner and the outer disks in the elaborate model with
$M=1.4M_{\odot}$ ({\em solid line}), and $M=2.0M_{\odot}$ ({\em
dashed line}). (b) {\em Right panel}: comparison of $\tilde{r}_{6}$
in different models with $M=1.4M_{\odot}$: (1) the elaborate model
({\em solid line}), (2) the simple model with $Y_{e}=1$ ({\em dashed
line}), and (3) the simple model with $Y_{e}=1/9$ ({\em dotted
line}).}\label{fig38}
\end{figure}
The left panel of Figure \ref{fig38} shows the size of the inner
disk in the elaborate model. We still choose the mass of the central
neutron star to be $M=1.4M_{\odot}$ and $M=2.0M_{\odot}$. From the
left panel of Figure 6 we can see that the size of the inner disk
$\tilde{r}$ decreases with increasing the accretion rate, and
reaches a minimum at $\dot{M} \sim 0.1 M_{\odot}s^{-1}$. Then the
value of $\tilde{r}$ increases with increasing the accretion rate.
This result is well consistent with what we have found in the simple
model in \S 3.3. The physical reason for this result has been
discussed in \S 3.3.3.2 and \S 3.3.4. Figure \ref{fig36}b shows the
solution of the inner disk size both in the simple model and the
elaborate model. We fix the central neutron star $M=1.4M_{\odot}$.
In the simple model of $\S$3.3, we take $Y_{e}$ as a free parameter
and plot two $\dot{m}_{d}-\tilde{r}_{6}$ lines with $Y_{e}=1$ and
$Y_{e}=1/9$, while in the elaborate model we only plot one line
since $Y_{e}$ can be directly determined through $\beta$-equilibrium
in the disk. From the right panel of Figure \ref{fig38} we conclude
that the solutions of the two models are basically consistent with
each other.

\begin{figure}
\resizebox{\hsize}{!} {\includegraphics{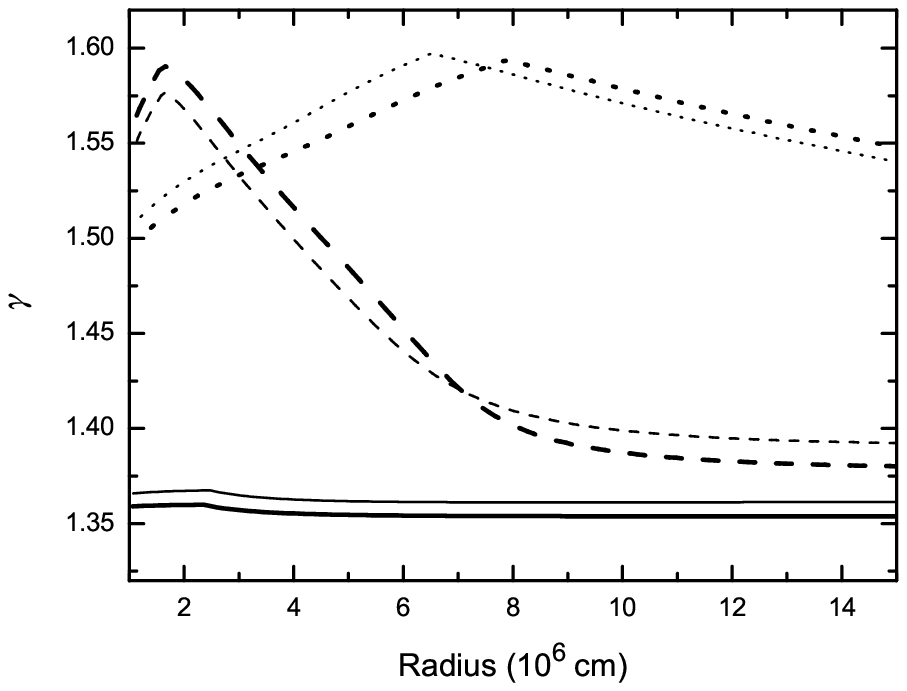}
\includegraphics{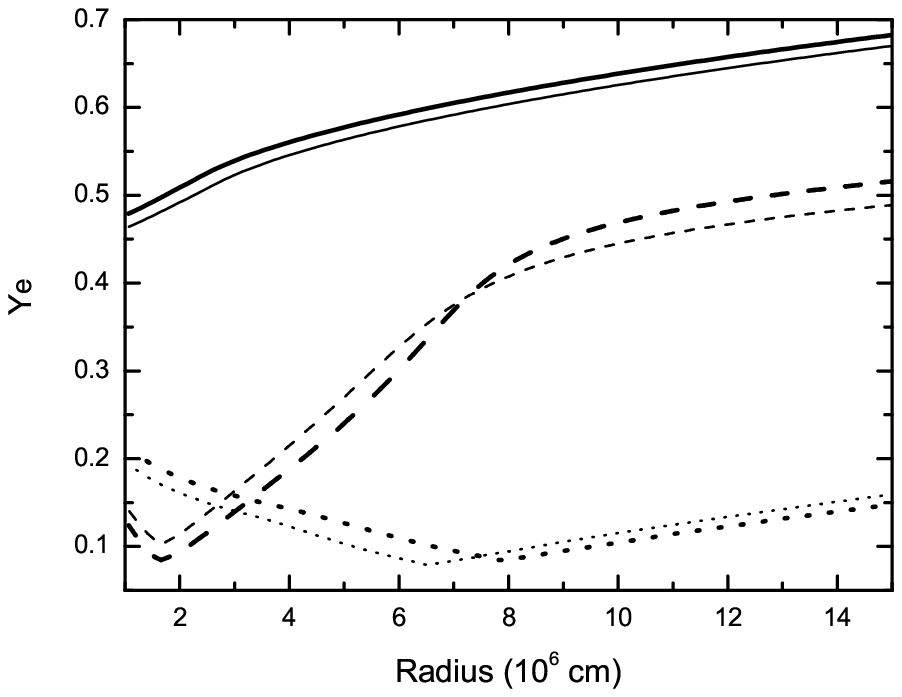}}
\caption{(a) {\em Left panel}: the ``equivalent" adiabatic index
$\gamma$ in the elaborate model. (b) {\em Right panel}: the electron
fraction $Y_{e}$ in the elaborate model as a function of radius. The
profiles are shown for three values of the accretion rate and two
values of the central neutron star mass: (1) $M=1.4M_{\odot}$ and
$\dot{M}=0.01M_{\odot}\,{\rm s}^{-1}$ ({\em thin solid line}), (2)
$M=1.4M_{\odot}$ and $\dot{M}=0.1M_{\odot}\,{\rm s}^{-1}$ ({\em thin
dashed line}), (3) $M=1.4M_{\odot}$ and $\dot{M}=1.0M_{\odot}\,{\rm
s}^{-1}$ ({\em thin dotted line}), (4) $M=2.0M_{\odot}$ and
$\dot{M}=0.01M_{\odot}\,{\rm s}^{-1}$ ({\em thick solid line}), (5)
$M=2.0M_{\odot}$ and $\dot{M}=0.1M_{\odot}\,{\rm s}^{-1}$ ({\em
thick dashed line}), and (6) $M=2.0M_{\odot}$ and
$\dot{M}=1.0M_{\odot}\,{\rm s}^{-1}$ ({\em thick dotted
line}).}\label{fig39}
\end{figure}
In Figure \ref{fig39}, we shows the ``equivalent" adiabatic index
$\gamma$ and the electron fraction $Y_{e}$ in the entire disk for
three values of the accretion rate $\dot{M}=0.01M_{\odot}s^{-1}$,
$0.1M_{\odot}s^{-1}$ and $1.0M_{\odot}s^{-1}$, and for two values of
the mass of the central neutron star $M=1.4M_{\odot}$ and
$2.0M_{\odot}$. The ``equivalent" adiabatic index $\gamma$ increases
with increasing the accretion rate in most region of the disks,
since gas will take over electrons and radiation to be the dominant
pressure when the accretion rate is high enough. In addition,
$\gamma$ decreases as the radius decreases in the inner disk. These
are consistent with the results of the simple model (see Fig.
\ref{fig33}). From the right panel of Figure \ref{fig39}, we can see
that $Y_{e}\sim 1$ when the accretion rate is low, and $Y_{e}\ll 1$
when the accretion rate becomes sufficiently high. This result is
consistent with Kohri et al. (2005, their Fig. 6b). Chen \&
Beloborodov (2007) and Liu et al. (2007) showed the electron
fraction $Y_{e}\leq 0.5$ in the disk, since they supposed that
initial neutrons and protons come from photodisintegration of
$\alpha$-particles at some large radius far from a central black
hole. However, since the hyperaccretion disk around a neutron star
we discuss has a size smaller than that of a black-hole disk, we
consider the mass fraction of free nucleons $X_{\rm nuc}=1$ in the
entire disk. Therefore, the fraction of protons can be slightly
higher and it is possible that the protons are richer than neutrons
in the disk or $Y_{e}\geq 0.5$ if the accretion rate is low enough.

\begin{figure}
\resizebox{\hsize}{!} {\includegraphics{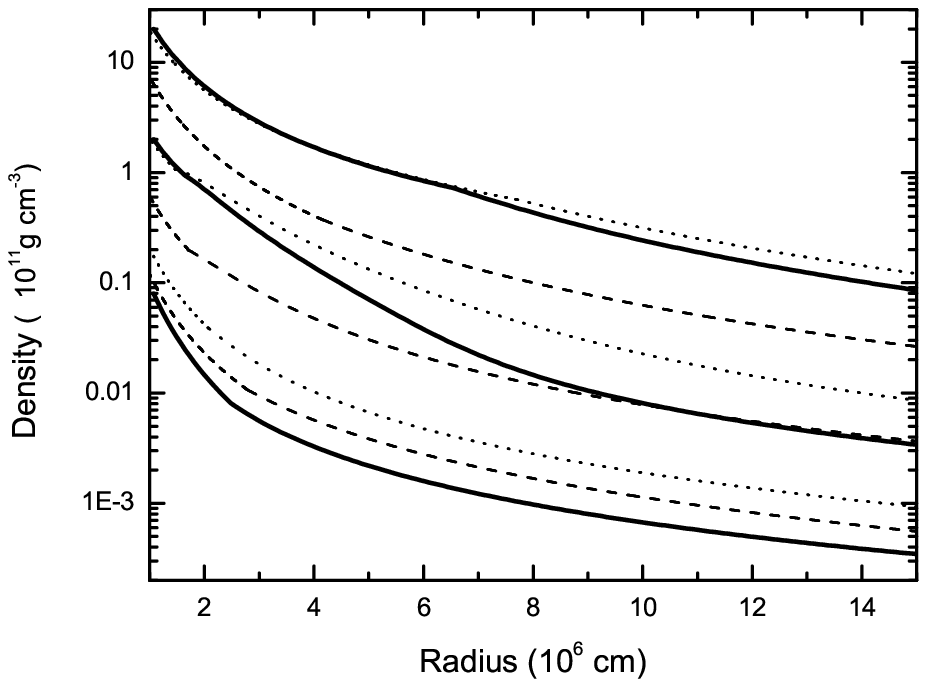}
\includegraphics{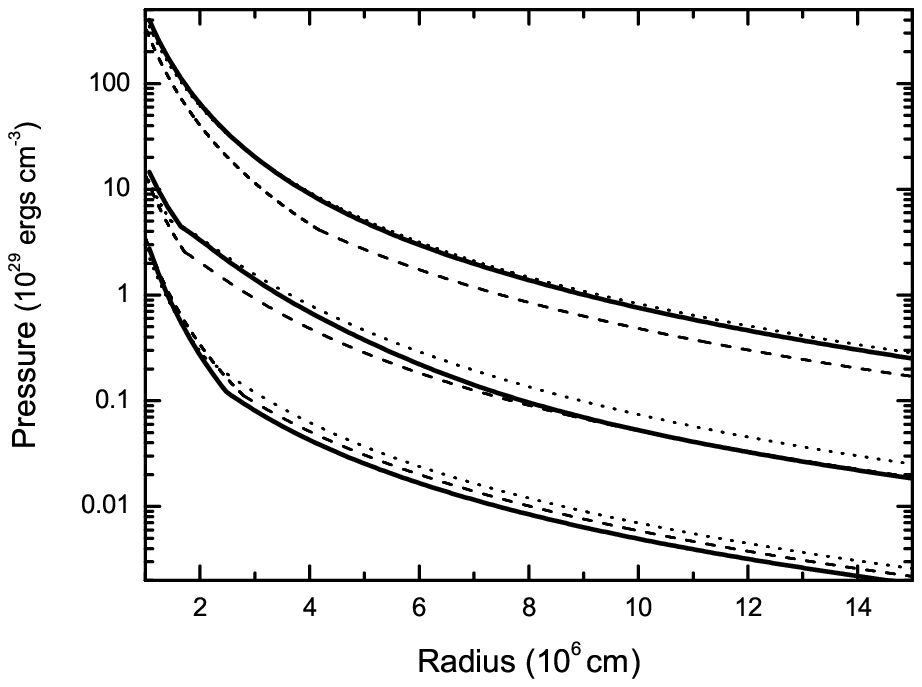}}
\\\resizebox{\hsize}{!} {\includegraphics{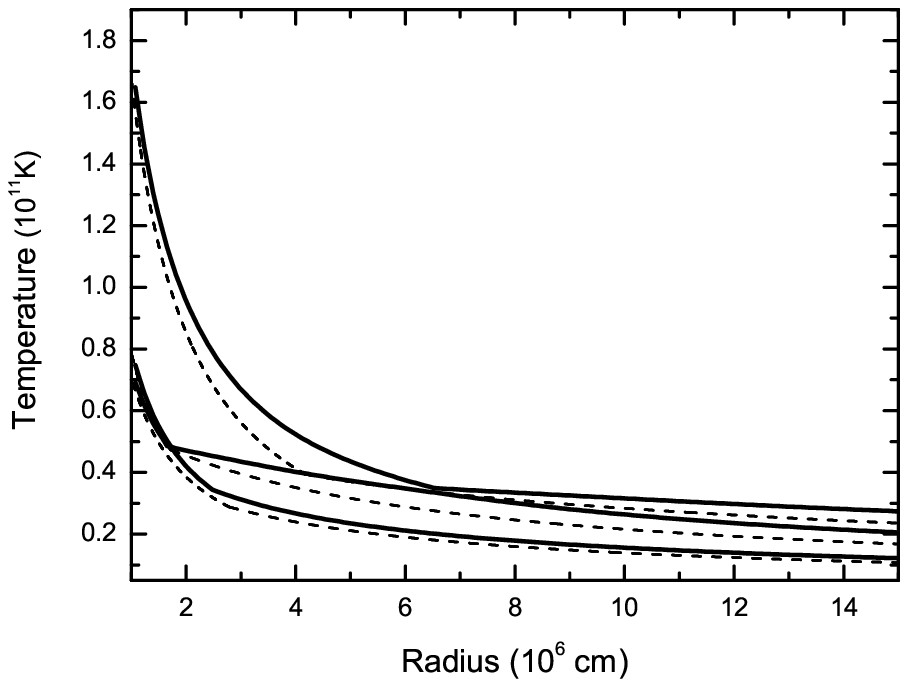}
\includegraphics{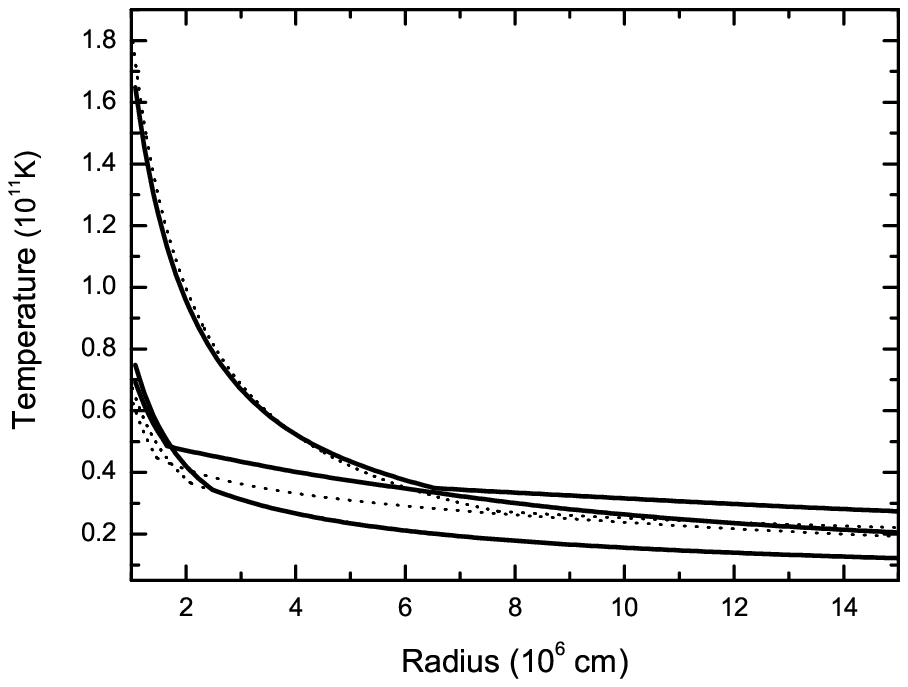}}
\caption{The density (in units of $10^{11}$g cm$^{-3}$), pressure
(in units of $10^{11}$ ergs cm$^{-3}$) and temperature (in units of
$10^{11}$K) of the entire disk in two models for $M=1.4M_{\odot}$.
The profiles include three groups of lines and each group also
includes three lines, which are shown for three values of the
accretion rate $\dot{M}=0.01M_{\odot}\,{\rm s}^{-1}$,
$0.1M_{\odot}\,{\rm s}^{-1}$ and $1.0M_{\odot}\,{\rm s}^{-1}$ from
bottom to top in these figures. The solution in the simple model
with $Y_{e}=1$ is shown by the thin dashed line, and $Y_{e}=1/9$ by
the thin dotted line. The solution in the elaborate model is shown
by the thick solid line. }\label{fig310}
\end{figure}

\begin{figure}
\resizebox{\hsize}{!}
{\includegraphics{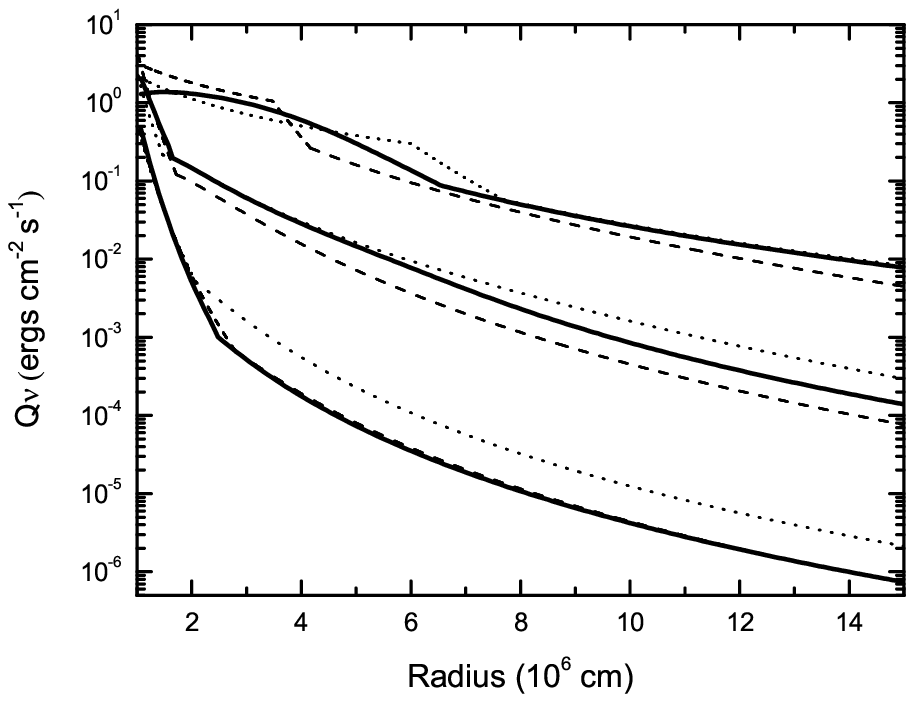}\\\includegraphics{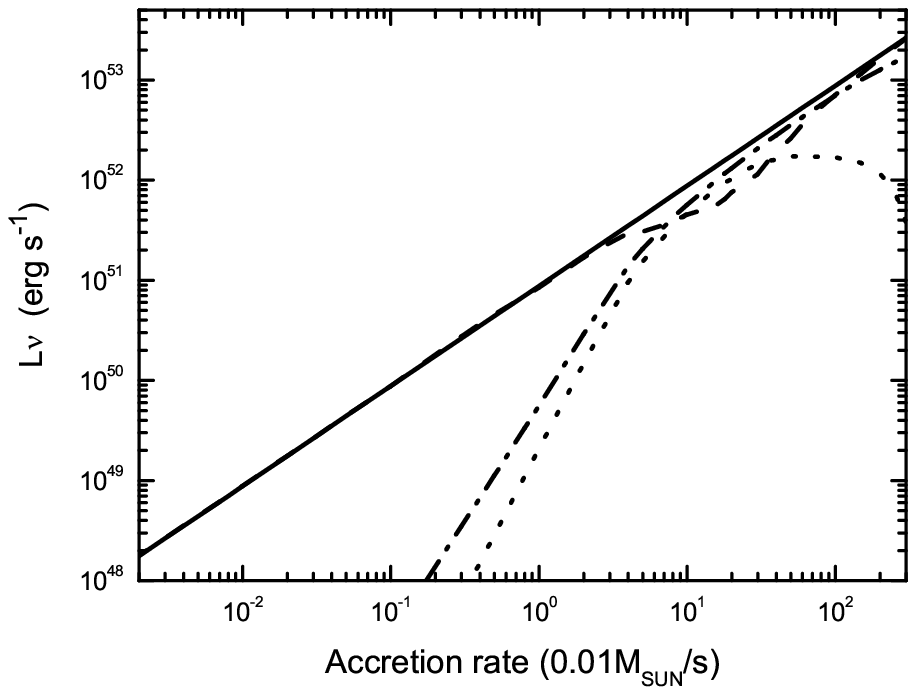}}
\caption{(a) {\em Left panel}: the neutrino luminosity per unit area
(in units of $10^{39}\,{\rm ergs}\,{\rm cm}^{-2}\,{\rm s}^{-1}$) in
both the simple model and the elaborate model. The meaning of
different lines are the same as those in Fig. 10. (b) {\em Right
panel}: the neutrino luminosity from the disk in the elaborate
model. The meanings of different lines are the same as those in Fig
.\ref{fig37}.}\label{fig311}
\end{figure}


Figure \ref{fig310} and the left panel of Figure \ref{fig311} show
the density, temperature, pressure and the neutrino luminosity per
unit area in the entire disk with three different accretion rates.
We fix $M=1.4M_{\odot}$, and also plot two other curves of solutions
in the simple model with $Y_{e}=1$ and $Y_{e}=1/9$. The density
$\rho$ and pressure $P$ in the elaborate model are smaller than
those in the simple model when the accretion rate is low
($\dot{m}_{d}=1.0$), but are similar to the solution of the simple
model with $Y_{e}=1/9$ in the high accretion rate
($\dot{m}_{d}=100$). In addition, $\rho$ and $P$ in the elaborate
model change from one solution ($Y_{e}=1$) to another solution
($Y_{e}=1/9$) in the simple model for an intermediate accretion rate
(e.g., $\dot{m}_{d}\sim 10$), since $Y_{e}\sim 0.5$ in the outer
edge of the disk and $Y_{e}\ll 1$ in the inner disk. The
distribution of the neutrino cooling rate $Q^{-}_{\nu}$ (luminosity
per unit area) in the elaborate model is almost the same as that in
the simple model with $Y_{e}=1$ and low accretion rate or
$Y_{e}=1/9$ and a high accretion rate. However, the value of
$Q^{-}_{\nu}$ is still different in these two models for the region
that is optically thick to neutrino emission in the disk.

We also plot the the total neutrino emission luminosity of the
entire disk, the outer and inner disks as functions of accretion
rate with the central neutron star mass of $1.4M_{\odot}$, and
compare the total neutrino luminosity with that of the black-hole
disk (Figure \ref{fig311}, \textit{right}). The results are similar
to what we have found in the simple model (Figure \ref{fig37}).

\section{Conclusions and Discussions}
In this chapter we have studied the structure, energy advection and
conversion, and neutrino emission of a hyperaccretion disk around a
neutron star. We considered a quasi-steady disk model without any
outflow. Similar to the disk around a black hole, the neutron star
disk with a huge mass accretion rate is extremely hot and dense,
opaque to photons, and thus is only cooled via neutrino emission, or
even optically thick to neutrino emission in some region of the disk
if the accretion rate is sufficiently high. However, a significant
difference between black hole and neutron star disks is that the
heat energy of the disk can be advected into the event horizon if
the central object is a black hole, but if the central object is a
neutron star, the heat energy should be eventually released from a
region of the disk near the stellar surface. As a result, the
neutrino luminosity of the neutron star disk should be much larger
than that in black hole accretion. We approximately took the disk as
a Keplerian disk. According to the Virial theorem, one half of the
gravitational energy in such a disk is used to heat the disk and the
other half to increase the rotational kinetic energy of the disk. We
assumed that most of the heat energy generated from the disk is
still cooled from the disk via neutrino emission and the rotational
energy is used to spin up the neutron star or is released on the
stellar surface via neutrino emission.

In a certain range of hypercritical accretion rates, depending on
the mechanisms of energy heating and cooling in the disk, we
considered a two-region, steady-state disk model. The outer disk is
similar to the outer region of a black hole disk. We used the
standard viscosity assumption, Newtonian dynamics and vertically
integrated method to study the structure of the outer disk. Since
the radial velocity of the disk flow is always subsonic, no stalled
shock exists in the disk and thus we considered that physical
variables in the disk change continuously when crossing the boundary
layer between the inner and outer disks. The inner disk, which
expands until a heating-cooling balance is built, could satisfy a
self-similar structure as shown by equation (\ref{3en02}).

In this chapter we first studied the disk structure analytically. To
do this, we adopted a simple disk model based on the analytical
method. We took the pressure as a summation of several extreme
contributions and simple formulae of neutrino cooling. And we took
the electron fraction $Y_{e}$ as a parameter in the simple model. We
used an analytical method to find the dominant-pressure distribution
(Table 2) and the radial distributions of the density, temperature
and pressure (solutions \ref{3s11}, \ref{3s12}, \ref{3e302},
\ref{3s2}, \ref{3e3031}) in the outer disk. Then we used the
equation of energy balance between heating and neutrino cooling to
calculate the size of the inner disk in four different cases:
whether the advection-dominated outer disk is radiation or gas
pressure-dominated, and whether the neutrino-cooled outer disk is
optically thin or thick to neutrino emission (Table \ref{tab33} to
\ref{tab36}). Subsequently, we numerically calculated the size of
the inner disk, the structure, and energy conversion and emission of
the entire disk in the simple model (Fig. \ref{fig32} to Fig.
\ref{fig37}) and compared the numerical results with the analytical
results. The numerical results are consistent with the analytical
ones from the simple model.

When the accretion rate is sufficiently low, most of the disk is
advection-dominated, the energy is advected inward to heat the inner
disk, and eventually released via neutrino emission in the inner
disk. In this case, the inner disk is very large, and quite
different from a black hole disk, which advects most of the energy
inward into the event horizon. If the accretion rate is higher, then
physical variables such as the density, temperature and pressure
become larger, the disk flow becomes NDAF, the advected energy
becomes smaller, and heating of the inner disk becomes less
significant. As a result, the size of the inner disk is much
smaller, and the difference between the entire disk and the black
hole disk becomes less significant. Furthermore, if the accretion
rate is large enough to make neutrino emission optically thick, then
the effect of neutrino opacity becomes important so that the
efficiency of neutrino emission from most of the disk decreases and
the size of the inner disk again increases until the entire disk
becomes self-similar. Besides, a different mass of the central star
or a different electron-nucleon ratio also makes physical variables
and properties of the disk different. However, the accretion rate
plays a more significant role in the disk structure and energy
conversion, as it varies much wider than the other parameters.

The simple model is based on the early works such as Popham et al.
(1999) and Narayan et al. (2001). We found that the simple model in
fact gives us a clear physical picture of the hyperaccretion disk
around a neutron star, even if we used some simplified formulae in
thermodynamics and microphysics in the disk. In $\S$3.4 we
considered an elaborate model, in which we calculated the pressure
of electrons and positrons by using the Fermi-Dirac distribution and
replaced the factor $11/12$ by $1/3$ in the radiation-pressure
equation. We adopted more advanced expressions of the neutrino
cooling rates, including the effect of all three types of neutrinos
and the electron degeneracy. Moreover, we considered
$\beta$-equilibrium in the disk to calculate the electron fraction
$Y_{e}$. Then, in the elaborate model, we also calculated the the
size of the inner disk (Fig. \ref{fig38}), the radial distributions
of the ``equivalent" adiabatic index $\gamma$, the electron fraction
(Fig. \ref{fig39}), the density, temperature and pressure in the
disk (Fig. \ref{fig310}), the neutrino cooling rate distribution of
the disk (Fig. \ref{fig311}, \textit{left}), the neutrino emission
luminosity from the inner and outer disks, and the total neutrino
luminosity of a neutron-star disk compared with that of a black-hole
disk (Fig. \ref{fig311}, \textit{right}).

The electron fraction $Y_{e}$ was also determined in the elaborate
model. We found that $Y_{e}$ drops with increasing the accretion
rate in the outer disk. $Y_{e}$ can be greater than 0.5 at a large
radius if the accretion rate is sufficiently low, and $Y_{e}\ll 1$
in the disk when the accretion rate is high enough. If we put these
results of $Y_{e}$ in the elaborate model into the simple model
correctly (i.e. $Y_{e}\sim 1$ for low accretion rate and $Y_{e}\ll
1$ for high accretion rate), we find that they are basically
consistent with each other (see Fig. \ref{fig38}, Fig. \ref{fig39},
Fig. \ref{fig310}, Fig. \ref{fig311}), and also consistent with most
of the early works (see the discussion in the end of \S3.3.1).

\begin{figure}
\centering\resizebox{0.7\textwidth}{!}
{\includegraphics{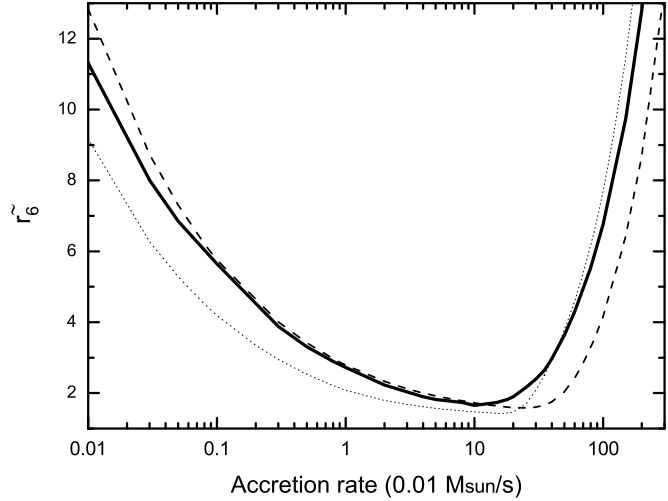}} \caption{Comparison of
$\tilde{r}_{6}$ in two models with $M=1.4M_{\odot}$. (1) The third
model discussed in \S 5 ({\em thick solid line}), (2) the simple
model with $Y_{e}=1$ ({\em dashed line}), and (3) the simple model
with $Y_{e}=1/9$ ({\em dotted line}). The solution of
$\tilde{r}_{6}$ in the third model is even more consistent with the
simple model than that of the elaborate model discussed in
\S3.4.}\label{fig312}
\end{figure}

A main difference in the structure between the simple and elaborate
models is caused by different expressions of pressure adopted in the
two models. In order to see it clearly, we introduce the third model
here. We still keep $P_{\rm rad}=11aT^{4}/12$ as the simple model
but just change the relativistic degeneracy pressure term of
electrons (the second term in equation \ref{3e32}) to the
Fermi-Dirac distribution (formula \ref{3z02} for $P_{e^{-}}$), and
use all the other formulae in $\S$3.4. In other words, the third
model is introduced by only changing one pressure term in the
elaborate model. We can find that the results from the third model
are even more consistent with those of the simple model than the
elaborate model in \S 3.4. Take Figure \ref{fig312} as an example.
We compare the solution of the inner disk of the third model with
that of the simple model. From Fig. \ref{fig312} we can see that if
the accretion rate is low and $Y_{e}\sim 1$, the thick solid line is
much closer to the thin dashed line, which results from the simple
model with $Y_{e}=1$; and if the accretion rate is high enough and
$Y_{e}\ll 1$, the solid line is much closer to the thick dotted
line, which is the result from the simple model with $Y_{e}=1/9$.
Compared with Fig. 8a, this result is even more consistent with the
simple model. The values of density $\rho$ and pressure $P$ in Fig.
10, for a low accretion rate, are smaller than those of the simple
model, which is also due to different expressions of the radiation
pressure used in these two models in \S 3.3 and \S 3.4. Therefore,
we conclude that the main difference of the results between the
simple model in \S 3.3 and the elaborate model in \S 3.4 come from
different expressions of the pressure in disks. However, we believe
that the pressure formulae given in the elaborate model are more
realistic since $11aT^{4}/12$ is only an approximated formula for
the pressure of a plasma of photons and $e^{+}e^{-}$ pairs. On the
other hand, as what has been pointed out by Lee et al. (2005) and
Liu et al. (2007), formulae $P_{\rm rad}=aT^{4}/3$ and (\ref{3z02})
in \S 3.4.1.1 are better and can automatically take relativistic
$e^{+}e^{-}$ pairs into account in the expression of $P_{e}$.

The different expression of the neutrino cooling rate $Q^-_{\nu}$
makes the neutrino luminosity distribution different in the region
where is optically thick to neutrino emission. A more advanced
expression of neutrino cooling rate $Q_{\nu}^{-}$ gives better
results of the neutrino luminosity per unit area than that given by
the rough expression $\frac{7}{8}\sigma_{B}T^{4}/\tau$ in \S3.

In this chapter \S 3 we studied the disk without any outflow, which
may exist if the disk flow is an ADAF. However, it is still unclear
whether an outflow or neutrino cooling plays a more important role
since the size of the disk is quite small. The other case that we
ignored is that, if the radius of the central neutron star is
smaller than that of the innermost stable circular orbit of the
accretion disk, the accreting gas eventually falls onto the neutron
star freely. In this case, a shock could form in the region between
the innermost stable circular orbit and neutron star surface
(Medvedev 2004). This effect can be studied if other effects such as
the equation of state of a differentially-rotating neutron star and
its mass-radius relation are together involved.

Neutrinos from a hyperaccretion disk around a neutron star will be
possibly annihilated to electron/positron pairs, which could further
produce a jet. It would be expected that such a jet is more
energetic than that from the neutrino-cooled disk of a black hole
with same mass and accretion rate as those of the neutron star
(Zhang \& Dai 2009a, see the next chapter). This could be helpful to
draw the conclusion that some GRBs originate from neutrino
annihilation rather than magnetic effects such as the
Blandford-Znajek effect.

We considered a central neutron star with surface magnetic field
weaker than $B_{s,{\rm cr}}\sim 10^{15}-10^{16}$ G for typical
hyperaccretion rates in \S 3. For magnetars (i.e., neutron stars
with ultra-strongly magnetic fields of $\sim B_{s,{\rm cr}}$),
however, the magnetic fields could play a significant role in the
global structure of hyperaccretion disks as well as underlying
microphysical processes, e.g., the quantum effect (Landau levels) on
the electron distribution and magnetic pressure in the disks could
become important. Thus, the effects of an ultra-strongly magnetic
field on hyperaccretion disks around neutron stars are an
interesting topic, which deserves a detailed study.


\chapter{Hyperaccreting Neutron-Star Disks and Neutrino Annihilation}
\label{chap4} \setlength{\parindent}{2em}

\section{Introduction}
The hyperaccreting disk surrounding a stellar-mass black hole
possibly formed by the merger of a compact object binary or the
collapse of a massive star has been argued to be a candidate for
central engines of gamma-ray bursts (GRBs) (e.g. Eichler et al.
1989; Narayan et al. 1992; Woosley 1993; Paczy\'{n}ski 1998; Popham
et al. 1999; MacFadyen \& Woosley 1999; Narayan et al. 2001). The
typical mass of the debris dense torus or disk is about $0.01-1
M_{\odot}$ with large angular momentum as well as high accretion
rate up to $\sim 1.0 M_{\odot}\,{\rm s}^{-1}$. Although the optical
depth of the accreting matter in the disk is enormous, the disk can
be cooled or partly cooled via neutrino emission. A number of
studies have investigated the structure and energy transfer of the
neutrino-cooled disk around a black hole both in steady-state and
time-dependent considerations over last several years (Popham et al.
1999; Narayan et al. 2001; Kohri \& Mineshige 2002; Di Matteo et al.
2002; Kohri et al. 2005; Gu et al. 2006; Chen \& Beloborodov 2007;
Liu et al. 2007; Janiuk et al. 2007; Metzger et al. 2008).

An alternative model of central engines of GRBs is newly, rapidly
rotating neutron stars or magnetars (Usov 1992; Klu\'{z}niak \&
Ruderman 1998; Dai \& Lu 1998b; Ruderman et al. 2000; Wheeler et al.
2000). In recent years, newborn neutron stars have also been
suggested as an origin of some GRBs and their afterglows. For
example, Dai et al. (2006) argued that the X-ray flares discovered
by {\em Swift} can be explained as being due to magnetic instability
and reconnection-driven events from highly-magnetized millisecond
pulsars; the shallow decay phase of X-ray afterglows is considered
to be due to energy injection to a forward shock by a relativistic
pulsar wind (Dai 2004; Yu \& Dai 2007); a newly-formed neutron star
rather than a black hole is expected to explain the light curve of
SN 2006aj associated with GRB 060218 (Mazzali et al. 2006; Soderberg
et al. 2006). Moreover, simulations on the merger of a compact
object binary show that it is possible to form a hypermassive
neutron star, at least a transiently existing neutron star after the
merger, depending on initial conditions of the binary, equations of
state of neutron matter and the structure of magnetic fields
(Shibata et al. 2003; Shibata 2003; Lee \& Ramirez-Ruiz 2007;
Anderson et al. 2008; Liu et al. 2008). Therefore, the
hyperaccreting disk around a neutron star can also be considered as
possible central engines for some GRBs. Based on these motivations,
we have studied the structure of the hyperaccretion disk around a
neutron star using both analytic and numerical methods (Zhang \& Dai
2008a, hereafter ZD08). We found that the neutron-star disk can cool
more efficiently and produce a higher neutrino luminosity than the
black-hole disk.

In ZD08, the quasi-steady disk around a neutron star is
approximately divided into two regions --- inner and outer disks,
depending on the energy transfer and emission in the disk. For the
outer disk, the heating energy rate $Q^{+}$ is mainly due to local
dissipation ($Q^{+}=Q_{\rm vis}^{+}$), and the structure of the
outer disk is very similar to the black-hole disk. On the other
hand, the heating energy in the inner disk includes both the energy
generated by itself and the energy advected from the outer region
($Q^{+}=Q_{\rm vis}^{+}+Q_{\rm adv}^{+}$), so the inner disk has to
be dense with a high pressure. We approximately take $Q^{+}=Q^{-}$
and the entropy-conservation self-similar condition $ds$=0 to
describe the inner disk. The size of the inner disk is determined by
the global energy equation of the inner disk. However, we need to
point out that the entropy-conversation structure is not the only
possible structure of the inner disk, which depends on the detailed
form of energy and mass transfer. In the case where $Q^{-}<Q^{+}$ in
the inner disk, we should take the advection-dominated self-similar
structure to describe the inner disk.

The net gravitational binding energy of the accreting matter is
proposed to drive a relativistic outflow or jet by two general
mechanisms that could provide energy for GRBs: neutrino annihilation
and magnetohydrodynamical mechanisms such as the Blandford-Znajek
effect. The mechanism of neutrino annihilation is easy to understand
and could be calculated based on the structure and neutrino
luminosity in the disk (Ruffert et al. 1997, 1998; Popham et al.
1999; Asano \& Fukuyama 2000, 2001; Di Matteo et al. 2002; Miller et
al. 2003; Birkl et al. 2007; Gu et al. 2006; Liu et al. 2007).
However, the annihilation rate due to neutrino emission from the
black-hole disk may not be able to produce a sufficiently high
luminosity to explain some energetic GRBs (Di Matteo et al. 2002).
Gu et al. (2006) and Liu et al. (2007) showed that the annihilation
luminosity can reach $10^{52}$ergs s$^{-1}$ even for an accretion
rate $\sim 10M_{\odot}$ s$^{-1}$. However, such an accretion rate is
too large for high-energy long GRBs, since this requires an
unreasonable massive accretion disk around a compact object. As the
neutron-star disk structure and neutrino luminosity are different
from the black-hole disk, it is interesting to calculate the
neutrino annihilation rate above the neutron-star disk and to
consider whether the annihilation energy rate and luminosity above a
neutron-star disk are high enough to produce energetic GRBs.

On the other hand, we do not consider any outflow from the disk in
ZD08, which may play a significant role in the structure and energy
transfer in the disks around neutron stars. A nonrelativistic or
subrelativistic ouflow or ``wind" from the disk can be considered as
an energy source of supernovae (MacFadyen \& Woosley 1999; Kohri et
al. 2005). This theoretical model becomes more attractive after the
discovery of the connection between some GRBs and supernovae (e.g.
Galama et al. 1998, Stanek et al. 2003, Prochaska et al. 2004,
Campana et al. 2006), while the GRB component is considered from a
relativistic jet produced by neutrino annihilation. As a result,
both outflow ejection and neutrino annihilation could be important
in the events of GRB-SN connections within the framework of the
collapsar model (Woosley \& Bloom 2006). However, an outflow from
the black-hole disk is expected and becomes important whenever the
accretion flow is an advection-dominated accretion flow (ADAF)
(Narayan \& Yi 1994, 1995), while the neutrino luminosity is
relatively low for ADAF. In other words, neutrino emission may not
provide a sufficiently high amount of energy for GRBs associated
with supernovae if a thermally-driven outflow is produced from the
advection-dominated disk at the same time. Therefore, we need to
calculate the neutrino luminosity and annihilation efficiency of the
advection-dominated disk with outflow around a neutron star if the
neutrino luminosity is much higher for the advection-dominated
neutron-star disk than the black-hole disk.

In this chapter we still consider the case in which the central
object is a neutron star rather than a black hole. Our purpose is to
further study the structure of a hyperaccreting neutron-star disk
following ZD08, and calculate the neutrino annihilation rate above
such a disk. This chapter is organized as follows. In \S4.2 we
introduce basic equations of the neutrino-cooled disk. We discuss
the properties of the inner disk in \S4.3 based on the two-region
disk scenario introduced in ZD08. We study the disk with different
values of the viscosity parameter $\alpha$ and the energy parameter
$\varepsilon$ (some quantities are given in this chapter in Table
\ref{tab41}), and then study the disk structure with an outflow. Two
models of an outflow driven from the neutron-star disk are
introduced in \S4.3.4. In \S4.4, we calculate the neutrino
annihilation rate and luminosity above the neutron-star disk in
various cases, and compare the results with the black hole disk. We
discuss the effect of the neutrino luminosity at the neutron star
surface boundary layer on the annihilation rate. In \S4.5, we
particularly focus on an astrophysical application of the
neutron-star disk in GRBs and GRB-SN connections. Conclusions are
presented in \S4.6.
\begin{table}
\begin{center}
Notation and definition of some quantities in this chapter.
\begin{tabular}{|l|l|r|r|r|r|}
\hline
notation & definition & \S/Eq. \\
\hline
 $\varepsilon$ & energy parameter& \S4.2.1, eq.(\ref{4a09})\\
 $\gamma$ & adiabatic index of the accreting matter& \S4.2.2, eq.(\ref{4b07})\\
 $s$ & outflow index& \S4.3.3, eq.(\ref{4e01})\\
 $\zeta$ & toroidal velocity difference between outflow and accretion disk& \S4.3.3, eq.(\ref{4e02})\\
 $\eta_{s}$ & efficiency factor to measure the surface emission& \S4.4.1, eq.(\ref{4g02})\\
 $\eta_{s,ADAF}$ & efficiency factor to measure the energy advected from the disk& \S4.4.1, eq.(\ref{4g06})\\
\hline
\end{tabular}
\caption{}
\end{center}\label{tab41}
\end{table}

\section{Basic Equations}
\subsection{Conservation equations}
In this chapter, all quantities are used as their usual meanings
(see ZD08). We adopt the cylindrical coordinates ($r, \varphi, z$)
to describe the disk. $v_{r}$, $v_{\varphi}$ are the radial and
rotational velocity, $\Omega$ is the angular velocity, and
$\Omega_{K}$ is the Keplerian angular velocity. $\Sigma=2\rho H$ is
the disk surface density with $\rho$ as the density and $H$ as the
half-thickness of the disk. Vertical hydrostatic equilibrium gives
$H=c_{s}/\Omega_{K}$, where the isothermal sound speed is
$c_{s}=(P/\rho)^{1/2}$ with $P$ to be the gas pressure. The
$\nu_{k}=\alpha c_{s}H$ is the kinematic viscosity coefficient in
the disk with $\alpha$ to be the viscosity parameter.

The mass continuity equation is
\begin{equation}
\frac{1}{r}\frac{d}{dr}\left(r\Sigma
v_{r}\right)=2\dot{\rho}H,\label{4a01}
\end{equation}
where $\dot{\rho}$ is the mass-loss term. If the outflow of the disk
is weak, the mass accretion rate $\dot{M}$ can be considered as a
constant and we have the accretion rate,
\begin{equation}
\dot{M}=-4\pi r\rho v_{r}H\equiv -2\pi r\Sigma v_r.\label{4a02}
\end{equation}
In \S 3.3 we will also discuss the disk structure with outflows.

The angular momentum conservation reads
\begin{equation}
\Sigma rv_{r}\frac{d(rv_{\varphi})}{dr}=\frac{d}{dr}\left(\Sigma
\alpha
\frac{c_{s}^{2}}{\Omega_{K}}r^{3}\frac{d\Omega}{dr}\right)+\frac{d}{dr}\dot{J}_{ext},\label{4a03}
\end{equation}
with $\dot{J}_{ext}$ as the external torque acted on the disk, such
as the torque acted by the outflow from the disk. The angular
momentum flows into the central compact star or the coupling exerted
by the star on the inner edge of the disk is
$C=-\dot{M}(GMr_{*})^{1/2}$ with $r_{*}$ being the neutron star
radius (Frank et al. 2002). Therefore, for a weak outflow, combined
with equation (\ref{4a01}) and the above boundary condition,
equation (\ref{4a03}) is integrated as
\begin{equation}
\nu
\Sigma=\frac{\dot{M}}{3\pi}\left(1-\sqrt{\frac{r_{*}}{r}}\right),\label{4a04}
\end{equation}
where $r_{*}$ is the  neutron star radius. Here we adopted the
standard assumption that the torque is zero at the inner boundary of
the disk $r_{*}+b$ with $b\ll r_{*}$. In \S 4.3.3 we will discuss
the angular momentum equation with outflow.

The energy equation of the disk is
\begin{equation}
\Sigma v_{r}T\frac{ds}{dr}=Q^{+}-Q^{-},\label{4a05}
\end{equation}
where $T$ is the temperature in the disk and $s$ is the local
entropy per unit mass, $Q^{+}$ and $Q^{-}$ are the heating and
cooling energy rates in the disk. In the outer disk, the energy
input is mainly due to the local viscous dissipation,
\begin{equation}
Q^{+}=Q^{+}_{\rm vis}=\frac{3GM\dot{M}}{8\pi
r^{3}}\left(1-\sqrt{\frac{r_{*}}{r}}\right).\label{4a06}
\end{equation}
The left term of equation (\ref{4a05}) can be taken as the energy
advection term $Q_{\rm adv}^{-}$. We can obtain (ZD08)
\begin{equation}
Q_{\rm adv}^{-}=\Sigma
v_{r}T\frac{ds}{dr}=v_{r}T\frac{\Sigma}{2r}\left[\frac{R}{2}\left(1+Y_{e}\right)+\frac{4}{3}g_{*}\frac{aT^{3}}{\rho}
\right],\label{4a07}
\end{equation}
where $R=8.315\times10^{7}$ergs mole$^{-1}$ K$^{-1}$ is the gas
constant, $Y_{e}$ is the ratio of electron to nucleon number
density, the free degree factor $g_{*}$ is 2 for photons and 11/2
for a plasma of photons and relativistic $e^{-}e^{+}$ pairs.

The energy cooling rate $Q^{-}$ is mainly due to neutrino emission,
i.e., $Q^{-}\approx Q^{-}_{\nu}$ (Popham \& Narayan 1995; Di Matteo
et al. 2002),
\begin{equation}
Q_{\nu}^{-}=\sum_{i=e,\mu,\tau}\frac{(7/8)\sigma_{B}T^{4}}{(3/4)[\tau_{\nu_{i}}/2+1/\sqrt{3}+1/(3\tau_{a,\nu_{i}})]}
.\label{4a08}
\end{equation}
The three types of neutrino cooling rate per unit volume are
\begin{equation}
\dot{q}_{\nu_{e}}=\dot{q}_{\rm eN}+\dot{q}_{e^{-}e^{+}\rightarrow
\nu_{e}\bar{\nu}_{e}}+\dot{q}_{\rm brem}+\dot{q}_{\rm
plasmon},\label{4a081}
\end{equation}
\begin{equation}
\dot{q}_{\nu_{\mu}}=\dot{q}_{\nu_{\tau}}=\dot{q}_{e^{-}e^{+}\rightarrow
\nu_{\tau}\bar{\nu}_{\tau}}+\dot{q}_{\rm brem}, \label{4a082}
\end{equation}
where $\dot{q}_{\rm eN}$, $\dot{q}_{e^{-}e^{+}\rightarrow
\nu_{i}\bar{\nu}_{i}}$, $\dot{q}_{\rm brem}$, and $\dot{q}_{\rm
plasmon}$ are the electron-positron pair capture rate, the
electron-positron pair annihilation rate, the nucleon bremsstrahlung
rate, and the plasmon decay rate. Following Kohri et al. (2005),
Janiuk et al. (2007) and Liu et al. (2007), we calculate the
absorption and scattering optical depth for three types of neutrinos
$\tau_{a,\nu_{i}(e,\mu,\tau)}$ and $\tau_{s,\nu_{i}(e,\mu,\tau)}$ as
well as the neutrino cooling rates. For hyperaccretion disks, the
electron-positron pair capture rate plays the most important role
among several types of neutrino cooling rates.

Moreover, besides the local energy equation (\ref{4a05}), we need
the global energy conservation equation of the inner disk in order
to decide the size of the inner disk. The maximum power that the
inner disk an release is estimated as (ZD08)
\begin{eqnarray}
L_{\nu,max}&\approx&\frac{3GM\dot{M}}{4}\left\{\frac{1}{3r_{*}}-\frac{1}{r_{\rm
out}}\left[1-\frac{2}{3}\left(\frac{r_{*}}{r_{\rm
out}}\right)^{1/2}\right]\right\}\nonumber\\
&&-\bar{f}_{\nu}\frac{3GM\dot{M}}{4}\left\{\frac{1}{\tilde{r}}
\left[1-\frac{2}{3}\left(\frac{r_{*}}{\tilde{r}}\right)^{1/2}\right]-
\frac{1}{r_{\rm out}}\left[1-\frac{2}{3}\left(\frac{r_{*}}{r_{\rm
out}}\right)^{1/2}\right]\right\}\nonumber\\
&&\approx\frac{3GM\dot{M}}{4} \left\{{\frac{1}{3
r_{*}}-\frac{\bar{f}_\nu}{\tilde{r}}\left[1-\frac{2}{3}
\left(\frac{r_{*}}{\tilde{r}}\right)^{1/2}\right]}\right\},
\end{eqnarray}
where $\tilde{r}$ is the radius between inner and outer disks (i.e.,
the size of the inner disk), the average neutrino cooling efficiency
$\bar{f}_{\nu}$ is determined by
\begin{equation} \bar{f}_{\nu}=\frac{\int_{\tilde{r}}^{r_{\rm
out}}Q_{\nu}^{-}2\pi r dr}{\int_{\tilde{r}}^{r_{\rm out}}Q^{+}2\pi r
dr}.\label{4a10}
\end{equation}
Thus we derive
\begin{equation}
\int_{r_{*}}^{\tilde{r}}Q_{\nu}^{-}2\pi rdr=\varepsilon
L_{\nu,max}=\varepsilon \frac{3GM\dot{M}}{4} \left\{{\frac{1}{3
r_{*}}-\frac{\bar{f}_\nu}{\tilde{r}}\left[1-\frac{2}{3}
\left(\frac{r_{*}}{\tilde{r}}\right)^{1/2}\right]}\right\}\label{4a09}
\end{equation}
with the energy parameter $\varepsilon$ being introduced to measure
the neutrino cooling efficiency of the inner disk. When the outer
disk flow is mainly an ADAF, we have $\bar{f}_\nu\sim 0$ and the
maximum energy release rate of the inner disk to be
$GM\dot{M}/4r_{*}$. When the outer disk flows is an efficiently
NDAF, then $\bar{f}_\nu\simeq 1$ and the energy release of the inner
disk mainly results from the heat energy generated by itself. The
values of $\bar{f}_\nu$ are calculated analytically in Zhang (2009,
Fig. 3.6). In ZD08, we simply set $\varepsilon=1$ and use the
entropy-conservation self-similar structure to describe the inner
disk. In \S3, we also discuss the case of $\varepsilon<1$ with
different structures of the inner disk. In addition, if we consider
an outflow ejected from the disk, equation (\ref{4a09}) should be
modified. We will discuss the modification in \S3.3.

\subsection{Pressure and $\beta$-equilibrium}
The total pressure in the disk is the summation of four terms:
nucleons, radiation, electrons (including $e^{+}e^{-}$ pairs) and
neutrinos,
\begin{equation}
P=P_{\rm nuc}+P_{\rm rad}+P_{e}+P_{\nu},\label{4b01}
\end{equation}
where the pressures of nucleons, radiation and electrons are
\begin{equation}
P_{\rm nuc}=\frac{\rho k_{B}T}{m_{B}},\label{4b02}
\end{equation}
\begin{equation}
P_{\rm rad}=\frac{1}{3}aT^{4},\label{4b03}
\end{equation}
\begin{equation}
P_{e^{\pm}}=\frac{1}{3}\frac{m_{e}^{4}c^{5}}{\pi^{2}\hbar^{3}}
\int^{\infty}_{0}\frac{x^{4}}{\sqrt{x^{2}+1}}\frac{dx}{e^{(m_{e}c^{2}\sqrt{x^{2}+1}\mp\mu_{e})/k_{B}T}+1},\label{4b05}
\end{equation}
and
\begin{equation}
P_{e}=P_{e^{-}}+P_{e^{+}}.\label{4b04}
\end{equation}
We use the Fermi-Dirac distribution to calculate the pressure of
electrons, where $\mu_{e}$ is the chemical potential of electron
gas, and $k_{B}$ is the Boltzmann constant.

The ratio of the neutrino pressure to the total pressure becomes
noticeable only in very opaque regions of the disk (e.g., Kohri et
al. 2005, their Fig. 6; Liu et al. 2007, their Fig. 3). The neutrino
pressure is
\begin{equation}
P_{\nu}=u_{\nu}/3, \label{4b06}
\end{equation}
where $u_{\nu}$ is the energy density of neutrinos. We adopt the
expression of $u_{\nu}$ from previous work (e.g., Di Matteo et al.
2002).

The adiabatic index of the accreting matter is important to
determine the size of the inner disk when it satisfies the
entropy-conservation condition. It can be written as
\begin{equation}
\gamma=1+(P_{\rm nuc}+P_{\rm rad}+P_{e}+P_{\nu})/(u_{\rm nuc}+u_{\rm
rad}+u_{e}+u_{\nu}). \label{4b07}
\end{equation}

Moreover, we need the equation of charge neutrality
\begin{equation}
n_{p}=\frac{\rho Y_{e}}{m_{B}}=n_{e^{-}}-n_{e^{+}},\label{4b08}
\end{equation}
and the chemical equilibrium equation
\begin{eqnarray}
n_{p}(\Gamma_{p+e^{-}\rightarrow
n+\nu_{e}}+\Gamma_{p+\bar{\nu}_{e}\rightarrow
n+e^{+}}+\Gamma_{p+e^{-}+\bar{\nu}_{e}\rightarrow n}) \nonumber\\=
n_{n}(\Gamma_{n+e^{+}\rightarrow
p+\bar{\nu}_{e}}+\Gamma_{n\rightarrow
p+e^{-}+\nu_{e}}+\Gamma_{n+\nu_{e}\rightarrow p+e^{-}})\label{4b09}
\end{eqnarray}
to determine the matter components in the disk, where $n_{p}$,
$n_{e^{-}}$ and $n_{e^{+}}$ are the number densities of protons,
electrons and positrons, and the various weak interaction rates
$\Gamma_{p\rightarrow n}$ ($\Gamma_{n\rightarrow p}$) can be
calculated following Janiuk et al. (2007; see also Kawanaka \&
Mineshige 2007). When neutrinos are perfectly thermalized, we derive
the $\beta$-equilibrium distribution in the disk
\begin{equation}
\textrm{ln}\left(\frac{n_{n}}{n_{p}}\right)
=f(\tau_{\nu})\frac{2\mu_{e}-Q}{k_{B}T}+[1-f(\tau_{\nu})]\frac{\mu_{e}-Q}{k_{B}T},\label{4b10}
\end{equation}
with $Q=(m_{n}-m_{p})c^{2}$, and the factor $f(\tau_{\nu})=\rm
exp(-\tau_{\nu_{e}})$ combines the formula from the
neutrino-transparent limit case with the the neutrino-opaque limit
case of the $\beta$-equilibrium distribution. However, we should
keep in mind that the $\beta$-equilibrium is established only if the
neutronization timescale $t_{n}$ is much shorter than the accretion
timescale $t_{a}$ in the disk $t_{n}<t_{a}$. Beloborodov (2003)
found that the equilibrium requires the accretion rate $\dot{M}$ to
satisfy
\begin{equation}
\dot{M}>\dot{M}_{eq}=2.24\times10^{-3}(r/10^{6}\textrm{cm})^{13/10}(\alpha/0.1)^{9/5}
(M/M_{\odot})^{-1/10}M_{\odot}s^{-1}.\label{4b11}
\end{equation}
When the accretion rate is sufficiently low, the electron fraction
$Y_{e}$ would freeze out from weak equilibrium, while the disk
becomes advection-dominated (e.g., Metzger et al. 2008b, 2009). In
this case, the chemical composition of the disk is determined by its
initial condition before its evolution. Metzger et al. (2009), for
example, showed that the hyperaccreting disk around a black hole
generically freeze out with the fixed $Y_{e}\sim 0.2-0.4$. We will
discuss the effect of chemical equilibrium on neutron-star disks
more detailedly in the next section. The $\beta$-equilibrium
assumption can be approximately adopted in our calculations even for
the ADAF case.

\section{Properties of the Disk}
The two-region disk scenario in ZD08 allows the gravitational energy
of the neutron-star disk system to be released in three regions:
outer disk, inner disk and neutron-star surface. The inner disk
region is formed due to the prevention effect of the neutron-star
surface, i.e., most advection energy generated in the disk still
need to be released in a region near the neutron-star surface.
Moreover, a difference between the angular velocity  of the
neutron-star surface and that of the disk inner boundary layer leads
to neutrino emission in the surface boundary layer. In ZD08, we
assume all the advected energy to be released in the disk and
furthermore all the advected energy from the outer region to be
released in the inner disk. Actually, it is possible that a part of
the advected energy can be transferred onto the neutron-star surface
and finally cooled by neutrino emission from the surface boundary
layer rather than the inner disk. Local microphysics quantities such
as the inner energy density, neutrino cooling rate, heating
convection and conduction properties as well as the advection and
cooling timescales should be calculated in order to simulate the
inner disk formation and the cooling efficiency of the steady-state
inner disk. In this chapter, however, we adopt a simple method to
determine the inner disk structure, i.e., we use the global energy
equation (\ref{4a09}) with the global parameter $\varepsilon$
instead of the local energy equation (\ref{4a05}). The inner disk
can release all the advected energy transferred inward for
$\varepsilon=1$, while a part of the advected energy can still be
transferred onto the neutron-star surface for $\varepsilon<1$.
Moreover, we approximately take $\varepsilon$ as a constant in the
inner disk, and adopt the self-similar treatment to calculate the
inner disk structure. We take $Q^{-}=Q^{+}$ or the
entropy-conservation condition $ds=0$ for $\varepsilon=1$ in the
entire inner disk, and $Q^{-}=\varepsilon Q^{+}$ for $\varepsilon<1$
with the advection-dominated self-similar structure.

With the energy parameter $\varepsilon$ in the global energy
equation (\ref{4a09}) and the self-similar treatment, the inner disk
model can be simplified and calculated by assuming the accretion
rate, the mass of the central compact object, and the self-similar
structure of the inner disk. We discussed the disk structure with
the entropy-conservation condition in ZD08, and in this section we
further discuss the properties of the neutron-star disk in various
cases. Furthermore, we need to consider the effect of an outflow
from the neutron-star disk.

\subsection{Entropy-Conservation Inner Disk with Different $\alpha$}
The viscosity parameter $\alpha$ was first used by Shakura \&
Sunyaev (1973) to express the relation between viscous stress
$t_{r\theta}$ and the pressure $P$ in the disk as
$t_{r\theta}=\alpha P$. Another formula introduces the turbulent
kinematic viscosity is $\nu_{k}=\alpha c_{s}H$ (Frank et al. 2002).
MHD instability simulations show a wide range of $\alpha$ from 0.6
to about 0.005 or less (Hawley et al. 1995; Brandenburg et al. 1995;
Balbus \& Hawley 1998; King et al. 2007). King et al. (2007)
summarized observational and theoretical estimates of the disk
viscosity parameter $\alpha$. They showed that there is a large
discrepancy between the typical values of $\alpha$ from the best
observational evidence ($\alpha\sim 0.1-0.4$ for fully ionized thin
disks) and those obtained from numerical MHD simulations
($\alpha\leq 0.02$ and even considerably smaller). More elaborate
numerical simulations should be carried out for resolving this
problem. For neutrino-cooled hyperaccreting disks, many previous
papers choose $\alpha=0.1$ as the most typical value. The disk
structure with $\alpha$ from 0.01 to 0.1 was discussed in Chen \&
Beloborodov (2007). On the other hand, hyperaccreting disks with
very low $\alpha$ have also been discussed. For example, Chevalier
(1996) studied the neutrino-cooled disk with an extremely small
$\alpha\sim 10^{-6}$. In this section, we discuss the neutron-star
disk with different $\alpha$. We choose the value of $\alpha$ from
0.001 to 0.1. The size of the inner disk alters with different
$\alpha$, and we still adopt the entropy-conversation self-similar
condition of the inner disk to study the effects of the viscosity
parameter.

As discussed in ZD08, from equation (\ref{4a09}), if
$\varepsilon\simeq 1$, the heating energy advected from the outer
disk together with the energy generated in the inner disk is totally
released in the inner disk, and the energy balance can be
established between heating and cooling in the inner disk, i.e.,
$Q^{+}=Q^{-}$, or from equation (\ref{4a05}), $Tds/dr=0$. In the
case where $v_{r}\ll v_{K}$ and $\Omega\sim \Omega_{K}$ but
$|\Omega-\Omega_{K}|\geq v_{r}/r$, we can obtain the
entropy-conservation self-similar structure of the inner disk
(Medvedev \& Narayan 2001; ZD08) by
\begin{equation}
\rho \propto r^{-1/(\gamma-1)},\,\, P \propto
r^{-\gamma/(\gamma-1)},\,\, v_r \propto
r^{(3-2\gamma)/(\gamma-1)}.\label{4c01}
\end{equation}
Since the adiabatic index of the accretion matter is not constant,
we modify expression (\ref{4c01}) as
\begin{eqnarray}
\frac{\rho(r)}{\rho(r+dr)}=\left(\frac{r}{r+dr}\right)^{-1/(\gamma(r)-1)},
\frac{P(r)}{P(r+dr)}=\left(\frac{r}{r+dr}\right)^{-\gamma(r)/(\gamma(r)-1)},
\nonumber\\\frac{v(r)}{v(r+dr)}=\left(\frac{r}{r+dr}\right)^{(3-2\gamma(r))/(\gamma(r)-1)}.\label{4c02}
\end{eqnarray}

\begin{figure}
\centering\resizebox{0.7\textwidth}{!} {\includegraphics{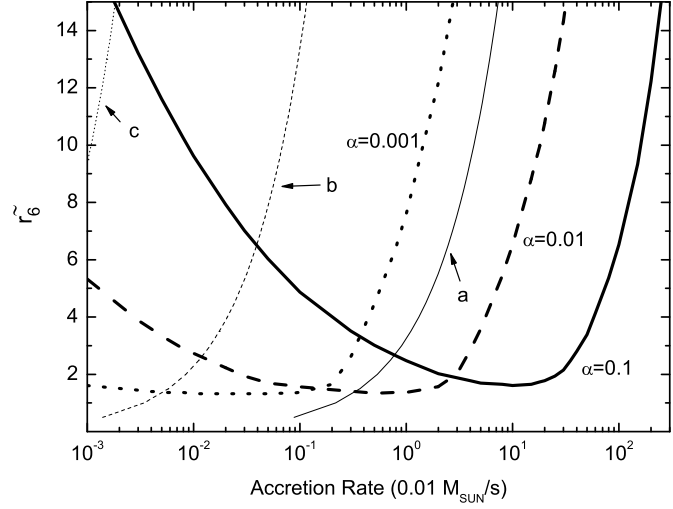}}
\caption{The
radius $\tilde{r}$ ($\tilde{r}_{6}$, in units of $10^{6}$ cm) of the
boundary layer between the inner and outer disks as a function of
accretion rate with different values of the viscosity parameter
$\alpha$=0.1 (thick solid line), 0.01 (thick dashed line) and 0.001
(thick dotted line). The three thin lines labeled ``a", ``b", and
``c" are the characteristic curves of $\beta$-equilibrium with three
values of $\alpha$=0.1, 0.01 and 0.001 respectively. Each
characteristic curve divides the $\dot{M}-\tilde{r}$ parameter plane
into two regions with a chosen $\alpha$, and $\beta$-equilibrium can
be established in the right region respectively.}\label{fig41}
\end{figure}
The size of the inner disk $\tilde{r}$ with various $\alpha$ can be
determined by equation (\ref{4a09}) in \S 4.2.1. Figure \ref{fig41}
shows the size of the inner disk as a function of accretion rate
with different viscosity parameter $\alpha$=0.1, 0.01 and 0.001.
Same as in ZD08, the outer edge radius of the inner disk $\tilde{r}$
decreases with increasing the accretion rate for a low accretion
rate when most part of the disk is advection-dominated. $\tilde{r}$
reaches its minimum value at $\dot{M} \sim 0.1 M_{\odot}s^{-1}$ for
$\alpha=0.1$, and then increases with increasing the accretion rate.
However, the size of the inner disk becomes smaller for lower
viscosity parameter $\alpha$ and would expand dramatically for a
higher accretion rate. The characteristic rate $\dot{M}_{0}$ which
minimizes the size of inner disk $\tilde{r}$ can be approximated by
$\dot{M}_{\rm ch}\sim \alpha M_{\odot}$ s$^{-1}$. The value of
accretion rate $\dot{M}_{\rm ch}$ is between those of characteristic
rates $\dot{M}_{\rm ign}$ (rate of ignition) and $\dot{M}_{\rm
opaque}$ (rate of transparency) in Chen \& Beloborodov (2007).
\begin{figure}
\centering\resizebox{1.0\textwidth}{!} {\includegraphics{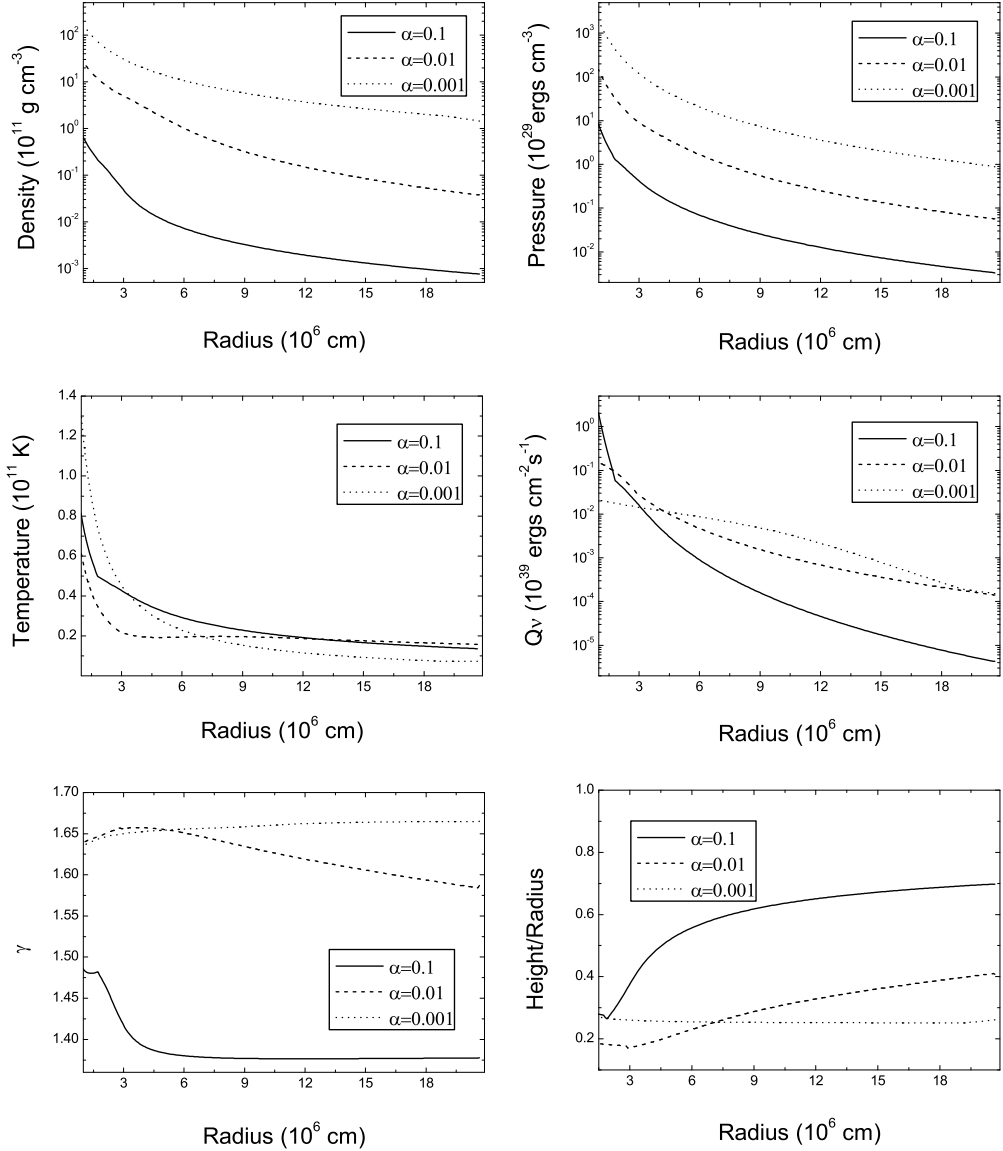}}
\caption{The
density, pressure, temperature, neutrino luminosity per unit area,
adiabatic index $\gamma$ and height (half-thickness of the disk) of
the entire disk with different values of the viscosity parameter
$\alpha$=0.1 (solid line), 0.01 (dashed line) and 0.001 (dotted
line) with a fixed accretion rate $\dot{M}=0.04M_{\odot}$ s$^{-1}$.}\label{fig42}
\end{figure}

Actually, from ZD08, we derive an approximate analytic equation of
the radius $\tilde{r}$ between the innner and outer disks as
\begin{equation}
\tilde{r}^{\frac{5(5-3\gamma)}{4(\gamma-1)}}\left(1-\sqrt{\frac{r_{*}}{\tilde{r}}}\right)^{3/4}
\left(r_{*}^{\frac{3\gamma-8}{2(\gamma-1)}}-
\tilde{r}^{\frac{3\gamma-8}{2(\gamma-1)}}\right) \propto
\dot{M}^{-3/2}\alpha_{-1}^{5/2}\label{4c03}
\end{equation}
for a radiation-pressure-dominated ADAF outer disk, and
\begin{equation}
\left(1-\sqrt{\frac{r_{*}}{\tilde{r}}}\right)^{-31/11}\tilde{r}^{(\frac{-1}
{\gamma-1}+\frac{3}{22})}\left(r_{*}^{\frac{2-3\gamma}{\gamma-1}}-
\tilde{r}^{\frac{2-3\gamma}{\gamma-1}}\right)^{-1}
\propto\dot{M}^{10/11}\alpha_{-1}^{-21/11}\label{4c04}
\end{equation}
for a gas-pressure-dominated ADAF. In both cases, the inner disk
size delines as the accretion rate increases or the viscosity
parameter $\alpha$ decreases. For a neutrino-dominated disk, the
inner disk size $\tilde{r}$ reaches its minimum value
\begin{equation}
\left(\frac{\gamma-1}{3\gamma-2}\right)
\tilde{r}^{\frac{2\gamma-1}{\gamma-1}}
\left(1-\sqrt{\frac{r_{*}}{\tilde{r}}}\right)\left(r_{*}^{\frac{2-3\gamma}{\gamma-1}}-
\tilde{r}^{\frac{2-3\gamma}{\gamma-1}}\right)
\propto\left\{\frac{1}{r_{*}}-\frac{3\bar{f}_{\nu}}{\tilde{r}}\left[1-\frac{2}{3}
\left(\frac{r_{*}}{\tilde{r}}\right)^{1/2}\right]\right\}.\label{4c05}
\end{equation}
The solution of minimum $\tilde{r}$ declines as $\alpha$ decreases
because the lower-$\alpha$ disk has a higher value of
$\bar{f}_{\nu}$. As a result, the inner disk size $\tilde{r}$ always
decrease with decreasing $\alpha$ in the ADAF case. This conclusion
is consistent with that of Figure \ref{fig41}.

For simplicity, we here adopt an unified model introduced in \S 4.2
to calculate the structure of the disk both in the ADAF and NDAF
cases. In particular, we assume that the disk is always in the
$\beta$-equilibrium state. Let us focus on this equilibrium
assumption. Following Beloborodov (2003, i.e., equation [\ref{4b11}]
in our work), we plot the characteristic curves of equilibrium with
different values of $\alpha$ in the $\dot{M}-\tilde{r}$ plane of
Figure \ref{fig41}. The $\beta$-equilibrium state can only be
established in the right region divided by the corresponding curve.
In the left region, as mentioned at the end of \S 4.2.2, the weak
interaction timescale become longer than the disk evolutionary
timescale, and the electron fraction $Y_{e}$ freezes out with a
fixed value. However, based on the analytic and numerical arguments
in ZD08, we found that the solutions of disk structure are
relatively insensitive to the value of $Y_{e}$, and the main results
of Figure \ref{fig41} can still be unchanged for various $Y_{e}$. In
ZD08, we fixed the value of $Y_{e}=1/9$ and $Y_{e}=1$ as the two
limits. The inner disk size increases slightly with increasing
$Y_{e}$ in the case of ADAF, and the main result (i.e., the
``U"-shape curve in the $\dot{M}-\tilde{r}$ plane as the solution of
inner disk size) is still kept for both $Y_{e}=1/9$ and $Y_{e}=1$.
Moreover, although larger $Y_{e}$ leads to slightly lower value of
density, temperature and pressure for ADAF, the physical properties
of the ADAF disk with low accretion rate beyond equation
(\ref{4b11}) is close to each other for the cases of $Y_{e}=1/9$ and
1 (ZD08, Fig. 8). Furthermore, if we adopt the equilibrium
assumption in the ADAF case, the value of $Y_{e}$ will actually not
deviate dramatically from $0.5$ (i.e., ZD08, the right panel of Fig.
7). Therefore, we always take $\beta$-equilibrium as an
approximation in our calculation.

Figure \ref{fig42} shows the structure of the disk for a chosen
accretion rate $0.04 M_{\odot}s^{-1}$ as a function of radius for
three values of $\alpha$. The disk with lower $\alpha$ is denser and
thinner with higher pressure and larger adiabatic index, and has a
brighter neutrino luminosity in most part of the disk except for a
part of the inner disk region, which satisfies the self-similar
structure. A low-$\alpha$ accretion flow with less kinematic
viscosity coefficient $\nu_{k}$ requires a higher surface density
$\Sigma$ for a fixed accretion rate compared to a high-$\alpha$
accretion flow. We have listed approximate analytic solutions of
accretion flows in various cases in ZD08. We obtain $\rho\propto
\alpha^{-1}$, $P\propto \alpha^{-1}$, $H\propto\alpha^{0}$ for a
radiation-pressure-dominated ADAF,
$\rho\propto(1+Y_{e})^{-12/11}\alpha^{-8/11}$,
$P\propto(1+Y_{e})^{-4/11}\alpha^{-10/11}$, $H\propto
(1+Y_{e})^{4/11}\alpha^{-1/11}$ for a gas-pressure-dominated ADAF,
$\rho\propto(1+Y_{e})^{-9/5}\alpha^{-13/10}$,
$P\propto(1+Y_{e})^{-3/5}\alpha^{-11/10}$, $H\propto
(1+Y_{e})^{3/5}\alpha^{1/10}$ for gas-pressure-dominated NDAF. The
density and pressure always increase with decreasing $\alpha$. These
results are consistent with those shown in Figure \ref{fig42}. The
disk region where is radiation-pressure-dominated is extremely small
for low $\alpha$ (eqs. [22] to [25] in ZD08). Also, as for a
low-$\alpha$ disk, the electron fraction $Y_{e}$ is also low, so the
disk is thinner compared to the high-$\alpha$ disk for
gas-pressure-dominated ADAF and NDAF, although the viscosity
parameter contributes an increasing factor $\alpha^{-1/11}$ for the
low-$\alpha$ disk with gas-dominated ADAF.

\subsection{Advection-Dominated Inner Disks}
The entropy-conservation self-similar structure has been used by
Medvedev \& Narayan (2001) and ZD08 to discuss the global accretion
disk structure. However, such a structure is not the only structure
for the neutron-star inner disk, as the entropy-conservation
condition $Q^{+}=Q^{-}$ can be satisfied only for
$\varepsilon\simeq1$ in equation (\ref{4a09}). In the case where
$\varepsilon<1$, i.e., the inner disk can only partly release the
heating energy generated by itself and advected from the outer
region, some part of the heating energy in the disk should be still
advected onto the neutron star surface and released from the surface
area, and thus the inner disk cannot satisfy the
entropy-conservation self-similar structure. In this case, a part of
the heating energy is still advected into the inner region until it
is released around the neutron star surface. We can approximately
take $Q^{-}=\varepsilon Q^{+}$ in the inner disk for
$\varepsilon\lesssim 1$, and thus the structure of the inner disk
can be described by the ADAF self-similar structure (Spruit et al.
1987, Narayan \& Yi 1994):
\begin{equation}
\rho \propto r^{-3/2},\,\, P \propto r^{-5/2},\,\, v_r \propto
r^{-1/2},\label{4d01}
\end{equation}

\begin{figure}
\centering\resizebox{0.7\textwidth}{!} {\includegraphics{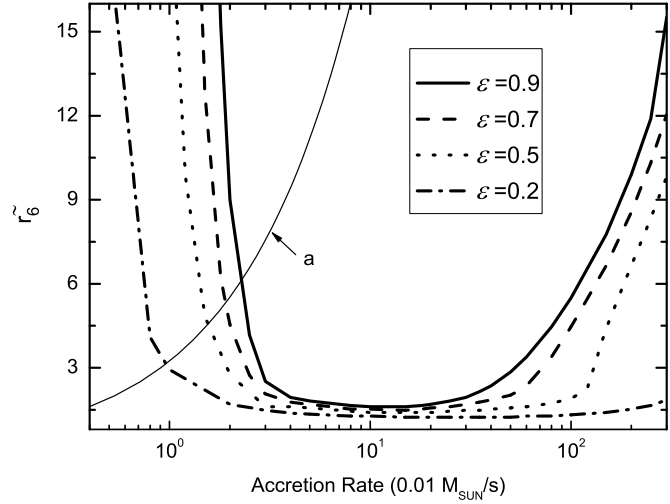}}
\caption{The
radius $\tilde{r}$ (in units of $10^{6}$ cm) between the inner and
outer disks as a function of accretion rate with different values of
the energy parameter $\varepsilon$=0.9, 0.7, 0.5 and 0.2. The thin
line labeled ``a" is the characteristic curve of equilibrium as in
Fig. 1 with $\alpha$=0.1.}\label{fig43}
\end{figure}
In Figure \ref{fig43}, we show the inner disk size for four values
of the energy parameter $\varepsilon$=0.9, 0.7, 0.5 and 0.2. We
still fix $\alpha$=0.1 in this chapter in order to see the effects
of advection in the inner disk with $\varepsilon<1$. This is because
the size of the inner disk becomes smaller for lower $\varepsilon$,
as more heating energy can be advected onto the neutron star
surface, and the inner disk size is small enough to keep energy
balance between heating and cooling in the disk.

Compared with the entropy-conservation self-similar structure, the
size of the advection-dominated inner disk is much larger for a low
accretion rate when most part of the disk is advection-dominated.
When the accretion rate is low, the adiabatic index of the accreting
matter is $\gamma\simeq 4/3$ and the entropy-conservation
self-similar structure (\ref{4c01}) can be approximately taken as
$\rho\propto r^{-3}$ and $P\propto r^{-4}$, which requires a more
dramatic change of density and pressure than those of the
advection-dominated inner disk $\rho\propto r^{-3/2}$ and $P\propto
r^{-5/2}$. This difference in structure between entropy-conservation
and advection-dominated inner disks makes the size of the inner
disks be different with each other.

Finally, what we should point out is that the structure of the
advection-dominated self-similar inner disk even with
$\varepsilon\rightarrow 1$ is different from the
entropy-conservation disk with $\varepsilon=1$, since these two
types of self-similar structure are based on different sets of
conservation equations. The advection-dominated structure is based
on the mass continuity, radial momentum and angular momentum
equations, while we do not consider the local energy equation in
which $Q^{+}=Q^{+}_{\rm vis}+Q^{+}_{\rm adv}$ with $Q^{+}_{\rm adv}$
to be difficult to determine locally, and we only consider the
global energy equation (\ref{4a09}) to calculate the size and
structure of the inner disk. On the other hand, under the
energy-conservation condition $Tds=0$, we can establish the relation
$P\propto \rho^{\gamma}$ from the local energy equation and obtain
the self-similar structure (\ref{4c01}) with a combination of the
mass continuity and the radial momentum and local energy equations
(ZD08). However, the relation $P\propto \rho^{\gamma}$ and the
integrated angular momentum equation (\ref{4a04}) cannot be both
satisfied in the entropy-conservation solution at the same time. In
other words, the angular momentum transfer in the inner disk with
the structure $P\propto \rho^{\gamma}$ cannot be merely due to the
viscosity. We should consider the external torque acted on the disk
or the angular momentum redistribution in the inner disk. We will
discuss the entropy-conservation structure in more details in \S4.5.

\subsection{Inner Disks with Outflows}
In \S4.3.1 and \S4.3.2, we do not consider outflows, which may have
important effects on the structure and energy flux distribution of
the entire disk in some cases. Following Narayan \& Yi (1994, 1995)
and Medvedev (2004), if the adiabatic index $\gamma<3/2$, then the
Bernoulli constant of the accretion flow is positive and a
thermally-driven wind or outflow can be produced from the disk.
Therefore, the outflow component can be important in ADAFs. On the
other hand, since the accretion rate of the neutron-star disk is
very large, it is reasonable to assume that the neutron star, which
has a solid surface and is different from the black hole, cannot
accumulate all of the accreting matter at once, and thus an outflow
could be produced near the neutron star surface and exist in the
inner region of the disk.

In this section, we consider the disk structure and neutrino
emission in the disk with an outflow. We consider two models
depending on two mechanisms. In the first model (hereafter model
O1), an outflow is mainly produced in the process of disk matter
accreting onto the surface of a neutron star. We assume that only
the inner disk produces an outflow and the accretion rate of the
outer disk can still be considered as a constant. We take
\begin{equation}
\dot{M}(r)=\dot{M}_{0}(r/\tilde{r})^{s}\label{4e01}
\end{equation}
for the inner disk with $s$ to be the outflow index and
$\dot{M}_{0}$ to be the constant accretion rate in the outer disk.
Second, if an outflow is produced by the accretion process in the
disk, then we consider that the outflow is produced in the entire
disk (hereafter model O2 in this section), i.e.,
\begin{equation}
\dot{M}(r)=\dot{M}_{0}(r/r_{out})^{s}\label{4e011}
\end{equation}
Strictly speaking, a thermally-driven outflow from the entire disk
is expected in the advection-dominated disk but not in the
neutrino-dominated disk. The winds from the disks which emit a
sufficient high neutrino luminosity are considered to be driven due
to neutrino irradiation (Metzger et al. 2008). We adopt equation
(\ref{4e011}) for all the disks with a wide range of the accretion
rate in model O2 for simplicity, and we also take the index $s$ of
the outflow as a constant. The angular momentum equation with an
outflow can be written as
\begin{equation}
\Sigma
\nu=\frac{1+2s\zeta}{1+2s}\frac{\dot{M}}{3\pi}\left(1-\sqrt{\frac{r_{*}}{r}}\right)\label{4e02}
\end{equation}
where $\zeta$ describes a difference between outflow velocity
$v_{outflow,\phi}$ and accretion-disk velocity $v_{\phi}$:
$v_{\phi}-v_{outflow,\phi}=\zeta v_{\phi}$. From equation
(\ref{4e02}), we see that if $s=0$ or $\zeta=1$, i.e., there is no
outflow or the toroidal velocity of the outflow is zero and no
angular momentum is taken away by the outflow, then the angular
momentum equation (\ref{4e02}) switches back to the common case of
equation (\ref{4a04}). In this section, we take $\zeta=0$, i.e.,
$v_{outflow,\phi}\approx v_{acc,\phi}$\footnote{Here we do not
consider the effects of magnetic fields. In fact, the differences in
azimuthal velocity and angular momentum between the outflow and
accretion inflow could also exist when the accretion flow is
governed by a magnetic field ($B_{\phi},B_{z}$) without strong
poloidal component $B_{r}$ (Xie \& Yuan 2008; Bu et al. 2009). The
poloidal magnetic field, however, would cause the outflow to
co-rotate with the disk out to the Alfv\'{e}n radius above the disk
surface. In this section, we take the outflow to co-rotate with the
accretion inflow $v_{outflow,\phi}\approx v_{acc,\phi}$.}.

With the outflow in the inner disk, the self-similar structure
becomes
\begin{equation}
\rho \propto r^{s-3/2},\,\, P \propto r^{s-5/2},\,\, v_r \propto
r^{-1/2}.\label{4e03}
\end{equation}
The energy conservation equation of the inner disk can be rewritten
as
\begin{equation}
\int_{r_{*}}^{\tilde{r}}Q_{\nu}^{-}2\pi
rdr=\varepsilon\left[\int_{r_{*}}^{\tilde{r}}\frac{9}{8}\nu\Sigma\frac{GM}{r^{3}}2\pi
rdr+(1-\bar{f}_\nu)\int_{\tilde{r}}^{r_{out}}\frac{9}{8}\nu\Sigma\frac{GM}{r^{3}}2\pi
rdr\right],\label{4e04}
\end{equation}
where we still keep the advection parameter $\varepsilon$. For model
O1 that the outflows merely exist in the inner disk due to the
neutron star surface, using equations (\ref{4e01}) and (\ref{4e02}),
we derive the energy conservation equation (\ref{4e04}) as
\begin{eqnarray}
\frac{1}{\varepsilon}\left(\int_{r_{*}}^{\tilde{r}}Q_{\nu}^{-}2\pi
rdr\right)& = &
\frac{3}{4}\left(\frac{1+2s\zeta}{1+2s}\right)\frac{GM\dot{M}_{0}}{\tilde{r}^{s}}\nonumber
\\ & & \times\left[\frac{1}{1-s}\left(\frac{1}{r_{*}^{1-s}}-\frac{1}{\tilde{r}^{1-s}}\right)
-\frac{r_{*}^{1/2}}{3/2-s}\left(\frac{1}{r_{*}^{3/2-s}}-\frac{1}{\tilde{r}^{3/2-s}}\right)\right]\nonumber
\\ & & +(1-\bar{f}_\nu)\frac{3GM\dot{M}_{0}}{4\tilde{r}}\left[1-\frac{2}{3}
\left(\frac{r_{*}}{\tilde{r}}\right)^{1/2}\right]\label{4e05}
\end{eqnarray}
for $s<1$, and
\begin{eqnarray}
\frac{1}{\varepsilon}\left(\int_{r_{*}}^{\tilde{r}}Q_{\nu}^{-}2\pi
rdr\right)& = &
\frac{3}{4}\left(\frac{1+2s\zeta}{1+2s}\right)\frac{GM\dot{M}_{0}}{\tilde{r}}\left\{
\textrm{ln}\left(\frac{\tilde{r}}{r_{*}}\right)
-2\left[1-\left(\frac{r_{*}}{\tilde{r}}\right)^{1/2}\right]\right\}\nonumber
\\ & & +(1-\bar{f}_\nu)\frac{3GM\dot{M}_{0}}{4\tilde{r}}\left[1-\frac{2}{3}
\left(\frac{r_{*}}{\tilde{r}}\right)^{1/2}\right]\label{4e06}
\end{eqnarray}
for $s=1$. Here we always consider the outflow index $s\leq1$.

If the outflows exist in the entire disk (model O2), then we have
\begin{eqnarray}
\frac{1}{\varepsilon}\left(\int_{r_{*}}^{\tilde{r}}Q_{\nu}^{-}2\pi
rdr\right)& = &
\frac{3}{4}\left(\frac{1+2s\zeta}{1+2s}\right)\frac{GM\dot{M}_{0}}{r_{out}^{s}}\nonumber
\\ & & \times\left[\frac{1}{1-s}\left(\frac{1}{r_{*}^{1-s}}-\frac{\bar{f}_\nu}{\tilde{r}^{1-s}}\right)
-\frac{r_{*}^{1/2}}{3/2-s}\left(\frac{1}{r_{*}^{3/2-s}}-\frac{\bar{f}_\nu}{\tilde{r}^{3/2-s}}\right)\right]
\label{4e07}
\end{eqnarray}
for $s<1$ and
\begin{eqnarray}
\frac{1}{\varepsilon}\left(\int_{r_{*}}^{\tilde{r}}Q_{\nu}^{-}2\pi
rdr\right)& = &
\frac{3}{4}\left(\frac{1+2s\zeta}{1+2s}\right)\frac{GM\dot{M}_{0}}{r_{out}}\nonumber
\\&&\times\left\{\textrm{ln}\left(\frac{r_{out}}{r_{*}}\right)-\bar{f}_{\nu}
\textrm{ln}\left(\frac{r_{out}}{\tilde{r}}\right)
+2\left[\bar{f}_{\nu}+\left(\frac{r_{*}}{\tilde{r}}\right)^{1/2}-2\right]\right\}
\label{4e08}
\end{eqnarray}
for $s=1$. We should point out that, since we derive the energy
equations (\ref{4e05})-(\ref{4e08}) in the disk with outflow using
the angular momentum equation (\ref{4e02}) rather than the
entropy-conservation  expression (\ref{4c02}), we still use the
self-similar structure (\ref{4e03}) to calculate the properties of
the inner disk and the entire disk in the case of $\varepsilon\simeq
1$ both in models O1 and O2.

\begin{figure}
\resizebox{\hsize}{!} {\includegraphics{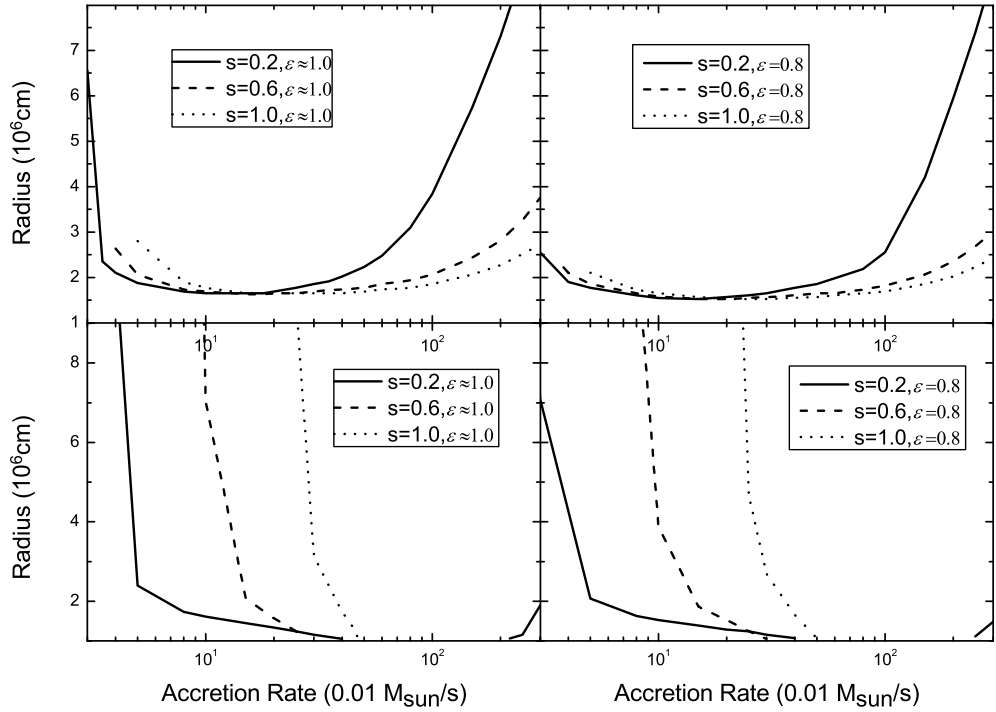}} \caption{(a) {\em
Top panels}: the radius $\tilde{r}$ between the inner and outer disk
in model O1 with different parameter sets of the outflow index
$s$=0.2, 0.6, 1 and the energy parameter $\varepsilon\simeq$ 1, 0.8.
(b) {\em Bottom panels}: the radius $\tilde{r}$ in model O2 with the
same parameter sets ($s, \varepsilon$) as in the top two panels.}\label{fig44}
\end{figure}
The top two panels in Figure \ref{fig44} show the size of the inner
disk with different values of the advection parameter $\varepsilon$.
In model O1, the inner disk is larger for a stronger outflow (larger
$s$) with low accretion rate ($<0.2M_{\odot}s^{-1}$); but for a high
accretion rate ($>0.2M_{\odot}s^{-1}$), the size of the inner disk
decreases with increasing the outflow index $s$. In fact, if the
accretion rate is low, the flow of the outer disk is mainly ADAF and
$\bar{f}_{\nu}\sim 0$. From equations (\ref{4e05}) and (\ref{4e06}),
most of the energy generated in the outer disk is advected into the
inner region, and the inner disk size is mainly determined by the
self-similar structure (\ref{4e03}), which requires a less dramatic
change of density and pressure as functions of radius for higher $s$
or stronger outflow. As a result, the size of the inner disk becomes
larger for a stronger outflow. On the other hand, if the accretion
rate is high, we have $\bar{f}_{\nu}\sim 1$ in the outer disk, the
inner disk size is mainly determined by the heating energy generated
in the inner disk, i.e., the first terms of the right-hand side in
equations (\ref{4e05}) and (\ref{4e06}), and a stronger outflow
carries away more energy from the disk and allows a smaller size of
the inner disk when the accretion rate is sufficiently high.

The bottom two panels in Figure \ref{fig44} show the inner disk size
in model O2 in which an outflow exists in the entire disk. The inner
disk structure cannot exist for a high accretion rate
($>0.4M_{\odot}s^{-1}$) except for the weak outflow case ($s=0.2$)
where the inner disk also exists in some range of a high accretion
rate. The change of the inner disk size is more significant for a
large outflow index $s$ and even the entire accretion disk can
satisfy the self-similar structure (\ref{4e02}) for a sufficiently
low accretion rate. For a high accretion rate
($>0.5M_{\odot}s^{-1}$), the outflow can take away enough heating
energy from the disk, and a balance between heating and cooling in
the entire disk can be built even without an inner disk. In this
range of a high accretion rate, the structure of the neutron-star
disk in model O2 is very similar to the black hole disk with an
outflow.

\begin{figure}
\resizebox{\hsize}{!} {\includegraphics{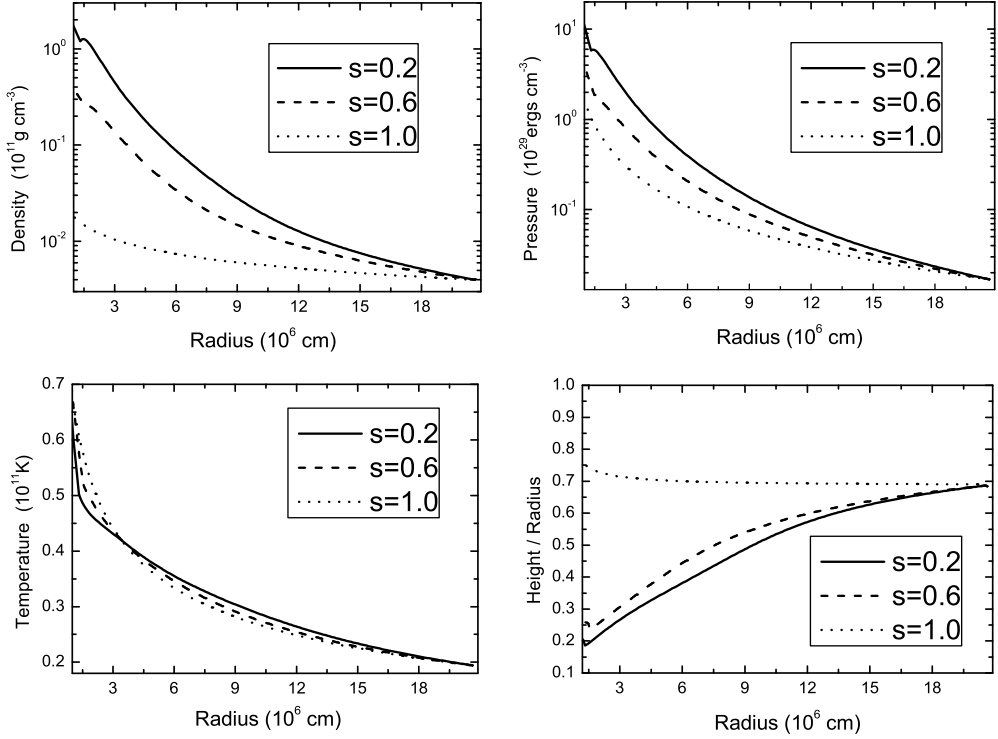}}
\caption{Density, pressure, temperature and height (half-thickness) as functions of
radius for different values of the outflow index $s$=0.2, 0.6, 1,
where we take the accretion rate as 0.2$M_{\odot}$ $\rm s^{-1}$ and
the energy parameter $\varepsilon$=0.8.}\label{fig45}
\end{figure}
Figure \ref{fig45} shows the structure of the disk with different
values of the outflow index $s$=0.2, 0.6 and 1 in model O2. We
choose the accretion rate $\dot{M}=0.2M_{\odot}$ s$^{-1}$ and the
energy parameter $\varepsilon$=0.8. The density and pressure in the
disk decrease with increasing the outflow strength in the entire
disk, and the disk becomes thicker for a stronger outflow. The
change of the temperature is not as obvious as the density and
pressure. Most part of the outer region of the disk will be cooler
for a stronger outflow, but the inner disk region can be hotter in
order to release the heating energy advected from the outer region
of the disk.

\begin{table}
\begin{center}
Self-similar structure and size of the inner disk in various cases
\footnotesize{\begin{tabular}{lcccc} \hline \hline
Cases & Inner disk self-similar structure & Inner disk size & \\
\hline
Case 1: $\varepsilon\simeq$1, no outflow & entropy-conservation structure (\ref{4c02})  & Figure 4.1\\
Case 2: $\varepsilon<$1, no outflow & advection-dominated structure (\ref{4d01})  & Figure 4.3\\
Case 3: outflow model O1 & outflow structure (\ref{4e03})  & Figure 4.4 top panels\\
Case 4: outflow model O2 & outflow structure (\ref{4e03})  & Figure 4.4 bottom panels\\
\hline \hline
\end{tabular}
\caption{We discuss case 1 in \S 4.3.1, case 2 in \S 4.3.2, case 3
and case 4 in \S 4.3.3 with different inner disk structures based on
energy and mass transfer in the entire disk.}\label{tab42}}
\end{center}
\end{table}
We make a brief summary in the end of this section. We propose the
inner disk to satisfy the self-similar structure. Table \ref{tab42}
shows the main results in this section. The inner disk satisfies the
entropy-conservation self-similar structure as in ZD08 for the
energy parameter $\varepsilon\simeq1$, while it becomes an
advection-dominated self-similar region for $\varepsilon<1$. In
outflow model O1, an outflow is produced in the inner disk due to
the prevention effect of the neutron star surface, and model O2
suggests that an outflow exists in the entire disk as in Kohri et
al. (2005). We discuss the size of the inner disk depending on
different structures of the inner disk and outflow and different
values of the viscosity parameter.

\section{Neutrino Annihilation}
\subsection{Calculation Method and Surface Boundary Condition}
Hyperaccreting black hole disks can convert some fraction of the net
accretion energy into the energy of a relativistic outflow or wind
by two general mechanisms: neutrino annihilation and
magnetohydrodynamical effects such as the Blandford-Znajek mechanism
or magnetic instabilities. However, for hyperaccretion disks
surrounding neutron stars, the energy conversion mechanism is mainly
due to the neutrino annihilation for magnetic fields at the neutron
star surface $\leq 10^{15}-10^{16}$ G. We consider the process of
electron capture $\nu_{i}+\bar{\nu}_{i}\rightarrow e^{+}+e^{-}$ as
the most important interaction for energy production. As the
neutron-star disk is denser, hotter with higher pressure, and has a
brighter neutrino luminosity compared with the black-hole disk in
most cases, the neutrino annihilation efficiency of the neutron-star
disk should be higher than that of the black-hole disk. Moreover,
the surface boundary of the neutron star, which carries away
gravitational-binding energy by neutrino emission, also makes the
neutrino annihilation luminosity of the neutron-star disk be higher
than that of the black-hole counterpart.

In this section, we follow the approximate method used by Ruffert et
al. (1997, 1998) and Popham et al. (1999) to calculate the neutrino
annihilation luminosity, i.e., the vertically-integrated disk is
modeled as a grid of cells in two dimensions ($r, \varphi$). The
symbols $\epsilon_{\nu_{i}}^{k}$ and $l_{\nu_{i}}^{k}$ are the mean
neutrino energy and neutrino radiation luminosity with three
different types of neutrino ($i=e,\tau,\mu$) in the cell $k$, and
$d_{k}$ is the distance from the cell $k$ to a certain spatial
point. The neutrino annihilation at any point above the disk is
\begin{eqnarray}
l_{\nu\bar{\nu}}=\sum_{i=e,\mu,\tau}A_{1,i}\sum_{k}\frac{l_{\nu_{i}}^{k}}{d_{k}^{2}}\sum_{k'}\frac{l_{\bar{\nu}_{i}}^{k'}}{d_{k'}^{2}}
(\epsilon_{\nu_{i}}^{k}+\epsilon_{\bar{\nu}_{i}}^{k'})(1-\cos\theta_{kk'})^{2}\nonumber
\\+\sum_{i=e,\mu,\tau}A_{2,i}\sum_{k}\frac{l_{\nu_{i}}^{k}}{d_{k}^{2}}\sum_{k'}\frac{l_{\bar{\nu}_{i}}^{k'}}{d_{k'}^{2}}
\frac{\epsilon_{\nu_{i}}^{k}+\epsilon_{\bar{\nu}_{i}}^{k'}}{\epsilon_{\nu_{i}}^{k}\epsilon_{\bar{\nu}_{i}}^{k'}}(1-\cos\theta_{kk'}),\label{4g01}
\end{eqnarray}
where the values of the neutrino cross section constants $A_{1,i}$
and $A_{2,i}$ can be seen in Popham et al. (1999). The total
neutrino annihilation luminosity above the disk can be integrated as
\begin{equation}
L_{\nu\bar{\nu}}=2\pi\int_{r_{*}}^{\infty}dr\int_{H}^{\infty}l_{\nu\bar{\nu}}rdz.\label{4g02}
\end{equation}

For a neutron-star disk, we should consider both its different
structure compared with the black-hole disk and the boundary
condition of the neutron star surface layer. The neutrino
annihilation luminosity is not only contributed by neutrinos emitted
from the disk but also from the neutron star surface layer. The
luminosity available to be radiated by the boundary layer at neutron
star surface is (Frank et al. 2002)
\begin{equation}
L_{s}=\frac{1}{4}\dot{M}r_{*}^{2}(\Omega_{K}^{2}-\Omega_{*}^{2})-G_{*}r_{*}\simeq
\frac{GM\dot{M}}{4r_{*}}\left(1-\frac{\Omega_{*}}{\Omega_{K}}\right)^{2},\label{4g03}
\end{equation}
where $\Omega_{*}$ is the angular velocity of the neutron star
surface, and $G_{*}\simeq
\frac{1}{2}\dot{M}r_{*}^{2}(\Omega_{K}-\Omega_{*})$ is the viscous
torque acted on the accretion disk. Here we only study the
vertically integrated disk over a half-thickness (height) $H$. As a
result, the luminosity is a function of accretion rate $\dot{M}$ and
neutron star surface angular velocity $\Omega_{*}$. In the case
where the inner disk satisfies the entropy-conservation self-similar
structure, we introduce the efficiency factor $\eta_{s}$ to measure
the energy emitting from the neutron star surface and rewrite
equation (\ref{4g03}) as
\begin{equation}
L_{s}=\eta_{s}\frac{GM\dot{M}}{4}\left(\frac{1}{r}-\frac{1}{r_{out}}\right).\label{4g02}
\end{equation}
If $\Omega_{*}\sim \Omega_{K}$, we have $\eta_{s}\sim 0$ and there
is no emission from the surface layer; or if $\Omega_{*}\sim 0$, we
have $\eta_{s}\sim 1$, which satisfies the Virial condition. We
consider the energy released from the surface layer is mainly
carried away by neutrino emission, and thus the neutrino emission
rate and the temperature at the layer are related by
\begin{equation}
Q_{\nu,s}=\frac{\eta_{s}}{2\pi
r_{*}H_{*}}\frac{GM\dot{M}}{4}\left(\frac{1}{r_{*}}-\frac{1}{r_{out}}\right)\sim
\frac{7}{8}\sigma_{B}T^{4}.\label{4g04}
\end{equation}

In the case where the self-similar inner disk is
advection-dominated, we obtain the gravitational energy released by
neutrino emission as
\begin{equation}
L_{s}=\eta_{s}\frac{GM\dot{M}}{4}\left(\frac{1}{r}-\frac{1}{r_{out}}\right)
+(1-\varepsilon)\frac{GM\dot{M}}{4}\left\{\frac{1}{r_{*}}-\bar{f}_\nu\left[1-\frac{2}{3}
\left(\frac{r_{*}}{\tilde{r}}\right)^{1/2}\right]\right\},\label{4g05}
\end{equation}
where the second term in the right side of equation (\ref{4g05}) is
the heating energy advected from the inner disk to the surface
boundary layer, and we can further write equation (\ref{4g05}) as
\begin{equation}
L_{s}=(\eta_{s}+\eta_{s,ADAF})\frac{GM\dot{M}}{4}\left(\frac{1}{r}-\frac{1}{r_{out}}\right),\label{4g06}
\end{equation}
where we take the equivalent factor $\eta_{s,ADAF}$ to measure the
energy advected from the advection-dominated inner disk to the
surface boundary. Table \ref{tab43} shows examples of
$\eta_{s,ADAF}$ with different inner disk structures.
\begin{table}
\begin{center}
Equivalent factor $\eta_{s,ADAF}$ with different inner disk
structures
\begin{tabular}{rccccrc}
\hline \hline $\eta_{s,ADAF}$ & 0.03 & 0.1 & 0.3 & 0.5 & 1.0 & 3.0\\
&($M_{\odot}$ s$^{-1}$)\\
\hline
$\varepsilon$=0.9 & 9.07e-2 & 5.25e-2 & 4.22e-2 & 5.56e-2 & 7.75e-2 & 9.56e-2\\
$\varepsilon$=0.5 & 0.419 & 0.196 & 0.159 & 0.149 & 0.180 & 0.446\\
$\varepsilon$=0.2 & 0.675 & 0.496 & 0.219 & 0.178 & 0.218 & 0.382\\
\hline
O1: $s$=0.2 & 0.228 & 0.113 & 7.84e-2 & 8.72e-2 & 0.107 & 0.202\\
$s$=0.6 & -- & 0.215 & 0.115 & 0.113 & 0.119 & 0.209\\
$s$=1.0 & -- & 0.315 & 0.199 & 0.168 & 0.203 & 0.336\\
\hline
O2: $s$=0.2 & 0.238 & 0.104 & 7.95e-3 & -- & -- & --\\
$s$=0.6 & -- & 0.428 & 5.79e-4 & -- & -- & --\\
$s$=1.0 &-- & -- & 0.639 & 4.36e-4 & -- & --\\
\hline \hline
\end{tabular}
\caption{Equations (\ref{4g05}) and (\ref{4g06}) give the value of
$\eta_{s,ADAF}$. We choose cases of the advection-dominated inner
disk with $\varepsilon$=0.9, 0.5, 0.2, and models O1 and O2 with
$s$=0.2, 0.6 1.0 to calculate $\eta_{s,ADAF}$.}\label{tab43}
\end{center}
\end{table}

When an outflow exists in the disk, the luminosity at the boundary
surface is dimmer since the accretion rate near the neutron star
surface is lower due to the outflow. We can modify equation
(\ref{4g05}) by changing the accretion rate $\dot{M}$ to be
$\dot{M}_{0}(r_{*}/\tilde{r})^{s}$ for model O1 in \S3.3 and
$\dot{M}_{0}(r_{*}/r_{out})^{s}$ for model O2.

\subsection{Results of Annihilation Luminosity}
\begin{figure}
\centering\resizebox{0.8\textwidth}{!}
{\includegraphics{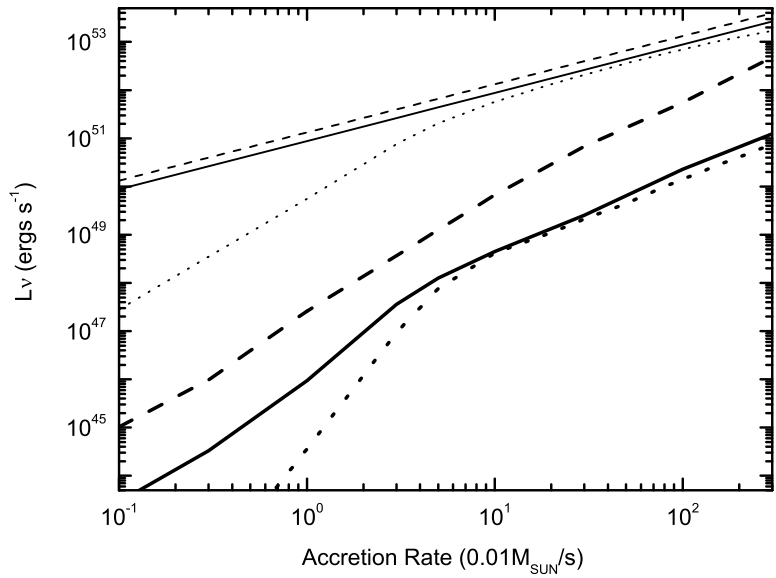}}\caption{Neutrino
annihilation luminosity $L_{\nu\bar{\nu}}$ ({\em thick lines}) and
total neutrino emission luminosity $L_{\nu}$ ({\em thin lines}) as
functions of accretion rate. The solid lines correspond to the
neutron-star disk with the entropy-conservation inner disk and the
boundary layer emission efficiency $\eta_{s}=0$, the dashed lines to
the neutron-star disk with the boundary condition $\eta_{s}=0.5$,
and the dotted lines to the black-hole disk.}\label{fig46}
\end{figure}

\begin{figure}
\resizebox{\hsize}{!} {\includegraphics{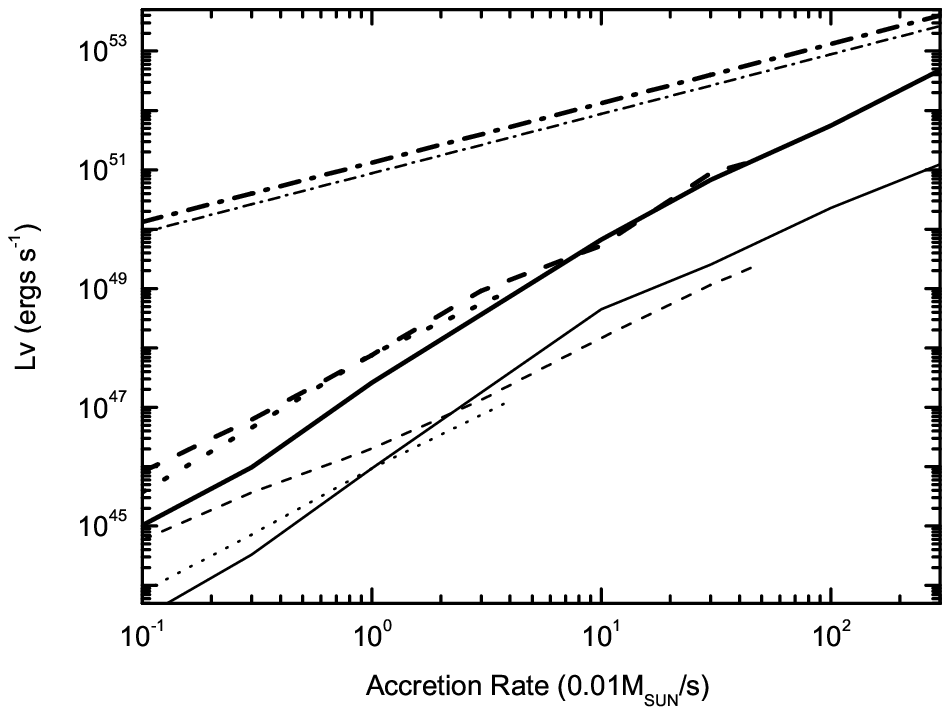}
\includegraphics{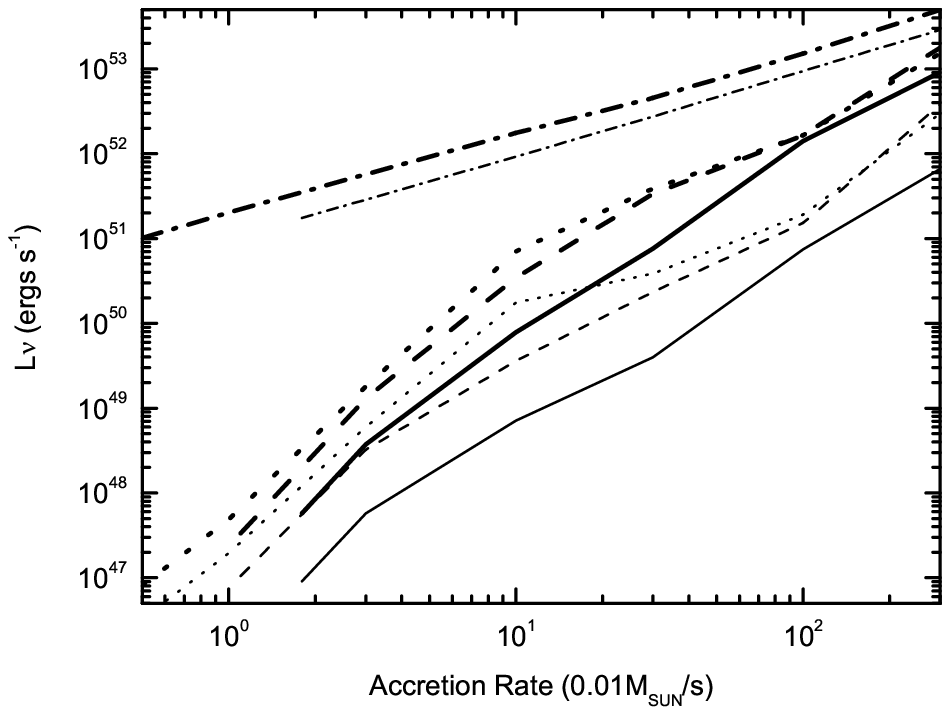}}
\caption{Neutrino annihilation luminosity $L_{\nu\bar{\nu}}$ and the
total neutrino luminosity $L_{\nu}$ with different values of the
viscosity parameter $\alpha$ and energy parameter $\varepsilon$. The
thin lines correspond to $\eta_{s}=0$ and thick lines to
$\eta_{s}=0.5$. (a) {\em Left panel}: $L_{\nu\bar{\nu}}$ with
$\alpha$=0.1 ({\em solid line}), 0.01 ({\em dashed line}), 0.001
({\em dotted line}) and total luminosity $L_{\nu}$ ({\em dash-dotted
line}), where we take $\varepsilon=1$ and the entropy-conservation
inner disk structure. (b) {\em Right panel}: $L_{\nu\bar{\nu}}$ of
the neutron-star disk with advection-dominated inner disk and
$\varepsilon$=0.9 ({\em solid line}), 0.5 ({\em dashed line}), 0.1
({\em dotted line}) and the total luminosity $L_{\nu}$ ({\em
dash-dotted line}).}\label{fig47}
\end{figure}

\begin{figure}
\resizebox{\hsize}{!} {\includegraphics{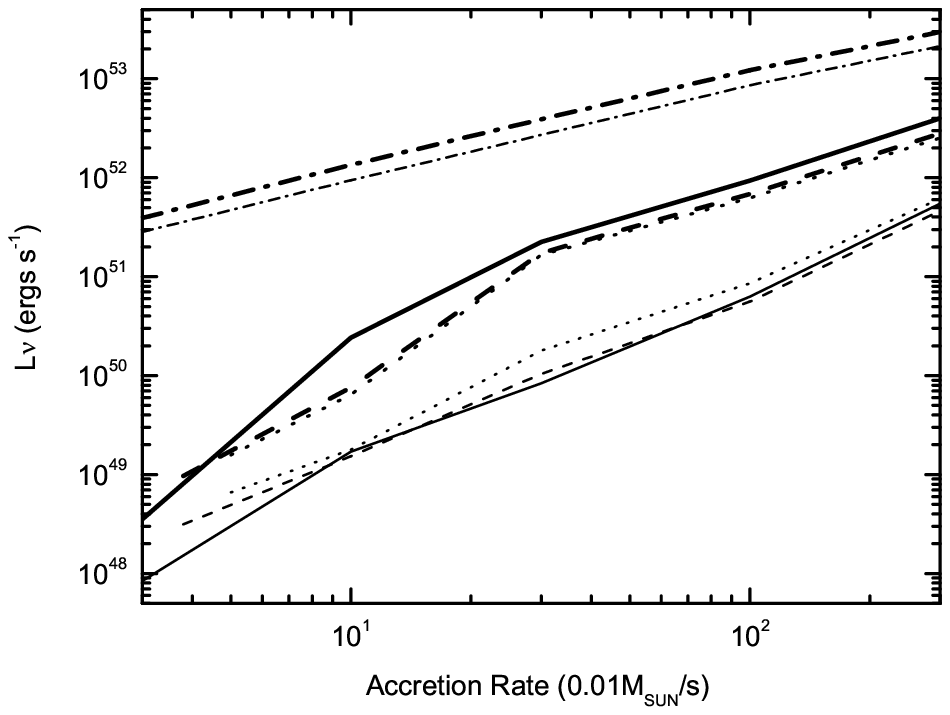}
\includegraphics{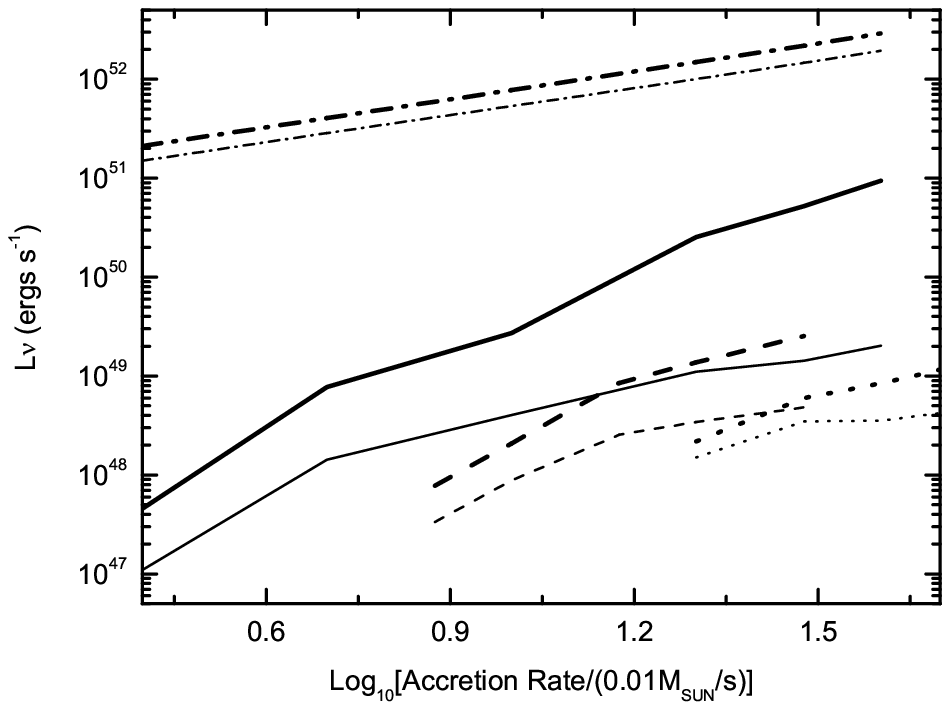}}
\caption{Neutrino annihilation luminosity $L_{\nu\bar{\nu}}$ of the
neutron-star disk with outflow index $s$=0.2 ({\em solid line}), 0.6
({\em dashed line}), 1.0 ({\em dotted line}) and the maximum value
of total luminosity $L_{\nu}$ in the case of $s$=0.2 ({\em
dash-dotted line}), where we take $\varepsilon$=0.8. Left panel
shows the results of model O1 and right panel to model O2. The thin
lines correspond to the luminosity to $\eta_{s}=0$ and thick lines
to $\eta_{s}=0.5$.}\label{fig48}
\end{figure}

\begin{figure}
\resizebox{\hsize}{!} {\includegraphics{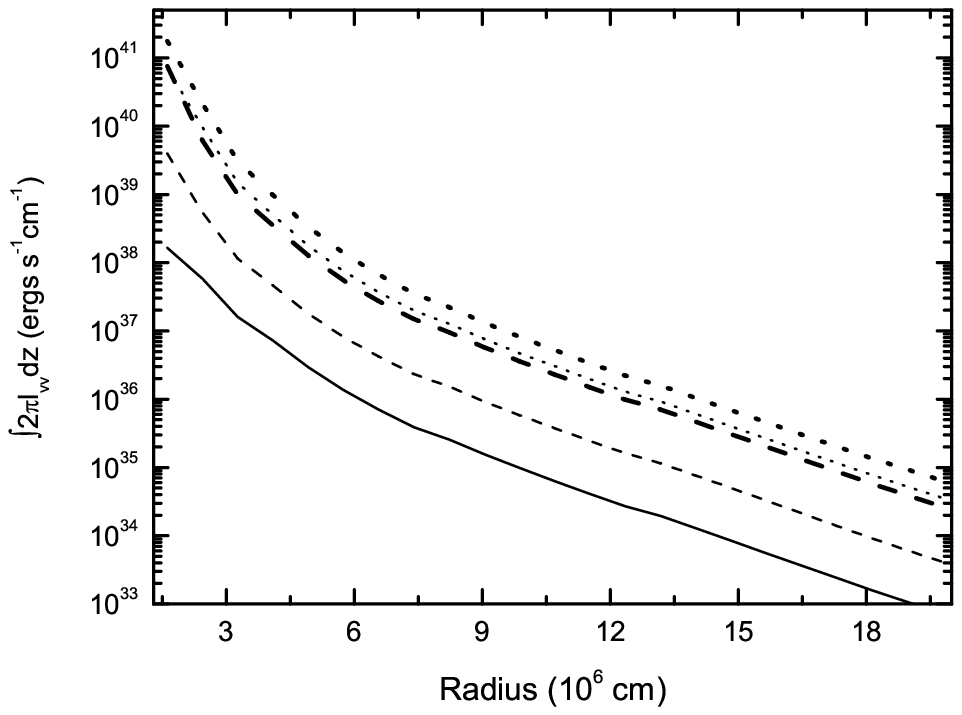}
\includegraphics{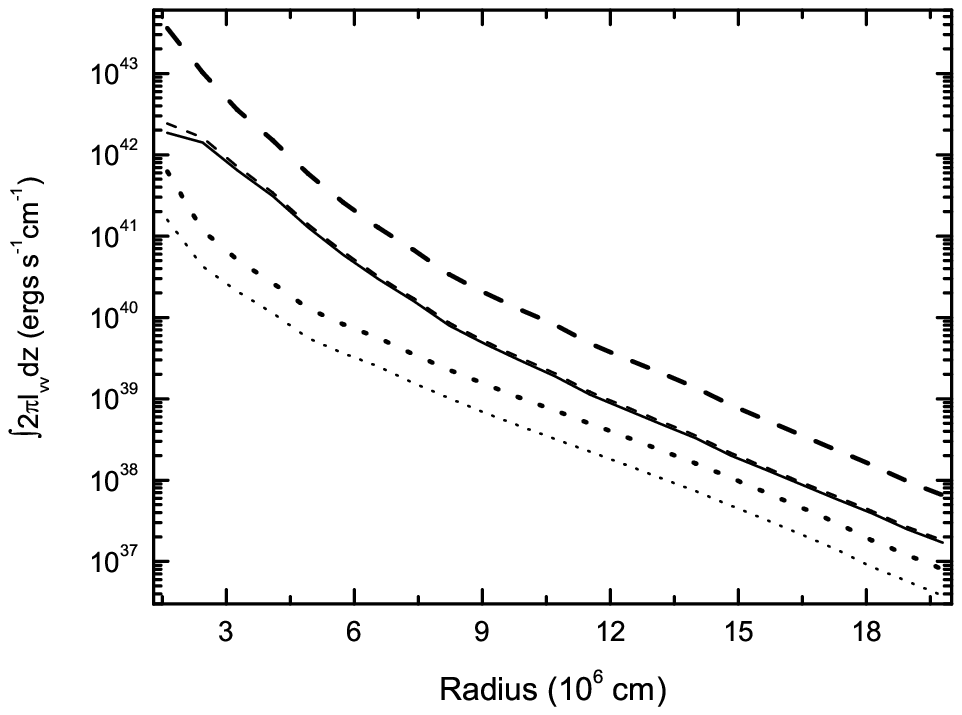}}
\caption{Neutrino annihilation luminosity per cm distribution as a
function of radius for $\dot{M}$=0.01$M_{\odot}$ s$^{-1}$ ({\em left
panel}) and $\dot{M}$=0.1$M_{\odot}$ s$^{-1}$ ({\em right panel}).
(a) {\em Left panel}: annihilation luminosity per cm for a
black-hole disk ({\em thin solid line}), neutron-star disks with
entropy-conservation inner disk and $\eta_{s}$=0 ({\em thin dashed
line}), and $\eta_{s}$=0.5 ({\em thick dashed line}), and
neutron-star disks with advection-dominated inner disk
$\varepsilon$=0.2 and $\eta_{s}$=0 ({\em thin dotted line}), and
$\varepsilon$=0.2 and $\eta_{s}$=0.5 ({\em thick dotted line}). (b)
{\em Right panel}: annihilation luminosity per cm for a neutron-star
disk with outflow (model O2) $\varepsilon$=1, $s$=0.6, $\eta_{s}$=0
({\em thin dotted line}), $\varepsilon$=1, $s$=0.6, $\eta_{s}$=0.5
({\em thick dotted line}). The thin solid line, thin dashed line and
thick dashed line are the same as the left panel.}\label{fig49}
\end{figure}

We calculate the neutrino annihilation luminosity $L_{\nu\bar{\nu}}$
and the total neutrino luminosity $L_{\nu}$ emitted from the disk
and neutron star surface. The results of $L_{\nu\bar{\nu}}$ and
$L_{\nu}$ depend on the value of the viscosity parameter $\alpha$,
the detailed structure of the inner disk, the strength of the
outflow, as well as the neutron star surface boundary condition. We
discuss the effects of these various factors in Figure \ref{fig46}
to Figure \ref{fig49}.

Figure \ref{fig46} shows the total neutrino annihilation luminosity
$L_{\nu\bar{\nu},NS}$ and the emission luminosity $L_{\nu,NS}$ of
the neutron-star disk with different surface boundary layer
conditions ($\eta_{s}$=0 and 0.5), and we compare them with the
black-hole disk. In this figure we take the neutron-star inner disk
structure to satisfy the entropy-conservation self-similar structure
(\ref{4c02}). The total luminosity and annihilation luminosity of
the neutron-star disk are brighter than those of a black-hole disk
with the same mass and accretion rate. If we study the neutrino
annihilation from the entire disk without surface boundary emission
($\eta_{s}$=0), the difference between $L_{\nu\bar{\nu},NS}$ and
$L_{\nu\bar{\nu},BH}$ is more significant for a low accretion rate
than for a high accretion rate. We have mentioned the reason in ZD08
that a larger inner disk for a neutron-star disk with a low
accretion rate makes the neutrino luminosity much brighter than its
black hole counterpart, and the annihilation rate also becomes
higher for the neutron-star disk. For a high accretion rate
($>$0.5$M_{\odot}$ s$^{-1}$), the effect of neutrino opacity on
$L_{\nu,BH}$ and $L_{\nu\bar{\nu},BH}$ also be less than that on
$L_{\nu,NS}$ and $L_{\nu\bar{\nu},NS}$. On the other hand, neutrino
emission from the neutron star surface boundary layer
($\eta_{s}$=0.5) makes the annihilation luminosity be more than one
order of magnitude higher than that without boundary emission
($\eta_{s}$=0). $L_{\nu\bar{\nu}}$ reaches 10$^{50}$ ergs s$^{-1}$
when $\dot{M}\sim 1M_{\odot}$ s$^{-1}$ for a black-hole disk or
neutron-star disk with $\eta_{s}$=0, but only needs $\dot{M}\sim
0.1M_{\odot}$ s$^{-1}$ for a neutron-star disk with the boundary
condition $\eta_{s}$=0.5. Therefore, a lower-spin neutron star with
hyperaccreting disk around it could have an obviously higher
annihilation efficiency than that of a higher-spin neutron star. We
will discuss the neutrino annihilation luminosity of a neutron-star
disk in more details in \S 4.5.

In Figure \ref{fig47}, we show the total neutrino annihilation
luminosity of a neutron star disk with different values of the
viscosity parameter $\alpha$=0.1, 0.01, 0.001 and the energy
parameter $\varepsilon$=0.9, 0.5, 0.1. The disk with a moderate
viscosity parameter ($\alpha$=0.01) has the highest annihilation
efficiency and luminosity for a low accretion rate, and the
annihilation luminosity from a high-$\alpha$ disk ($\alpha$=0.1)
becomes the brightest for an accretion rate $\dot{M}\geq 0.05
M_{\odot}$ s$^{-1}$. As discussed in Figure \ref{fig42}, a
low-$\alpha$ disk has a brighter neutrino luminosity $Q_{\nu}$, but
it is thinner than a high-$\alpha$ disk. These two competitive
factors lead to the annihilation results shown in Figure
\ref{fig47}a. Figure \ref{fig47}b shows that the annihilation
efficiency increases with increasing $\varepsilon$. This is because
the disk with lower $\varepsilon$ means more heating energy in the
disk to be advected onto the neutron star surface and increases the
neutrino luminosity of the surface layer, and the value of
$\eta_{s,ADAF}$, which plays an important role in increasing the
annihilation efficiency of the entire disk.

Figure \ref{fig48} shows the total neutrino annihilation luminosity
$L_{\nu\bar{\nu}}$ of the disk with an outflow. We consider the
results of model O1 in which an outflow only exists in the inner
disk due to the neutron star surface and of model O2 in which an
outflow exists in the entire disk. The difference in annihilation
luminosity with different values of the outflow index $s$ but the
same surface boundary condition is not obvious in model O1. However,
$L_{\nu\bar{\nu}}$ becomes much dimmer for a high outflow index $s$
in model O2, which means a strong outflow in the entire disk and
decreases the neutrino annihilation efficiency significantly.

Furthermore, we study the spatial distribution of the neutrino
annihilation luminosity. Figure \ref{fig49} illustrates the
integrated annihilation luminosity per cm distribution
\begin{equation}
2\pi r\int_{H}^{\infty}l_{\nu\bar{\nu}}dz\label{4h01}
\end{equation}
for two accretion rates $\dot{M}$=0.01$M_{\odot}$ s$^{-1}$ and
0.1$M_{\odot}$ s$^{-1}$ with different physical structures of the
disk. We find that the integrated annihilation luminosity drops
dramatically along the disk radius, and a majority of the
annihilation energy is ejected from the cylindric region above the
disk with $r<3\times10^{6}$cm. The difference of the integrated
luminosity per cm between the black-hole disk and neutron-star disk
is more significant for a low accretion rate when the inner
entropy-conservation disk is sufficiently large. The luminosity of
the neutron-star disk with an advection inner disk can be about four
orders of magnitude higher than its black-hole counterpart to
produce a jet with higher energy, while an outflow makes the
annihilation luminosity above the neutron-star disk be lower than
that of the black-hole disk.

Compared with the black-hole disk, the neutron-star disk could
produce a brighter neutrino flux and more powerful annihilation
luminosity. However, the mass-loss rate driven by neutrino-on-baryon
absorption reactions (Qian \& Woosley 1996) along the polar axis of
the neutron-star disk is also significantly higher than its
black-hole counterpart. This raises the problem of whether
simultaneously more powerful annihilation luminosity and higher mass
loss rate could work together to produce a relativistic jet required
for the GRB phenomena. For example, a heavily baryon-loaded wind
from a new-born neutron star within 100 ms prevents any production
of a relativistic jet (Dessart et al. 2009). However, a weaker wind
above the polar region and spherical asymmetry of the outflow at
late times could make production of a relativistic jet above the
stellar pole become possible. We will discuss this issue in more
details in \S 4.5.3.

\section{Discussions}

\subsection{Size of the Inner Disk}
In \S 4.3 we study various self-similar structures of the inner
disk. In this section we first want to discuss the
entropy-conservation structure. The inner disk can be determined by
the self-similar structure (\ref{4c02}) and the energy equation
(\ref{4a09}). However, the inner disk with the entropy-conservation
structure cannot satisfy the integrated angular momentum equation
(\ref{4a04}). There are two explanations. First, we can consider the
term $d\dot{J}_{ext}/dr\neq 0$ in equation (\ref{4a03}) for a
steady-state disk since the angular momentum redistributes in the
inner disk before the entire disk becomes steady-state. As mentioned
in ZD08, because the neutron star surface prevents heating energy
from being further advected inward, the inner disk is formed to
balance the heating and cooling energy in the entire disk. As a
result, energy and angular momentum could redistribute in the inner
disk. Second, besides this consideration on angular momentum
redistribution, we can also discuss another type of inner disk with
the entropy-conservation structure discussed by Medvedev \& Narayan
(2000), which still satisfies the angular momentum equation
(\ref{4a04}) with $v_{r}\ll v_{K}$ and $\Omega\simeq $constant. The
angular momentum transfer and heating energy generation due to
viscosity can be neglected in the inner disk, as $\Omega$ is
approximately equal to an constant, and the heating energy in the
inner disk is merely the energy advected from the outer disk, i.e.,
$Q^{+}=Q^{+}_{\rm adv}$ in the inner disk. As a result, the global
inner-disk energy conservation equation (\ref{4a09}) should be
modified as
\begin{equation}
\int_{r_{*}}^{\tilde{r}}Q_{\nu}^{-}2\pi rdr= \frac{3GM\dot{M}}{4}
\left\{{\frac{1}{3
\tilde{r}}-\frac{\bar{f}_\nu}{\tilde{r}}\left[1-\frac{2}{3}
\left(\frac{r_{*}}{\tilde{r}}\right)^{1/2}\right]}\right\},\label{4h01}
\end{equation}
for $\Omega\simeq {\rm constant}=\Omega(\tilde{r})$ in the inner
disk. Compared with equation (\ref{4a09}), equation (\ref{4h01})
requires a smaller value of $\tilde{r}$ with the same neutron star
mass and accretion rate.
\begin{figure}
\centering\resizebox{0.7\textwidth}{!}{\includegraphics{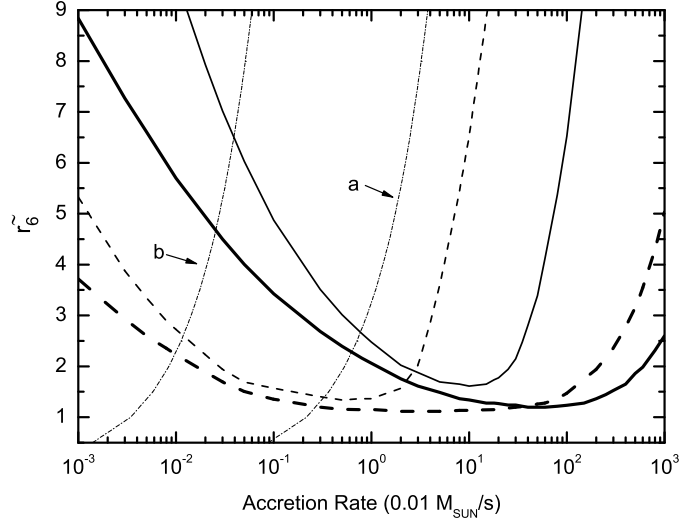}}\caption{The
radius $\tilde{r}$ (in units of $10^{6}$ cm) between the inner and
outer disks with the viscosity parameter $\alpha$=0.1 ({\em solid
line}) and 0.01 ({\em dashed line}). The thick lines show the
results based on the consideration $\Omega\simeq$ const in the inner
disk in \S 5, and the thin lines based on the angular velocity
distribution $\Omega\propto r^{-3/2}$ as in \S 3.}\label{fig410}
\end{figure}
Figure \ref{fig410} shows the value of $\tilde{r}$ for $\alpha$=0.1
and 0.01 with $\Omega\simeq$constant and $\Omega\propto r^{-3/2}$.
The constant angular velocity in the inner disk decreases the inner
disk size compared to the case of $\Omega\propto r^{-3/2}$. However,
in order to get a unified scenario of the entire disk in various
cases, we adopt $\Omega\propto r^{-3/2}$ for all the self-similar
structures in \S 4.3.

\subsection{Annihilation Results and Disk Geometry}

Several previous studies have been performed to calculate the
neutrino annihilation efficiency above the disk around a black hole
with the effects of disk geometry, gravitational bending, rotation
of central black holes and so on (e.g., Ruffert et al. 1997, 1998;
Popham et al. 1999; Asano \& Fukuyama 2000, 2001; Miller et al.
2003; Birkl et al. 2007). The simulations based on general
relativity show the effects of general relativity such as the Kerr
black holes and bending neutrino geodesics in spacetime increase the
total annihilation rate  by a factor of a few. Also, compared with
the spherical and torus neutrinosphere, the disk neutrinosphere with
the same temperature and neutrino luminosity distribution usually
have the highest annihilation efficiency (Birkl et al. 2007). In
this chapter, we still adopt the calculation approach on
annihilation based on Ruffert et al. (1997, 1998) and Popham et al.
(1999), and consider the vertically-integrated Newtonian disk. The
most important results of annihilation calculation in our work is
that the neutron-star disk produce more energetic annihilation
luminosity compared to the black-hole counterpart. Moreover, we
consider neutrinos emitted from the stellar surface region, and the
neutrino emission concentrated in this surface region plays a
significant role in increasing the annihilation luminosity to
produce relativistic ejecta formed by $e^{+}e^{-}$ plasma. On the
other hand, we should note that the effect of surface boundary
condition emission could be reduced if the emission region becomes
larger than we consider in this chapter due to outflow or the other
cooling mechanisms rather than neutrino cooling at the surface. Some
other works focusing on neutrino annihilation in supernovae
discussed neutrino emission from the entire spherical neutron star
surface (Cooperstein et al. 1986; Goodman et al. 1987; Salmonson \&
Wilson 1999). Therefore, a further work should be done to study the
effects of boundary emission and cooling based on more elaborate
considerations on cooling mechanisms and energy transfer at the
boundary around neutron-star disk. However, as the neutron-star disk
and the surface emission increase the annihilation luminosity more
significantly than the general relativity effects, we conclude our
main results would maintain for more elaborate simulations based on
advanced calculations on neutrino annihilation above the disk.

\subsection{Application to GRB Phenomena}

Energy can be deposited in the polar region of black-hole and
neutron-star disks by neutrino annihilation and MHD processes. We
focus on the annihilation process in this chapter. In the black-hole
case, the environment along the polar axis (i.e., the rotation axis
of the disk around neutron star) can be baryon-free. For example,
GRB can commence after the initial collapse for $\sim 15$ s in the
collapsar scenario, when the accretion process in the polar region
becomes sufficiently weak to produce a relatively clean environment
with the mass density $\leq10^{6}$ g cm$^{-3}$ (MacFadyen \& Woosley
1999). In the neutron-star case, however, the enormous neutrino
luminosity of a neutron star would drive appreciable mass-loss from
its surface during the first 20 s of its life (Qian \& Woosley
1996). The wind material will feed the polar region, where a large
amount of annihilation energy is deposited. Dessart et al. (2009)
showed that a newly formed neutron star from the neutron star binary
merger will develop a powerful neutrino-driven wind in the polar
funnel in a few milliseconds after its formation. The mass-loss
associated with the neutrino-driven wind is on the order of
$10^{-3}M_{\odot}$ s$^{-1}$, preventing energy outflow from being
accelerated to relativistic speed and producing a GRB. Their
numerical simulations stops at $t\sim100$ ms, while they considered
that the neutron star will collapse to a black hole quickly.
However, a stable neutron star with much longer lifetime $\gg100$ ms
has also been proposed as the GRB central engine. A rapidly rotating
neutron star formed by the neutron star binary merger (Gao \& Fan
2006) or the merger of a white dwarf binary (King et al. 2001) could
produce extended emission (Metzger et al. 2008b) or X-ray flares
(Dai et al. 2006) about 10-100 s afetr the neutron star birth, and
last for tens of seconds. In the collapsar scenario, a newly formed
neutron star or a magnetar could also form after the initial
collapse or associated supernova explosion, and the lifetime of the
neutron star before collapsing to a black hole can reach as long as
several months to yrs (Vietri \& Stella 1998). Therefore, many GRB
models lead to a longtime ($\gg100$ ms) neutron star, which can
produce other phenomena accompanying the GRB prompt emission. On the
other hand, the accretion timescale can last for longer than 10s in
the collapse process of a massive star, or for several seconds after
the merger of a compact star binary with the disk viscosity
parameter $\alpha<0.1$ (Narayan et al. 2001). Thus we can discuss
the steady-state scenario of a neutron star with lifetime $\gg100$
ms surrounded by a hyperaccreting disk.

\begin{figure}
\centering\resizebox{0.8\textwidth}{!}{\includegraphics{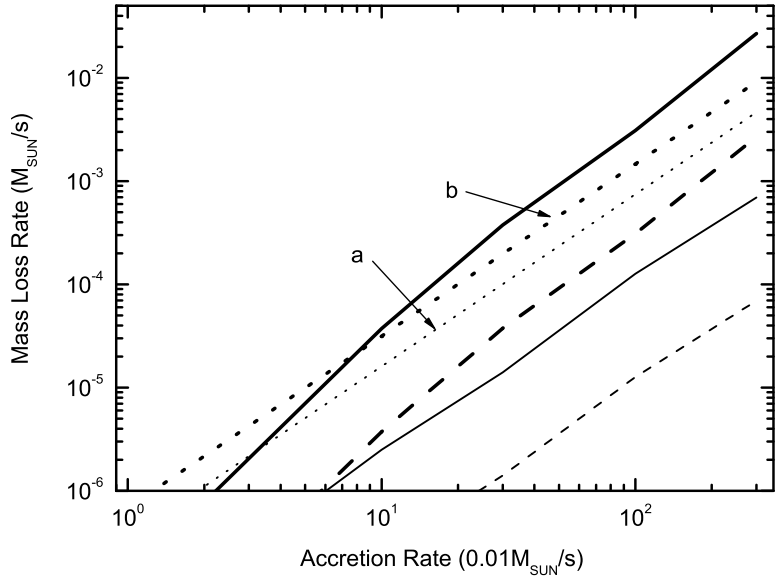}}
\caption{Upper limit of mass-loss rate due to neutrino-driven wind from the neutron
star surface for the outflow being accelerated to $\Gamma=10$
(moderately relativistic) with the surface emission boundary
condition $\eta_{s}=0$ (\textit{thin solid line}) or $\eta_{s}=0.5$
(\textit{thin dashed line}); or $\Gamma=100$ (ultrarelativistic)
with the surface emission boundary condition $\eta_{s}=0$
(\textit{thick solid line}) or $\eta_{s}=0.5$ (\textit{thick dashed
line}). The dotted lines ``a" and ``b" correspond to the strength of
a thermally neutrino-driven wind from the stellar surface for the
boundary condition $\eta_{s}=0$ and $\eta_{s}=0.5$ respectively.}\label{fig411}
\end{figure}

For a longtime neutron star, the strengthen of a neutrino-driven
winds above the stellar polar region would drop quickly at late
times, as the total mass-loss rate $\dot{M}\sim t^{-5/3}$, and the
neutrino-driven wind become weaker above the polar region than that
from the low latitudes and midplane region of the neutron star. It
is difficult to calculate the spatial distribution of Lorentz factor
of the outflow material above the polar region precisely in our
present work, because we only use the approximate disk geometry to
calculate the steady-state spatial distribution of the neutrino
annihilation efficiency. We do not simulate the dynamical evolution
of the neutrino-driven wind. However, we can estimate the speed of
outflow material using semi-analytic methods as follows. The mass
loss rate for a thermal neutrino-driven wind can be approximately
given by (Qian \& Woosley 1996)
\begin{equation}
\dot{M}_{\rm
wind}\approx1.14\times10^{-10}C^{5/3}L_{\bar{\nu}_{e},51}\epsilon_{\nu_{e},
MeV}^{10/3}(r/10^{6}\textrm{cm})^{5/3}(M/1.4M_{\odot})^{-2}M_{\odot}\,\textrm{s}^{-1},\label{4wind01}
\end{equation}
where $10^{51}$$L_{\bar{\nu}_{e},51}$ergs s$^{-1}$ is the luminosity
of the $\bar{\nu}_{e}$ emission,
$\epsilon_{\nu_{e}}$=1MeV$\epsilon_{\nu_{e}, MeV}$ is the mean
neutrino energy of the neutron star surface, and
$C\approx1+0.733(r/10^{6}\textrm{cm})^{-1}(M/1.4M_{\odot})$. Thus
the typical steady state spherical mass-loss rate due to thermal
neutrino absorption reactions for a neutron star with 10km radius is
on the order of a few $10^{-6}-10^{-5}M_{\odot}$ s$^{-1}$, depending
on the initial configuration of the material above the neutron star
surface (i.e., Qian \& Woosley 1996, Tab. 1). The mass-loss rate
depends on the neutrino luminosity and temperature above the neutron
star surface sensitively. When the neutron star with weak magnetic
field $<10^{15}$ G is surrounded by a hyperaccreting disk, the
neutron star surface should have a higher temperature near the star
midplane than its polar region, as mentioned in the last section \S
4. Thus the neutron star should produce a stronger wind above the
midplane region than above its poles. As a result, the mass-loss
rate via neutrino absorption above the star polar region is
estimated as
\begin{equation}
\dot{M}_{\rm polar}=\dot{M}_{\rm
wind}\left(\frac{\Delta\Omega}{2\pi}\right)f_{asy},\label{4wind03}
\end{equation}
where $\Delta\Omega=\int \textrm{cos}\varphi d\varphi d\phi$ is the
solid angle of the polar funnel, and $f_{asy}<1$ is used to measure
the degree of spherical asymmetry. Moreover, the bulk Lorentz factor
of the outflows from the neutron star surface is
\begin{equation}
\Gamma\geq\frac{L_{\nu\bar{\nu}}f_{k}}{\dot{M}_{\rm
polar}c^{2}}=\frac{L_{\nu\bar{\nu}}}{\dot{M}_{\rm
wind}c^{2}}\left(\frac{2\pi f_{k}}{\Delta\Omega
f_{asy}}\right),\label{4wind04}
\end{equation}
where $f_{k}$ is the fraction of deposited annihilation energy which
provides kinetic energy of the neutron star wind above the polar
region. We can estimate whether the outflow material from the polar
region of the neutron star surface can be accelerated to a
relativistic speed using equation (\ref{4wind04}). Here we take the
bulk Lorentz factor $1<\Gamma<10$ as a mildly relativistic speed,
$10\leq\Gamma<100$ as a moderately relativistic speed and
$\Gamma>100$ as an ultrarelativistic speed. If we take the bulk
Lorentz factor $\Gamma$ as a parameter, the solid angle of the polar
region $\Delta\Omega/2\pi\sim 0.1$, and $f_{k}\sim f_{asy}$, then we
can estimate the upper limit of total thermal mass-loss rate from
the neutron star surface with particular boundary layer conditions,
and compare such limits with the actual mass-loss rate of thermal
neutrino-driven winds calculated by equation (\ref{4wind01}). In
Figure \ref{fig411} we explore the possibility of producing a
relativistic jet. We show the maximum allowed strength of total
mass-loss for the wind material above the polar region being
accelerated to $\Gamma=10$ and $\Gamma=100$ with the chosen boundary
layer condition $\eta_{s}=0$ and $\eta_{s}=0.5$ as in \S 4.4. The
neutrino emission from the neutron star surface layer increases the
neutrino annihilation luminosity and efficiency, while it increases
the stellar surface mass-loss simultaneously.
From Figure \ref{fig411}, we find that moderately relativistic
outflows above the neutron star polar region are possible under some
cases, e.g., the wind material can be accelerated to $10<\Gamma<100$
above the polar region for the disk accretion rate
$\dot{M}\geq0.08M_{\odot}$ s$^{-1}$ with $\eta_{s}=0.5$. This result
can still be kept for other values of $f_{k}/f_{asy}$ around unit.
However, annihilation process can never produce any winds with bulk
Lorentz factor $\Gamma\geq100$, as the heavily mass-loaded
neutron-star wind precludes ultrarelativistic speed even for very
high annihilation efficiency with sufficiently bright neutrino
emission from the innermost disk radius. Zhang et al. (2003, 2004)
found that relativistic jets formed in the accreting black hole
systems can be collimated by their passage through the stellar
envelope; moderately and even mildly relativistic jets can be partly
accelerated to an ultrarelativistic speed after they break out in
the massive star, because the jets' internal energy can be converted
to kinetic energy after jet breakout. Such jet-stellar-envelope
interactions also happen around the hyperaccreting neutron star
systems, although the neutron-star disk systems can be surrounded by
a cavity $\sim 10^{9}$ cm inside the progenitor stars with the
stellar radius of several $10^{10}$ cm (e.g., Bucciantini et al.
2008, 2009). In the compact star binary merger scenario, on the
other hand, a moderately relativistic jet is possible to produce a
short-duration GRB.

As a result, the neutron-star disk can produce a sufficiently larger
annihilation luminosity than its black-hole counterpart. However, a
neutrino-driven outflow from the newly formed neutron star at early
times ($\sim100$ ms) is so heavily mass-loaded that it can in no way
be accelerated to relativistic speed. For a longtime neutron star,
however, mass-loss becomes weaker above the stellar pole than above
the low latitudes and midplane. Thus a moderately relativistic jet
can be produced in a hyperaccreting neutron-star system with
sufficiently high disk accretion rate and bright boundary emission
(e.g., the accretion rate $\dot{M}\geq0.08M_{\odot}$ s$^{-1}$ for
$\eta_{s}=0.5$). Also, the jet can be collimated by
jet-stellar-envelope interactions, and partly accelerated to an
ultrarelativistic speed after jet breakout if the neutron star forms
through stellar collapse. Therefore, some hyperaccreting
neutron-star system can produce GRBs, which are more energetic than
those from the black-hole systems. We know that some long-duration
GRBs can reach a peak luminosity of $\sim 10^{52}$ ergs s$^{-1}$
(e.g. GRB 990123, Kulkarni et al. 1999), which requires a very high
accretion rate $\sim 10 M_{\odot}$ s$^{-1}$ as well as a much more
massive disk ($\geq 10 M_{\odot}$) around a black hole compared to
the typical disk or torus mass $0.01-1 M_{\odot}$ if the energy is
provided by neutrino annihilation above the black-hole disk.
However, if we consider the neutron-star disk with the surface
boundary emission (e.g. $\eta_{s}\sim 0.5$), we only need an
accretion rate $\sim 1 M_{\odot}$ s$^{-1}$ onto the neutron star,
which is one order of magnitude less than that of the black-hole
disk, and the disk mass can be $\sim 1 M_{\odot}$ for a long burst
with the peak luminosity $10^{52}$ ergs s$^{-1}$ in a time of about
1s. However, other reasons or mechanisms such as the jet effect to
reduce the total burst energy or the magnetic mechanism rather than
neutrino annihilation to provide the GRB energy have been introduced
into the GRB central engine models. Therefore, it is necessary to
search new important observational evidence to show the existence of
a central neutron star rather than black hole surrounded by a
hyperaccreting disk. X-ray flares after GRBs may be a piece of
evidence, which shows that an activity of the central neutron star
after the burst may be due to magnetic instability and reconnection
effects in differentially-rotating pulsars (Dai et al. 2006).
However, more studies should be done to compare other models of
X-ray flares (King et al. 2005; Proga \& Zhang 2006; Perna et al.
2005; Lee \& Ramires-Ruiz 2007, Lazzati et al. 2008) with the
differential-rotating pulsar model and show more effects of such a
magnetized pulsar. For example, the spin-down power of a magnetar
probably explains the peculiar optical to X-ray integrated
luminosity of GRB 060218 (Soderberg et al. 2006).

Furthermore, we propose that other GRB-like events may be produced
by hyperaccreting neutron star systems, if winds fail to reach a
proper relativistic speed. As mentioned by MacFadyen et al. (2001),
a mildly or moderately relativistic jet may lead to X-ray flashes,
which are less energetic than normal GRBs. This will also happen in
the neutron-star case. Another possible result is that a heavily
mass-loaded wind with nonrelativistic speed could produce a bright
SN-like optical transient event (Kohri et al. 2005; Metzger et al.
2008a).

\subsection{GRB-SN Association}
Besides GRBa and GRB-like phenomena, we think a nonrelativistic or
mildly relativistic outflow from a hyperaccreting disk and neutron
star surface may feed a supernova explosion associated with a GRB.
The discovery of connection between some GRBs and supernovae has
inspired studies of the origin of GRB-SN association (e.g., Iwamota
et al. 1998; MacFadyen \& Woosley 1999; Zhang et al. 2004; Nagataki
et al. 2006, 2007; Mazzali et al. 2006). As discussed in Kohri et
al. (2005), an outflow from the hyperaccreting disk is possible to
provide a successful supernova with both prompt explosion or delayed
explosion. On the other hand, neutrino annihilation above the disk
is considered as one of the possible mechanisms to produce GRBs.
Thus, it is reasonable to propose a general scenario for the origin
of GRB-SN connection: hyperaccreting disks with outflows around
compact objects are central engines of GRBs companied by supernovae.
The outflow energy from the disk provides a part of or even a
majority of the kinetic energy of a supernova, and neutrino
annihilation from the disk provides the energy of a GRB. Now we use
the results in \S 4.3.3 and \S 4.4.2 to discuss the energy of the
outflow from the disk, and compare it with the neutrino annihilation
luminosity and energy above the disk. The upper limit of the energy
rate carried away by the outflow can be estimated by subtracting the
heating energy rate generated in the disk from the ideal heating
energy rate of the disk without outflow, i.e., the maximum energy
rate of the outflow is
\begin{equation}
\dot{E}_{o, max}\sim \frac{3GM\dot{M}}{4} \left\{{\frac{1}{3
r_{*}}-\frac{1}{r_{out}}\left[1-\frac{2}{3}
\left(\frac{r_{*}}{r_{out}}\right)^{1/2}\right]}\right\}-\int_{r_{*}}^{r_{out}}\frac{9}{8}\nu\Sigma\frac{GM}{r^{3}}2\pi
rdr,\label{4h02}
\end{equation}
and the total actual outflow energy can be considered to be $\sim
0.1-1$ fraction of the maximum outflow energy.
\begin{table}
\begin{center}
Comparison of energy rates between outflow and neutrino annihilation
\footnotesize{\begin{tabular}{rccccrc} \hline \hline Cases & disk
heating rate & max outflow energy& $L_{\nu\bar{\nu}}$ ($\eta_{s}$=0)
& $L_{\nu\bar{\nu}}$ ($\eta_{s}$=0.5)& \\& ($10^{51}$ergs s$^{-1}$)&
rate ($10^{51}$ergs
s$^{-1}$) & ($10^{51}$ergs s$^{-1}$) & ($10^{51}$ergs s$^{-1}$)\\
\hline
model O1: $s$=0.2 & 25.3 & 1.09 & 8.37e-2 & 2.24\\
$s$=0.6 & 24.7 & 1.72 & 0.104 & 1.74\\
$s$=1.0 & 18.4 & 7.95 & 0.179 & 1.65\\
\hline
model O2: $s$=0.2 & 14.6 & 11.8 & 1.42e-2 & 0.522\\
$s$=0.6 & 6.50 & 19.9 & 4.84e-3 & 2.54e-2\\
$s$=1.0 & 4.07 & 22.3 & 3.48e-3 & 5.99e-3\\
\hline \hline
\end{tabular}}
\caption{We consider model O1 and O2 with the outflow index $s$=0.2,
0.6, 1.0, the accretion rate $\dot{M}=0.3M_{\odot}$ s$^{-1}$ and the
energy parameter $\varepsilon$=0.8.}\label{tab44}
\end{center}
\end{table}
In Table \ref{tab44} we list various energy rates, including the
heating energy rate in the disk, the maximum energy injection rate
to an outflow and the neutrino annihilation rate above the disk. We
choose the accretion rate to be 0.3$M_{\odot}$ s$^{-1}$ in both
models O1 and O2 in \S 3.3. Compared with model O1, the neutron-star
disk with an outflow from the entire disk (model O2) produces higher
outflow energy but less neutrino annihilation rate. If the disk mass
around a neutron star is $\sim 1M_{\odot}$, then the maximum outflow
energy is $\sim 10^{51}$ ergs in model O1, and $\sim 10^{52}$ ergs
in model O2, but the annihilation luminosity in model O2 would be
one or two orders of magnitude less than that in model O1. Besides
the case of a neutron star, the black-hole disk with an outflow can
provide the same order of outflow energy ejecta as in model O2, but
even dimmer annihilation luminosity than model O2. Therefore, we
think that further studies of an energy relation between GRBs and
supernovae in GRB-SN events could distinguish between the
neutron-star disk models (model O1 and O2) and the black-hole disk
model.

\subsection{Effects of Magnetic Fields}
In this chapter, we do not consider the effect of magnetic fields.
As mentioned in ZD08, the high magnetic fields of central neutron
stars or magnetars $> 10^{15}-10^{16}$G could play a significant
role in the global structure of the disk as well as various
microphysical processes in the disk. Moreover, compared to the
neutron star surface, which produces $e^{+}e^{-}$ jets and outflows,
the central magnetars could be considered as a possible source to
produce magnetically-dominated outflows and collimated jets with
ultrarelativistic bulk Lorentz factors (e.g., Usov 1992; Lyutikov
2006; Uzdensky \& MacFadyen 2007; Tchekhovskoy et al. 2008; Metzger
et al. 2008a, Bucciantini et al. 2008, 2009). For example,
Bucciantini et al. (2008, 2009) modeled the interaction between the
wind from a newly formed rapidly rotating protomagnetar and the
surrounding progenitor. The free-flowing wind from protomagnetar is
not possible to achieve simultaneously collimation and acceleration
to high Lorentz factor. However, a bubble of relativistic plasma and
a strong toroidal magnetic field created by the magnetar wind
shocking on the surrounding stellar envelope can work together to
drive a relativistic jet, which is possible to produce a long GRBs
and the associated Type Ic supernova. Metzger et al. (2008a) showed
that protomagnetars are capable to produce neutron-rich long GRB
outflows for submillisecond rotation period $P\leq0.8$ ms. Besides
the study on the central magnetar, the properties of magnetized NDAF
disk has also been studied. Recently Lei et al. (2009) investigated
the properties of the NDAF with the magnetic torque acted between
the central black hole and the disk. The neutrino annihilation
luminosity can be increased by one  order of magnitude higher for
accretion rate $\sim0.5M_{\odot}$\,s$^{-1}$, and the disk becomes
thermally and viscously unstable in the inner region. Therefore, it
is also interesting to study the effects of ultra-highly magnetic
fields of neutron stars or magnetars on hyperaccreting disks.

\section{Conclusions}
In this chapter we have studied the structure of a hyperaccreting
disk around a neutron star based on the two-region scenario, and
calculate the neutrino annihilation luminosity above the disks with
different structures. The neutron-star disk model is still
Newtonian, vertically-integrated with one-dimensional variable
radius $r$, and based on the $\alpha$-prescription. The accretion
rate $\dot{M}$ is the basic quantity to determine the properties of
the disk and the annihilation luminosity, and we also discuss the
effects of the energy parameter $\varepsilon$ of the inner disk
(eqs. [4.13], [4.35] to [4.39]), the viscosity parameter $\alpha$,
the outflow structure and strength (eqs. [4.31], [4.32]), the
neutrino emission for the stellar surface layer to increase the
total annihilation energy rate (eqs. [4.42] to [4.44]).

We adopt the self-similar structure to describe the inner disk, in
which the heating mechanism is different from the outer disk. Table
\ref{tab42} shows the inner disk structure and size in various cases
depending on the value of $\varepsilon$ and the properties of the
outflow. We introduce two outflow models in \S 4.3.3. In model O1,
we consider the outflow is mainly from the inner disk, while model
O2 suggests that an outflow exists in the entire disk. In \S 4.5, we
also discuss the other possibilities of the entropy-conservation
self-similar structure of the inner disk.

Compared to the high-$\alpha$ disk ($\alpha\sim 0.1$), the size of
the low-$\alpha$ is smaller for a low accretion rate ($\leq 0.1
M_{\odot}$ s$^{-1}$) and increases dramatically with increasing
accretion rate (Fig. \ref{fig41}). A low-$\alpha$ disk is denser,
thinner with higher pressure and larger adiabatic index, and has a
brighter neutrino luminosity (Fig. \ref{fig42}). The size of the
inner disk which satisfies the advection-dominated self-similar
structure for $\varepsilon<1$ becomes smaller for lower
$\varepsilon$, and is much larger compared to the
entropy-conservation inner disk for a low accretion rate ($\leq 0.1
M_{\odot}$ s$^{-1}$, Fig. \ref{fig43}). In outflow model O1, the
inner disk is larger for a stronger outflow with low accretion rate,
but its size decreases with increasing the outflow index $s$ for a
high accretion rate. Moreover, the inner disk would not exist in
outflow model O2 when the accretion rate $\geq 0.5M_{\odot}$
s$^{-1}$ (Fig. \ref{fig44}). The outflow in the entire disk
decreases the density and pressure, but increases the thickness of
the disk (Fig. \ref{fig45}).

The neutrino annihilation luminosity above the neutron-star disk
$L_{\nu\bar{\nu},NS}$ is higher than the black-hole disk
$L_{\nu\bar{\nu},BH}$, and the difference between
$L_{\nu\bar{\nu},NS}$ and $L_{\nu\bar{\nu},BH}$ is more significant
for a low accretion rate due to the different disk structure and
neutrino luminosity between them (Fig. \ref{fig46}). The neutrino
emission from the neutron star surface boundary layer is produced in
the process of disk matter accreting onto the surface, and the
boundary emission can increase the total neutrino annihilation rate
above the disk significantly for about one order of magnitude higher
than the disk without boundary emission (Fig. \ref{fig46}, Fig.
\ref{fig47}, Fig. \ref{fig48}). The disk with an advection-dominated
inner disk could produces the highest neutrino luminosity while the
disk with an outflow from the entire disk (model O2) produces the
lowest annihilation luminosity (Fig \ref{fig47}, Fig \ref{fig48},
Fig \ref{fig49}). We show that the annihilation luminosity can reach
$10^{52}$ergs s$^{-1}$ when the accretion rate $\sim 1 M_{\odot}$
s$^{-1}$ for neutron-star disks, while the higher accretion rate
$\sim 10 M_{\odot}$ s$^{-1}$ is needed to reach $10^{52}$ ergs
s$^{-1}$ for black-hole disks. Therefore, the neutron-star disk can
produce a sufficiently large annihilation luminosity than its
black-hole counterpart. Although a heavily mass-loaded outflow from
the neutron star surface at early times of neutron star formation
prevents the outflow material from being accelerated to a high bulk
Lorentz factor, an energetic relativistic jet can be produced above
the stellar polar region at late times if the disk accretion rate
and the neutrino emission luminosity from the surface boundary are
sufficiently high. Such relativistic jet may be further accelerated
by jet-stellar-envelope interaction and produce the GRB or GRB-like
events such as X-ray flashes (XRFs).

The outflow from the advection-dominated disk and low latitudes of
the neutron star surface can be considered to provide the energy and
sufficient $^{56}$Ni for successful supernova explosions associated
with GRBs or XRFs in some previous works. However, the energy
produced via neutrino annihilation above the advection-dominated
black-hole disk is usually not sufficient for relativistic ejecta
and GRBs. On the other hand, the advection-dominated disk around a
neutron star can produce a much higher annihilation luminosity
compared to the black-hole disk. Outflow model O2 produces a higher
neutrino annihilation rate but less outflow energy compared to model
O1, while the black-hole disk could provide the same order of
outflow energy but even less annihilation energy compared to model
O2. Therefore, observations on GRB-SN connection would further
constrain these models between hyperaccreting disks around black
holes and neutron stars with outflows.


\chapter{Self-similar structure of magnetized ADAFs and CDAFs}
\label{chap5} \setlength{\parindent}{2em}

\section{Introduction}
Rotating accretion flows with viscosity and angular momentum
transfer can be divided into several classes, depending on different
structures and energy transfer mechanisms in the flows:
advection-dominated accretion flows (ADAFs), advection-dominated
inflow-outflows (ADIOs), convection-dominated accretion flows
(CDAFs), neutrino-dominated accretion flows (NDAFs) and
magnetically-dominated accretion flows (MDAFs).

ADAFs were introduced by Ichimaru (1977) and then have been widely
studied over thirty years. The opically-thick ADAFs with
super-Eddington accretion rates were discussed by Abramowicz et al.
(1988) in details (see also Begelman et al. 1982; Eggum et al.
1988). The optically-thin ADAFs with low, sub-Eddington accretion
rates were discussed by Rees et al. (1982) and Narayan \& Yi (1994,
1995a, 1995b) (see also Ichimaru 1977; Abramowicz et al. 1995;
Gammie \& Popham 1998; Popham \& Gammie 1998; Wang \& Zhou 1999). In
particular, Narayan \& Yi (1994) introduced self-similar solutions
for ADAFs with the fixed ratio of the advective cooling rate to the
viscous heating rate in the disk. Wang \& Zhou (1999) solved
self-similar solutions for optically-thick ADAFs. The effects of
general relativity were considered in Gammie \& Popham (1998) and
Popham \& Gammie (1998).

CDAFs were presented in details in Narayan et al. (2000, hereafter
NIA). They discussed the effects of convection on angular momentum
and energy transport, and presented the relations between the
convective coefficient $\alpha_{c}$ and the classical viscosity
parameter $\alpha$. A non-accreting solution can be obtained when
convection moves angular momentum inward and the viscosity parameter
$\alpha$ is small. Later, a series of works have been published to
discuss the disk structure, the MHD instability, the condition of
angular momentum transport in CDAFs (e.g., Igumenshchev et al. 2000,
2002, 2003; Quataert \& Gruzinov 2000; Narayan et al. 2002;
Igumenshchev 2002; Lu et al. 2004; van der Swaluw et al. 2005).

The effects of a magnetic field on the disk were also studied (see
Balbus \& Hawley 1998; Kaburiki 2000; Shadmehri 2004; Meier 2005;
Shadmehri \& Khajenabi 2005, 2006; Akizuki \& Fukue 2006; Ghanbari
et al. 2007). Balbus \& Hawley (1998) discussed the MHD turbulence
initiated by magnetorotational instability (MRI) and its effects on
the angular momentum transportation. Kaburaki (2000) considered an
analytic model to describe the ADAFs with a global magnetic field
and Meier (2005) considered how a turbulent and magnetized disk
creates a global well-ordered magnetic field, and introduced a
magnetically-dominated flow. Shadmehri (2004) and Chanbari et al.
(2007) discussed the self-similar structure of the magnetized ADAFs
in spherical polar coordinates. Moreover, Shamehri \& Khajenabi
(2005, 2006, hereafter SK05, SK06) presented self-similar solutions
of flows based on the vertically integrated equations. They
discussed the relations between magnetic fields components in
different directions, and mainly focused on the effects of the
magnetic field on the disk structure. Akizuki \& Fukue (2006,
hereafter AF06), different from SK05 and SK06, emphasized an
intermediate case where the magnetic force is comparable to other
forces by assuming the physical variables in the disk only as
functions of radius. However, they merely discussed a global
toroidal magnetic field in the disk.

In this chapter \S 5, we first extend the work of AF06 by
considering a general large-scale magnetic field in all the three
components in cylindrical coordinates ($r, \varphi, z$) and then
discuss effects of the global magnetic field on the flows with
convection. We adopt the treatment that the flow variables are
functions of the disk radius, neglect the different structure in the
vertical direction except for the $z$-component momentum equation.
We also discuss magnetized accretion flows with convection, and
compare our results with those in NIA, in which a large-scale
magnetic field is neglected.

This chapter \S 5 is organized as follows: basic equations are
presented in \S2. We obtain self-similar solutions in \S5.3 and
discuss the effects of a general large-scale magnetic field on the
disk flow. In \S5.4 we investigate the structure and physical
variables in magnetized CDAFs, and present the relation of the
convective parameter $\alpha_{c}$ and the classical viscosity
parameter $\alpha$. We adopt a more realistic form of the kinematic
viscosity in \S5.5. Our conclusions are presented in \S5.6.

\section{Basic Equations}

In this chapter \S 5, we use all quantities with their usual
meanings: $r$ is the radius of the disk, $v_{r}$ and $v_{\varphi}$
are the radial and rotation velocity, $\Omega=v_{\varphi}/r$ is the
angular velocity of the disk, $\Omega_{K}=(GM/r^{3})^{1/2}$ is the
Keplerian angular velocity, $\Sigma=2\rho H$ is the disk surface
density with $\rho$ to be the disk density and $H$ to be the
half-thickness, and $c_{s}=(p/\rho)^{1/2}$ is the isothermal sound
speed with $p$ to be the gas pressure in the disk.

Moreover, we consider a large-scale magnetic field in the disk with
three components $B_{r}$, $B_{\varphi}$ and $B_{z}$ in the
cylindrical coordinates ($r, \varphi, z$). The total magnetic field
$\hat{B}_{i=r,\varphi,z}$ is contributed by both large-scale
magnetic field $B_{i=r,\varphi,z}$ and the turbulence local-scale
field $b_{i=r,\varphi,z}$. We have $\hat{B}_{i}=B_{i}+b_{i}$ and
$\nabla\cdot \hat{\textbf{B}}=0$. However, since we only consider
the vertically-integrated disk, we can approximately take the
average magnetic fields $<\hat{B}_{i}>\approx B_{i}$, and neglect
the local turbulence field in our self-similar calculations. Thus we
define the Alfv\'{e}n sound speeds $c_{r}$, $c_{\varphi}$ and
$c_{z}$ in three directions of the cylindrical coordinates as
$c_{r,\varphi,z}^{2}=B_{r,\varphi,z}^{2}/(4\pi \rho)$. We consider
that all flow variables are only functions of radius $r$, and write
basic equations, i.e., the continuity equation, the three components
($r,\varphi,z$) of the momentum equation and the energy equation:

\begin{equation}
\frac{1}{r}\frac{d}{dr}\left(r\Sigma
v_{r}\right)=2\dot{\rho}H,\label{e10}
\end{equation}

\begin{equation}
v_{r}\frac{dv_{r}}{dr}=\frac{v_{\varphi}^{2}}{r}-\frac{GM}{r^{2}}-\frac{1}{\Sigma}\frac{d}{dr}\left(\Sigma
c_{s}^{2}\right)-\frac{1}{2\Sigma}\frac{d}{dr}\left(\Sigma
c_{z}^{2}+\Sigma
c_{\varphi}^{2}\right)-\frac{c_{\varphi}^{2}}{r},\label{e11}
\end{equation}

\begin{equation}
\frac{v_{r}}{r}\frac{d(rv_{\varphi})}{dr}=\frac{1}{\Sigma
r^{2}}\frac{d}{dr}\left(\Sigma \alpha
\frac{c_{s}^{2}}{\Omega_{K}}r^{3}\frac{d\Omega}{dr}\right)+\frac{c_{\varphi}c_{r}}{r}
+\frac{c_{r}}{\sqrt{\Sigma}}\frac{d}{dr}\left(\sqrt{\Sigma
}c_{\varphi}\right),\label{e12}
\end{equation}

\begin{equation}
\Omega_{K}^{2}H-\frac{1}{\sqrt{\Sigma}}c_{r}\frac{d}{dr}\left(\sqrt{\Sigma}
c_{z}\right)=\frac{c_{s}^{2}+\frac{1}{2}\left(c_{\varphi}^{2}+c_{r}^{2}\right)}{H},\label{e13}
\end{equation}

\begin{equation}
\frac{v_{r}}{\gamma-1}\frac{dc_{s}^{2}}{dr}-v_{r}\frac{c_{s}^{2}}{\rho}\frac{d\rho}{dr}=f\frac{\alpha
c_{s}^{2}r^{2}}{\Omega_{K}}\left(\frac{d\Omega}{dr}\right)^{2}.\label{e131}
\end{equation}
Here we consider the height-integrated equations using the classical
$\alpha$-prescription model with $\alpha$ to be the viscosity
parameter, and use the Newtonian gravitational potential. In the
mass continuity equation, we also consider the mass loss term
$\partial \rho/\partial t$. In the energy equation we take $\gamma$
to be the adiabatic index of the disk gas and $f$ to measure the
degree to which the flow is advection-dominated (NY94), and neglect
the Joule heating rate.

In AF06, a general case of viscosity
$\eta=\rho\nu=\Omega_{K}^{-1}\alpha p_{\rm gas}^{\mu}(p_{\rm
gas}+p_{\rm mas})^{1-\mu}$ with $\mu$ to be a parameter is
mentioned. If the ratio of the magnetic pressure to the gas pressure
is constant (as assumed in the self-similar structure), the solution
of the basic equations can be obtained with replacing $\alpha$ by
$\alpha(1+\beta)^{1-\mu}$. In our section \S 5, however, we first
adopt the classical form $\nu=\alpha c_{s}^{2}/\Omega_{K}$ for
simplicity in \S 5.3 and \S 5.4, in which we mainly focus on the
effects of a magnetic field on the variables $v_{r}$, $v_{\varphi}$
and $c_{s}$. A more realistic model requires $\nu=\alpha c_{s}H$
with both $c_{s}$ and $H$ as functions of the magnetic field
strength. We discuss this model in \S 5.5 and compare it with the
results in \S 5.3 and \S 5.4.

Our equations are somewhat different from those in SK05 and SK06,
since we only consider the disk variables as functions of radius
$r$, while SK05 and SK06 discuss the magnetic field structure in the
vertical direction. More details about the basic equations are
discussed in Appendix A. When $B_{r}=0$ and $B_{z}=0$, our equations
switch back to the equations in AF06, in which only the toroidal
magnetic field is considered and all the variables are taken to
depend merely on radius $r$.

In addition, we need the the three-component induction equations to
measure the magnetic field escaping rate:
\begin{equation}
\dot{B_{r}}\approx 0,\label{e14}
\end{equation}

\begin{equation}
\dot{B_{\varphi}}=\frac{d}{dr}\left(v_{\varphi}B_{r}-v_{r}B_{\varphi}\right),\label{e15}
\end{equation}

\begin{equation}
\dot{B_{z}}=-\frac{d}{dr}\left(v_{r}B_{z}\right)-\frac{v_{r}B_{z}}{r}.\label{e16}
\end{equation}

\section{Self-Similar Solutions for ADAF}
If we assume the parameters $\gamma$ and $f$ in the energy equation
are independent of radius $r$, then we can adopt a self-similar
treatment similar to NY94 and AF06,
\begin{equation}
v_{r}(r)=-c_{1}\alpha\sqrt{\frac{GM}{r}},\label{e201}
\end{equation}

\begin{equation}
v_{\varphi}(r)=c_{2}\sqrt{\frac{GM}{r}},\label{e202}
\end{equation}

\begin{equation}
c_{s}^{2}(r)=c_{3}\frac{GM}{r},\label{e203}
\end{equation}

\begin{equation}
c_{r,\varphi,z}^{2}(r)=\frac{B_{r,\varphi,z}^{2}}{4\pi
\rho}=2\beta_{r,\varphi,z}c_{3}\frac{GM}{r},\label{e204}
\end{equation}
where the coefficients $c_{1}$, $c_{2}$ and $c_{3}$ are similar to
those in AF06, and $\beta_{r}$, $\beta_{\varphi}$ and $\beta_{z}$
measure the ratio of the magnetic pressure in three directions to
the gas pressure, i.e., $\beta_{r,\varphi,z}=p_{{\rm
mag},r,\varphi,z}/p_{\rm gas}$. Following AF06, we also denote the
structure of the surface density $\Sigma$ by
\begin{equation}
\Sigma(r)=\Sigma_{0}r^{s}.\label{e205}
\end{equation}
The half-thickness of the disk still satisfies the relation
$H\propto r$ and we obtain
\begin{equation}
H(r)=H_{0}r.\label{e206}
\end{equation}

Substituting self-similar relations (\ref{e201})--(\ref{e206}) to
equations (\ref{e11}), (\ref{e12}) and (\ref{e131}), we can obtain
the algebraic equations of $c_{1}$, $c_{2}$ and $c_{3}$:
\begin{equation}
-\frac{1}{2}c_{1}^{2}\alpha^{2}=c_{2}^{2}-1-[(s-1)+\beta_{z}(s-1)+\beta_{\varphi}(s+1)]c_{3},\label{e207}
\end{equation}

\begin{equation}
-\frac{1}{2}c_{1}c_{2}\alpha=-\frac{3}{2}\alpha(s+1)c_{2}c_{3}+c_{3}(s+1)\sqrt{\beta_{r}\beta_{\varphi}},\label{e208}
\end{equation}

\begin{equation}
c_{2}^{2}=\frac{4}{9f}\left(\frac{1}{\gamma-1}+s-1\right)c_{1}.\label{e209}
\end{equation}

If in the cylindrical coordinates we assume the three components of
magnetic field $B_{r,\varphi,z}>0$, then $v_{\varphi}(r)$ can be
either positive or negative, depending on the detailed magnetic
field structure in the disk. In a particular case where $B_{r}=0$ or
$B_{\varphi}=0$, we can only obtain the value of $|c_{2}|$, but in a
general case where $B_{r}B_{\varphi}\neq0$, we are able to determine
the value of $c_{2}$.

\begin{figure}
\resizebox{\hsize}{!} {\includegraphics{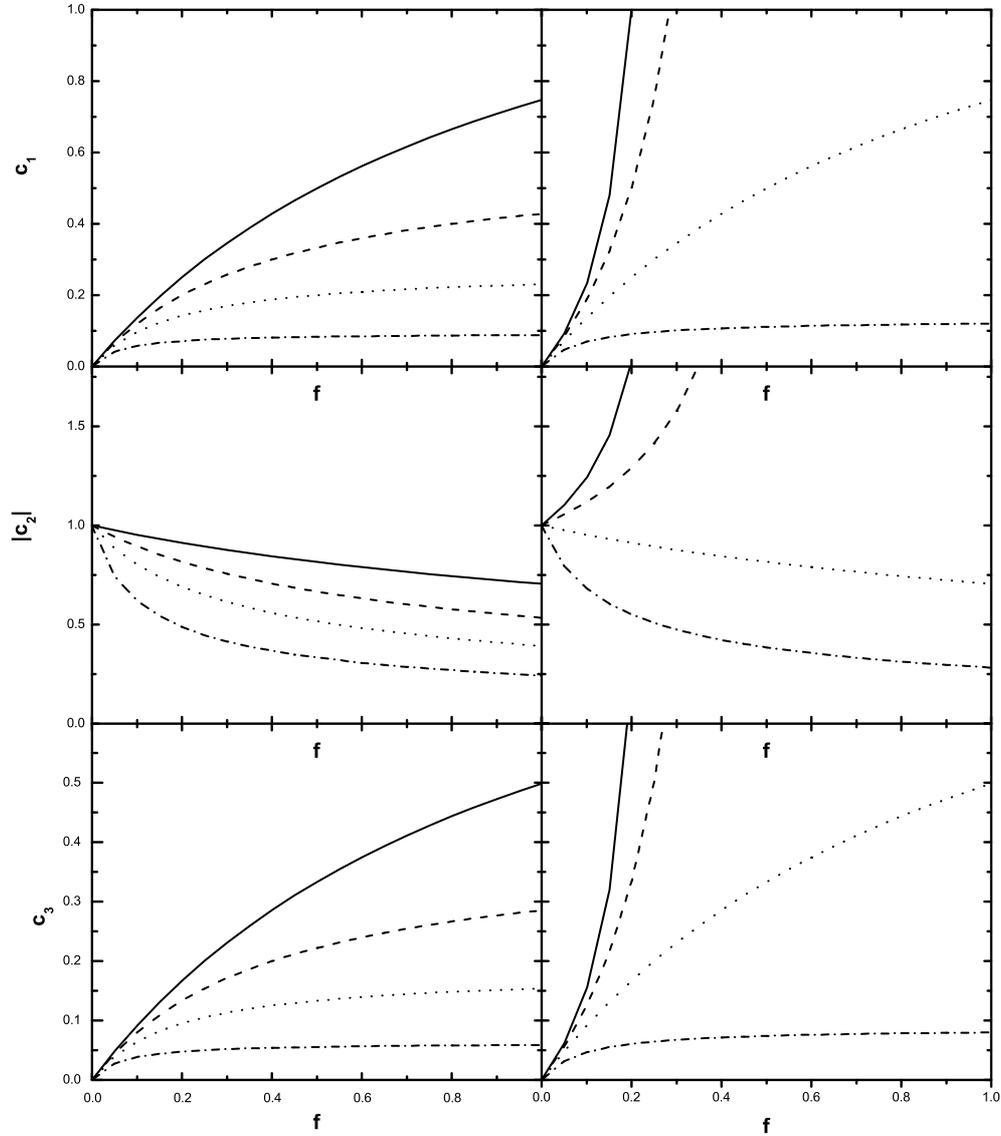}} \caption{The
self-similar coefficients $c_{1}$, $|c_{2}|$, and $c_{3}$ as
functions of the advection parameter $f$ for different sets of
parameters $\beta_{r}$, $\beta_{\varphi}$ and $\beta_{z}$. We take
$\alpha$=0.1, $\gamma=4/3$ and $s=-1/2$. The left three panels
correspond to ($\beta_{r}$, $\beta_{\varphi}$ $\beta_{z}$)$=(0,1,0)$
({\em solid lines}), $(0,1,1)$ ({\em dashed lines}), $(0,1,3)$ ({\em
dotted lines}), and (0,1,10) ({\em dash-dotted lines}). The right
three panels correspond to ($\beta_{r}$, $\beta_{\varphi}$
$\beta_{z}$)$=(0,10,0)$ ({\em solid lines}), $(0,10,1)$ ({\em dashed
lines}), $(0,10,3)$ ({\em dotted lines}), and $(0,10,10)$ ({\em
dash-dotted lines}).}\label{fig51}
\end{figure}
\begin{figure}
\resizebox{\hsize}{!} {\includegraphics{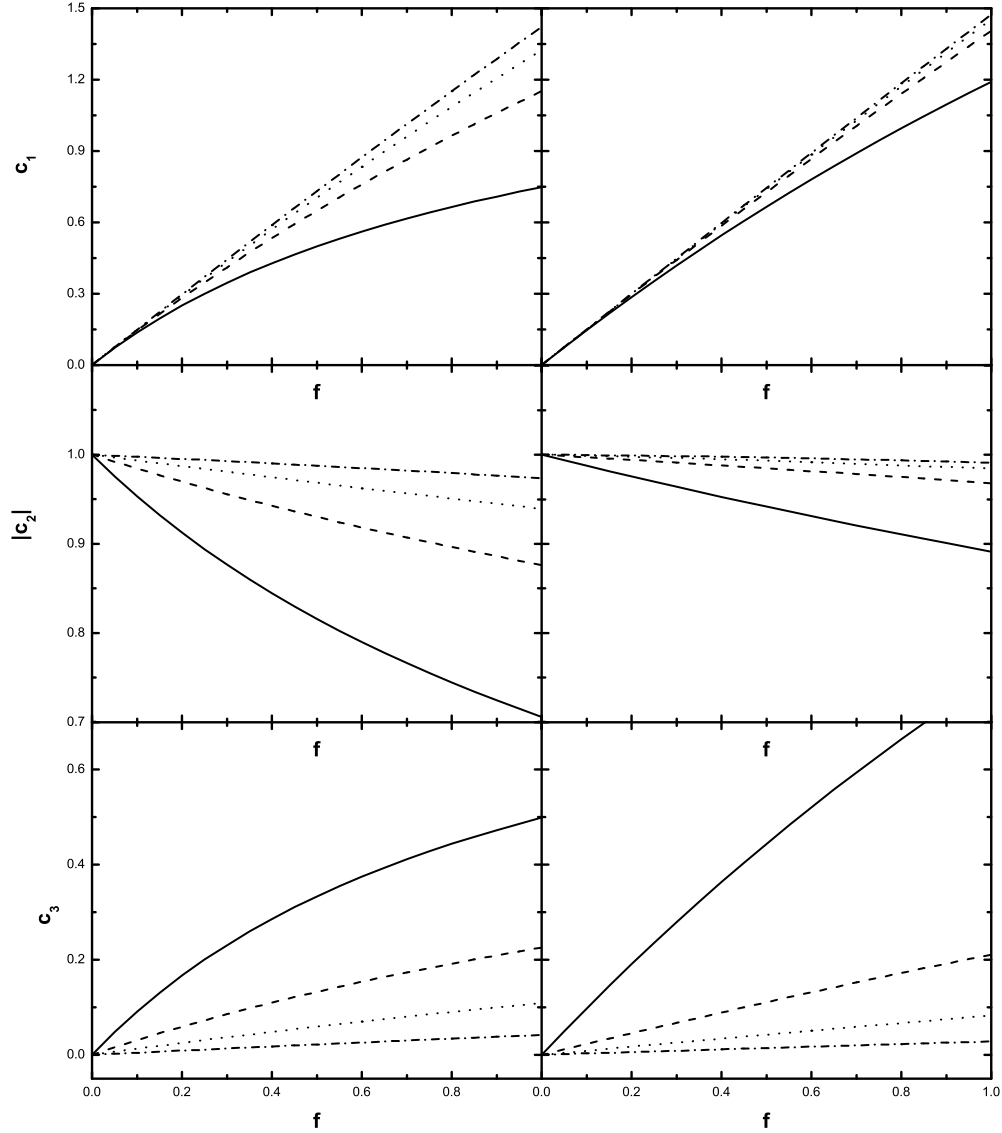}} \caption{The
self-similar coefficients $c_{1}$, $|c_{2}|$, and $c_{3}$ as
functions of the advection parameter $f$ for different sets of
parameters $\beta_{r}$, $\beta_{\varphi}$ and $\beta_{z}$ with
$\alpha$=0.1, $\gamma=4/3$ and $s=-1/2$. The left three panels
correspond to ($\beta_{r}$, $\beta_{\varphi}$ $\beta_{z}$)$=(0,1,0)$
({\em solid lines}), $(0.1,1,0)$ ({\em dashed lines}), $(1,1,0)$
({\em dotted lines}), and $(10,1,0)$ ({\em dash-dotted lines}). The
right three panels correspond to ($\beta_{r}$, $\beta_{\varphi}$
$\beta_{z}$)=$(0,2.5,0)$ ({\em solid lines}), $(0.1,2.5,0)$ ({\em
dashed lines}), $(1,2.5,0)$ ({\em dotted lines}), and $(10,2.5,0)$
({\em dash-dotted lines}).}\label{fig52}
\end{figure}

Figures \ref{fig51} and \ref{fig52} show the self-similar
coefficients $c_{1}$, $|c_{2}|$ and $c_{3}$ as functions of the
advection parameter $f$ with different $(\beta_{r}, \beta_{\varphi},
\beta_{z})$. We consider the disk to be radiation dominated with
$\gamma=4/3$, and take $s=-1/2$ (i.e., $\rho\propto r^{-3/2}$ as the
common case) and the viscosity parameter $\alpha=0.1$, which is the
widely used value.

Figure \ref{fig51} shows changes of the coefficients $c_{1}$,
$|c_{2}|$, $c_{3}$ with $\beta_{z}$ and $\beta_{\varphi}$. We
neglect the radial magnetic fields $B_{r}$, and take the parameters
$(\beta_{r}, \beta_{\varphi}, \beta_{z})$ in the left three panels
in Figure \ref{fig51} for (0, 1, 0), (0, 1, 1), (0, 1, 3) and (0, 1,
10), and then take the value of $\beta_{\varphi}$ to be 10 in the
right three panels. As $\beta_{r}=0$, we can only obtain $|c_{2}|$
without needing to determine the direction of $v_{\varphi}$. The
coefficients $c_{1}$ and $c_{3}$ increase with increasing the
advection parameter $f$, but $|c_{2}|$ decreases monotonously as a
function of $f$ except for a strong toroidal magnetic field.
Moreover, with the fixed ratio $\beta_{z}$, an increase of
$\beta_{\varphi}$ makes all the coefficients $|c_{i}|$ become
larger. Oppositely, $|c_{i}|$ decreases with increasing $\beta_{z}$.
In fact, with a small radial magnetic field $\beta_{r}\approx 0$, we
can obtain an analytical solution of $c_{i}$ from equations
(\ref{e207})-(\ref{e209}), which are similar to expressions
(27)-(29) in AF06, but we should replace $(1-s)/(1+s)$ by
$(1-s)(1+\beta_{z})/(1+s)$ and $\beta$ by $\beta_{\varphi}$ in those
expressions instead. Also, we have $c_{2}^{2}\propto f^{-1}c_{1}$,
$c_{3}\propto c_{1}$, and $c_{1}\propto \beta_{\varphi}$ for large
$\beta_{\varphi}$ and $c_{1}\propto \beta_{z}^{-1}$ for large
$\beta_{z}$, all of which are consistent with the results in Figure
\ref{fig51}.

As a result, from Figure \ref{fig51}, we first find that a strong
toroidal magnetic field leads to an increase of the infall velocity
$|v_{r}|$, rotation velocity $|v_{\varphi}|$ and isothermal sound
speed $c_{s}$, and $|v_{r}|$ and $c_{s}$ are large in the case where
the disk flow is mainly advection-dominated, but the rotation
velocity $|v_{\varphi}|$ increases with increasing $f$ only in the
case where the toroidal magnetic field is large enough. This
conclusion is consistent with the case 1 in AF06. Second, the high
ratio $\beta_{z}$ decreases the value of $|v_{r}|$, $|v_{\varphi}|$
and $c_{s}$, which means that a strong magnetic pressure in the
vertical direction prevents the disk matter from being accreted, and
decreases the effect of gas pressure as accretion proceeds.

Figure \ref{fig52} shows how $c_{1}$, $|c_{2}|$ and $c_{3}$ change
with $\beta_{r}$ and $\beta_{\varphi}$, where we neglect the
vertical magnetic $B_{z}$. For a small value of $\beta_{r}$, the
coefficients $c_{1}$, $|c_{2}|$ and $c_{3}$ also increase with
increasing $\beta_{\varphi}$. However, a change of $c_{i}$ is not
obvious for a large value of $\beta_{r}$. We are able to calculate
the limiting value of $c_{i}$ in the extreme case where $\beta_{r}$
is large enough and $\beta_{r}\beta_{\varphi}\neq 0$ using an
analytical method. From equations (\ref{e207})-(\ref{e209}), we can
obtain $c_{1}=2/[\epsilon''+\sqrt{(\epsilon'')^{2}+2\alpha^{2}}]$
and $c_{2}^{2}=\epsilon''c_{1}$ for large $\beta_{r}$, where
$\epsilon''=\frac{4}{9f}\left\{(\gamma-1)^{-1}+s-1\right\}$. If
$\epsilon''\gg \alpha$, we have $c_{1}\sim (\epsilon'')^{-1}\propto
f$ and $|c_{2}|\sim 1$, which means that the infall velocity
$|v_{r}|$ increases with advection parameter $f$ linearly, and the
radial velocity $|v_{\varphi}|$ is nearly the Keplerian velocity, no
matter whether the disk is efficiently cooled or not. Also, for a
large value of $\beta_{r}$, equation (\ref{e208}) becomes
$-(c_{1}c_{2}\alpha)/2\sim
c_{3}(s+1)\sqrt{\beta_{r}\beta_{\varphi}}$. Since $c_{1},c_{3}>0$ in
the accretion disk, we obtain $c_{2}<0$, which means that the
direction of rotation in the disk is opposite to the toroidal
magnetic field $B_{\varphi}$. Actually, in the case where
$\beta_{r}$ is sufficient large and $\beta_{r}\beta_{\varphi}\neq
0$, the angular momentum transported due to the magnetic field
stress is dominated over that due to the viscosity, and balances
with the advection angular momentum. As we take $B_{r,\varphi,z}>0$,
from equation (\ref{e12}), we obtain that the large angular momentum
due to the magnetic field stress makes the value of the advection
angular momentum ($\dot{M}v_{\varphi}r$, where $\dot{M}$ is the mass
accretion rate) increase in the disk, which requires $v_{\varphi}<0$
in the self-similar structure\footnote{In some previous works (e.g.,
Wang 95, Lai 98, SK05 and SK06), the rotation velocity is taken to
be positive and the toroidal magnetic field $B_{\varphi}$ to be
negative. The advection transports angular momentum inward, while
the magnetic stress transports angular momentum outward instead.
This previous result is consistent with our result here if we change
the cylindrical coordinate used above from ($r,\varphi,z$) to
($r,-\varphi,z$). In this chapter \S 5 it is convenient for us to
take $B_{r,\varphi,z}>0$ and to obtain s series of self-similar
solutions about magnetized flows in many different cases.}.

From the mass-continuity equation (\ref{e10}) and the induction
equations (\ref{e14})-(\ref{e16}) as well as the solved coefficients
$c_{i}$, we can solve the self-similar structure of the mass loss
and magnetic field escaping rate with forms of
$\dot{\rho}=\dot{\rho}_{0}r^{s-5/2}$ and
$\dot{B}_{r,\varphi,z}=\dot{B}_{r_0,\varphi_0,z_0}r^{(s-5)/2}$ where
$\dot{\rho}_{0}$ satisfies

\begin{equation}
\dot{\rho}_{0}=-\left(s+\frac{1}{2}\right)\frac{c_{1}\alpha\Sigma_{0}\sqrt{GM}}{2H_{0}}.\label{e210}
\end{equation}
As mentioned in AF06, when $s=-1/2$, i.e., $\Sigma\propto r^{-1/2}$
or $\rho\propto r^{-3/2}$, there is no wind in the disk, and thus we
can use the formula $-2\pi rv_{r}\Sigma =\dot{M}$ to determine the
surface density $\Sigma$. In addition, in the region of the disk
where is adiabatic with $\rho \propto r^{-1/(\gamma-1)},\,\,
p\propto r^{-\gamma/(\gamma-1)},\,\, v_r \propto
r^{(3-2\gamma)/(\gamma-1)}$ (i.e., $s=(\gamma-2)/(\gamma-1)$), we
obtain the self-similar solution of $c_{1}\alpha=\sqrt{2}, c_{2}=0$
and $c_{3}=0$, which describe the Bondi accretion. However, if the
disk region satisfies the entropy-conservation condition with $f=0$,
we can still obtain an accretion-disk solution beyond the
self-similar treatment. For a CDAF with $\rho\propto r^{-1/2}$
(NIA), a steady disk without wind requires $c_{1}=0$ or $v_{r}=0$.

$\dot{B}_{r_0,\varphi_0,z_0}$ satisfy
\begin{equation}
\dot{B}_{r_0}\approx 0,\label{e211}
\end{equation}

\begin{equation}
\dot{B}_{\varphi_0}=\left(\frac{s-3}{2}\right)GM
\left\{c_{2}\sqrt{\frac{4\pi\beta_{r}c_{3}\Sigma_{0}}{H_{0}}}+c_{1}
\alpha\sqrt{\frac{4\pi\beta_{\varphi}c_{3}\Sigma_{0}}{H_{0}}}\right\},\label{e2111}
\end{equation}

\begin{equation}
\dot{B}_{z_0}=\left(\frac{s-1}{2}\right)c_{1}\alpha(GM)\sqrt{\frac{4\pi\beta_{z}\Sigma_{0}c_{3}}{H_{0}}},\label{e212}
\end{equation}
where $H_{0}$ in the expression (\ref{e206}) can be obtained from
the hydrostatic equilibrium equation (\ref{e13}), that is,
\begin{equation}
H_{0}=\frac{1}{2}\left[(s-1)c_{3}\sqrt{\beta_{r}\beta_{z}}+\sqrt{c_{3}^{2}(s-1)^{2}\beta_{r}\beta_{z}
+4(1+\beta_{\varphi}+\beta_{r})c_{3}}\right],\label{e213}
\end{equation}
and thus we obtain the half-thickness of the disk
$H=H_{0}c_{s}/(\sqrt{c_{3}}\Omega_{K})$. We will discuss the effects
of the magnetic field on $H$ later in \S 5.5.

\section{Self-Similar Solutions for CDAF}

In CADFs, both advection and convection play contributions to the
angular momentum and energy transportation. We propose a CDAF model
in a global magnetic field in order to compare it with
non-globally-magnetized CDAFs. We follow the idea of NIA in this
section and consider the effect of a magnetic field\footnote{We
adopt the ($\alpha, \alpha_{c}$)-prescription following NIA. The MHD
simulations beyond this prescription can be seen in Igumenshchev et
al. (2002, 2003), Hawley \& Balbus (2002) and so on. Moreover,
Quataert \& Gruzinov (2000) also develop an analytical model for
CDAFs. }. The viscosity angular momentum flux is
\begin{equation}
\dot{J}_{v}=-\alpha\frac{c_{s}^{2}}{\Omega_{K}}\rho
r^{3}\frac{d\Omega}{dr},\label{e301}
\end{equation}
and the convection angular momentum flux can be written as
\begin{equation}
\dot{J}_{c}=-\alpha_{c}\frac{c_{s}^{2}}{\Omega_{K}}\rho
r^{3(1+g)/2}\frac{d}{dr}\left(\Omega
r^{3(1-g)/2}\right),\label{e3011}
\end{equation}
where $\alpha_{c}$ is the dimensionless coefficient to measure the
strength of convective diffusion, $g$ is the parameter to determine
the condition of convective angular momentum transport. Convection
transports angular momentum inward (or outward) for $g<0$ (or $>0$).

The energy equation of a CADF is
\begin{equation}
\rho
v_{r}T\frac{ds}{dr}+\frac{1}{r^{2}}\frac{d}{dr}\left(r^{2}F_{c}\right)=Q^{+}=f\frac{(\alpha+g\alpha_{c})
\rho
c_{s}^{2}r^{2}}{\Omega_{K}}\left(\frac{d\Omega}{dr}\right)^{2},\label{e302}
\end{equation}
where the convective energy flux $F_{c}$ is
\begin{equation}
F_{c}=-\alpha_{c}\frac{c_{s}^{2}}{\Omega_{K}}\rho
T\frac{ds}{dr},\label{e303}
\end{equation}
where we still consider the general energy equation without vertical
integration as in NIA, and still neglect the Joule heating rate.

Using the angular momentum equation, the energy equation of the CDAF
and the self-similar structure (\ref{e201})-(\ref{e206}), we can
obtain the relevant algebraic equations of self-similar structure of
the CDAF,
\begin{equation}
-\frac{1}{2}c_{1}c_{2}\alpha=-\frac{3}{2}(\alpha+g\alpha_{c})(s+1)c_{2}c_{3}
+c_{3}(s+1)\sqrt{\beta_{r}\beta_{\varphi}},\label{e304}
\end{equation}

\begin{equation}
\left(s-\frac{1}{2}\right)\left(\frac{1}{\gamma-1}+s-1\right)c_{3}\alpha_{c}+\left(\frac{1}{\gamma-1}+s-1\right)c_{1}\alpha
=(\alpha+g\alpha_{c})\frac{9f}{4}c_{2}^{2}.\label{e305}
\end{equation}
The radial momentum equation is still the same as that in ADAF.
Combining equation (\ref{e207}), (\ref{e304}) and (\ref{e305}), we
can finally solve the coefficients $c_{1}$, $c_{2}$, and $c_{3}$ in
the case of CDAF and compare them with those in ADAF. The
dimensionless coefficient $\alpha_{c}$ can be calculated using the
mixing length theory, and we adopt equation (15) in NIA, who
describes the relation of $\alpha_{c}$ with $s$, $c_{3}$ and
$\gamma$ (in NIA, they used the symbols $a$ and $c_{0}$, where
$a=1-s$, and $c_{0}^{2}=c_{3}$ in our section \S 5).

To simplify the problem, we first prefer using a simpler treatment
with a fixed $\alpha_{c}$ to discussing the solutions of
(\ref{e207}), (\ref{e304}) and (\ref{e305}), i.e., we take
$\alpha_{c}$ as a free parameter rather than a calculated variable,
since $\alpha_{c}$ does not dramatically change in many cases. Then
we can adopt a similar treatment as in \S3 to solve equations
(\ref{e207}), (\ref{e304}) and (\ref{e305}). Similarly as in \S 3,
we first use the analytical method to discuss some particular cases.

When $\beta_{r}\beta_{\varphi}\sim 0$ (which implies that the radial
or toroidal magnetic fields are weak), we can obtain an analytical
solution similar to that in \S5.3 (see Appendix B for more details
about the calculation). We discuss two cases. One is that the
toroidal magnetic field $B_{\varphi}$ is dominated and the radial
magnetic field is weak ($B_{r}\approx 0$). Then we can have an
approximate solution,
\begin{equation}
c_{1}\alpha\sim \frac{2\beta_{\varphi}}{3(\alpha+g\alpha_{c})},
\label{e306}
\end{equation}

\begin{equation}
c_{2}^{2}=
\frac{c_{1}\alpha}{\alpha+g\alpha_{c}}\left[\epsilon''-|\xi''|\frac{\alpha_{c}}{3(\alpha+g\alpha_{c})(s+1)}\right],
\label{e3071}
\end{equation}

\begin{equation}
c_{3}\sim \frac{2\beta_{\varphi}}{9(\alpha+g\alpha_{c})^{2}(s+1)}.
\label{e3072}
\end{equation}
where $\xi''=\frac{4}{9f}(s-\frac{1}{2})(\frac{1}{\gamma-1}+s-1)$.
From these equations, we know the coefficients $c_{1}$ and $c_{3}$
increase with increasing the convective parameter $\alpha_{c}$ for
$g<0$ (i.e., convection transports angular momentum inward), but we
cannot obtain the relation between $\alpha_{c}$ and $|c_{2}|$ unless
the values of $s$, $g$ and $\gamma$ are given in detail. The other
case is that the vertical magnetic field $B_{z}$ is dominated. In
this case, we obtain
\begin{equation}
c_{1}\alpha\sim \frac{3(\alpha+g\alpha_{c})(1+s)}{(1-s)\beta_{z}},
\label{e3073}
\end{equation}

\begin{equation}
c_{2}^{2}=
\frac{3(1+s)}{(1-s)\beta_{z}}\left[\epsilon''-|\xi''|\frac{\alpha_{c}}{3(\alpha+g\alpha_{c})(s+1)}\right],
\label{e3074}
\end{equation}

\begin{equation}
c_{3}\sim \frac{1}{(1-s)\beta_{z}}. \label{e3072}
\end{equation}
We find that the coefficients $c_{1}$ and $|c_{2}|$ decrease with
increasing $\alpha_{c}$ for $g<0$ while the value of $c_{3}$ is more
or less the same for a fixed $\beta_{z}$.

Another analytical solution can be obtained when $\beta_{r}$ is
large and $\beta_{\varphi}\neq 0$, and we have the relations
\begin{equation}
c_{1}\propto
\left[\frac{\alpha}{\alpha+g\alpha_{c}}\epsilon''+\sqrt{\left(\frac{\alpha}{\alpha+g\alpha_{c}}\right)^{2}\epsilon''^{2}
+2\alpha^{2}}\right]^{-1},\label{e308}
\end{equation}

\begin{equation}
|c_{2}|\propto
\left[\epsilon''+\sqrt{\epsilon''^{2}+2(\alpha+g\alpha_{c})^{2}}\right]^{-1/2},
\label{e309}
\end{equation}

\begin{equation}
c_{3}\propto -c_{1}c_{2}. \label{e3091}
\end{equation}
From formulae (\ref{e308}) and (\ref{e309}),  we again find changes
of $c_{1}$ and $|c_{2}|$ with $\alpha_{c}$, which is similar to the
former case. From equation (\ref{e3091}) and $c_{1,3}>0$, we get
$c_{2}<0$, which has also been obtained in \S 3.

\begin{figure}
\resizebox{\hsize}{!} {\includegraphics{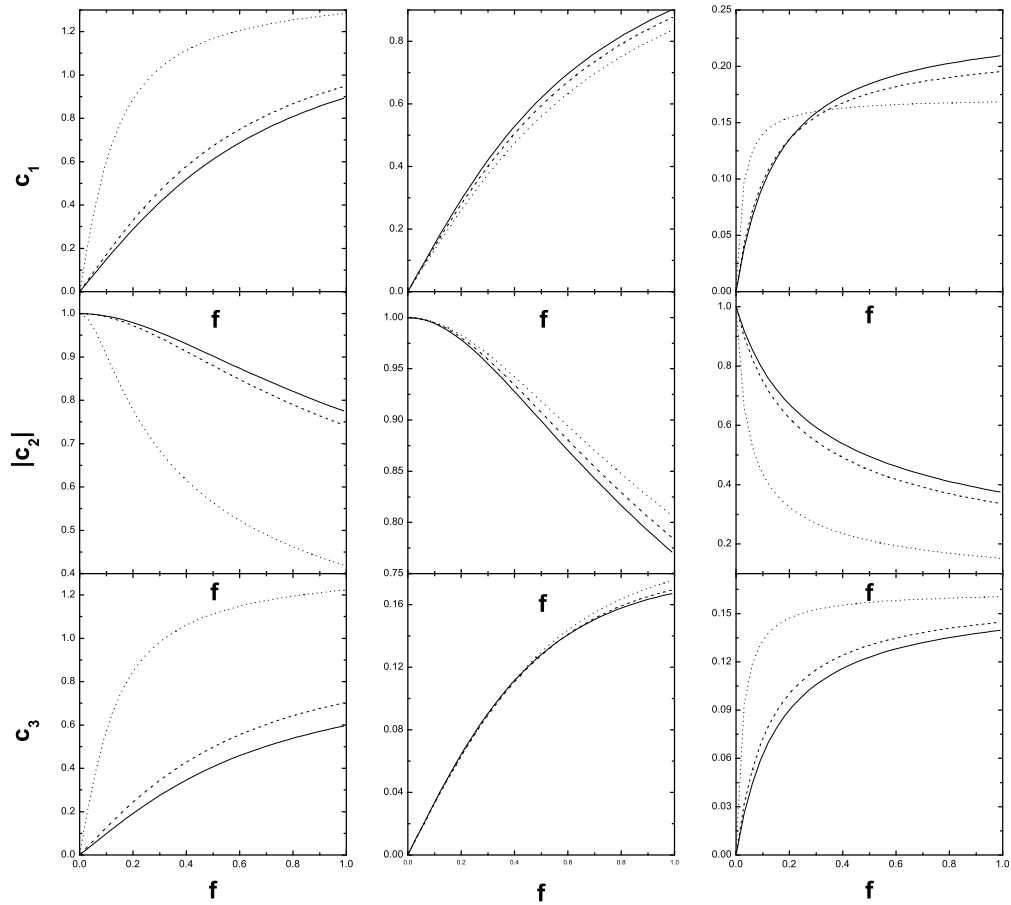}} \caption{The
coefficients $c_{1}$, $|c_{2}|$, and $c_{3}$ as functions of $f$
with different sets of parameters $\alpha_{c}$ and ($\beta_{r}$,
$\beta_{\varphi}$ $\beta_{z}$). We take $\alpha$=1, $\gamma=4/3$,
$s=-1/2$ and $g=-1/3$. The left three panels correspond to
($\beta_{r}$, $\beta_{\varphi}$ $\beta_{z}$)$=(0,3,0)$, the middle
three panels to ($\beta_{r}$, $\beta_{\varphi}$
$\beta_{z}$)$=(3,3,0)$, and the right panels to ($\beta_{r}$,
$\beta_{\varphi}$ $\beta_{z}$)$=(0,0,3)$. Different lines refer to
$\alpha_{c}=0$ ({\em solid lines}), $\alpha_{c}=0.3$ ({\em dashed
lines}) and $\alpha_{c}=0.9$ ({\em dotted lines}).}\label{fig53}
\end{figure}
Figure \ref{fig53} shows some examples of the effect of the
convection parameter $\alpha_{c}$ on the three coefficients $c_{i}$.
In order to see the results clearly, we take $\alpha=1$ and change
the value of $\alpha_{c}$ from 0 to 0.9 with several sets of
magnetic field parameters ($\beta_{r}$, $\beta_{\varphi}$,
$\beta_{z}$)$=(0,3,0)$, $(3,3,0)$ and $(0,0,3)$. Also we set
$\gamma=4/3$, $s=-1/2$ and $g=-1/3$. The basic results in Figure 3
are consistent with the above discussion using the analytical
method. In particular, we notice that the three coefficients do not
change dramatically in the case of $\beta_{r}\sim \beta_{\varphi}$,
since the magnetic field gives a contribution to the angular
momentum rather than the viscosity, and reduces the effect of
convection on the disk.

Next we want to obtain the relation between $\alpha_{c}$ and
$\alpha$ following NIA, i.e., we consider $\alpha_{c}$ to be the
variable as a function of $s$, $\gamma$ and $c_{3}$. Using the
treatment in NIA based on the mixing length theory and equations
(\ref{e207}), (\ref{e304}) and (\ref{e305}), we can establish the
$\alpha_{c}$-$\alpha$ relation. NIA discussed such a relation with
$g=1$ and $g=-1/3$, and find that the solution with $s=-1/2$ is
available only for $\alpha$ greater than a certain critical
$\alpha_{\rm crit}$ when the isothermal sound speed reaches its
maximum value, and the value of $\alpha_{c}$ decreases monotonously
as $\alpha$ increases. However, our results are quite different from
those in NIA for two reasons. First, we keep the term
$v_{r}dv_{r}/dr$ in the radial momentum equation, while in NIA this
term is neglected. As a result, in many cases, the sound speed to
determine the actual critical $\alpha$ for available solutions does
not reach exactly its maximum value. Second and more importantly, we
consider the effect of the large-scale magnetic field on the disk.

From equation (\ref{e207}), we obtain
\begin{equation}
\frac{1}{2}c_{1}^{2}\alpha^{2}+[(1-s)(1+\beta_{z})-\beta_{\varphi}(1+s)]c_{3}-1
<0.\label{e3092}
\end{equation}
If the radial magnetic field is weak, then we have
$c_{3}<[(1-s)\beta_{z}]^{-1}$ for a large vertical magnetic field
and $c_{3}<2\beta_{\varphi}/[9(\alpha+g\alpha_{c})^{2}(1+s)]$ for a
large toroidal magnetic field. On the other hand, from NIA, we have
$c_{3}>\gamma/[(2-s)(2+s\gamma-s)]$ for the convective process to be
available. Therefore, the structure of flows with convection cannot
be maintained for a large vertical magnetic field. Moreover, if the
term $\beta_{r}\beta_{\varphi}$ is large, we still obtain a small
value of $c_{3}\sim
c_{1}|c_{2}|\alpha/(2\sqrt{\beta_{r}\beta_{\varphi}})$ with a small
value of $\alpha_{c}$, which is almost independent of the variation
of $\alpha$.

\begin{figure}
\resizebox{\hsize}{!} {\includegraphics{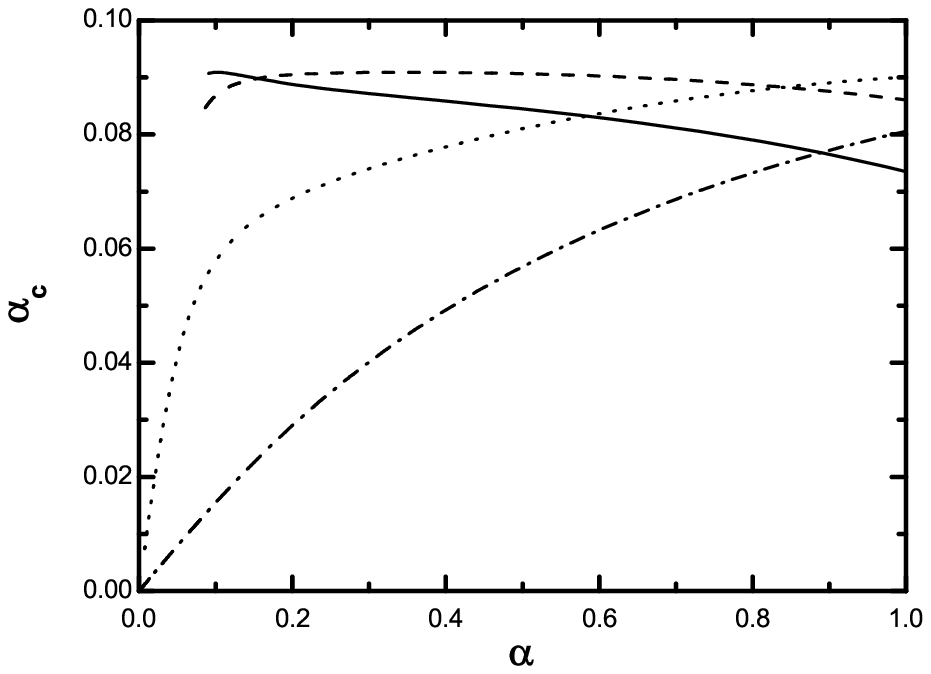}
\includegraphics{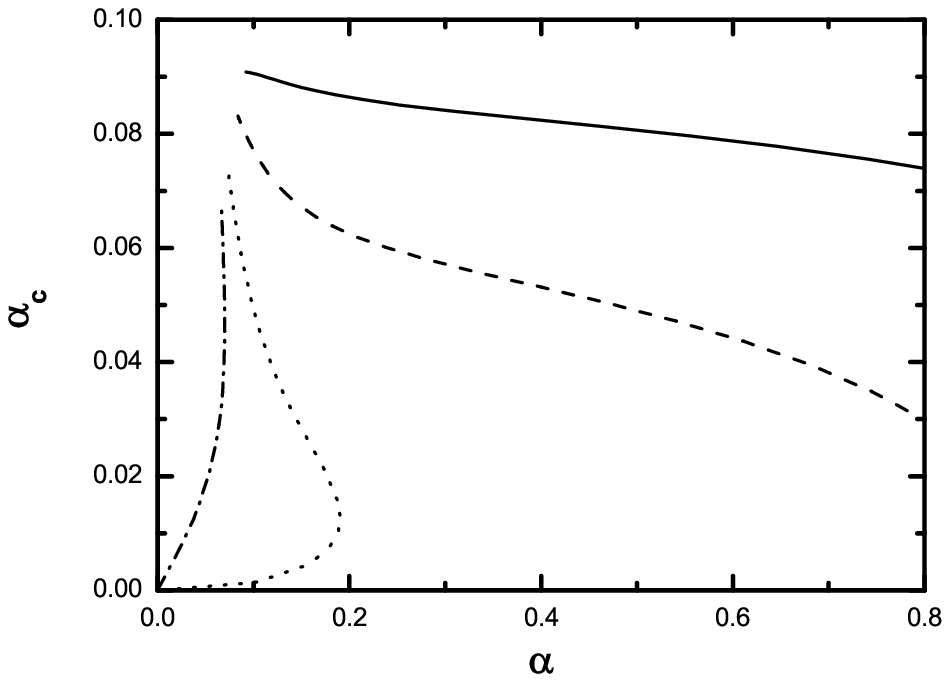}}
\caption{The convective coefficient $\alpha_{c}$ as a function of
viscosity parameter $\alpha$ with $s=-1/2$, $\gamma$=1.4, $f=1$ and
different sets of parameters ($\beta_{r}$,
$\beta_{\varphi}$,$\beta_{z}$). (a) {\em Left panel}: ($\beta_{r}$,
$\beta_{\varphi}$,$\beta_{z}$)$=(0, 0, 0)$ ({\em solid line}), $(0,
1, 0)$ ({\em dashed line}), $(0, 3, 0)$ ({\em dotted line}) and $(0,
5, 0)$ ({\em dash-dotted line}); (b) {\em Right panel}:
($\beta_{r}$, $\beta_{\varphi}$,$\beta_{z}$)$=(0, 0, 0.1)$ ({\em
solid line}), $(0, 0, 0.5)$ ({\em dashed line}), $(0, 0, 0.7)$ ({\em
dotted line}) and $(0, 0, 0.8)$ ({\em dash-dotted
line})}\label{fig54}
\end{figure}
Figure \ref{fig54} shows examples of the $\alpha_{c}-\alpha$
relation with different magnetic field structures. We take
$\gamma=1.4$ and $s=-1/2$. The left panel shows the
$\alpha_{c}-\alpha$ relation with different values of
$\beta_{\varphi}$. When the magnetic field is small
($\beta_{\varphi}$=0 and 1 in this panel), $\alpha_{c}$ decreases
with increasing $\alpha$, and $\alpha$ has its critical (minimum)
value for the solution to be available. These results are basically
consistent with those in NIA. However, when $\beta_{\varphi}$
becomes large, $\alpha_{c}$ increases as the viscosity parameter
$\alpha$ increases, and the critical value of $\alpha$ becomes
extremely small or even disappears. The right panel of Figure
\ref{fig54} shows the $\alpha_{c}-\alpha$ relation with different
values of $\beta_{z}$. When $\beta_{z}$ becomes large, the critical
value of $\alpha$ also disappears, but $\alpha$ has its maximum
value. This result is quite different from NIA, who found that only
the minimum value of $\alpha$ exists and becomes important.

From the above discussion, we conclude that a strong vertical
magnetic field or large $\beta_{r}\beta_{\varphi}$ prevents the
convective process in flows, while a moderate vertical magnetic
field is available for small $\alpha$. A strong toroidal magnetic
field with weak radial field makes the convective process become
important even for large $\alpha$ in flows.

In NIA, a self-similar convection-driven non-accreting solution with
$s=1/2$ (i.e. $\Sigma\propto r^{1/2}$) was given for
$\alpha+g\alpha_{c}=0$ when $\alpha$ is smaller than the critical
value $\alpha_{\rm crit}$. However, the relation
$\alpha+g\alpha_{c}=0$ cannot be satisfied if
$\beta_{r}\beta_{\varphi}\neq 0$ for magnetized CDAFs. In fact, we
are still able to get a self-similar structure for magnetized CDAFs
when the $\alpha_{c}-\alpha$ relation mentioned is no longer
satisfied (i.e., inequality (\ref{e3092}) is not satisfied). For
$\beta_{r}\beta_{\varphi}\neq 0$, the zero infall velocity (i.e.
$c_{1}=0$) requires $s=-1$ ($\rho\propto r^{-2}$) from equation
(\ref{e304}), and $\alpha$ as a function of $c_{3}$,
\begin{equation}
\alpha=\alpha_{c}\left(|g|-\frac{|\xi''|c_{3}}
{c_{2}^{2}}\right)<\alpha_{c}|g|,\label{e311}
\end{equation}
with the maximum value of $\alpha_{c}$ to be
\begin{equation}
\alpha_{c,\rm
crit2}=\frac{(1+\beta_{z})}{9\sqrt{2}}\sqrt{\frac{9-(2\beta_{z}+5)\gamma}{2\gamma(1+\beta_{z})}}.\label{e312}
\end{equation}

Furthermore, if we turn the radial momentum equation from its
vertical integration to its general from, we can still have the
relation (\ref{e311}) but $s=0$ ($\rho\propto r^{-1}$), and the
maximum value of $\alpha_{c}$ to be
\begin{equation}
\alpha_{c,\rm
crit2}=\frac{(1+\beta_{z})}{4\sqrt{2}}\sqrt{\frac{2-(1+\beta_{z})\gamma}{\gamma(1+\beta_{z})}}.\label{e313}
\end{equation}
Such a structure of $\rho\propto r^{-1}$ was also obtained by
Igumenshchev et al. (2003), who explained the structure as a result
of vertical leakage of convective energy flux from the disk. In our
model, however, we show that this structure is due to the
inefficient angular momentum transfer by viscosity and the zero
Lorentz force in the $\varphi$-direction.

As a result, we obtained a self-similar solution for magnetized
CDAFs with $c_{1}=0$, $s=-1$ or $s=0$ (for the general form) and
$\alpha+g\alpha_{c}< 0$. This solution is adopted when the normal
self-similar solutions mentioned above for convective flows cannot
be satisfied.

\section{A more realistic form of kinematic viscosity}
In the above sections \S5.3 and \S5.4, we assume the kinematic
viscosity $\nu=\alpha c_{s}^{2}/\Omega_{K}$ and take the viscosity
parameter $\alpha$ as a constant in our discussion for simplicity. A
more realistic model based on the physical meaning of the viscosity
parameter is $\nu=\alpha c_{s}H$ with $H\neq c_{s}/\Omega_{K}$ in
the magnetized disk. In this section we consider the effect of
different forms of kinematic viscosity $\nu$. In order to compare
with the results in \S5.3 and \S5.4, we replace $\alpha$ in the last
two sections by $\alpha'$ and take $\nu=\alpha
c_{s}H=\alpha'c_{s}^{2}/\Omega_{K}$ in this section. Also, we still
adopt the definition of $c_{1}$ using equation (\ref{e201}). From
formula (\ref{e213}), we are able to obtain
\begin{equation}
\alpha'=\alpha\left\{\left[\left(\frac{1-s}{2}\right)^{2}\beta_{r}\beta_{z}c_{3}
+(1+\beta_{r}+\beta_{\varphi})\right]^{1/2}
-\left(\frac{1-s}{2}\right)(\beta_{r}\beta_{z}c_{3})^{1/2}\right\}.\label{e601}
\end{equation}
When $\beta_{r,z}=0$, we have
$\alpha'=\alpha\sqrt{1+\beta_{\varphi}}$ and equation (\ref{e601})
switches back to the case of $\mu=1/2$ in AF06. For large
$\beta_{r}$ or $\beta_{\varphi}$ and small $\beta_{z}$, we have
$\alpha'\sim \alpha\sqrt{1+\beta_{r}+\beta_{\varphi}}$ and
$\alpha'\gg \alpha$. For large $\beta_{z}$, we obtain $\alpha'\sim
\alpha(1+\beta_{r}+\beta_{\varphi})/(1-s)\sqrt{\beta_{r}\beta_{z}c_{3}}$
and $\alpha'\ll \alpha$. This result can be explained as being due
to the fact that a large toroidal or radial magnetic field makes the
disk half-thickness $H$ become large and increase the kinematic
viscosity (since $\nu\propto H$), but a large vertical field reduces
the height $H$ and decreases the kinematic viscosity.

\begin{figure}
\resizebox{\hsize}{!} {\includegraphics{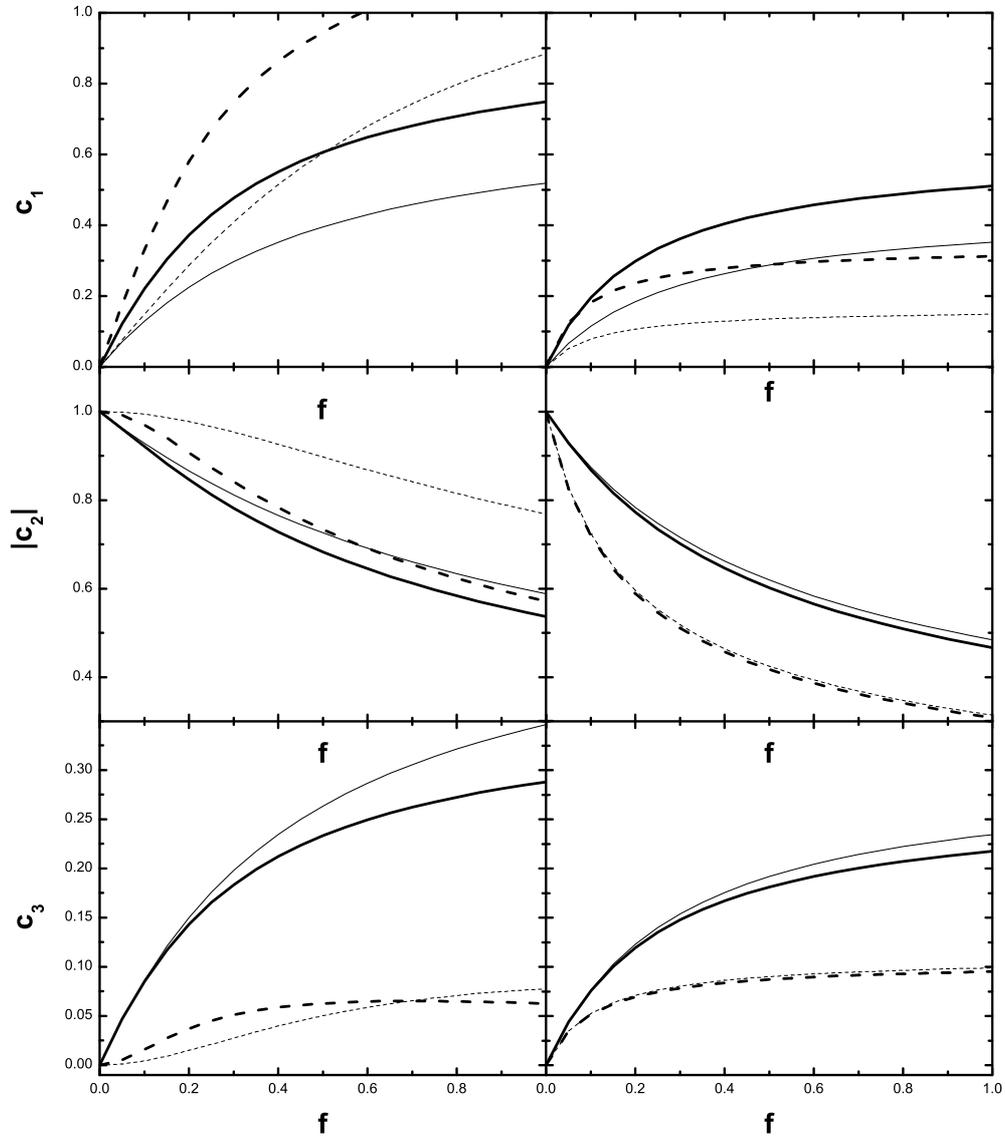}}
\caption{Comparison between two forms of the kinematic viscosity
$\nu$, while the results of $\nu=\alpha c_{s}^{2}/\Omega_{K}$ are
shown by thin lines, and those of $\nu=\alpha c_{s}H$ are shown by
thick lines. We adopt $s=-1/2$, $\gamma=4/3$ and $\alpha=1$. (a)
{\em Left panels}: ($\beta_{r}$, $\beta_{\varphi}$ $\beta_{z}$)$=(2,
0, 0)$ ({\em solid lines}) and $(2, 2.5, 0)$ ({\em dashed lines});
(b) {\em Right panels}: ($\beta_{r}$, $\beta_{\varphi}$
$\beta_{z}$)$=(2, 0, 1)$ ({\em solid lines}) and $(2, 0, 5)$ ({\em
dashed lines}). }\label{fig55}
\end{figure}

Figure \ref{fig55} shows the effect of a modified kinematic
viscosity on the three coefficients $c_{1}$, $|c_{2}|$ and $c_{3}$
in ADAFs. A more realistic expression of $\nu$ increases the infall
velocity, but decreases the radial velocity and the isothermal sound
speed. However, a difference between these two cases of kinematic
viscosity is obvious for a large toroidal magnetic field rather than
a large vertical field. In fact, if the toroidal magnetic field
$B_{\varphi}$ is strong and dominated in ($B_{r}, B_{\varphi},
B_{z}$), we can adopt a similar solution of (49)-(51) in AF06 for
$\mu=1/2$, and find that $c_{1}$ increases but $|c_{2}|$ and $c_{3}$
reaches their limiting values with increasing $\beta_{\varphi}$. If
the radial magnetic field $B_{r}$ is strong and dominated, we can
obtain $c_{1}\sim {\rm const}$, $|c_{2}|\propto \beta_{r}^{-1/4}$
and $c_{3}\propto \beta_{r}^{-1/2}$, which are different from \S3 in
which $|c_{2}|\sim 1$ for large $\beta_{r}$. Furthermore, if
$\beta_{z}$ is large enough, we have the limiting value $c_{1} \sim
3(1+\beta_{r}+\beta_{\varphi})(1-s)^{-3/2}\beta_{z}^{-1}\beta_{r}^{-1/2}$,
$c_{3}\sim (1-s)^{-1}\beta_{z}^{-1}$ and
$c_{2}^{2}=3\epsilon''(s+1)c_{3}$, and the values of $|c_{2}|$ and
$c_{3}$ are more or less the same, no matter what the form of
kinematic viscosity is.

For flows with convection, it is convenient for us to adopt the
general definition of $\alpha_{c}$ from NIA, which measures a degree
of convection in the flows. We find that the conclusions in \S4 are
not basically changed if we replace $\alpha$ in \S5.4 by $\alpha'$.
The $\alpha'-\alpha_{c}$ relation can be turned back to the
$\alpha-\alpha_{c}$ relation using equation (\ref{e601}). However,
there is no dramatic change between these two relations except for
extremely strong magnetic fields.

\section{Conclusions}
In this chapter \S 5 we have studied the effects of a global
magnetic field on viscously-rotating and vertically-integrated
accretion disks around compact objects using a self-similar
treatment. Our conclusions are listed as follows:

(1) We have extended Akizuki and Fukue's self-similar solutions
(2006) by considering a three-component magnetic field $B_{r}$,
$B_{\varphi}$, and $B_{z}$ in ADAFs. If we set the kinematic
viscosity $\nu=\alpha c_{s}^{2}/\Omega_{K}$ as its classical form,
then with the flow to be advection-dominated, the infall velocity
$|v_{r}|$ and the isothermal sound speed $c_{s}$ increase, and even
the radial velocity $|v_{\varphi}|$ can exceed the Keplerian
velocity with a strong toroidal magnetic field. The strong magnetic
field in the vertical direction prevents the disk from being
accreted, and decreases the effect of the gas pressure. For a large
radial magnetic field, $v_{r}$, $v_{\varphi}$ and $c_{s}$ can reach
their limiting values, and the direction of radial velocity is
actually negative, since the angular momentum transfer due to the
magnetic field stress in this case is dominated over that due to the
viscosity in the disk, and makes the value of advection angular
momentum increase inward.

(2) If the convective coefficient $\alpha_{c}$ in flows is set as a
free parameter, $|v_{r}|$ and $c_{s}$ increase with increasing
$\alpha_{c}$ for large $B_{\varphi}$ and weak $B_{r}$. Also,
$|v_{r}|$ becomes smaller and $|v_{\varphi}|$ becomes larger (or
smaller) with increasing $\alpha_{c}$ for a strong and dominated
radial (or vertical) magnetic field.

(3) The $\alpha_{c}-\alpha$ relation in the magnetized disk is
different from that in the non-magnetized disk. For large
$B_{\varphi}$ and weak $B_{r}$, $\alpha_{c}$ increases with
increasing $\alpha$, the critical value $\alpha_{\rm crit}$ to
determine different cases of the $\alpha_{c}-\alpha$ relation
disappears, and $\Sigma\propto r^{-1/2}$ can be satisfied for any
value of $\alpha$. A moderate vertical magnetic field is available
for small $\alpha$. The large $B_{z}$ or $B_{r}B_{\varphi}$, on the
other hand, prevents the convective process in flows.

(4) The self-similar convection envelope solution in NIA should be
replaced by $c_{1}=0$, $\alpha+g\alpha_{c}<0$ and $s=-1$
($\rho\propto r^{-2}$) for the vertical integration form of angular
equations and  $s=0$ ($\rho\propto r^{-1}$) for the general form in
magnetized CDAFs. This solution can be adopted in the region that
does not satisfy the normal self-similar solutions for flows with
convection and $\alpha_{c}<\alpha_{c,\rm crit2}$.

(5) The magnetic field increases the disk height $H$ for large
$B_{r}$ and $B_{\varphi}$, but decreases it for large $B_{z}$ in the
magnetized disk. A more realistic model of the kinematic viscosity
$\nu=\alpha c_{s}H$ makes the infall velocity in ADAFs increase and
the sound speed and toroidal velocity decrease compared with the
simple case when the form $\nu=\alpha c_{s}^{2}/\Omega_{K}$ is
assumed.


\section{Appendix A} The momentum equation of accretion flows can
be written as (Frank et al. 2002)
\begin{equation}
(\textbf{v}\cdot\nabla)\textbf{v} =-\frac{1}{\rho}\nabla
P-\nabla\Phi+\Omega^{2}\textbf{r}+(\nabla\cdot\sigma)+\frac{1}{\rho
c}\textbf{j}\times \textbf{B},\label{e401}
\end{equation}
where $\sigma$ is the viscosity stress tensor, $\textbf{j}\times
\textbf{B}/(\rho c)$ is the density Lorentz force. Also, the
Amp\`{e}re's law and the induction equation (Faraday's law) are
\begin{equation}
\textbf{j}=\frac{c}{4\pi}\left(\nabla\times
\textbf{B}\right).\label{e402}
\end{equation}

\begin{equation}
\frac{\partial \textbf{B}}{\partial t}=\nabla\times(\textbf{v}\times
\textbf{B})+\eta_{m}\nabla^{2}\textbf{B}.\label{e403}
\end{equation}
where $\eta_{m}=c^{2}/(4\pi\sigma_{e})$ is the magnetic diffusivity
and $\sigma_{e}$ is the electrical conductivity. For simplicity, we
consider the extreme case that $\sigma_{e}\rightarrow \infty$ and
$\eta_{m}\approx 0$ , and then neglect the second term in the right
side of the induction equation (\ref{e403}). Combining equations
(\ref{e401}) and (\ref{e402}), we can obtain the three components of
the momentum equation. In particular, the three components of the
Lorentz force in the cylindrical coordinates are
\begin{equation}
\frac{4\pi}{c}(\textbf{j}\times
\textbf{B})_{r}=-\frac{1}{2}\frac{\partial}{\partial
r}(B_{z}^{2}+B_{\varphi}^{2})+B_{z}\frac{\partial B_{r}}{\partial
z}-\frac{B_{\varphi}^{2}}{r},\label{e404}
\end{equation}

\begin{equation}
\frac{4\pi}{c}(\textbf{j}\times
\textbf{B})_{\varphi}=\frac{1}{r}B_{\varphi}B_{r}+B_{r}\frac{\partial
B_{\varphi}}{\partial r}+B_{z}\frac{\partial B_{\varphi}}{\partial
z},\label{e405}
\end{equation}

\begin{equation}
\frac{4\pi}{c}(\textbf{j}\times
\textbf{B})_{z}=-\frac{1}{2}\frac{\partial}{\partial
z}(B_{r}^{2}+B_{\varphi}^{2})+B_{r}\frac{\partial B_{z}}{\partial
r}.\label{e406}
\end{equation}

Based on the consideration that all flow variables including the
magnetic field are mainly functions of radius $r$, we can conclude
$v_{z}=0$ and $\partial/\partial z=0$. Or a more realistic
consideration requires $\partial/\partial z\sim (H/r)
\partial/\partial r\ll \partial/\partial r$. Also, we take $\partial/\partial \varphi=0$ for the axisymmetric disk.
We rewrite the Lorentz force using the Alfven sound speed as
\begin{equation}
\frac{1}{\rho c}(\textbf{j}\times
\textbf{B})_{r}=-\frac{1}{2\rho}\frac{\partial}{\partial
r}[\rho(c_{z}^{2}+c_{\varphi}^{2})]-\frac{c_{\varphi}^{2}}{r},\label{e407}
\end{equation}

\begin{equation}
\frac{1}{\rho c}(\textbf{j}\times
\textbf{B})_{\varphi}=\frac{1}{r}c_{\varphi}c_{r}+\frac{c_{r}}{\sqrt{\rho}}\frac{\partial}{\partial
r}(\sqrt{\rho}c_{\varphi}),\label{e408}
\end{equation}

\begin{equation}
\frac{1}{\rho c}(\textbf{j}\times
\textbf{B})_{z}=\frac{c_{r}}{\sqrt{\rho}}\frac{\partial}{\partial
r}(\sqrt{\rho}c_{z}),\label{e409}
\end{equation}

These expressions are different from SK05 and SK06, who considered
the magnetic field structure as a function of both radius $r$ and
height $z$: $B_{r}(r,z)=z(B_{r})_{H}/H$,
$B_{\varphi}(r,z)=z(B_{\varphi})_{H}/H$ with $H$ to be the
half-thickness of the disk, and $B_{z}(r,z)=B_{z}(r)$. However, in
this chapter \S 5 we take the magnetic field to be homogeneous in
the vertical direction and neglect the term of $\partial/\partial z$
as mentioned above except for the $z$-component equation, in which
we take the total pressure as $P_{\rm tot}=P_{\rm
gas}+(B_{\varphi}^{2}+B_{r}^{2})/8\pi$ in the vertical direction,
and adopt $\partial P_{\rm tot}/\partial z\sim-P_{\rm tot}/H$ to
estimate the value of $H$. In \S2, we use the height-integration
equations.

\section{Appendix B}
Expressions (29)-(37) in \S4 can be derived as follows:

When $\beta_{r}=0$ or $\beta_{\varphi}=0$, we obtain equations for
the three coefficients $c_{i}$ in CDAFs as
\begin{equation}
-\frac{1}{2}c_{1}^{2}\alpha^{2}=c_{2}^{2}-1-[(s-1)(1+\beta_{z})+(1+s)\beta_{\varphi}]c_{3},\label{e501}
\end{equation}
\begin{equation}
c_{1}\alpha=3(\alpha+g\alpha_{c})(s+1)c_{3},\label{e502}
\end{equation}
\begin{equation}
c_{2}^{2}=\epsilon''\frac{\alpha
c_{1}}{\alpha+g\alpha_{c}}+\xi''\frac{\alpha_{c}c_{3}}{\alpha+g\alpha_{c}},\label{e503}
\end{equation}
with $\epsilon''=\frac{4}{9f}(\frac{1}{\gamma-1}+s-1)$ and
$\xi''=\frac{4}{9f}(s-\frac{1}{2})(\frac{1}{\gamma-1}+s-1)$. Then we
can write the equation for $c_{1}$ as
\begin{equation}
\frac{1}{2}c_{1}^{2}\alpha^{2}+c_{1}\alpha\left\{\epsilon''
+\xi''\frac{\alpha_{c}}{3(s+1)(\alpha+g\alpha_{c})}
+\frac{1}{3}\left[\left(\frac{1-s}{1+s}\right)(1+\beta_{z})-\beta_{\varphi}\right]\right\}-1=0.\label{e504}
\end{equation}
When $\beta_{z}$ is large, the above equation can be simplified as
\begin{equation}
\frac{1}{2}c_{1}^{2}\alpha^{2}+\frac{c_{1}\alpha}{3}\left(\frac{1-s}{1+s}\right)
\frac{\beta_{z}}{\alpha+g\alpha_{c}}-1=0,\label{e505}
\end{equation}
and we obtain
\begin{equation}
c_{1}\alpha\sim
\frac{3(\alpha+g\alpha_{c})(1+s)}{(1-s)\beta_{z}}\label{e506}.
\end{equation}
Similarly, we can get the solution for large $\beta_{\varphi}$ and
small $\beta_{r}$.

On the other hand, if the radial magnetic field is strong and
dominated and $\beta_{\varphi}\neq0$, then equation (\ref{e502})
should be replaced by
\begin{equation}
-\frac{1}{2}c_{1}c_{2}\alpha=(s+1)c_{3}\sqrt{\beta_{r}\beta_{\varphi}},\label{e507}
\end{equation}
and we obtain an equation in the extreme case,
\begin{equation}
\frac{1}{2}c_{1}^{2}\alpha^{2}+\epsilon''\frac{\alpha
c_{1}}{\alpha+g\alpha_{c}}-1=0\label{e508},
\end{equation}
and get
\begin{equation}
\alpha
c_{1}=2(\alpha+g\alpha_{c})\left[\epsilon''+\sqrt{\epsilon''^{2}+2(\alpha+g\alpha_{c})^{2}}\right]^{-1}.\label{e509}
\end{equation}


\chapter{Remarks and Future Prospects}
I give a detailed review of the GRB progenitor and central engine
models, and introduce the accreting and hyperaccreting BH systems.
After than, I discuss the scenario of hyperaccreting neutron-star
disks with their neutrino emission and annihilation. Then I study
the structure of magnetized self-similar ADAFs and CDAFs. In this
chapter, I mention a few problems which may be important and
attractive in the next decade. I also mention some points of my
future work.

Does the GRB bimodality classification based on $T_{90}$ duration
actually reflect their different progenitors and central engines? It
seems that a better classification can be done based on
environmental properties as well as prompt emission properties
(e.g., Zhang 2006). Prompt emission properties include $T_{90}$
duration, spectrum hard or soft condition and spectral lag
condition. Environmental properties includes the host galaxy type,
the stellar population in the host, the existence of an accompanying
SN and so on. Thus the duration of $T_{90}$ is only one factor for
an more elaborate classification. Nakar (2007) mentioned several
exceptions for the long and short-duration bimodality classification
(e.g., GRB 050416A, 051221A, 060121, 060505, 060614). The
environmental properties are important to understand GRBs with their
progenitors, but they could become the basis for classification only
if they reflect the nature of GRB production. The best
classification, of course, should be based on the GRB central
engines and their progenitors. Although in current state it is
difficult to probe the real types of progenitor and central engines
of each GRB, but we should understand the very observational
characters of each type. For example, GRB produced by a magnetar
formed after massive star collapse is different from a GRB produced
by the accreting BH system formed in the collapse, as they have
different jet formation and jet-star interaction processes. We list
several possible types of GRBs based on their energy sources.
\begin{table}
\begin{center}
\begin{tabular}{cll}
\hline\hline
Type & Progenitor & Central Engine \\
\hline
I & collapsar & hyperaccreting BH system \\
II & collapsar & magnetar and neutron star \\
III & collapsar & supranova scenario\\
IV & compact binary merger & hyperaccreting BH system \\
V & compact binary merger & magnetar and neutron star\\
VI & magnetar & magnetar\\
VII & neutron star & strange star\\
\hline \hline
\end{tabular}
\end{center}
\caption{}\label{tab61}
\end{table}

Can we obtain an entire consistent collapsar scenario, from the
initial massive star collapse process to the jet formation and
propagation? First of all, the collapse mechanism could be
completely understood only if we understand the mechanism of
supernova (SN) explosions. The collapsar scenario, which was once
purposed as the ``failed supernova" scenario, could actually produce
SNe associated with GRBs. Thus, it is important to study the GRB and
SN phenomena as a whole. Here we farther list several special
questions which should be answered in future study:
\begin{itemize}
\item
What are the energy sources in the collapsars that might transport
energy and lead to an SN explosion? As we know, the neutrino energy
deposition by itself fails to laugh and sustain an outbound shock
with sufficient energy for a supernova explosion by current computer
simulations. Perhaps we need to add some additional ``key" physics
factors such as rotation and magnetic fields for producing a
successful SN. In the collapsar scenario, however, the core-collapse
SN might be driven or partially driven by outflow energy from the
accretion disk around collapsed center BH. Moreover, since the
central accreting BH and the jet from a collapsar could not produce
sufficient $^{56}\textrm{N}_{\rm i}$ for the observed SN, most
$^{56}\textrm{N}_{\rm i}$ might be from the accretion disks. Thus
the questions are, could the outflows from the central accreting BH
system produce a successful SN together with a GRB phenomenon? Are
there any differences between normal SNe and the SNe from collapsars
(i.e., hypernova)? It is possible that the explosion mechanisms
between them are different. Furthermore, it is also interesting to
consider the effects of magnetic fields, since previous work only
consider the neutrino processes and thermal winds by rotation for SN
explosion. Are MHD processes significant for generating a SN
explosion in normal and collapsar cases? We will also discuss the
field effects on GRBs later.

\item
Can we consistently obtain the conclusion that collapsars are not
necessary to connect with SNe, since observations have shown that at
least some long-duration bursts are not associated SNe (e.g., GRB
060505 and GRB 060614)? If the progenitors of most long-duration
GRBs are collapsars, why some collapsars could also produce SNe, but
others cannot? We can give an artificial parameterized explain for
the possible connection between GRBs and SNe in the collapsar model,
but the most important thing is to understand the actual main
mechanisms for SNe and related GRBs.

\item
Are GRBs and XRTs associated with other types of SNe besides Type
Ibc? Today most SN events associated with GRBs are Type Ic, i.e.,
the SNe lose their hydrogen and helium envelopes. However, since the
sample of GRB-SN is still small (only five confirmed events), it is
still not clear which types of SNe are favored by collapsars. For
example, short-hard GRB 970514 and GRB 980910 were reported to be
possible associated with Type II SN 1997cy and SN 1997E
respectively, though these reports are still controversial. However,
we would like to ask a reversed question: if Type Ibc SNe are most
likely to associated with GRB and XRT phenomena, what GRB-like
phenomena can be produced in Type II SN progenitors? MacFayden et
al. (2001) discussed blue and red supergiants would produce both
XRTs and Type II SNe. Moreover, if some Type II SNe can be confirmed
to be associated with short-hard bursts, the model of short GRB
progenitors would be challenged. Furthermore, we collect the
questions from Woosley \& Bloom (2006): does the unified model of
GRBs and XRTs in the collapsar scenario work? Are XRTs the results
of GRBs seen off axis? Are XRTs the more common form than GRBs in
the universe? As we have already discuss these topics in theoretical
model, the answers of these questions should based on future
observations.

\item
Are collapsars which can effectively produce GRBs favored by binary
environments, as mentioned in Fryer et al. (2007)? More work both
observational and theoretical should be done to understand the
collapsar scenario.

\end{itemize}

Another questio is that, what is the functions of magnetic fields on
producing a GRB? This is also a broad question, we discussion it in
several aspects:
\begin{itemize}
\item
The evolution of magnetic fields in a collapsar. As mentioned by
Herger, Woosley \& Spruit (2005) and Woosley \& Herger (2006),
the collapsars which could produce GRBs successfully are favored
by weak magnetic fields and low metallicity. However, on the
other hand, a strong magnetic field $\sim 10^{14}-10^{15}$ G
(e.g., Popham et al. 1999) is required near the BH for forming a
relativistic jet via MHD process and directly produce a GRB. In
some collapsars (such as Type II collapsar), neutrino
annihilation mechanism is much less effective to produce a jet
than MHD process, and recent simulations (e.g., Nagataki 2009)
shows that the jet is Poynting-flux-dominated. Therefore, the
question is that, how could the magnetic fields be amplified so
significantly for $\sim10$ s? In the collapsar scenario, we need
to consider that the magnetic torques are negligible in order to
give GRB progenitors necessary angular momentum, but in the MHD
process of jet formation, we need a strong magnetic field near
the BH. Thus we need to describe the mechanism of field
amplification and field evolution during the whole collapse
process about 10 s.

\item
The role of magnetic fields played in compact object binary mergers
and HMNS collapses. GRMHD simulations are needed to show the effects
of magnetic fields. As mentioned by Liu et al. (2008), it is
possible that the effects of magnetic fields are significant, but
not dramatic. The role of magnetic fields is long term and for
secular evolution. However, farther works should be carried out. For
instance, what the effects of magnetic fields in WD-WD mergers? THe
problem is somewhat important, if we consider that magnetars can be
formed by double magnetized WD mergers. Is the magnetic field
structure near the new formed BH after mergers similar to that in
the collapsar scenario, and can we describe the magnetic fields
evolution during the whole merger process? Different from the
collapsars, whose jets can be collimated by their passage through
the stellar envelope, the jets formed after compact star mergers
could only be collimated by the magnetic fields near the accreting
BHs.

\item
The differences among Poynting-flux-dominated jets, baryonic jets
and the intermediate case--magnetized fireball jets. As mentioned by
Piran (2005) and Woosley \& Bloom (2006), this issue is also
important for AGNs, microquasars, pulsars as well as GRBs. The
polarization of afterglows and the strength of the optical
afterglows may be the observational diagnostics. However, it is
still difficult to detect the obvious polarization results in
current state. Can we find other determined methods to probe the
properties of the jets based on current observation technology, and
then indirectly know the process of jet formation?

\end{itemize}

Finally I also mention some of my further work after I finished this
thesis:
\begin{itemize}

\begin{figure}
\centering\resizebox{1.0\textwidth}{!} {\includegraphics{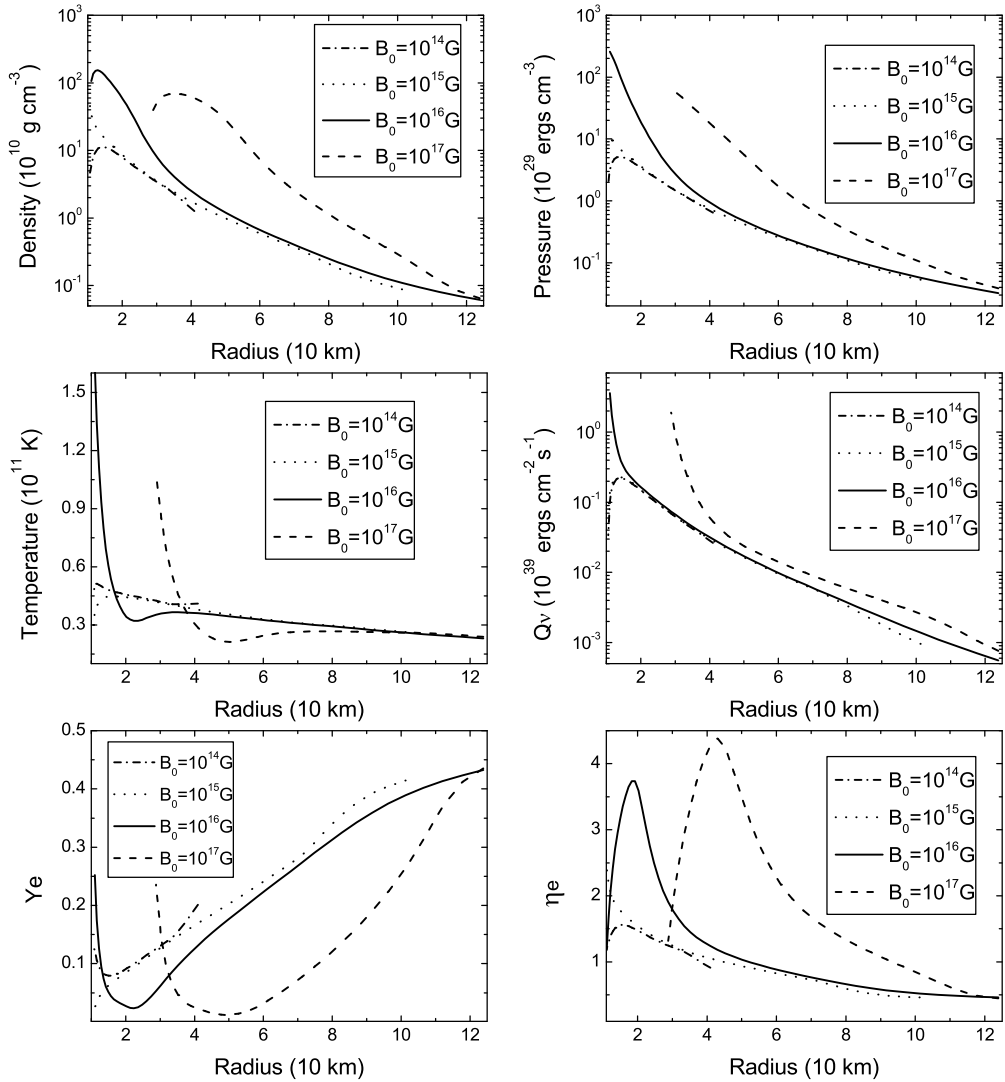}}
\caption{Disk structure with $\dot{M}=0.1M_{\odot}$ s$^{-1}$ for the magnetar
surface vertical field $B_{0}=10^{14},10^{15},10^{16}$ and $10^{17}$
G, where the magnetic field in the disk is in an open configuration.}\label{fig61}
\end{figure}
\item
What are the effects of high strong magnetic fields
$\sim10^{15}-10^{17}$ G on the hyperaccreting disk? The magnetic
fields are generated in the new formed central magnetar. We need
to consider the neutrino emission and annihilation processes in
strong magnetic fields, i.e., the intermediate case between
Poynting-flux-dominated jet and annihilation-induced jet. We
have discussed some of the microphysics equations in strong
magnetic fields in \S 2.5. Figure \ref{fig61} gives a first
result: the density, pressure, temperature, electron fraction
and degeneracy in the disk as a function of radius for a open
magnetic fields configuration with magnetar surface fields as
$10^{14}$, $10^{15}$, $10^{16}$ and $10^{17}$ G. More work can
see our recent paper ``Hyperaccreting Disks around Magnetars for
Gamma-Ray Bursts: Effects of Strong Magnetic Fields'', which is
not included in this thesis.

\end{itemize}

\newpage
\chapter*{Author's Papers List}
\markboth{Papers List}{Papers List}

\begin{enumerate}
\item {\bf Dong Zhang} and Zi-Gao Dai (2008)

  ``Hyperaccretion disks around Neutron Stars"

  {\em the Astrophysical Journal}, 683, 329

  (astro-ph/0712.0423)

\item {\bf Dong Zhang} and Zi-Gao Dai (2008)

  ``Self-similar structure of magnetized advection dominated accretion flows and convection dominated accretion
  flows"

  {\em Monthly Notices of the Royal Astronomical Society}, 388, 1409

  (astro-ph/0805.3254)

\item {\bf Dong Zhang} and Zi-Gao Dai (2008)

  ``Hyperaccreting Neutron Stars and Neutrino-cooled Disks"

  {\em AIP Conferenece Proceedings}, 1065, 294

\item {\bf Dong Zhang} and Zi-Gao Dai (2009)

  ``Hyperaccreting Neutron-Star Disks and Neutrino Annihilation"

  {\em the Astrophysical Journal}, 703, 461

  (astro-ph/0901.0431)

\item {\bf Dong Zhang} and Zi-Gao Dai (2009)

  ``Hyperaccreting Disks around Magnetars for Gamma-Ray Bursts:
  Effects of Strong Magnetic Fields"

  {\em Submitted to the Astrophysical Journal}

  (astro-ph/0911.5528)

\end{enumerate}


%
\end{document}